Daniel Gruss

# Software-based Microarchitectural Attacks

## PhD Thesis

Assessors: Stefan Mangard, Thorsten Holz

June 2017

Institute for Applied Information Processing and Communications
Graz University of Technology

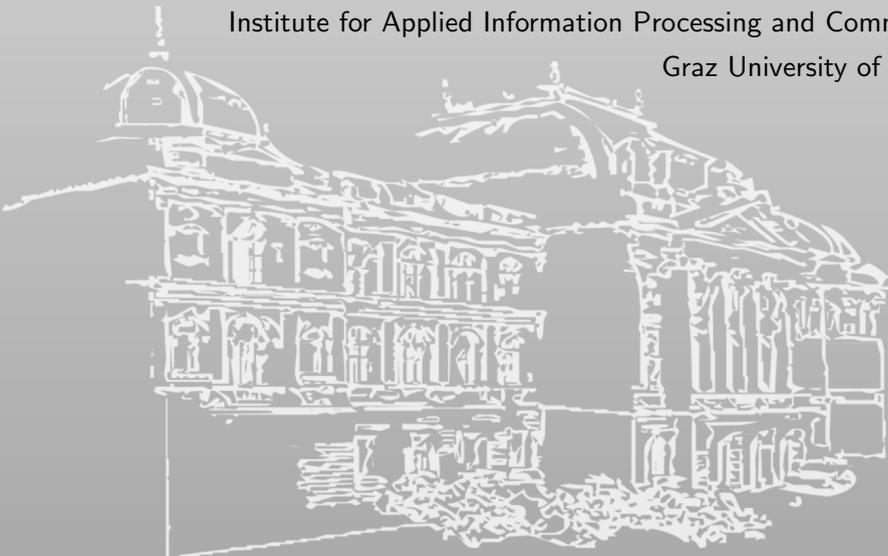

SCIENCE ▪ PASSION ▪ TECHNOLOGY

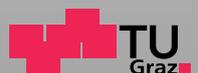

# Abstract

Modern processors are highly optimized systems where every single cycle of computation time matters. Many optimizations depend on the data that is being processed. Software-based microarchitectural attacks exploit effects of these optimizations. Microarchitectural side-channel attacks leak secrets from cryptographic computations, from general purpose computations, or from the kernel. This leakage even persists across all common isolation boundaries, such as processes, containers, and virtual machines.

Microarchitectural fault attacks exploit the physical imperfections of modern computer systems. Shrinking process technology introduces effects between isolated hardware elements that can be exploited by attackers to take control of the entire system. These attacks are especially interesting in scenarios where the attacker is unprivileged or even sandboxed.

In this thesis, we focus on microarchitectural attacks and defenses on commodity systems. We investigate known and new side channels and show that microarchitectural attacks can be fully automated. Furthermore, we show that these attacks can be mounted in highly restricted environments such as sandboxed JavaScript code in websites. We show that microarchitectural attacks exist on any modern computer system, including mobile devices (e.g., smartphones), personal computers, and commercial cloud systems.

This thesis consists of two parts. In the first part, we provide background on modern processor architectures and discuss state-of-the-art attacks and defenses in the area of microarchitectural side-channel attacks and microarchitectural fault attacks. In the second part, a selection of our papers are provided without modification from their original publications.[1] I have co-authored these papers, which have subsequently been anonymously peer-reviewed, accepted, and presented at renowned international conferences.

---

[1]Several of the original publications were in a two-column layout. Without any changes in content, we updated the layout for all included papers from the camera-ready versions to fit the layout and formatting of this thesis, including resizing some figures and tables, and changing the citation format.

# Acknowledgements

First and foremost, I want to thank my advisor Stefan Mangard. You constantly challenged my mind with ambitious ideas like the one of a cache-based key logger at the beginning of my PhD. You gave me the freedom to research anything I'm interested in and supported me in all endeavors. Thank you for your guidance throughout the last years.

I also want to thank my assessor Thorsten Holz for valuable comments and interesting discussions at past conferences.

I really enjoyed my time at IAIK and in the Secure Systems group – you were the best colleagues one can wish for and I owe thanks to all of you! You all helped me directly or indirectly throughout my PhD.

I owe special thanks to some of my colleagues. Thank you Raphael for helping me in the first months of my PhD and in my first steps on cache attacks. Thank you Clémentine. I've learned so much from you – without you, the amazing last years would not have been possible. It was a great time and a lot of fun working with you, independent of linguistic opinions! Thank you Michael and Moritz. You are two of the most brilliant minds I know and I enjoy every day working with you!

I also want to thank Anders Fogh for fierce competition, great collaborations, great ideas and lots of fun!

Thank you Peter Lipp and David Derler. I enjoyed teaching during my PhD so much and not least because it's great working in a team with you!

Finally, I want to thank my fiancée Maria Eichlseder, my friends, my family, and my fiancée's family for supporting me during my PhD. Thank you Maria for supporting me in everything I do, helping me with my numerous questions on cryptography and maths, and tolerating what I call a healthy work-job balance.

# Contents

# Part I.

# Introduction to Software-based Microarchitectural Attacks

# 1

# Introduction

The idea of learning the secret code for a safe by listening to the clicking sounds of the lock, is likely as old as safes are. The clicking sound is an inadvertent influence on the environment revealing secret information. In 1996, Kocher [Koc96] described side-channel attacks, a technique that allows to derive secret values used in a computation from inadvertent influences the computation has on its environment. This seminal work was the beginning of an entire area of research on side channels. Kocher performed what we now describe as a timing attack, an attack exploiting differences in the execution time of an algorithm. In the following years, side-channel attacks have been demonstrated based on virtually any measurable environmental change caused by various types of computations, such as power consumption [MOP08], electro-magnetic radiation [RR01; KOP09], temperature [HS13], photonic emission [Sch+12; CSW17], acoustic emissions [Bac+10], and many more. These attacks have in common that they require an attacker to have some form of physical access to the target device.

In contrast to side-channel attacks, which do not cause any damage to the target device, there are also fault attacks [BDL97; BS97]. In a fault attack an attacker tries to manipulate computations of a device to either evade security mechanisms of the device or to leak its secrets. For this purpose, the attacker manipulates the environment in a way that influences the target device. Typically such fault-inducing environments are at the border of or beyond the specification range of the target device. Like for side-channel attacks, different environment manipulations have been investigated, such as exposure to voltage glitching [Aum+02], clock glitching [SMC09], extreme temperatures [HS13], or photons [SA02]. Again, to perform a fault attack, some form of physical access to the target device is required.

*1. Introduction*

Modern computer systems are highly complex and highly optimized. Consequently, information leakage, the inadvertent influence of the environment in a secret-dependent way, is not only introduced on an algorithmic level. Optimizations are performed based on the specific data values that are processed, the location of the data, the frequency of accesses to locations, and many other factors. It is clear to see, that any adversary observing the effects of these optimizations through a side channel can make deductions on the specific cause of the optimizations. Through these deductions, the adversary learns information about the secret data values that are processed.

In this thesis, we investigate *software-based microarchitectural attacks*. Software-based microarchitectural side-channel attacks exploit timing and behavior differences that are (partially) caused through microarchitectural optimizations, *i.e.*, differences that are not architecturally documented. Software-based microarchitectural fault attacks induce faults through microarchitectural optimizations, *i.e.*, operate elements of modern computer systems at the border of or beyond their specification range. Generally, software-based microarchitectural attacks do not require physical access, but instead only some form of code execution on the target system.

Cache attacks are the most prominent class of software-based microarchitectural attacks. The possibility of timing differences induced through processor caches was first described by Kocher [Koc96]. Cache timing attacks have first mostly been applied on cryptographic algorithms in software-based attacks [Pag02; TSS03; Ber05; BM06].

Cache attacks in more recent works are usually instances of three generic cache attack techniques. These techniques have been used in targeted attacks on cryptographic algorithms [Ber05; Per05; GBK11] and were later on generalized by Osvik et al. [OST06] and Yarom et al. [YF14]. These generic techniques are independent of the specific cache and hardware on which they are performed. Osvik et al. [OST06] described two generalized cache attack techniques. First, *Evict+Time*, where an attacker measures how the execution time of an algorithm is influenced by evicting a chosen cache set. Second, *Prime+Probe*, where an attacker measures whether a victim computation influences how long it takes to access every way of a chosen cache set.

In both attacks the attacker learns that the chosen cache set was used by the victim. Yarom et al. [YF14] introduced the third generalized attack technique, *Flush+Reload*. In a *Flush+Reload* attack, the attacker flushes a shared memory location from the cache and subsequently measures how long it takes to reaccess it. If the victim loaded the shared memory location back into the cache in the meantime, the reaccess is faster. In a *Flush+Reload* attack the attacker does not only learn which cache set was used by the victim, but even the specific memory location (at the granularity of cache lines).

Based on these three attack primitives various computations have been attacked, for instance cryptographic algorithms [YF14; Liu+15], web server function calls [Zha+14], user input [GSM15; Gru+16b; Ore+15], kernel addressing information [HWH13; Gru+16a].

Software-based fault attacks are considerably more difficult to build in practice as faults must be induced in hardware. Hence, software has to move the system component that is targeted to the border of or beyond its specification range. Only in 2014 software-based fault attacks have been found to be practical, in the so-called Rowhammer attack [Kim+14; SD15]. In concurrent work, Karimi et al. [Kar+15] demonstrated a second software-based fault attack. They showed that a carefully crafted instruction stream can deteriorate the processor stability and cause severe permanent damage to the processor if executed continuously for weeks. Rowhammer attacks have by now been demonstrated in JavaScript [GMM16; Bos+16], on supposedly safe DDR4 [Pes+16], on co-located virtual machines [Raz+16; Xia+16], and on mobile devices [Vee+16].

To develop and evaluate potential countermeasures against software-based microarchitectural attacks, it is necessary to map and understand the attack surface in detail. In this thesis, we aim to improve the general understanding of the attack surface of software-based microarchitectural attacks and to provide novel insights to software-based microarchitectural attacks and attack vectors. Our research includes the minimization of requirements, the automation of previous attacks, and the identification of previously unknown side channels. Figure 1.1 gives an overview how the papers relate to each other and where they are located in this exploration space.

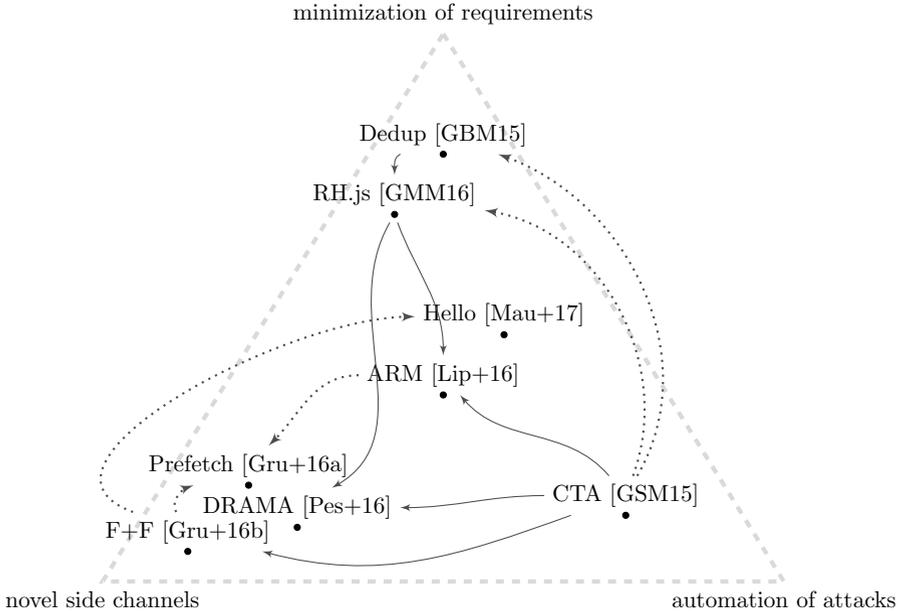

Figure 1.1.: Overview of the relation of the papers to each other. The dotted arrows illustrate where ideas from one paper facilitated the research conducted in the other. The continuous arrows illustrate where concepts from one paper where more directly applied in the other paper.

## 1.1. Main Contributions

We started the work on software-based microarchitectural attacks by enhancing the *Flush+Reload* cache attack technique [YF14] by an automated method to find and exploit vulnerabilities. This generic technique called Cache Template Attacks has been published at the USENIX Security 2015 conference [GSM15] in collaboration with Raphael Spreitzer and Stefan Mangard. We demonstrated that Cache Template Attacks can be used to identify and exploit leakage in old implementations of cryptographic algorithms, or automatically spy on user input events such as keystrokes or mouse movements. This publication is included as Chapter 5 of this thesis.

While caches buffer the comparably slow DRAM, the DRAM itself buffers the even slower hard disk. Hence, side-channel attacks are also possible on the DRAM level. Suzaki et al. [Suz+11] demonstrated a side-channel attack on page deduplication, as performed by the operating system or hypervisor, which reveals whether specific data can be found in memory. We demonstrated that such attacks can even be performed from JavaScript integrated into a website. Our results have been published at the ESORICS 2015 conference [GBM15] in collaboration with David Bidner and Stefan Mangard. This publication is included as Chapter 6 of this thesis.

Based on these two works we investigated the possibility of Rowhammer attacks [Kim+14; SD15] from JavaScript integrated into websites. For such attacks to work it was necessary to evict data from caches using regular memory accesses fast enough to replace the comparably fast clflush instruction. Our investigations showed that cache eviction can be performed fast enough for such attacks to successfully be mounted. In our proof-of-concept implementation we were able to trigger bit flips in exploitable memory locations from JavaScript. The results of this research have been published at the DIMVA 2016 conference [GMM16] in collaboration with Clémentine Maurice, and Stefan Mangard. This publication is included as Chapter 7 of this thesis.

Related to the *Flush+Reload* attack we developed a new cache attack called *Flush+Flush*. This technique makes cache attacks faster and stealthier as it does not perform memory accesses itself. We evaluated the performance and stealthiness of the attack in comparison to other cache attacks such as *Flush+Reload* and *Prime+Probe* as well as to Rowhammer attacks. The results of our work are also published at the DIMVA 2016 conference [Gru+16b] in collaboration with Clémentine Maurice, Klaus Wagner, and Stefan Mangard. This publication is included as Chapter 8 of this thesis.

Our work on fast cache eviction motivated investigations on fine-grained access-driven cache attacks ARM Cortex-A systems, which usually has no user-space flush instruction. We demonstrated that all cache attack techniques can be performed on ARM Cortex-A systems as well. Based on these attack primitives we demonstrated that Cache Template Attacks provide a powerful means to find and exploit cache leakage on mobile devices. Our results have been published at the USENIX Security 2016 conference [Lip+16] in collaboration with Moritz Lipp, Raphael Spreitzer, Clémentine Maurice, and Stefan Mangard. This publication is included as Chapter 9 of this thesis.

Besides the unified cache hierarchy there is also a second cache hierarchy in modern processors for page table entries. We found that prefetch instructions have a different execution time based on the state of these page translation caches. Even worse, the x86 prefetch instructions allow unprivileged processes to prefetch privileged memory into the cache. We exploited these observations in order to defeat kernel address space-layout randomization (KASLR). Our results have been published at the CCS 2016 conference [Gru+16a] in collaboration with Clémentine Maurice, Anders Fogh, Moritz Lipp, and Stefan Mangard. This publication is included as Chapter 10 of this thesis.

## 1.2. Other Contributions

While working on this thesis, several contributions to other works that are not included as a part of this thesis. Nonetheless, we discuss them here to draw the complete picture of all contributions.

While working on the Rowhammer attack we observed timing differences caused by so-called row hits and row conflicts in the DRAM module. To get a better understanding of these timing differences we developed a fully automated method to reverse-engineer the mapping of physical addresses to DRAM cells in software. Using these reverse-engineered mappings reduces the runtime of Rowhammer attacks significantly. Investigating the timing differences in more detail, we found significant side-channel leakage that is comparable to that of cache attacks. These novel DRAM side-channel attacks have been published at the USENIX Security 2016 conference [Pes+16] in collaboration with Peter Pessl, Clémentine Maurice, Michael Schwarz, and Stefan Mangard.

Our previous works on Rowhammer, cache eviction on ARM Cortex-A, and DRAM reverse-engineering systems, sparked the idea of performing Rowhammer attacks on Android devices. In a collaboration with Victor van der Veen, Yanick Fratantonio, Martina Lindorfer, Clémentine Maurice, Giovanni Vigna, Herbert Bos, Kaveh Razavi, Cristiano Giuffrida we published our results on Rowhammer attacks on mobile devices at the ACM CCS 2016 conference [Vee+16].

Related work on software-based microarchitectural side channels typically discusses the capacity of a side channel based on the raw capacity of a covert channel built on top of it. Due to the nature of side channels, these covert channels are not error-free. Previous work claimed that straightforward application of error correcting codes is sufficient to eliminate all errors. Thus, to provide realistic estimates the error rate is taken into account to compute a real-world capacity for the channel. Investigating how realistic these estimates are, we built an entirely error-free covert channel. We found that the application of error correcting codes is possible but has to be combined with other error detection techniques in a non-trivial way. Our channel is so reliable that we can even tunnel an SSH connection through it. Our results have been published at the NDSS 2017 conference [Mau+17] in collaboration with Clémentine Maurice, Manuel Weber, Michael Schwarz, Lukas Giner, Carlo Alberto Boano, Kay Römer, and Stefan Mangard.

Many microarchitectural attacks could generally run in JavaScript, but require high-precision timers. We investigated high-precision timing sources in JavaScript and found techniques which allow to mount reliable attacks. We demonstrate this by building a covert channel through DRAM between JavaScript running on a website and an unprivileged application running inside a virtual machine. Our results have been published at the Financial Crypto 2017 conference [Sch+17c] in collaboration with Michael Schwarz, Clémentine Maurice, and Stefan Mangard.

A new feature in modern Intel processors is Intel SGX, an environment for secure execution on untrusted hardware and operating systems. SGX enclaves are highly secure and can generally not be inspected or monitored by the operating system. However, they are also restricted environments, which cannot perform any system calls directly. We investigated whether it is possible to exploit the security features to protect malicious software running inside an SGX enclave. We built cache side-channel attacks extracting cryptographic keys from the host or from co-located SGX enclaves. Our results will be published at the DIMVA 2017 conference [Sch+17b] in collaboration with Michael Schwarz, Samuel Weiser, Clémentine Maurice, and Stefan Mangard.

We investigated possible countermeasures against attacks on address-translation caches (cf. Chapter 10). Our solution called KAISER is a practical extension for the Linux kernel, which eliminates the leakage entirely while having a low performance overhead on modern processors. Our results will be published at the ESSoS 2017 conference [Gru+17b] in collaboration with Moritz Lipp, Michael Schwarz, Richard Fellner, Clémentine Maurice, and Stefan Mangard.

Finally, we also investigated generic countermeasures against cache side-channel attacks. Modern Intel processors implement hardware-transactional memory on top of the cache hierarchy. Through creative instrumentation we can use hardware-transactional memory to abort upon conflicting memory operations and cache misses. This effectively eliminates the leakage which is exploited in cache attacks. Our results will be published at the Usenix Security 2017 conference [Gru+17a] in collaboration with Julian Lettner, Felix Schuster, Olga Ohrimenko, Istvan Haller, and Manuel Costa.

## 1.3. Thesis Outline

This thesis consists of two parts. In the first part (Chapter 2 – Chapter 4), a we provide an overview on software-based microarchitectural attacks.

Chapter 2 explains the relevant background. We first detail in Section 2.1 how processors are organized. We explain the basic concept of virtual memory in Section 2.2 and discuss the idea of caching in Section 2.3. In Section 2.4, we provide an overview of how DRAM works.

Chapter 3 discusses the state of the art in microarchitectural attacks. We discuss software-based microarchitectural side-channel attacks in Section 3.1. We describe how cache attacks developed in the past decades to the current state of the art and provide a discussion of attacks on various caches. In Section 3.2 we introduce the concept of software-based microarchitectural fault attacks. Finally, we discuss countermeasures and defense mechanisms in Section 3.3.

Chapter 4 discusses future work and draws conclusions.

In the second part of this thesis (Chapter 5 – Chapter 10), a list of all publications is provided, together with transcripts for a selection of papers constituting this thesis.

Chapter 5 consists of our USENIX Security 2015 conference paper [GSM15] on Cache Template Attacks. Chapter 6 consists of our ESORICS 2015 conference paper [GBM15] on page deduplication attacks in JavaScript. Chapter 7 consists of our DIMVA 2016 conference paper [GMM16] on Rowhammer.js. Chapter 8 consists of our DIMVA 2016 conference paper [Gru+16b] on *Flush+Flush*. Chapter 9 consists of our USENIX Security 2016 conference paper [Lip+16] on cache attacks on ARM-based mobile devices. Chapter 10 consists of our CCS 2016 conference paper [Gru+16a] on prefetch side-channel attacks.

# 2

# Background

In this chapter, we provide background which is necessary to understand and discuss microarchitectural attacks. First of all, these attacks target microarchitectural elements, *i.e.*, elements of specific processor families and models. Hence, we discuss how processors are organized in Section 2.1. With virtual memory the processor provides process isolation for modern operating systems. It is deeply rooted in today's processor architectures and influences how various microarchitectural elements work. We discuss virtual memory in Section 2.2. Cache attacks are the most important class of microarchitectural attacks. They typically exploit timing differences introduced by the specific organization of the cache. We discuss caches in detail in Section 2.3. Finally, we discuss how the memory controller interacts with DRAM and exposes timing differences caused by the DRAM to software in Section 2.4.

## 2.1. Processor organization

Modern processors are highly parallelized machines operating at extremely high speeds. With shrinking process technology sizes, energy demands have decreased allowing to increase processor clock frequencies. Processor clock frequencies stayed roughly at the same level in the past decade. However, besides the processor clock frequency, there are other ways to improve the processing speed. Many optimizations reduce the execution time of specific instructions by a few cycles, sometimes depending on the data or the processor state.

**Pipelining.** Pipelining is one of the main contributors to performance improvements. Pipelining splits instructions into several stages. The num-ber of stages and their purpose varies between processors. Modern processors have many pipelining stages, most importantly

- a fetch stage, loading the instruction opcode into the processor,
- a decode stage, decoding the instruction opcode to an internal representation of the instruction,
- and an execute stage, executing the instruction.

Hence, the processor can execute multiple instructions at once, each with a slight interleaving to the previous and the next. Modern processors can have multiple identical stages to perform certain computations in parallel and further improve performance.

**Multi-core.**  Instead of optimizing the processing speed of a single execution core it is also possible to increase the number of execution cores. Especially in server environments multiple processors are installed in a single machine to multiply the performance for parallelizable tasks. Naturally, if a task is not parallelized, there is no performance difference. Hence, depending on the workload, a multi-processor system can provide a significant performance improvement. Workloads on personal computers have changed over the past decades. Today, these systems run several hundred tasks in parallel all the time and thus benefit from multiple processors. For this reason, multi-core processors have been introduced. These processors combine multiple execution cores into a single processor. Each of these execution cores has some private resources, e.g., registers and the execution pipeline, and some shared resources, e.g., the main memory interface.

**Instruction Stream Optimization.**  Another idea to increase the processor performance is to execute branches speculatively. With this optimization the processor makes a guess and execute a branch before it knows whether this branch will be executed. If the guess was correct, the processor already has the information ready when it is required. Otherwise, the processor just discards the result. Another idea is to execute instructions out of order. The execution of instructions with pending data dependencies is delayed and other instructions without pending data dependencies are executed before, to optimize the CPU throughput.

## 2.2. Virtual memory

As multi-processing became more popular, proper isolation between different processes became more important. As a first step processor manufacturers introduced coarse-grained forms of *virtual memory*, such as Intel's x86 segmentation. Virtual memory regions are mapped to *physical memory* regions in large coherent blocks. At this level of virtual memory, the operating system can specify access privileges, an offset, and a length for each segment. Different processes use different segments and thus work on different physical memory. Consequently, we distinguish between two types of addresses: *virtual addresses*, which are specific to a process, and *physical addresses*, which are valid system-wide but not directly accessible for processes.

Instead of segmentation, modern processors employ a better form of virtual memory, called "paging", which works at a granularity of "pages", which are memory blocks of a fixed size. The entire virtual memory is sliced into virtual pages and the entire physical memory is sliced into physical pages. By numbering the pages we obtain *page numbers*. Paging can be seen as a process-specific map from virtual page numbers to physical page numbers. Modern processors typically support multiple page sizes where the smallest page size is often 4 KB or 1 KB. Larger page sizes are always multiples of smaller page sizes. Pages are aligned in physical memory and in virtual memory to their own size, *i.e.*, 4 KB pages are aligned to 4 KB boundaries in virtual and physical address space.

With 64-bit processors the address-width was increased significantly from 32 bits to 48 bits and an extension to 57 bits is already planned. Address translation is extremely critical to performance and the processor must be able to handle the data structure. Hence, the data structure must be similar to a simple array. An array that could possibly map every 4 KB page of the 48-bit virtual address space to a physical page would already consume 512 GB of physical memory just to store the map, at a map-entry size of 64 bits. Clearly, this is not a practical solution. The idea behind multi-level translation tables is that the virtual address space is usually mapped sparsely to physical memory. With multiple levels, the size of the map can be reduced to a negligible overhead.

To translate a virtual address into a physical address the processor first locates the top-level translation table by reading its address from a processor register. This register value is exchanged upon a context switch to enable multi-processing. This is the basis for the process isolation we mentioned before. Each process has its own virtual address mappings and can only access to its own address space.

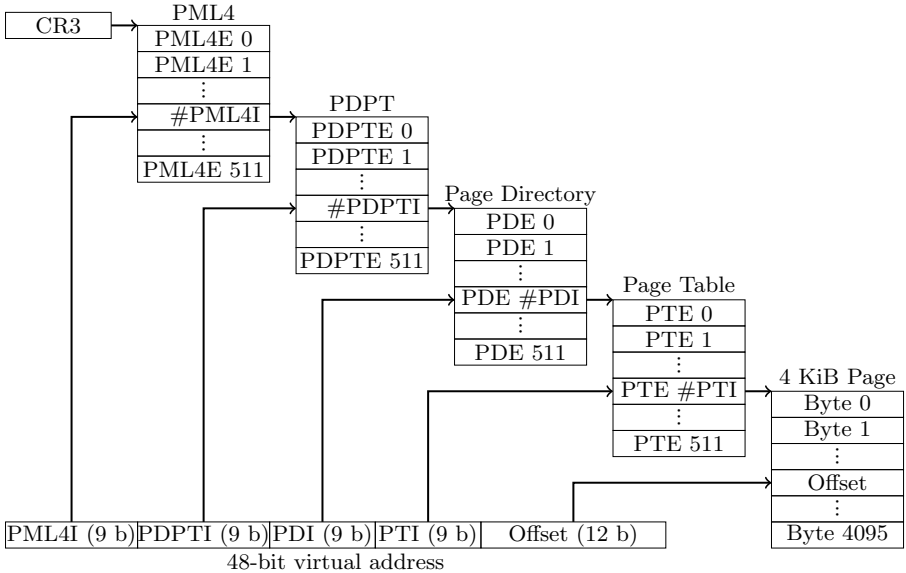

Figure 2.1.: Address translation for 4 KB pages on x86-64 processors. Starting with the PML4 base address from the CR3 register, the processor determines the physical address by gradually using parts of the virtual address.

Modern Intel processors have 4 levels of translation tables as shown in Figure 2.2. The top-most level is the page map level 4 (PML4). It divides the 48-bit virtual address space into 512 memory regions of each 512 GB (PML4 entries). PML4 entries always map to page directory pointer tables (PDPT), *i.e.*, there is no possibility to map a 512 GB page. Each PML4 entry defines properties of the corresponding 512 GB memory region, e.g., whether the memory is mapped to physical memory, whether it is readable, writable, and whether it is accessible to user space. The lower levels are organized in the same way. Each PDPT again has 512 entries, with each entry defining the properties of a 1 GB virtual memory region. This 1 GB virtual page can directly be mapped to a so-called 1 GB page or to a page directory (PD). Modern operating systems use 1 GB virtual pages for instance for the large direct-physical mapping in kernel space which allows working on physical address directly although running in virtual addressing mode. Each PD again has 512 entries, defining the properties of a 2 MB virtual memory region. Modern operating systems commonly use 2 MB pages to map files or large arrays. Alternatively, the PD entry can map a page table (PT). Each PT again has 512 entries, each controlling a 4 KB page. This is the default page size for most use cases.

For any operation the processor performs, one or more virtual addresses have to be translated into physical addresses. Consequently, the address translation latency must be very small. With translation tables being located in the main memory, this is not the case. Consequently, address translation caches have been introduced to hide the DRAM latency as we will see in the next section.

## 2.3. Caches

In this section we discuss caches and cache organization in detail. We will discuss the general organization of caches and the basic concepts in Section 2.3.1. Subsequently, we discuss cache replacement policies in Section 2.3.2. We describe the relation between virtual and physical addresses and caches in Section 2.3.3. Finally, we discuss how caches on modern Intel processors work in Section 2.3.4.

### 2.3.1. Cache Organization

The plain computation speed of processors was the bottleneck for a long time. However, with increasing processor frequencies, the latency of physical memory (DRAM) increasingly became a new bottleneck. While the bandwidth of DRAM has increased over the past decades, the latency is still very high. Processors caches are small and fast buffers intended to hide the latency of the slow DRAM. Modern processors have multiple cache hierarchies for different purposes with each multiple levels of different sized caches. Some caches are private to one execution core while other caches are shared among all cores.

Generally, all memory accesses go through the cache. If a memory access is served from the cache it is called a *cache hit*. Otherwise, it is a *cache miss* causing a fetch from the slow main memory.

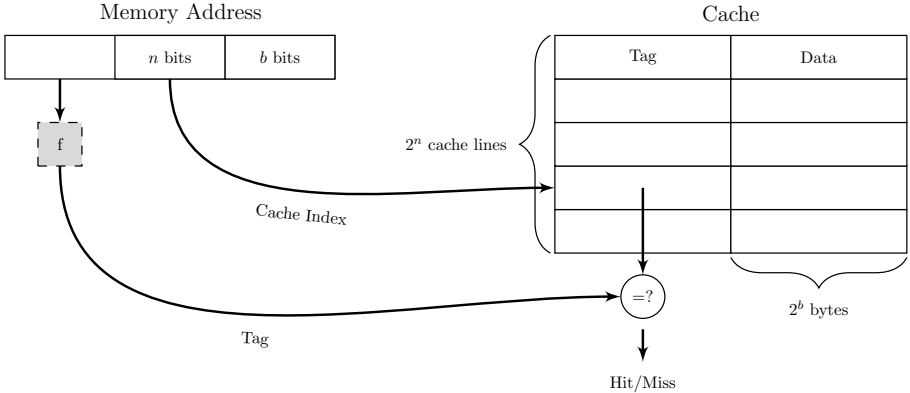

Figure 2.2.: A directly-mapped cache. Based on the middle $n$ bits the cache index is computed to choose a cache line. The tag is used to check whether an address is cached. If it is cached (cache hit), the $2^b$ bytes data are returned to the processor.

**Directly-mapped Caches.** The most simple form of a cache is a *directly-mapped cache*, as illustrated in Figure 2.2. A cache consists $2^n$ *cache lines*, each consisting of a *tag* and $2^b$ bytes of associated data. The tag is computed from the corresponding memory address that is buffered in this cache line. It is used to determine whether or not a cache line currently buffers a specific memory address. The lowest $b$ bits of the address are used as an offset within the cache line data. Most modern processors have a *cache line size* of 64 bytes, *i.e.*, $b = 6$. The middle $n$ bits of the memory address are used as a *cache index*, telling the processor in which cache line to look for corresponding data. The size of the cache determines how many bits are used, *i.e.*, how many indices there are. Addresses with the same middle $n$ bits are *congruent*, as they map to the same cache line. A significant problem of directly-mapped caches is that they can only store a single cache line out of all congruent cache lines. Hence, if the processor needs to work on two or more congruent cache lines, a directly-mapped cache would experience cache misses most of the time.

**Fully-associative Caches.** The congruency problem does not exist in *fully-associative caches*, as illustrated in Figure 2.3. Fully-associative caches do not have cache indices and thus they do not have any cache lines. Instead they have multiple *cache ways* to store data. The tag is now used to determine whether the corresponding memory address is cached and which cache way contains the associated data. Fully-associative caches are increasingly expensive with the number of ways. Hence, they are typically restricted to a small number of ways, e.g., translation-lookaside buffers with 64 ways can be found in several modern processors.

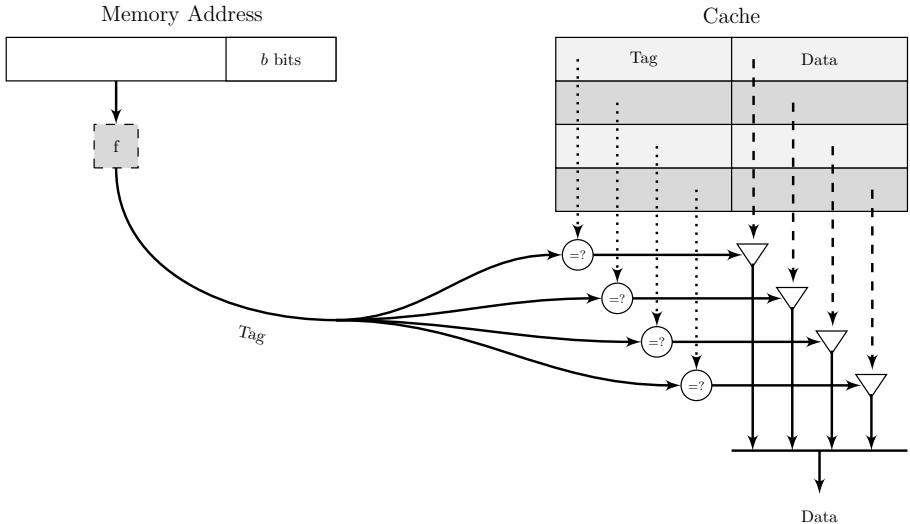

Figure 2.3.: A fully-associative cache. All cache ways are checked in parallel using the tag. The corresponding data value is selected based on the tag comparison.

**Set-associative Caches.** An elegant compromise are *set-associative caches*, which have *cache sets* instead of cache lines. These caches are widely used in modern processors, often referred to as $m$-way set-associative caches. Figure 2.4 shows an abstract model of a 2-way set-associative cache. The cache is divided into $2^n$ cache sets. The *cache set index* is determined from the middle $n$ bits of the memory address. Each cache set has $m$ ways to provide storage locations for $m$ congruent addresses. Note that cache sets can also be seen as tiny fully-associative caches with $m$ ways, for the set of congruent addresses. Hence, the tag is again used to determine which cache way buffers a specific memory address.

Data and instruction caches today are typically implemented as $m$-way set-associative caches, but processor manufacturers are also transitioning some address translation caches from fully-associative to set-associative.

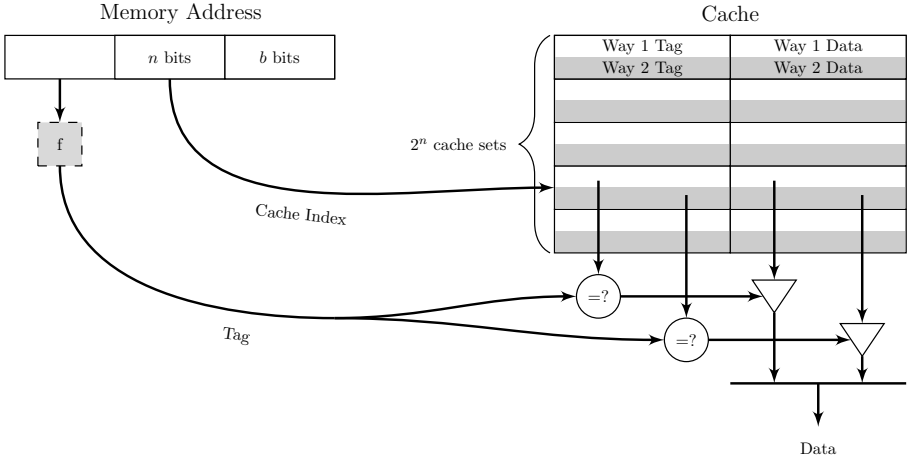

Figure 2.4.: A 2-way set-associative cache. The middle $n$ bits are the cache index, selecting the cache set. The tag is used to check all ways simultaneously. The data in the matching cache way is returned to the execution core.

## 2.3.2. Cache Replacement Policies

Inevitably, with a limited number of ways and many congruent addresses mapping to the same cache set, the processor constantly has to evict cache ways to replace their content with newly requested data fetched from the main memory. Hence, a replacement policy is implemented to decide which cache way is replaced next when loading new data into a cache set. Processor manufacturers keep the details of their replacement policies mostly secret as they are a relevant contributing factor for the overall processor performance.

A wide-spread replacement policy commonly used by Intel for different caches is least-recently used (LRU). With an LRU replacement policy, every cache way has some form of last-usage timestamp. This is often an approximation to reduce the implementation complexity. Whenever the processor has to load new data into a cache set, LRU replaces the cache way with the oldest last-usage timestamp. LRU fails in cases where the processor works on a set of congruent addresses larger than the number of cache ways. In this case an LRU replacement policy yields a worst-case performance as every single memory access will be a cache miss.

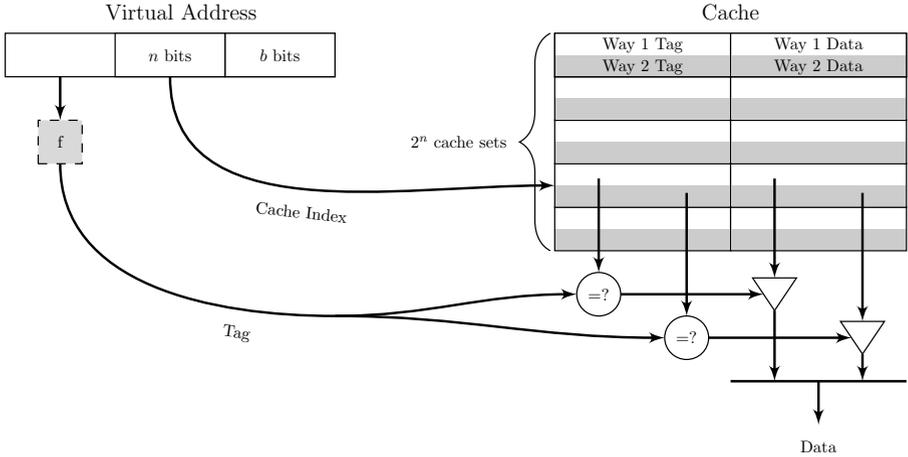

Figure 2.5.: A virtually-indexed virtually-tagged (VIVT) cache. The virtual address is used to compute both index and tag. The processor does not have to translate any addresses.

ARM processors commonly employs a pseudo-random replacement policy. Here the next cache way to be replaced is determined by a pseudo-random number generator. Random replacement policies are the easiest to implement in hardware [Sez93]. They additionally have the advantage of being energy-efficient [PS05]. In practice, random replacement policies have shown to deliver a high performance.

To overcome the limitations of simpler algorithms, since Ivy Bridge, Intel uses a bimodal insertion policy where the CPU can switch between the two strategies to achieve optimal cache usage [Qur+07]. For a group of cache sets the processor can either use a quad-age LRU strategy or a strategy that replaces more recent cache lines first unless they are accessed multiple times. This yields a significantly better performance when using slightly more congruent addresses than would fit in a cache set.

### 2.3.3. Addressing Modes

Caches can use either virtual addresses or physical addresses to compute the cache index and tag. Three designs have found their way into real-world processors.

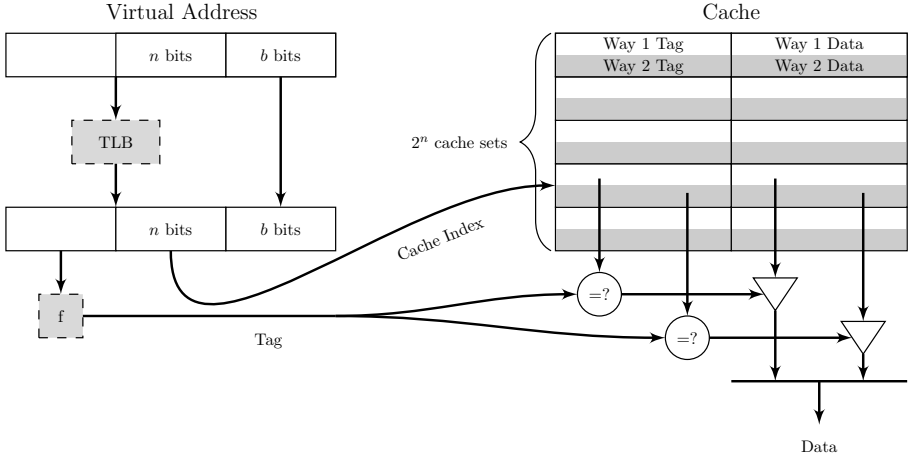

Figure 2.6.: A physically-indexed physically-tagged (PIPT) cache. The physical address is used to compute both index and tag. The processor has to translate the virtual address before the cache set lookup.

Virtually-indexed virtually-tagged (VIVT) caches (cf. Figure 2.5) use the virtual address for both index and tag. Consequently, they don't have to translate addresses at all and have the advantage of a low latency. However, this comes at the price that shared memory might not be shared in the cache, causing an unnecessary increase in cache utilization. Furthermore, upon a context switch it may be necessary to invalidate entries because the virtual tag is not unique. VIVT caches have been used for the smallest and fastest data and instruction caches in some ARM processors. Furthermore, address translation caches are typically VIVT caches.

Physically-indexed physically-tagged (PIPT) caches (cf. Figure 2.6) use the physical address for both index and tag. Consequently, they have a significantly higher latency than VIVT caches. However, shared memory will always be shared in the cache. Thus, there is no unnecessary cache utilization. Furthermore, the tag is physically unique and thus caches do not need to be invalidated upon context switches. Today, PIPT caches are widely used for data and instruction caches with the address translation latency mostly hidden in the address translation caches.

Virtually-indexed physically-tagged (VIPT) caches (cf. Figure 2.7) try to combine the advantages of both approaches by using the virtual address for the index which is required immediately. While the cache index is looked up the tag is computed. This hides the latency of the address translation mostly and allows using a physical tag.

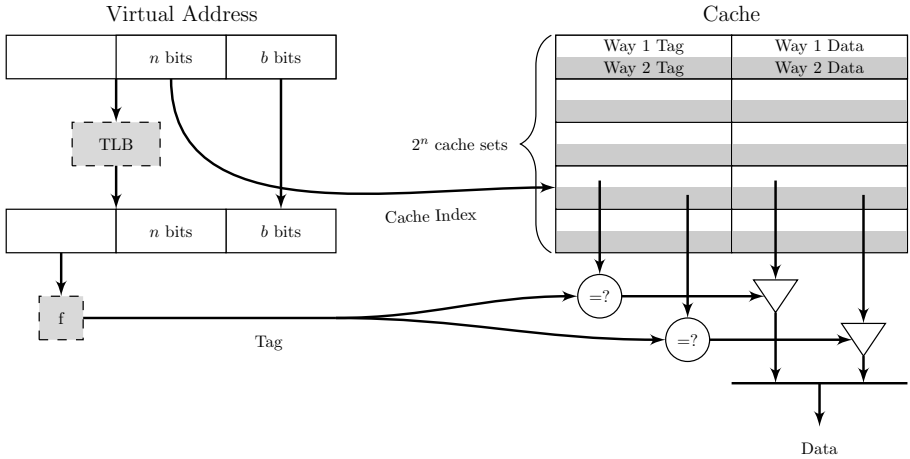

Figure 2.7.: A virtually-indexed physically-tagged (VIPT) cache. The physical address is used to compute both the tag, but the virtual address is used to compute the index. The cache set lookup is done in parallel to the address translation and tag computation.

To avoid the disadvantages of VIVT caches, the cache index should not use address bits that are not part of the page offset in the virtual address. With a page size of 4 KB and a cache line size of 64 bytes, there are 6 bits which can be used as a cache index. Most Intel x86 processors from the past decade integrate two 8-way set-associative VIPT L1 caches per processor core, one for instructions and one for data. Consequently, the size of Intel's L1 caches is $2^6 \cdot 64 \cdot 8 = 32$ KB for most processors from the past decade.

## 2.3.4. Caches in Modern Intel Processors

For instructions and data, Intel x86 processors have a cache hierarchy consisting of L1, L2, and L3 cache. The instruction and data L1 caches are the fastest and smallest caches in this hierarchy. They are private per-core caches, *i.e.*, they are not shared with other cores. The L2 cache is a unified cache, storing both instruction and data cache lines. There is no strict relation between L1 and L2 cache, *i.e.*, cache lines can be presented in none, in one of the two caches, or in both caches. The L3 cache is a unified cache which is shared among all CPU cores. It is also commonly referred to as the *last-level cache.* The L3 cache is inclusive to L1 and L2 caches, *i.e.*, all cache lines in the L1 and L2 caches must also be present in the L3 cache. Both L2 and L3 cache are PIPT caches, enabling to share cache lines based on the physical address.

| Address Bit | | 37 | 36 | 35 | 34 | 33 | 32 | 31 | 30 | 29 | 28 | 27 | 26 | 25 | 24 | 23 | 22 | 21 | 20 | 19 | 18 | 17 | 16 | 15 | 14 | 13 | 12 | 11 | 10 | 09 | 08 | 07 | 06 |
|---|---|---|---|---|---|---|---|---|---|---|---|---|---|---|---|---|---|---|---|---|---|---|---|---|---|---|---|---|---|---|---|---|---|
| 2 cores | $o_0$ | | ⊕ | ⊕ | | | ⊕ | | ⊕ | | ⊕ | ⊕ | ⊕ | ⊕ | ⊕ | | ⊕ | | ⊕ | | ⊕ | ⊕ | ⊕ | | ⊕ | | ⊕ | | ⊕ | | | | ⊕ |
| 4 cores | $o_0$ | | ⊕ | ⊕ | | | ⊕ | | ⊕ | | ⊕ | ⊕ | ⊕ | ⊕ | ⊕ | | ⊕ | | ⊕ | | ⊕ | ⊕ | ⊕ | | ⊕ | | ⊕ | | ⊕ | | | | ⊕ |
| | $o_1$ | ⊕ | | ⊕ | ⊕ | ⊕ | | ⊕ | | ⊕ | ⊕ | | ⊕ | | ⊕ | ⊕ | ⊕ | ⊕ | ⊕ | ⊕ | | ⊕ | | ⊕ | | ⊕ | | ⊕ | | | ⊕ | | |
| 8 cores | $o_0$ | | ⊕ | ⊕ | | | ⊕ | | ⊕ | | ⊕ | ⊕ | ⊕ | ⊕ | ⊕ | | ⊕ | | ⊕ | | ⊕ | ⊕ | ⊕ | | ⊕ | | ⊕ | | ⊕ | | | | ⊕ |
| | $o_1$ | ⊕ | | ⊕ | ⊕ | ⊕ | | ⊕ | | ⊕ | ⊕ | | ⊕ | | ⊕ | ⊕ | ⊕ | ⊕ | ⊕ | ⊕ | | ⊕ | | ⊕ | | ⊕ | | ⊕ | | | ⊕ | | |
| | $o_2$ | ⊕ | ⊕ | ⊕ | ⊕ | | | ⊕ | ⊕ | | ⊕ | ⊕ | | | ⊕ | ⊕ | | ⊕ | | | | | ⊕ | | ⊕ | | ⊕ | ⊕ | | | | ⊕ | |

Table 2.1.: Complex addressing functions extrapolated from [Mau+15a].

To enhance the performance, the last-level cache is divided into cache slices since the Intel Nehalem microarchitecture [Mau+15a]. On current Intel processors, each core has its own L3 cache slice. The slices are interconnected by a ring bus allowing all cores to access all L3 cache lines. The mapping from physical addresses to slices is not documented by Intel and referred is to as a *complex addressing* function. This function has recently been reversed-engineered by researchers [Mau+15a; Inc+15; Yar+15]. Table 2.1 lists the slices functions for different processors. Knowledge of the complex addressing functions facilitates cache side-channel attacks.

## 2.4. DRAM

The main memory of modern computer systems is typically DRAM. DRAM has a significantly higher latency than the various caches inside the processors. The reason for the latency is not only the low clock frequency of DRAM cells but also how DRAM is organized and how it is connected to the processor. As it is difficult to reduce the latency, hardware manufacturers instead focused on increasing the bandwidth of DRAM. The high bandwidth can be utilized to hide the latency, e.g., through speculative prefetching. Modern processors have an on-chip memory controller which communicates through the memory bus with the DRAM.

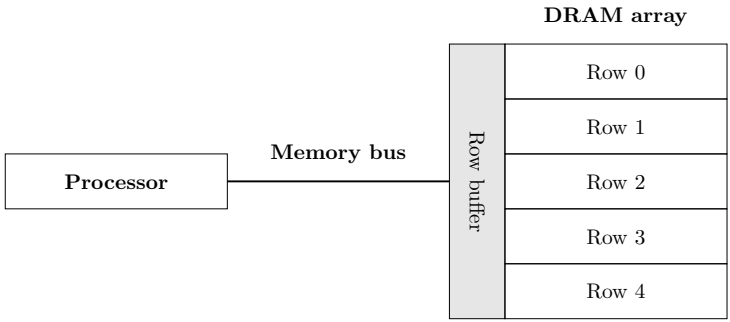

Figure 2.8.: A very simple computer system, with a single DRAM array, which is connected to the processor through a memory bus. The DRAM array consists of rows which are each 8 KB in size.

Figure 2.8 shows a very simple computer system which consists of a single DRAM array, which is connected to the processor through a memory bus. This DRAM array consists of DRAM *rows* and *columns* (typically 1024). Modern systems organize the memory in a way that a DRAM row typically has a *row size* of 8 KB. A row can either be *opened* or *closed*. If it is currently opened, the entire row is preserved in the *row buffer*.

To fetch a data value from DRAM, the processor sends a request to the integrated memory controller. The memory controller then determines a sequence of commands to send over the memory bus to the DRAM to retrieve the data. If the currently opened row contains the data to retrieve, the memory controller just fetches the data from the DRAM row buffer. As this situation is very similar to a cache hit, we call it a *row hit*. If the currently opened row does not contain the data to retrieve, we call it a *row conflict*. The memory controller then first closes the current row, *i.e.*, writes back the entire row to the actual DRAM cells. It subsequently activates the row with the data to retrieve, which is then loaded into the row buffer. Then the memory controller fetches the data value from the row buffer. Similar to cache misses, row conflicts incur an increased access latency.

We can see that the DRAM row buffer in our simple computer system (cf. Figure 2.8) behaves identically to a directly-mapped cache (cf. Figure 2.2). Just as congruent memory accesses constantly lead to cache misses in a directly-mapped cache, alternating row accesses constantly lead to row conflicts in DRAM. Unsurprisingly, similar approaches as in the case of caches have been implemented to increase DRAM performance and eliminate bottlenecks. Modern computer systems organize the DRAM into *channels*, *DIMMs* (Dual Inline Memory Modules), *ranks*, and *banks*.

| CPU | Ch. | DIMM | BA0 | BA1 | BA2 | BA3 | Rank | DIMM | Channel |
|-----|-----|------|-----|-----|-----|-----|------|------|---------|
| Sandy Bridge | 1 | 1 | 13, 17 | 14, 18 | 15, 19 | - | 16 | - | - |
| | 2 | 1 | 14, 18 | 15, 19 | 16, 20 | - | 17 | - | 6 |
| Ivy Bridge / Haswell | 1 | 1 | 13, 17 | 14, 18 | 16, 20 | - | 15, 19 | - | - |
| | 1 | 2 | 13, 18 | 14, 19 | 17, 21 | - | 16, 20 | 15 | - |
| | 2 | 2 | 14, 18 | 15, 19 | 17, 21 | - | 16, 20 | - | 7, 8, 9, 12, 13, 18, 19 |
| | 2 | 4 | 14, 19 | 15, 20 | 18, 22 | - | 17, 21 | 16 | 7, 8, 9, 12, 13, 18, 19 |
| Skylake | 2 | 1 | 7, 14 | 15, 19 | 17, 21 | 18, 22 | 16, 20 | - | 8, 9, 12, 13, 18, 19 |
| 2x Haswell-EP (interleaved) | 1 | 2 | 6, 22 | 19, 23 | 20, 24 | 21, 25 | 14 | 7, 17 | - |
| | 2 | 4 | 6, 23 | 20, 24 | 21, 25 | 22, 26 | 15 | 7, 17 | 8, 12, 14, 16, 18, 20, 22, 24, 26 |
| 2x Haswell-EP (non-interleaved) | 1 | 2 | 6, 21 | 18, 22 | 19, 23 | 20, 24 | 13 | - | - |
| | 2 | 4 | 6, 22 | 19, 23, | 20, 24 | 21, 25 | 14 | - | 7, 12, 14, 16, 18, 20, 22, 24, 26 |
| Exynos 7420 | 2 | - | 14 | 15 | 16 | 8, 13 | - | - | 7, 12 |

Table 2.2.: DRAM addressing functions from [Pes+16].

Our simple computer system (cf. Figure 2.8) had only 1 channel, 1 DIMM, 1 rank, and 1 bank. Modern DDR3 DRAM memory has 8 banks and DDR4 DRAM memory has even 16 banks (per rank). Each of these banks has an independent state and thus can have different rows opened at the same time. This reduces the chance of row conflicts for random memory accesses significantly. Modern DRAM modules typically have 1 to 4 ranks, multiplying the number of banks even further. Similarly, modern systems often allow to install multiple DIMM modules, again multiplying the number of banks with the number of DIMMs. Finally, the memory controller reorders memory accesses to reduce the number of row conflicts, e.g., accesses to the same row and bank are grouped together temporally.

This spatial parallelism reduces the number of row conflicts. Hence, it does not directly influence the bandwidth, but the average latency. To increase temporal parallelism and thus the bandwidth directly, modern computer systems employ multiple channels. Each channel is operated independently and in parallel over the DRAM bus. This effectively multiplies the bandwidth by the number of channels.

Only if two addresses map to the same DIMM, rank and bank they can be physically adjacent in the DRAM chip. In this case the two addresses are in the same bank. If two addresses map to the same bank number, but to a different rank or DIMM, they are not in the same bank and thus generally not physically adjacent in the DRAM chip.

Similar to the slice functions of the last-level cache there are functions mapping from physical addresses to channels, DIMMs, ranks, and banks.

While AMD publicly documents these addressing function [Adv13, p. 345], Intel does not. However, the addressing functions have recently been reverse-engineered for one architecture by Seaborn [Sea15] and with a generic software-based approach by us [Pes+16]. Table 2.2 lists the DRAM addressing functions for several common configurations. Knowledge of the DRAM addressing functions enables DRAM-based side-channel attacks.

# 3

# State of the Art

In this chapter, we discuss the state-of-the-art microarchitectural attacks and defenses. We discuss software-based microarchitectural side-channel attacks in Section 3.1. We discuss software-based microarchitectural fault attacks in Section 3.2. Finally, we discuss defenses against software-based microarchitectural attacks in Section 3.3.

## 3.1. Software-based Microarchitectural Side-Channel Attacks

In this section, we discuss the most important microarchitectural side-channel attacks. We show that imperfections of the hardware, introduced by optimizations on a microarchitectural level, undermine system security and software security. The hardware leaks part of its internal state including potentially secret information through differences in behavior and timing. Software-based microarchitectural side-channel attacks exploit these differences entirely from software.

With the constantly growing field of microarchitectural attacks and software-based side-channel attacks several works aim to provide a systematization of attacks and defenses in this area of research [AK; Ge+16b; Ge+16a; BWM16; Sze16; DK16; Spr+16; Zha+16].

We discuss state-of-the-art cache attacks in Section 3.1.1, including *Evict+Time*, *Prime+Probe*, and *Flush+Reload* and variants of these attacks. We discuss branch-prediction attacks in Section 3.1.2. We discuss attacks on page-translation caches in Section 3.1.3, including prefetch side-channel attacks. We discuss exception-based attacks in Section 3.1.4, including page deduplication attacks. We discuss DRAM-based attacks in Section 3.1.5.

Finally, we discuss other microarchitectural side-channel attacks in Section 3.1.6.

### 3.1.1. Cache Attacks

The idea of caches is to hide the latency of the comparably slow physical main memory. Fetching data from the cache has a significantly lower latency. Kocher [Koc96] described the possibility of exploiting these timing differences in so-called cache-timing attacks. The idea is to deduce information on cryptographic secrets from the influence of the cache on the execution time of a cryptographic implementation. Cache timing attacks have been studied in many works [Kel+00; Pag02; TSS03; Ber05; BM06].

The first cache attacks have been cache-based timing attacks. More recent cache attacks are usually categorized into instances of three generalized cache attack techniques, which have first been performed on cryptographic algorithms [Ber05; Per05; GBK11] and were later on generalized by Osvik et al. [OST06] (*Evict+Time* and *Prime+Probe*) and Yarom et al. [YF14] (*Flush+Reload*). We discuss the different attack techniques in detail in the following.

**Bernstein's attack.** Bernstein [Ber05] described a remote cache-timing attack on an AES T-table implementation. T-tables are preprocessed S-box computations that directly follow the AES design [Nat01; DR13]. The entire AES algorithm can then be implemented as a fast sequence of T-table lookups. The T-table entries are accessed based on an algorithmically defined scheme. For instance in the first encryption round, the algorithm accesses the T-table entries $T_j[p_i \oplus k_i]$, where $p_i$ is the $i$-th plaintext byte and $k_i$ is the $i$-th key byte, with $i \equiv j \mod 4$ and $0 \leq i < 16$. Bernstein observed that these accesses may be cached and depending on whether they are cached a timing difference can be observed. By observing the timing difference, the attacker can deduce which T-table entry was accessed and thus, learns the upper 4 bits of the result of $p_i \oplus k_i$. In a chosen-plaintext attack the attacker can eliminate $p_i$ from the equation and thus obtain the upper 4 bits for every key byte. Combining the information from not only the first round but from multiple rounds yields the full AES key. Bernstein's attack has been reproduced and evaluated in many works [NSW06; BM06; SP13b; Wei+14; SG14].

**Evict+Time.** Osvik et al. [OST06] described *Evict+Time* as a generic cache-timing attack technique. The attacker triggers several victim computations and measures the victim's execution time. To measure the influence of a specific cache set the attacker evicts the cache set before the computation for half of the victim runs. If there is a timing difference when evicting the cache set, the attacker can conclude that the cache set was used by the victim computation.

Similar as in Bernstein's attack, the timing differences observed this way yield which T-table entries were accessed. *Evict+Time* yields information on a cache-set granularity but suffers from various sources of noise on the execution time. Hence, several hundred thousand repetitions may be necessary to extract an entire AES key. *Evict+Time* requires the attacker to be able to measure the exact starting and end time of a victim computation. This might not be possible in asynchronous attacks where the attacker cannot trigger the computation or more generally in many cloud scenarios. An advantage of *Evict+Time* is that it does not require any shared memory. Complex addressing functions and replacement policies of modern processors make eviction more difficult and thus make *Evict+Time* attacks harder.

*Evict+Time* attacks have been investigated by Osvik et al. [OST06] in an attack on OpenSSL AES. Spreitzer et al. [SP13a; Lip+16] demonstrated that *Evict+Time* on OpenSSL AES is also applicable to mobile ARM-based devices. Hund et al. [HWH13] demonstrated that *Evict+Time* can be used to defeat kernel address space-layout randomization (KASLR).

**Prime+Probe.** The second technique presented by Osvik et al. [OST06] is much more powerful. In a *Prime+Probe* attack, the attacker continuously fills (primes) a cache set and measures how long it takes to refill the cache set, as illustrated in Figure 3.1. Osvik et al. described that the time it takes to refill the cache set is proportional to the number of cache ways that have been replaced by other processes. While this proportionality is not strictly present anymore for more recent microarchitectures, the general idea is still valid. A higher timing means that at least one cache way has been replaced. A lower timing means that likely no cache way has been replaced.

*Prime+Probe* has the same granularity as *Evict+Time*, *i.e.*, a cache set. The accuracy is higher than with *Evict+Time* as it measures the cache access times directly, whereas *Evict+Time* measures it indirectly through the execution time. *Prime+Probe* does not require any measurement of the execution time and thus allows performing asynchronous attacks. Analogous to *Evict+Time*, complex addressing functions and replacement policies of modern processors also make *Prime+Probe* attacks more difficult.

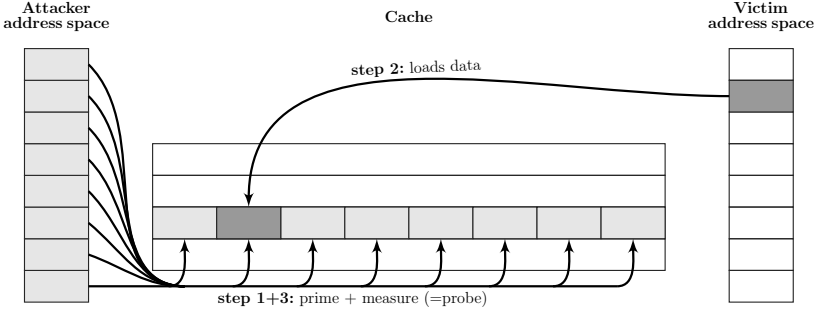

Figure 3.1.: A *Prime+Probe* attack illustrated in 3 steps. The attacker continuously primes a cache set using its own memory locations and measures the execution time of this step (step 1 and step 3). In step 2, the victim possibly accesses (non-shared) memory locations that map to the same cache set. If the victim accessed memory locations in the same cache set in step 2, the execution time of the priming (*i.e.*, the probe step) is high as one of the cache ways has been replaced. Otherwise, the execution time of the priming is low.

The first *Prime+Probe* attacks targeted the L1 cache. However, the reverse-engineering of the last-level cache [Mau+15a; Inc+15; Yar+15] opened up the possibility to perform cross-core *Prime+Probe* attacks through the inclusive last-level cache. The first *Prime+Probe* attacks on the L1 cache have first been demonstrated by Percival [Per05] on RSA. Neve et al. [NS06] attacked an AES implementation, Osvik et al. [OST06] demonstrated an an attack on OpenSSL AES, Aciicmez et al. [AK06; Acı07b; AS08a; AS07] demonstrated attacks on OpenSSL AES and RSA exploiting the L1 instruction cache, Bonneau et al. [BM06] attack OpenSSL AES exploiting internal collisions, Brumley and Hakala [BH09] demonstrated and attack on ECDSA, Aciicmez et al. [ABG10] demonstrated an attack on DSA exploiting the L1 instruction cache, Zhang et al. [Zha+12] demonstrated an attack on ElGamal.

The first *Prime+Probe* attacks on the last-level cache have been performed by Ristenpart et al. [Ris+09] and later by Zhang et al. [Zha+11] to detect co-location in the cloud and eavesdrop on co-located virtual machines. However, these attacks were performed on microarchitectures with a simpler organization, *i.e.*, before Intel Nehalem, without cache slices and without complex addressing functions [Mau+15a]. Maurice et al. [Mau+15b] presented a *Prime+Probe* covert channel through the last-level cache on a recent Intel processor. Liu et al. [Liu+15] demonstrated an attack on ElGamal, Irazoqui et al. [IES15] and later Kayaalp et al. [Kay+16] demonstrated attacks on AES. Oren et al. [Ore+15] performed a *Prime+ Probe* attack in the browser to eavesdrop on user activities, Inci et al. [Inc+15] attacked ElGamal. Our *Prime+Probe* attack on BouncyCastle AES was the first last-level *Prime+Probe* attack on ARM-based mobile devices [Lip+16]. Finally, we presented a fast and robust *Prime+Probe* covert channel on the Amazon EC2 cloud which employs synchronization techniques and error correction codes to achieve a 0% error rate [Mau+17].

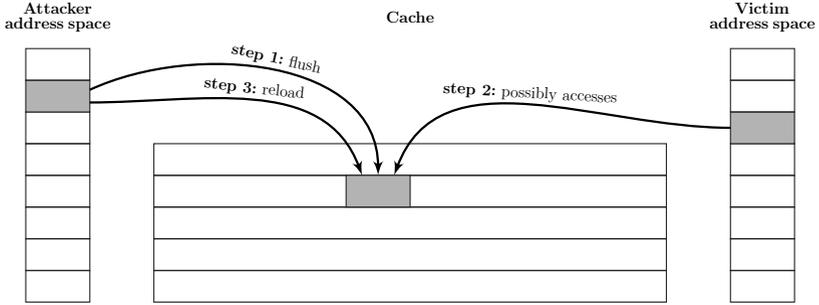

Figure 3.2.: A *Flush+Reload* attack illustrated in 3 steps. In step 1, the attacker flushes a shared memory location in the attacker virtual address space. In step 2, the victim possibly accesses the same shared memory location in the victim virtual address space. In step 3, the attacker reloads the shared memory location and measures the access latency. If the victim accessed the memory location in step 2, the access latency observed in the reload step is low. Otherwise, the access latency in the reload step is high.

Brasser et al. [Bra+17] showed that *Prime+Probe* attacks can also be performed by a malicious operating system on Intel SGX enclaves. Intel SGX enclaves aim to provide a secure execution environment where software can be executed securely even if the operating system is compromised. We showed that *Prime+Probe* attacks can even be performed from inside an enclave [Sch+17b]. They use the protection features provided by Intel SGX to effectively hide the attack from the operating system. They demonstrate that they can steal most key bits of an RSA private key from a single RSA exponentiation running on the host or a co-located enclave.

**Flush+Reload.** *Flush+Reload* is often considered the most powerful cache attack. They work on a single cache line granularity and even more, they reveal to the attacker whether a very specific memory location was cached or not. The attack works by frequently flushing a cache line using the `clflush` instruction. The attacker then measures the time it takes to reload the data. If the reload time is low the attacker learns that another process (*i.e.*, the victim) must have reloaded the very same memory location into the cache. If the reload time is high the attacker learns that likely no other process accessed the memory location in the meantime. This general attack flow is illustrated in Figure 3.2. *Flush+Reload* exploits the availability of shared memory and especially shared libraries between the attacker and the victim program. Hence, in scenarios where shared memory is not available, *Flush+Reload* cannot be applied and an attacker has to resort to *Prime+Probe* instead.

There are many variants of *Flush+Reload*, most prominently *Evict+Reload* [GSM15; Lip+16] and *Flush+Flush* [Gru+16b; Lip+16], both of which we introduced. In an *Evict+Reload* attack the `clflush` instruction is replaced by a cache eviction as in a *Prime+Probe* attack. This makes *Evict+Reload* applicable to architectures that do not expose a flush instruction. *Flush+Flush* exploits a timing difference in the `clflush` instruction to determine whether a memory location is cached. Hence, the attacker can omit the reload step from *Flush+Reload* and build a significantly faster and stealthier cache attack that does not perform a single memory access.

The first *Flush+Reload*-like attack has been demonstrated by Gullasch et al. [GBK11], attacking AES. Yarom and Falkner [YF14] demonstrated the first full *Flush+Reload* attack, attacking RSA. *Flush+Reload* attacks on AES have been demonstrated on Intel processors by Irazoqui et al. [IES15; Ira+15a] and Gülmezoğlu et al. [Gül+15]. We also demonstrated *Flush+Reload* attacks on AES on Intel processors [GSM15] and on mobile ARM-based devices [Lip+16]. *Flush+Reload* attacks have also been demonstrated on ECDSA by Benger et al. [Ben+14], van de Pol et al. [PSY15], and Yarom and Benger et al. [YB14]. Allan et al. [All+16] demonstrated that a combination with a denial-of-service attack to degrade the speed of the ECDSA algorithm can yield a more efficient *Flush+Reload* attack on ECDSA. Other attacks have been performed by Zhang et al. [Zha+14] on activities in co-located virtual machines and by Irazoqui et al. [Ira+15b] on TLS. Inci et al. [Inc+16] demonstrated that they can recover encryption keys used in co-located VMs on Amazon EC2 in a larger and automated scale. Bruinderink et al. [Gro+16] demonstrated an attack on the BLISS signature scheme. Irazoqui et al. [Ira+14; IES16] demonstrated cross-VM and cross-CPU variants of the *Flush+Reload* attack. Another variation of the *Flush+Reload* attack combines it with return-oriented programming to attack cache designs that are less straightforward to attack [ZXZ16].

Beyond *Flush+Reload* attacks on cryptographic implementations, we demonstrated *Flush+Reload* attacks on user input [GSM15; Lip+16].

## 3.1.2. Branch-Prediction Attacks

Another set of caches are used for branch prediction. The branch-pattern table stores past results on branches and uses them to predict the outcome of future branches. The branch-target buffer caches branch targets from past branches to predict targets of future branches. Both caches are virtually-indexed and thus an adversary can target these caches without knowledge of physical addresses.

Software-based side-channel attacks based on branch prediction hits and misses have first been demonstrated by Aciicmez et al. [ASK07; Acı07a] in attacks on the RSA implementation of OpenSSL. The attacker primes the branch-target buffer by executing a sequence of branches. If the victim experiences a branch misprediction, an entry of the branch-target buffer will be replaced. The attacker subsequently observes a higher execution time due to a misprediction for one of its branches.

Bhattacharya et al. [BM15] show that branch prediction attacks based on hardware performance counters can be used to extract RSA keys from exponentiations running in other processes. Evtyushkin et al. [EPA15] demonstrate a covert channel between two processes manipulating the branch predictor. Evtyushkin et al. [EPA16] also demonstrate that KASLR can be defeated using the branch-target buffer. They infer where code has been executed in the kernel based on the mapping from virtual addresses to the branch-target buffer cache lines. Lee et al. [Lee+16] show that a malicious operating system can reverse-engineer the control flow of SGX enclaves through branch-prediction analysis. Ge et al. [Ge+16a] analyze besides other side channels also the capacity and error rate of a branch-prediction-based covert channel.

### 3.1.3. Page-translation Cache Attacks

Hund et al. [HWH13] presented the first attack exploiting timing differences caused by page-translation caches. Triggering page faults on inaccessible memory regions reveals whether the memory region would be valid for the kernel, as the valid page-translation entries are cached, independent of the current privilege level. This allows recovering which addresses are valid and even which addresses are used by specific parts of the kernel, *i.e.*, it defeats kernel address-space-layout randomization (KASLR). To run the attack, multiple page faults are triggered. When processing the first page fault, the processor walks through the page translation tables, caching every valid entry. For every subsequent page fault on the same address, the translation table entries are already cached and thus the time until the page fault is triggered is significantly lower.

Jang et al. [JLK16] exploited TSX transaction aborts upon page faults to optimize this attack. Their attack is significantly faster and more reliable and allows defeating KASLR within seconds. Furthermore, they observed a timing difference when trying to execute inaccessible kernel addresses. For executable kernel addresses the latency until the TSX transaction abort is lower than for non-executable kernel memory. Simultaneously to their work, we demonstrated that prefetch instructions leak the same timing difference and can be used to defeat KASLR as well [Gru+16a].

Van Schaik et al. [Sch+17a] showed that timing attacks allow to reverse-engineer the size and structure of the page translation caches. Gras et al. [Gra+17] showed that this information can be used in *Evict+Reload* attacks to defeat ASLR from sandboxed JavaScript.

### 3.1.4. Exception-based Attacks

Exception-based attacks deduce information from processor exceptions they trigger. Typical exceptions are scheduler interrupts, instruction aborts, page faults, but also behavioral differences, e.g., instructions providing the user with an error code. Through this behavior the CPU leaks direct information (*i.e.*, the behavior itself) and indirect information (*i.e.*, timing differences due to the behavior).

We include exception-based attacks as microarchitectural attacks, as they exploit both architecturally defined and undefined behavior. Especially the implementation of instructions depends on the specific microarchitectural design. Unspecified cases may influence the processor state and operation on some microarchitectures whereas others ignore it.

Two decades ago, Warner et al. [War+96] presented the first covert channel based on page faults. Page faults can leak sensitive information in three ways: the location of the page fault, the execution time of the page fault, and the mere fact that a page fault occurred *i.e.*, the memory access was delayed or not successful. Page deduplication is a mechanism to share identical memory pages across boundaries of virtual machines to reduce the memory footprint of systems [Mil+09]. However, the fact that a page fault occurred reveals to a user process that a second copy of the same page exists somewhere on the same machine. Suzaki et al. [Suz+11] presented an attack exploiting this to detect programs running in co-located virtual machines. Owens et al. [OW11] demonstrated fingerprinting based on page deduplication attacks. Xiao et al. [Xia+12; Xia+13] demonstrated cross-VM covert channels based on the page deduplication side channel. We demonstrated page deduplication attacks from JavaScript running in a website [GBM15].

Xu et al. [XCP15] demonstrated that page faults can be used as a side-channel by a malicious operating system to spy on an application running in a secure enclave. Shinde et al. [Shi+16] later confirmed their results. The malicious operating system frequently changes the virtual memory mappings of the enclave from valid to invalid. This forces the enclave to experience page faults for almost every memory reference. Although the page fault address in Intel SGX is truncated to be page-aligned the operating system is able derive accurate information of what the application in the secure enclave is processing. Weichbrodt et al. [Wei+16] used a similar approach to interrupt the secure enclave frequently to exploit double-fetch vulnerabilities in enclaves reliably. Zhang et al. [ZW09] proposed an attack that uses system interrupts information to derive user input. Simon et al. [SXA16] demonstrated a similar attack on Android that allows an attacker to recover words and sentences.

## 3.1.5. DRAM-based Attacks

Modern cloud systems often have multiple processors installed in a multi-socket mainboard. The processor caches are kept coherent with an inter-processor coherency protocol. However, this only has an effect on shared memory cache lines. For co-located virtual machines that do not share memory, the cache does not provide a communication channel in this setup.

Wu et al. [WXW12; WXW14] found that timing differences caused by memory bus locking can be exploited to build a covert channel between co-located virtual machines in this setup. Their covert channel works through channel contention and has a raw capacity of 13.5 KB/s at an error rate of 0.75% on the Amazon EC2 cloud. Inci et al. [İnc+16] found that memory bus locking can be used to verify co-location in the Microsoft Azure cloud.

More recently, we found found that the DRAM itself can also be exploited [Pes+16]. They described two attack primitives as so-called DRAM addressing (DRAMA) attacks. The row-conflict attack primitive works similar to *Prime+Probe*, the row-hit attack primitive is a side channel that is comparable to a *Flush+Reload* attack. Both side channels work without requiring any shared memory. DRAMA attacks exploit the DRAM row buffer which acts similarly as a directly-mapped per-bank cache for the DRAM rows.

In case of the row-conflict attack, the attacker and victim share a DRAM bank, but no DRAM row. The attacker continuously opens the same row in a bank. Whenever the victim accesses a different row in the same bank, the attacker observes a higher latency. The row-conflict covert channel achieves a performance of 74.5 KB/s with an error rate of 0.4% in a cross-VM cross-processor cloud setup.

In a row-hit attack, attacker and victim share a DRAM row in hardware. The attacker loads another row into the row buffer, comparable to the flush-step in a *Flush+Reload* attack. If the victim accesses the shared row again, it will be loaded into the row buffer. The attacker then reloads a memory location from the shared row and measures the access latency. If the victim accessed the shared row in the meantime, the access latency is low. Otherwise, the access latency is high.

Bhattacharya et al. [BM16] exploited the row-conflict side channel to locate the DRAM bank of a cryptographic secret exponent. They use this side-channel information to perform a Rowhammer attack on the secret exponent, leading to faulty signatures which allow a full key recovery. We demonstrated that the timing differences of the row-conflict side channel are large enough to be measured from JavaScript [Sch+17c]. Based on this observation, we demonstrated a transmission from an unprivileged process inside a virtual machine with no network access to JavaScript code running inside a website. The covert channel achieves a raw capacity of 11 b/s and an error rate of 0%.

### 3.1.6. Other Microarchitectural Side-Channel Attacks

Besides these main categories of software-based microarchitectural side-channel attacks, some works have investigated other interferences in instructions and microarchitectural elements. These interferences originate in the throughput limitations of processors. Aciicmez et al. [AS08b] demonstrated that parallel execution of multiplication instructions can leak an RSA key used in a square-and-multiply exponentiation. Evtyushkin et al. [EP16] build a covert channel exploiting timing differences of the `rdseed` instruction depending on the state of the internal random number buffer.

## 3.2. Software-based Microarchitectural Fault Attacks

A common assumption in system security and software security is the security of the hardware and its error-free operation. However, hardware is not perfect and especially when operated outside the specification, faults can be induced by an attacker [BDL97; BS97; Aum+02; SMC09; HS13; SA02]. A unique feature of microarchitectural fault attacks is that they use effects caused by microarchitectural elements or operations implemented on a microarchitectural level. In the software-based variant these effects and operations are triggered from software.

The first software-based microarchitectural fault attack was the so-called Rowhammer bug. Kim et al. [Kim+14] found that it can be triggered from software and that this could have implications on system security. They execute a sequence of memory accesses and `clflush` operations to frequently open and close DRAM rows, as illustrated in Figure 3.3. If the DRAM rows are in physical proximity, a bit can flip in another DRAM row without accessing it. This other memory location might be inaccessible to the attacker and even belong to another security domain. Their work sparked a series of publications that investigated the security implications of the Rowhammer bug and the requirements to successfully trigger it.

In early 2015, Seaborn and Dullien [SD15] presented the first two practical Rowhammer exploits. The first exploit escapes from the Google NaCl sandbox by causing a bit flip in dynamically generated indirect jumps. They spray the memory with indirect jumps, to maximize the probability to flip a bit in one of them. After a bit flipped in the jump instruction, the attacker gains control over the jump and can thus escape from the NaCl sandbox. The second exploit escalates from user mode to kernel privileges by causing a bit flip in a page table. Again they spray the memory, but this time with page tables, by memory-mapping the same file over and over again. Every mapping requires page table entries, which consist to more than 80% of address bits. The file contents are only kept in memory once. Hence, they fill almost the entire memory with address bits. When a bit flips in an address bit, the user memory mappings change from the file contents to another page. Hence, the user program has very likely gained access to its own page tables and thus has gained kernel privileges.

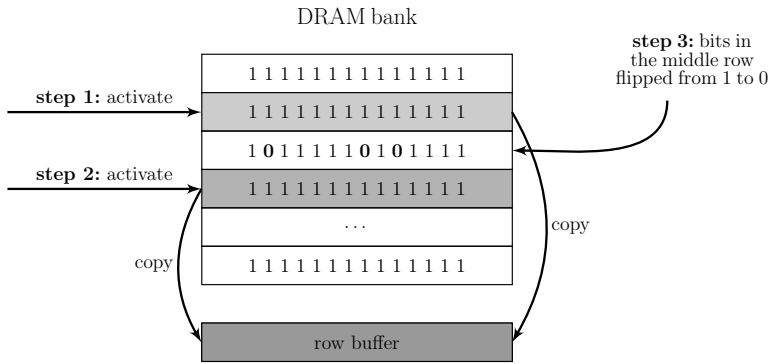

Figure 3.3.: To trigger the Rowhammer bug, memory locations in the same bank but in different rows are alternately accessed (steps 1 and 2) in a high frequency. Depending on the DRAM cell susceptibility to Rowhammer and the activation frequency, bits flip in step 3.

Seaborn and Dullien found 85% of all DDR3 modules they tested to be susceptible to the Rowhammer bug.

In July 2015, we demonstrated the first Rowhammer attack from sandboxed JavaScript [GMM16]. We triggered bit flips in page tables from JavaScript, by using cache eviction instead of the `clflush` instruction. We introduced a new hammering technique called amplified single-sided hammering, where an attacker hammers two DRAM rows in a 2 MB page to flip a bit in another 2 MB page.

Despite claims by DRAM vendors, we publicly reported the first bit flips on DDR4 memory in late 2015 and published our results in early 2016 [Pes+16]. For our attack we reverse-engineered the DRAM addressing functions for this purpose to perform an optimized attack. Our results on DDR4 were later on confirmed independently by Lanteigne [Lan16]. He observed a susceptibility rate of 67% of the DDR4 memory modules he tested. Aichinger [Aic15a; Aic15b] found that the Rowhammer bug also exists in ECC memory.

Also in 2016, Qiao and Seaborn [QS16] implemented a Rowhammer attack with non-temporal memory accesses, showing that prohibiting access to `clflush` is not sufficient. Bosman et al. [Bos+16] developed a reliable Rowhammer exploit in JavaScript exploiting page deduplication on Windows systems besides the Rowhammer bug. Bhattacharya et al. [BM16] demonstrated the first Rowhammer attack on cryptographic secrets. They trigger bit flips in an RSA key used in an exponentiation. Subsequently, they recover the RSA key through the Chinese remainder theorem as in first fault attacks on RSA [BDL97]. Xiao et al. [Xia+16] implemented a Rowhammer attack on Xen-PVM, triggering bit flips in hypervisor page tables and consequently obtaining hypervisor privileges.

Razavi et al. [Raz+16] demonstrated a Rowhammer attack on co-located tenants in the presence of page deduplication. They first scan the entire DRAM of the system for bit flips using the Rowhammer attack. Subsequently, they fill vulnerable pages with data that they suspect to be in a victim machine that will be co-located in the future. As soon as the victim is co-located, the identical memory pages are deduplicated. If the attacker now performs the Rowhammer attack again to trigger a bit flip in the page that is now also used by the victim. Through this attack they are able to manipulate installed certificates and update URLs used in the co-located machine to install malicious fake updates.

Besides personal computers and servers, mobile devices can also be attacked using the Rowhammer bug [Vee+16]. In our work with Van der Veen et al. we show that memory allocation techniques on Android devices expose uncached memory to user programs. Thus, the user can hammer DRAM rows efficiently and flip bits in kernel-level data structures.

In simultaneous work another software-based microarchitectural fault attack has been presented. Karimi et al. [Kar+15] demonstrated that software can artificially age circuits used in specific pipeline stages. A carefully crafted instruction stream increases the latency of the critical path of circuits if executed for several weeks. When the latency of the critical path exceeds the specification the subsequent pipeline stages work with incorrect values. Consequently, any further computations on the processor may have erroneous results.

## 3.3. Defenses Against Software-based Microarchitectural Attacks

Defenses against software-based microarchitectural attacks can be implemented on the user-space layer, system layer, or hardware layer. System layer and user-space layer would allow for protecting commodity systems. However, on a hardware level, defenses may induce a smaller performance overhead.

Most generic defenses try to reduce the amount of resource sharing to mitigate specific side-channel attacks. A countermeasure that has been proposed against same-core attacks is to schedule processes in different security domains only to different CPU cores [Per05; Mar+10]. Similarly, to mitigate cross-core attacks one could schedule such processes to different CPUs. For cross-CPU attacks one could perform the entire computation on different physical systems. However, the central idea of the cloud is to share resources and cloud providers will not eliminate multi-tenancy. Furthermore, information leakage through a remote interface might still be possible [Ber05].

Eliminating resource sharing does not work at all in the case of personal computers. Users deliberately want to execute third-party code such as native binaries or JavaScript on a website. Hence, it is important to find defense mechanisms that do not eliminate sharing but provide protection through other means. Furthermore, defenses against Rowhammer, as a microarchitectural fault attack, often follow similar approaches as defenses against microarchitectural side-channel attacks in many cases. In the following, we discuss possible state-of-the-art defenses grouped by the three layers: user-space layer, system layer, and hardware layer.

### 3.3.1. User-Space Layer

**Constant-time and data-obliviousness.** In his seminal work on cache timing attacks, Bernstein [Ber05] already proposed several mitigation tech-niques to protect AES computations. He emphasized that constant-time implementations are the most important protection mechanisms for cryptographic algorithms. Furthermore, he criticized the design of AES that suggests to perform data-dependent lookups in software. Similar defense mechanisms have been advised by later work [Cop+09; BLS12; Zha+12]. In his work on RSA cache timing attacks, Aciicmez et al. [Ac107a] proposed eliminating secret-dependent branches to defeat branch prediction attacks. Agosta et al. [Ago+07] suggested to either eliminate secret-dependent branches by replacing the branch by arithmetic operations or by converting branches to indirect jumps. Joye et al. [JT07] proposed an oblivious software exponentiation that performs no secret-dependent data or code accesses and thus is cache side-channel resistant.

Today, constant-time and oblivious computations are a standard countermeasure against attacks on cryptographic implementations. However, it has been found very hard to write truly constant-time implementations for some algorithms [BS13; YGH16; GBY16]. Still, microarchitectural side-channel attacks on cryptographic algorithms can generally be mitigated entirely in user-space implementations.

Preloading all data into the cache before running an algorithm has been investigated as a countermeasure against cache side-channel attacks. Hilton et al. [HLL16] demonstrated that this can improve performance of secure enclaves significantly while eliminating leakage based on cache misses. However, this approach is probabilistic and an attacker running on a second core might still be able to manipulate the cache state to restore the cache misses and thus, the leakage. However, experiments we performed showed that prefetching or preloading AES T-tables does not have any significant effect on asynchronous *Flush+Reload* and *Prime+ Probe* cross-core cache attacks.

### 3.3.2. System Layer

**Constant-time and data-obliviousness.** Andrysco et al. [And+15] developed a library that allows constant-time fixed-point numeric computations. Ohrimenko et al. [Ohr+16] developed a framework that allows making any algorithm data- and code-oblivious. Through conditional CPU operations the control flow is always the same and memory fetches are always performed. They demonstrated the practicality of their solution on commonly used machine learning algorithms run in Intel SGX enclaves. However, the framework does not protect against accidentally adding code that leaks side-channel information through timing or data accesses. From a more theoretical aspect Oblivious RAM (ORAM) could provide data-obliviousness in general. To perform an array lookup, a simple ORAM construction would access every element in the array. More complex ORAM constructions achieve a lower runtime overhead while still maintaining certain lower security bounds for distinguishing any two array lookups based on the access sequence. In practice, ORAM suffers from severe performance and latency penalties and is therefore not widely applicable.

**Manipulation of timing sources.** Most microarchitectural attacks require some form of accurate timing measurements. Consequently, simulating timing sources to reduce their usefulness for attackers has been proposed in several independent works [Avi+10a; Avi+10b; For12; Wu+15a; Wu+15b; LGR13; MDS12]. In modern cloud environments, this is typically already the case. Every virtual machine has its own timing offsets, including low-level timers like cycle counter registers. However, microarchitectural attacks seem widely unaffected from this [Ira+14; IES16; Liu+15].

Vattikonda et al. [VDS11] proposed to add additional fuzziness to timers to destroy any reliable timing information for the guest. However, microbenchmark measurements as used in modern cache attacks only require minimal timing differences [Gru+16b; Mau+17]. Furthermore, statistical methods allow to align traces and recover the secret information [Liu+15]. Furthermore, even in absence of any timing source an attacker can fall-back to a counting thread [Wra92; Lip+16] or build even more sophisticated and accurate timers [Sch+17c; Sch+17b]. Hence, it would be necessary to prevent any form of parallelism in the attacker process and also prevent any access to indirect timing sources such as interrupts from a preemptive scheduler [Dun+02; Avi+10a; Ste+13; Coc+14].

Timing differences in general can be made invisible to other processes, by making algorithms always consume the worst case time or by bucketing their computation time [KD09; AZM10; ZAM11; Coc+14]. This can be effective against timing attacks including the *Evict+Time* cache timing attack. However, it does not have a significant effect on other cache side-channel attacks such as *Flush+Reload* or *Prime+Probe*, as they can derive exactly when a memory access is performed and not only overall execution time differences. Many threat models, especially for multi-user environments, exclude direct measurement of the execution time of a specific algorithm. Hence, *Evict+Time* is not possible in such a threat model, whereas *Flush+Reload* and *Prime+Probe* are.

**Disabling cache-line sharing and shared memory.** Disabling resource sharing can be applied on every level for different resources with different granularities. As the last-level cache is typically physically-indexed and physically-tagged, cache lines can only be shared across processes if they are part of a shared memory region. Disabling cache-line sharing by avoiding shared memory has first been suggested by Yarom et al. [YF14] against *Flush+Reload* attacks. However, this would increase memory utilization significantly and also increase the execution time due to higher cache miss rates.

Another source of shared memory is page deduplication. Suzaki et al. [Suz+11] recommended to disable page deduplication in cloud environments to mitigate page deduplication attacks. The importance of disabling page deduplication in cloud environments has been substantiated with more sophisticated attacks being demonstrated [OW11; Xia+12; Xia+13; Bar+15; Raz+16]. In our work on page deduplication attacks, we also recommended disabling page deduplication to prevent attacks on personal computers [GBM15]. Bosman et al. [Bos+16] came to the same conclusion in order to prevent Rowhammer attacks on Windows systems.

While the large cloud providers have already disabled page deduplication, cloud providers generally have a large interest in keeping page deduplication enabled to optimize resource utilization and increase cost efficiency. Ning et al. [Nin+16] designed a system that enables page deduplication between virtual machines belonging to the same group, while preventing cross-group attacks.

Zhou et al. [ZRZ16] proposed "copy-on-access", a more dynamic approach to disable cache-line sharing. They developed a system where the operating system or hypervisor dynamically creates copies of pages that are used simultaneously by multiple programs. However, in practice it only increases the amount of measurements an attacker has to perform, but does not fully mitigate any cache side-channel attack.

To mitigate microarchitectural attacks on KASLR, we proposed to use separate paging structures for kernel space and user space to avoid sharing cache lines in paging structure caches [Gru+16a]. This was also proposed in concurrent work by Jang et al. [JLK16]. Gruss et al. [Gru+17b] showed that separating paging structures indeed eliminates the page-translation cache side channel.

**Avoiding cache-set sharing.** Assuming the problem of cache-line sharing is solved and thus *Flush+Reload* attacks are not possible anymore, cache-set sharing still allows performing *Prime+Probe* attacks. To prevent cache-set sharing, cache-coloring has been proposed as a countermeasure against side-channel attacks by Shi et al. [Shi+11]. Kim et al. [KPM12] implemented a protection mechanism for cryptographic implementations based on cache coloring on modern Intel CPUs. Godfrey [GZ14] implemented a similar protection mechanism in the Xen hypervisor. Cock [Coc+14] evaluated cache coloring on ARM-based devices. In all cases the authors measured only a small performance impact. However, the memory overhead is significant: colors are fixed to physical addresses and thus large portions of physical memory have to be assigned to the same virtual machine or process in order to provide strict cache coloring without other virtual machines or processes working in the same cache set. Costan et al. [CD16] proposed Sanctum, an alternative to Intel SGX that employs cache coloring to protect against cache side-channel attacks on enclaves.

The slices used in the last-level cache in modern Intel processors can be utilized for cache coloring [HWH13; Mau+15a; Inc+15; Yar+15]. They already implicitly make cache attacks more difficult as the attacker has to gain knowledge on how addresses map to slices. Additionally, slices facilitates implementing cache coloring schemes on a system level. With every slice the number of colors is multiplied by 2.

With Intel CAT (cache allocation technology) [Int14], system software can now directly control how the slices are used. Intel CAT allows restricting cores to a subset of slices of the last-level cache. By separating processes of different security domain to different cores and thus their data into different cache slices, any cache-set sharing is eliminated. Liu et al. [Liu+16] implemented CATalyst, a system that instruments Intel CAT to protect general purpose software and cryptographic algorithms. They use Intel CAT to pin cache lines in the cache by first restricting access to one slice to its core and subsequently removing all cores from this slice. Still, cached values are served from the cache, effectively pinning the values cached in this slice into the cache. A compiler could generate the code to protect secret-dependent operations with Intel CAT. The performance overhead for low-memory tasks is negligible.

Intel CAT likely can also be used to prevent DRAMA side-channel attacks. The attacker may not be able to flush or evict victim data. Consequently, the victim process has full control over all its memory accesses in terms of cache hits and cache misses and can avoid information leakage through cache misses.

Weiß et al. [Wei+14] developed a scheduler that reduces the amount of cache set sharing between different virtual machines. Moon et al. [MSR15] proposed frequent VM migration to avoid colocation for longer periods in time. A similar approach based on container migration has been proposed by Azab et al. [AE16].

**Avoiding spatial proximity.** To mitigate Rowhammer attacks it is not sufficient to avoid memory and cache-set sharing, but it is also necessary to avoid spatial proximity in DRAM. Brasser et al. [Bra+16] proposed to isolate processes running in different security domains in DRAM such that no security domain can flip bits in another security domain.

**Cache cleansing.** If we assume that attacker and victim cannot access any cache set simultaneously, the question is how to cope with leakage that remains in the cache after the victim was descheduled. Cache cleansing aims to protect against attacks in such a scenario. It tries to maintain the cache in a state that leaks no information to prevent cache attacks. Zhang et al. [ZR13] and Godfrey et al. [GZ14] proposed cache cleansing to prevent leakage in cloud scenarios. They flush the cache upon context switches to eliminate the secret information from the cache. Cock et al. [Coc13] implemented a scheduler that reduces the amount of flushes on context switches and while maintaining the same security properties. Varadarajan et al. [VRS14] proposed a minimum-runtime guarantee for virtual machines in the cloud to prevent frequent context switches between different virtual machines. They also use cache cleansing to mitigate leakage of sensitive data. While this does not degrade system performance significantly it increases latency for requests to other VMs by up to 17%. Braun et al. [BJB15] proposed to compile specially annotated functions to be constant-time and data-oblivious. These functions also do not access any shared cache sets. Furthermore, they employ cache cleansing before and after the secret-dependent execution.

With the advent of multi-core processors, cache cleansing lost some of its practical relevance. Although disabling hyperthreading might be viable, disabling multi-core or the last-level cache entirely is not a practical solution. Even without the last-level cache, coherency protocols can keep cache lines coherent across processors and reintroduce the supposedly eliminated timing differences.

**Detecting vulnerabilities.** A different branch of countermeasures are mechanisms to detect vulnerabilities in software. Detected vulnerabilities can be eliminated by patching the software.

With Cache Template Attacks [GSM15], we presented a way to scan software for vulnerabilities. Doychev et al. [Doy+15] proposed to detect potential leakage in applications using static analysis techniques. A similar system has been proposed by Irazoqui et al. [IES17] to detect microarchitectural attacks including DRAMA and Rowhammer. Reparaz et al. [RBV16] proposed black-box leakage detection for cryptographic implementations and other algorithms. Zankl et al. [ZHS16] advised to incorporate automated leakage detection in the deployment process for cryptographic libraries. In line with these works are also approaches to quantify cache leakage using detailed abstract models of the cache [Dem+12; DK16; Cha+16]. A developer can use this information to eliminate the leakage through source code improvements.

To mitigate Rowhammer attacks, Kim et al. [Kim+14] proposed to detect vulnerable DRAM rows and remap them to spare DRAM cells. Brasser et al. [Bra+16] proposed to disable these DRAM regions in the boot loader.

**Detecting and stopping ongoing attacks.** Another form of detection mechanisms aim at detecting ongoing attacks. Following the idea of virus and malware scanners, a software runs continuously checking the system for malicious activity and subsequently stop the attacking processes or virtual machines. Zhang et al. [Zha+11] proposed HomeAlone, a system using a *Prime+Probe* covert channel to detect colocation. Their system allows detecting when a virtual machine is co-located with other virtual machines on the same physical machine although being billed for a dedicated machine. With Cache Template Attacks [GSM15], we presented a way to search and detect ongoing cache attacks. However, both approaches increase the system load significantly.

Cardenas et al. [CB12] used performance counters to detect microarchitectural denial-of-service attacks in cloud environments. Demme et al. [Dem+13] proposed the use of performance counters to detect abnormal cache behavior to detect malware and Tang et al. [TSS14] enhanced this idea by evaluating performance counters using unsupervised learning.

Chouhan et al. [CH16] proposed to use bloom filters on the cache miss traces to detect yet unknown cache side-channel attacks. Hunger et al. [Hun+15] proposed detecting side-channel attacks through measuring the performance variations in a program that mimics a typical victim application.

Herath and Fogh [HF15] proposed to monitor cache misses to detect *Flush+Reload* attacks and Rowhammer. Similar approaches have been chosen by Chiappetta et al. [CSY15] and Zhang et al. [ZZL16]. Both built systems that use cache hit and miss traces to detect *Flush+Reload* attacks in native and cloud environments respectively. We showed that detection mechanisms through performance counters might be an insufficient solution [Gru+16b]. We showed that performance counters fail to detect all variants of cache attacks, such as *Flush+Flush* and slowed-down variants of *Flush+Reload*. In response, Fogh [Fog15] developed a mechanism that uses performance-monitor interrupts on `rdtsc` instructions. This mechanism checks the program code around the current instruction pointer for suspicious instructions like `clflush`. If suspicious instructions are found, the potentially malicious program is slowed down or terminated. Payer [Pay16] developed a system called HexPADS, which detects cache attacks and Rowhammer at runtime. HexPADS uses different performance events like cache references, cache misses, and page faults to evaluate whether a process is malicious. HexPADS can easily be applied to commodity operating systems.

Chen et al. [Che+17] designed a framework to detect ongoing controlled-channel attacks on SGX enclaves at runtime. They use TSX to built a trusted in-enclave counting thread. This counting thread is used to measure the execution time of code sections. If the execution time is too high an interrupt must have occurred and thus the untrusted operating system interrupted the enclave, likely to perform an attack.

To specifically detect and stop ongoing Rowhammer attacks, several works propose the usage of performance counters to detect whether to induce or wait for row refreshes [Kim+14; Cor16; Awe+16].

### 3.3.3. Hardware Layer

**Eliminating timing differences.** Naturally, timing differences introduced by the hardware could be eliminated in hardware [Pag03; Ber05; Per05]. The best example might be the `clflush` instruction, which has a small but exploitable timing difference when accessing cached and un-cached memory locations. Making this instruction constant-time would likely not be noticed in practice [Gru+16b]. The instruction is rarely used and a timing penalty of less than 10 cycles is negligible for rarely used instructions. Similarly, prefetch instructions leak through timing differences that could be eliminated without a significant performance penalty [Gru+16a].

Timing differences due to hardware modification only occur if the software runs through an instruction stream that runs into these hardware optimizations. Leakage introduced this way could be avoided by providing constant-time instructions as building blocks for more complex algorithms, e.g., cryptographic implementations. Indeed, processor manufacturers are incorporating and increasing number of constant-time instructions for cryptographic primitives [Int08; AMD09; ARM12; ARM13], most prominently Intel AES-NI. Instruction set extensions like AES-NI have been proposed as countermeasures against various attacks [Pag03; Ber05; Per05]. Today, many cryptographic libraries use these instructions typically by default [Ope].

Wang et al. [WFS14] proposed to change memory controllers to eliminate timing side-channels. Most importantly, they suggest changes to the row-buffer policy. Instead of keeping the row buffer open, they immediately close the row buffer again, leaving the DRAM in a pre-charged state. This introduces a lower latency penalty than a row conflict, but still a significant performance penalty as compared to current implementations. This countermeasure would likely eliminate DRAMA side-channel attacks [Pes+16].

Timing differences are no problem if the execution time does not vary depending on secret information. Wang et al. [WL07] proposed to eliminate cache side-channel leakage with their so-called partition-locked caches (PLcache) and random-permutation caches (RPcache). The PLcache allows locking cache lines in the cache and prevent their eviction. Hence, an attacker cannot observe any cache misses from the victim process as the victim operates entirely on the cache. The RPcache approach introduces a different mapping from physical addresses to cache sets for every process at runtime. Hence, each program has its own cache sets which may be overlapping or disjoint but are never identical. Consequently, the attacker cannot prime a cache set of the victim and thus the attacker cannot perform a cache attack on the victim.

Kong et al. [Kon+08] showed that both the PLcache and the RPcache protections can be circumvented by an attacker. They proposed informing loads as an extension to the RPcache to protect against cache side channels. Informed loads are special instructions that do not only perform the memory load but also re-randomize data structures [Kon+09]. Liu et al. [LL14] proposed a cache design where the mapping from addresses to cache sets is dynamically randomized at runtime. While the randomized address mapping does not prevent that attacker and victim share a cache set, it does effectively prevent that attacker and victim share a cache set over a longer period in time. Consequently, many attacks are mitigated. Fuchs et al. [FL15] proposed to use specialized prefetching algorithms to mitigate side-channel attacks.

**Disabling resource sharing.** Also on the hardware level it is possible to disable or reduce the amount of resource sharing. Page [Pag05] suggested partitioned caches to avoid cache-set sharing across processes. These per-process partitions are maintained dynamically, avoiding any static cache mapping and cache sharing. Wang et al. [WL08] proposed a new cache architecture to mitigate cache side-channel attacks. Their cache architecture would prevent cache-set sharing between attacker and victim using dynamic mappings between addresses and cache sets. Tiwari et al. [Tiw+09] proposed a mechanism to execute untrusted code with tight upper limits on the leakage in terms of time and side effects, with a moderate performance impact. Sanchez et al. [SK10] proposed a faster cache design which decouples cache ways and cache associativity. This cache design is also likely to impact the applicability of eviction-based attacks like *Prime+ Probe*. Domnitser et al. [Dom+11] proposed non-monopolizable caches as a defense against cache attacks. Non-monopolizable caches prevent that any process can allocate enough cache lines to observe cache collisions with another process. Domnitser et al. observed a low performance penalty for cryptographic algorithms. However, using non-monopolizable caches for larger parts of the software stack would severely impact the system performance, as it is equivalent to reducing the cache size per process.

To mitigate branch prediction attacks, Tan et al. [TWG14] proposed a new branch target buffer scheme which allows detecting potentially malicious activity. After detection, the hardware prevents that branch target buffer entries are shared with the suspected malicious processes.

**Rowhammer countermeasures in hardware.** Hardware countermeasures against Rowhammer are specific to the Rowhammer hardware defi-ciency. Kim et al. [Kim+14] proposed several solutions to the Rowhammer bug, including usage of ECC memory, building more reliable DRAM cells, and increasing the refresh rate. They also showed that increasing the refresh rate is not always effective unless it is increased by a factor of 7 or more. Instead, they propose PARA, a mechanism which probabilistically opens adjacent rows. As Rowhammer attacks require a huge number of accesses, the adjacent rows are very likely refreshed early enough and no bit flip occurs.

A different mechanism is target-row refresh (TRR). TRR refreshes lines after a certain number of accesses to adjacent lines. Although TRR has been announced for DDR4 modules, it was removed from the final DDR4 standard [GMM16].

# **4**

# Future Work and Conclusions

We can draw conclusions on four different axis from this thesis and the corresponding publications.

First, microarchitectural attacks can be widely automated. We have shown this with our work on Cache Template Attacks [GSM15], but automation also played a significant role in our other works [Lip+16; Mau+17]. Automation provides any unsophisticated user with the ability to perform microarchitectural attacks. It also enables more large scale attacks. Future work will likely investigate automation of microarchitectural attacks in further detail.

Second, unknown and novel side channels are very likely to exist and to be found. We showed that modern microarchitectures expose several previously unknown side channels, such as the `clflush` instruction [Gru+16b], the DRAM [Pes+16], or prefetch instructions [Gru+16a]. While we found several new side channel, it is more difficult to find all microarchitectural side channels. Hence, we can expect to find more microarchitectural side channels and especially find new side channels with every new microarchitecture. Furthermore, many microarchitectural side channels have not been investigated in detail yet. For instance, it is likely that the prefetch side channel [Gru+16a] contains information that has not been used in any published attacks. Also microarchitectural elements at low levels, closer to the execution core, should be investigated for new side channels. Future work should investigate whether these hardware components, such as graphic adapters, can be instrumented in attacks.

Third, it is possible to reduce and minimize requirements of known attacks to a point where they can be performed in highly-restricted and sandboxed environments. We have shown that this is the case in our work on Rowhammer attacks in JavaScript [GMM16] and in our work on page deduplication attacks in JavaScript [GBM15]. In terms of software-based microarchitectural fault attacks, we are just starting to investigate various hardware elements and how they can be accessed from unprivileged environments. Investigating the applicability of Rowhammer [Kim+14; SD15] and MAGIC [Kar+15] attacks in different scenarios will help to assess their risks. However, besides the DRAM and specific processor components, there is an abundance of other hardware elements in modern systems that could be attacked. Such novel attacks might reduce the requirements for fault attacks even further.

Fourth, constructing both effective and efficient countermeasures is a difficult task. Research often over-ambitiously aims to find universal countermeasures against microarchitectural attacks, ignoring that the various attacks have vastly different requirements and properties [Gru+16b; Pes+16]. At the core of microarchitectural attacks is usually a temporal or behavioral difference that is intended by the processor manufacturer to optimize the performance. Hence, it we cannot always find a universal countermeasure that does not degrade the performance as was the case for the prefetch side channel [Gru+16a]. Security and performance are contradicting each other to a growing extent. Countermeasures can only be practical if they provide useful and possibly dynamic trade-offs between security and performance. Instead of universal countermeasures, it may appear as a low-hanging fruit to protect specific scenarios, but it is also more likely to be practical. Especially cryptographic implementations are already being constantly improved to defend against new microarchitectural attacks.

# Part II.

# Publications

# List of Publications

During my thesis, I contributed to 13 publications in conference proceedings, 6 of which are included in this thesis as shown below.

## Publications in this Thesis

[GBM15a]     D. Gruss, D. Bidner, and S. Mangard. Practical Memory Deduplication Attacks in Sandboxed JavaScript. In: 20th European Symposium on Research in Computer Security (ESORICS'15). 2015.

[GMM16a]     D. Gruss, C. Maurice, and S. Mangard. Rowhammer.js: A Remote Software-Induced Fault Attack in JavaScript. In: DIMVA'16. 2016.

[Gru+16a]    D. Gruss, C. Maurice, A. Fogh, M. Lipp, and S. Mangard. Prefetch Side-Channel Attacks: Bypassing SMAP and Kernel ASLR. In: CCS'16. 2016.

[Gru+16b]    D. Gruss, C. Maurice, K. Wagner, and S. Mangard. Flush+Flush: A Fast and Stealthy Cache Attack. In: DIMVA'16. 2016.

[GSM15a]     D. Gruss, R. Spreitzer, and S. Mangard. Cache Template Attacks: Automating Attacks on Inclusive Last-Level Caches. In: USENIX Security Symposium. 2015.

[Lip+16a]    M. Lipp, D. Gruss, R. Spreitzer, C. Maurice, and S. Mangard. ARMageddon: Cache Attacks on Mobile Devices. In: USENIX Security Symposium. 2016.

## Other Contributions

[Gru+17a]    D. Gruss, J. Lettner, F. Schuster, O. Ohrimenko, I. Haller, and M. Costa. Strong and Efficient Cache Side-Channel Protection using Hardware Transactional Memory. In: USENIX Security Symposium. (to appear). 2017.

# 5

# Cache Template Attacks: Automating Attacks on Inclusive Last-Level Caches

## Publication Data

D. Gruss, R. Spreitzer, and S. Mangard. Cache Template Attacks: Automating Attacks on Inclusive Last-Level Caches. In: USENIX Security Symposium. 2015

## Contributions

Main author.

# Cache Template Attacks: Automating Attacks on Inclusive Last-Level Caches

Daniel Gruss, Raphael Spreitzer, and Stefan Mangard

Graz University of Technology, Austria

# Abstract

Recent work on cache attacks has shown that CPU caches represent a powerful source of information leakage. However, existing attacks require manual identification of vulnerabilities, *i.e.*, data accesses or instruction execution depending on secret information. In this paper, we present *Cache Template Attacks*. This generic attack technique allows us to profile and exploit cache-based information leakage of any program automatically, without prior knowledge of specific software versions or even specific system information. Cache Template Attacks can be executed online on a remote system without any prior offline computations or measurements.

Cache Template Attacks consist of two phases. In the profiling phase, we determine dependencies between the processing of secret information, e.g., specific key inputs or private keys of cryptographic primitives, and specific cache accesses. In the exploitation phase, we derive the secret values based on observed cache accesses. We illustrate the power of the presented approach in several attacks, but also in a useful application for developers. Among the presented attacks is the application of Cache Template Attacks to infer keystrokes and—even more severe—the identification of specific keys on Linux and Windows user interfaces. More specifically, for lower-case only passwords, we can reduce the entropy per character from $\log_2(26) = 4.7$ to 1.4 bits on Linux systems. Furthermore, we perform an automated attack on the T-table-based AES implementation of OpenSSL that is as efficient as state-of-the-art manual cache attacks.

---

# 1. Introduction

Cache-based side-channel attacks have gained increasing attention among the scientific community. First, in terms of ever improving attacks against cryptographic implementations, both symmetric [Pag02; Ber04; BM06; OST06; WHS12; GBK11] as well as asymmetric cryptography [YB14; Ben+14; Che+13; BH09], and second, in terms of developing counter-measures to prevent these types of attacks [LL14; MKS12]. Recently, Yarom and Falkner [YF14] proposed the Flush+Reload attack, which has been successfully applied against cryptographic implementations [Ben+14; Ira+14b; Gül+15]. Besides the possibility of attacking cryptographic implementations, Yarom and Falkner pointed out that their attack might also be used to attack other software as well, for instance, to collect keystroke timing information. However, no clear indication is given on how to exploit such vulnerabilities with their attack. A similar attack has already been suggested in 2009 by Ristenpart et al. [Ris+09], who reported being able to gather keystroke timing information by observing cache activities on an otherwise idle machine.

The limiting factor of all existing attacks is that sophisticated knowledge about the attacked algorithm or software is necessary, *i.e.*, access to the source code or even modification of the source code [BH09] is required in order to identify vulnerable memory accesses or the execution of specific code fragments manually.

In this paper, we make use of the Flush+Reload attack [YF14] and present the concept of *Cache Template Attacks*,[1] a generic approach to exploit cache-based vulnerabilities in any program running on architectures with shared inclusive last-level caches. Our attack exploits four fundamental concepts of modern cache architectures and operating systems.

1. Last-level caches are shared among all CPUs.
2. Last-level caches are inclusive, *i.e.*, all data which is cached within the L1 and L2 cache must also be cached in the L3 cache. Thus, any modification of the L3 cache on one core immediately influences the cache behavior of all other cores.
3. Cache lines are shared among different processes.

---

[1]The basic framework can be found at `https://github.com/IAIK/cache_template_attacks`.

4. The operating system allows programs to map any other program binary or library, *i.e.*, code and static data, into their own address space.

Based on these observations, we demonstrate how to perform Cache Template Attacks on any program automatically in order to determine memory addresses which are accessed depending on secret information or specific events. Thus, we are not only able to attack cryptographic implementations, but also any other event, e.g., keyboard input, which might be of interest to an attacker.

We demonstrate how to use Cache Template Attacks to derive keystroke information with a deviation of less than 1 microsecond from the actual keystroke and an accuracy of almost 100%. With our approach, we are not only able to infer keystroke timing information, but even to infer specific keys pressed on the keyboard, both for GTK-based Linux user interfaces and Windows user interfaces. Furthermore, all attacks to date require sophisticated knowledge of the attacked software and the executable itself. In contrast, our technique can be applied to any executable in a generic way. In order to demonstrate this, we automatically attack the T-table-based AES [Nat01; DR13] implementation of OpenSSL [Ope14].

Besides demonstrating the power of Cache Template Attacks to exploit cache-based vulnerabilities, we also discuss how this generic concept supports developers in detecting cache-based information leaks within their own software, including third party libraries. Based on the insights we gained during the development of the presented concept, we also present possible countermeasures to mitigate specific types of cache attacks.

**Outline**

The remaining paper is organized as follows. In Section 2, we provide background information on CPU caches, shared memory, and cache attacks in general. We describe Cache Template Attacks in Section 3. We illustrate the basic idea on an artificial example program in Section 4 and demonstrate Cache Template Attacks against real-world applications in Section 5. In Section 6, we discuss countermeasures against cache attacks in general. Finally, we conclude in Section 7.

# 2. Background and Related Work

In this section, we give a basic introduction to the concept of CPU caches and shared memory. Furthermore, we provide a basic introduction to cache attacks.

## 2.1. CPU Caches

The basic idea of CPU caches is to hide memory accesses to the slow physical memory by buffering frequently used data in a small and fast memory. Today, most architectures employ set-associative caches, meaning that the cache is divided into multiple cache sets and each cache set consists of several cache lines (also called ways). An index is used to map specific memory locations to the sets of the cache memory.

We distinguish between virtually indexed and physically indexed caches, which derive the index from the virtual or physical address, respectively. In general, virtually indexed caches are considered to be faster than physically indexed caches. However, the drawback of virtually indexed caches is that different virtual addresses mapping to the same physical address are cached in different cache lines. In order to uniquely identify a specific cache line within a cache set, so-called tags are used. Again, caches can be virtually tagged or physically tagged. A virtual tag has the same drawback as a virtual index. Physical tags, however, are less expensive than physical indices as they can be computed simultaneously with the virtual index.

In addition, there is a distinction between inclusive and exclusive caches. On Intel systems, the L3 cache is an inclusive cache, meaning that all data within the L1 and L2 caches are also present within the L3 cache. Furthermore, the L3 cache is shared among all cores. Due to the shared L3 cache, executing code or accessing data on one core has immediate consequences for all other cores. This is the basis for the Flush+Reload [YF14] attack as described in Section 2.3.

Our test systems (Intel Core i5-2/3 CPUs) have two 32 KB L1 caches—one for data and one for instructions—per core, a unified L2 cache of 256 KB, and a unified L3 cache of 3 MB (12 ways) shared among all cores. The cache-line size is 64 bytes for all caches.

## 2.2. Shared Memory

Operating systems use shared memory to reduce memory utilization. For instance, libraries used by several programs are shared among all processes using them. The operating system loads the libraries into physical memory only once and maps the same physical pages into the address space of each process.

The operating system employs shared memory in several more cases. First, when forking a process, the memory is shared between the two processes. Only when the data is modified, the corresponding memory regions are copied. Second, a similar mechanism is used when starting another instance of an already running program. Third, it is also possible for user programs to request shared memory using system calls like `mmap`.

The operating system tries to unify these three categories. On Linux, mapping a program file or a shared library file as a read-only memory with `mmap` results in sharing memory with all these programs, respectively programs using the same shared library or program binary. This is also possible on Windows using the `LoadLibrary` function. Thus, even if a program is statically linked, its memory is shared with other programs which execute or map the same binary.

Another form of shared memory is content-based page deduplication. The hypervisor or operating system scans the physical memory for pages with identical content. All mappings to identical pages are redirected to one of the pages while the other pages are marked as free. Thus, memory is shared between completely unrelated processes and even between processes running in different virtual machines. When the data is modified by one process, memory is duplicated again. These examples demonstrate that code as well as static data can be shared among processes, even without their knowledge. Nevertheless, page deduplication can enhance system performance and besides the application in cloud systems, it is also relevant in smaller systems like smartphones.

User programs can retrieve information on their virtual and physical memory using operating-system services like `/proc/<pid>/maps` on Linux or tools like `vmmap` on Windows. The list of mappings typically includes all loaded shared-object files and the program binary.

## 2.3. Cache Attacks

Cache attacks are a specific type of side-channel attacks that exploit the effects of the cache memory on the execution time of algorithms. The first theoretical attacks were mentioned by Kocher [Koc96] and Kelsey et al. [Kel+00]. Later on, practical attacks for DES were proposed by Page [Pag02] as well as Tsunoo et al. [Tsu+03]. In 2004, Bernstein [Ber04] proposed the first time-driven cache attack against AES. This attack has been investigated quite extensively [Nev06].

A more fine-grained attack has been proposed by Percival [Per05], who suggested to measure the time to access all ways of a cache set. As the access time correlates with the number of occupied cache ways, an attacker can determine the cache ways occupied by other processes. At the same time, Osvik et al. [OST06] proposed two fundamental techniques that allow an attacker to determine which specific cache sets have been accessed by a victim program. The first technique is Evict+Time, which consists of three steps. First, the victim program is executed and its execution time is measured. Afterwards, an attacker evicts one specific cache set and finally measures the execution time of the victim again. If the execution time increased, the cache set was probably accessed during the execution.

The second technique is Prime+Probe, which is similar to Percival's attack. During the Prime step, the attacker occupies specific cache sets. After the victim program has been scheduled, the Probe step is used to determine which cache sets are still occupied.

Later on, Gullasch et al. [GBK11] proposed a significantly more powerful attack that exploits the fact that shared memory is loaded into the same cache sets for different processes. While Gullasch et al. attacked the L1 cache, Yarom and Falkner [YF14] presented an improvement called Flush+Reload that targets the L3 cache.

Flush+Reload relies on the availability of shared memory and especially shared libraries between the attacker and the victim program. An attacker constantly flushes a cache line using the `clflush` instruction on an address within the shared memory. After the victim has been scheduled, the attacker measures the time it takes to reaccess the same address again. The measured time reveals whether the data has been loaded into the cache by reaccessing it or whether the victim program loaded the data into the cache before reaccessing. This allows the attacker to determine the memory accesses of the victim process. As the L3 cache is shared among all cores, it is not necessary to constantly interrupt the victim process. Instead, both processes run on different cores while still working on the same L3 cache. Furthermore, the L3 cache is a unified inclusive cache and, thus, even allows to determine when a certain instruction is executed. Because of the size of the L3 cache, there are significantly fewer false negative cache-hit detections caused by evictions. Even though false positive cache-hit detections (as in Prime+Probe) are not possible because of the shared-memory-based approach, false positive cache hits can still occur if data is loaded into the cache accidentally (e.g., by the prefetcher). Nevertheless, applications of Flush+Reload have been shown to be quite reliable and powerful, for example, to detect specific versions of cryptographic libraries [Ira+15a], to revive supposedly fixed attacks (e.g., Lucky 13) [Ira+15b] as well as to improve attacks against T-table-based AES implementations [Gül+15].

As shared memory is not always available between different virtual machines in the cloud, more recent cache attacks use the Prime+Probe technique to perform cache attacks across virtual machine borders. For example, Irazoqui et al. [IES15] demonstrated a cross-VM attack on a T-Table-based AES implementation and Liu et al. [Liu+15] demonstrated a cross-VM attack on GnuPG. Both attacks require manual identification of exploitable code and data in targeted binaries. Similarly, Maurice et al. [Mau+15] built a cache-index-agnostic cross-VM covert channel based on Prime+Probe.

Simultaneous to our work, Oren et al. [Ore+15] developed a cache attack from within sandboxed JavaScript to attack user-specific data like network traffic or mouse movements. Contrary to existing attack approaches, we present a general attack framework to exploit cache vulnerabilities automatically. We demonstrate the effectiveness of this approach by inferring keystroke information and, for comparison reasons, by attacking a T-table-based AES implementation.

# 3. Cache Template Attacks

Chari et al. [CRR02] presented template attacks as one of the strongest forms of side-channel attacks. First, side-channel traces are generated on a device controlled by the attacker. Based on these traces, the template—an exact model of signal and noise—is generated. A single side-channel trace from an identical device with unknown key is then iteratively classified using the template to derive the unknown key.

Similarly, Brumley and Hakala [BH09] described cache-timing template attacks to automatically analyze and exploit cache vulnerabilities. Their attack is based on Prime+Probe on the L1 cache and, thus, needs to run on the same core as the spy program. Furthermore, they describe a profiling phase for specific operations executed in the attacked binary, which requires manual work or even modification of the attacked software. In contrast, our attack only requires an attacker to know how to trigger specific events in order to attack them. Subsequently, Brumley and Hakala match these timing templates against the cache timing observed. In contrast, we match memory-access templates against the observed memory accesses.

Inspired by their work we propose *Cache Template Attacks*. The presented approach of Cache Template Attacks allows the exploitation of any cache vulnerability present in any program on any operating system executed on architectures with shared inclusive last-level caches and shared memory enabled. Cache Template Attacks consist of two phases: 1) a profiling phase, and 2) an exploitation phase. In the profiling phase, we compute a Cache Template matrix containing the cache-hit ratio on an address given a specific target event in the binary under attack. The exploitation phase uses this Cache Template matrix to infer events from cache hits.

Both phases rely on Flush+Reload and, thus, attack code and static data within binaries. In both phases the attacked binary is mapped into read-only shared memory in the attacker process. By accessing its own virtual addresses in the allocated read-only shared memory region, the attacker accesses the same physical memory and the same cache lines (due to the physically-indexed last level cache) as the process under attack. Therefore, the attacker completely bypasses address space layout randomization (ASLR). Also, due to shared memory, the additional memory consumption caused by the attacker process is negligible, *i.e.*, in the range of a few megabytes at most.

In general, both phases are performed online on the attacked system and, therefore, cannot be prevented through differences in binaries due to different versions or the concept of software diversity [Fra10]. However, if online profiling is not possible, e.g., in case the events must be triggered by a user or Flush+Reload is not possible on the attacked system, it can also be performed in a controlled environment. Below, we describe the profiling phase and the exploitation phase in more detail.

## 3.1. Profiling Phase

The profiling phase measures how many cache hits occur on a specific address during the execution of a specific event, *i.e.*, the *cache-hit ratio*. The cache-hit ratios for different events are stored in the Cache Template matrix which has one column per event and one row per address. We refer to the column vector for an event as a *profile*. Examples of Cache Template matrices can be found in Section 4 and Section 5.1.

An *event* in terms of a Cache Template Attack can be anything that involves code execution or data accesses, e.g., low-frequency events, such as keystrokes or receiving an email, or high-frequency events, such as encryption with one or more key bits set to a specific value. To automate the profiling phase, it must be possible to trigger the event programmatically, e.g., by calling a function to simulate a keypress event, or executing a program.

The Cache Template matrix is computed in three steps. The first step is the generation of the cache-hit trace and the event trace. This is the main computation step of the Cache Template Attack, where the data for the Template is measured. In the second step, we extract the cache-hit ratio for each trace and store it in the Cache Template matrix. In a third post-processing step, we prune rows and columns which contain redundant information from the matrix. Algorithm 1 summarizes the profiling phase. We explain the corresponding steps in detail below.

**Cache-Hit Trace and Event Trace**

The generation of the cache-hit trace and the event trace is repeated for each event and address for the specified duration (the while loop of Algorithm 1). The cache-hit trace $g_{a,e}^{(H)}$ is a binary function which has value 1 for every timestamp $t$ where a cache hit has been observed. The function value remains 1 until the next timestamp $t$ where a cache miss has been observed. We call subsequent cache hits a cache-hit phase. The event trace $g_{a,e}^{(E)}$ is a binary function which has value 1 when the processing of one specific event $e$ starts or ends and value 0 for all other points.

In the measurement step, the binary under attack is executed and the event is triggered constantly. Each address of the attacked binary is profiled for a specific duration $d$. It must be long enough to trigger one or more events. Therefore, $d$ depends only on the execution time of the event to be

---

**Algorithm 1:** Profiling phase.

---

**Input**: Set of events $E$, target program binary $B$, duration $d$
**Output**: Cache Template matrix $T$

Map binary $B$ into memory
**foreach** *event e in E* **do**
    **foreach** *address a in binary B* **do**
        **while** *duration d not passed* **do**
            **simultaneously**
            Trigger event $e$ and save event trace $g_{a,e}^{(E)}$
            Flush+Reload attack on address $a$
                and save cache-hit trace $g_{a,e}^{(H)}$
        **end**
        Extract cache-hit ratio $H_{a,e}$ from $g_{a,e}^{(E)}$
            and $g_{a,e}^{(H)}$ and store it in $T$
    **end**
**end**
Prune Cache Template matrix $T$

---

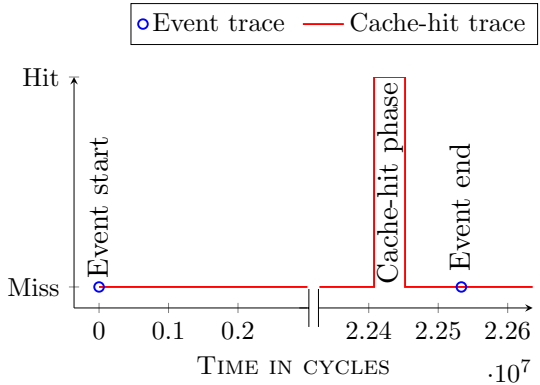

Figure 5.1.: Trace of a single keypress event for address `0x4ebc0` of `libgdk.so`.

measured. The more events triggered within the specified duration $d$, the more accurate the resulting profile is. However, increasing the duration $d$ increases the overall time required for the profiling phase.

The results of this measurement step are a cache-hit trace and an event trace, which are generated for all addresses $a$ in the binary and all events $e$ we want to profile. An excerpt of such a cache-hit trace and the corresponding event trace is shown in Figure 5.1. The start of the event is measured directly before the event is triggered. As we monitor library code, the cache-hit phase is measured before the attacked binary observes the event.

The generation of the traces can be sped up by two factors. First, in case of a cache miss, the CPU always fetches a whole cache line. Thus, we cannot distinguish between offsets of different accesses within a cache line and we can deduce the same information by probing only one address within each cache-line sized memory area.

Second, we reduce the overall number of triggered events by profiling multiple addresses at the same time. However, profiling multiple addresses on the same page can cause prefetching of more data from this page. Therefore, we can only profile addresses on different pages simultaneously. Thus, profiling all pages only takes as long as profiling a single page.

In case of low-frequency events, it is possible to profile all pages within one binary in parallel. However, this may lead to less accurate cache-hit traces $g_{a,e}^{(H)}$, *i.e.*, timing deviations above 1 microsecond from the real event, which is only acceptable for low-frequency events.

**Hit-Ratio Extraction**

After the cache-hit trace and the event trace have been computed for a specific event $e$ and a specific address $a$ (the while loop of Algorithm 1), we derive the cache-hit ratio for each event and address. The cache-hit ratio $H_{a,e}$ is either a simple value or a time-dependent ratio function. In our case it is the ratio of cache hits on address $a$ and the number of times the event $e$ has been triggered within the profiling duration $d$.

To illustrate the difference between a cache-hit ratio with time dependency and without time dependency, we discuss two such functions. The cache-hit ratio with time dependency can be defined as follows. The event traces contain the start and end points of the processing of one event $e$. These start and end points define the relevant parts (denoted as *slices*) within the cache-hit trace. The slices are stored in a vector and scaled to the same length. Each slice contains a cache-hit pattern relative to the event $e$. If we average over this vector, we get the cache-hit ratio function for event $e$.

The second, much simpler approach is to define the cache-hit ratio without time dependency. In this case, we count the number of cache hits $k$ on address $a$ and divide it by the number of times $n$ the event $e$ has been triggered within the profiling duration $d$. That is, we define $H_{a,e} = \frac{k}{n}$. In case of a low-noise side channel and event detection through single cache hits, it is sufficient to use a simple hit-ratio extraction function.

Like the previous step, this step is repeated for all addresses $a$ in the binary $b$ and all events $e$ to be profiled. The result is the full Cache Template matrix $T$. We denote the column vectors $\vec{p}_e$ as *profiles* for specific events.

**Pruning**

In the exploitation phase, we are limited regarding the number of addresses we can attack. Therefore, we want to reduce the number of addresses in the Cache Template. We remove redundant rows from the Cache Template matrix and merge events which cannot be distinguished based on their profiles $\vec{p}_e$.

As cache hits can be independent of an event, the measured cache-hit ratio on a specific address can be independent of the event, *i.e.*, code which is always executed, frequent data accesses by threads running all the time, or code that is never executed and data that is never accessed. In order to be able to detect an event $e$, the set of events has to contain at least one event $e'$ which does not include event $e$. For example, in order to be able to detect the event "user pressed key A" we need to profile at least one event where the user does not press key A.

The pruning happens in three steps on the matrix. First, the removal of all addresses that have a small difference between minimum and maximum cache-hit ratio for all events. Second, merging all similar columns (events) into one set of events, *i.e.*, events that cannot be distinguished from each other are merged into one column. The similarity measure for this is, for example, based on a mean squared error (MSE) function. Third, the removal of redundant lines. These steps ensure that we select the most interesting addresses and also allows us to reduce the attack complexity by reducing the overall number of monitored addresses.

We measure the reliability of a cache-based side channel by true and false positives as well as true and false negatives. Cache hits that coincide with an event are counted as true positive and cache hits that do not coincide with an event as false positive. Cache misses which coincide with an event are counted as true negative and cache misses which do not coincide with an event as false negative. Based on these four values we can determine the accuracy of our Template, for instance, by computing the F-Score, which is defined as the harmonic mean of the cache-hit ratio and the positive predictive value (percentage of true positives of the total cache hits). High F-Score values show that we can distinguish the given event accurately by attacking a specific address. In some cases further lines can be pruned from the Cache Template matrix based on these measures. The true positive rate and the false positive rate for an event $e$ can be determined by the profile $\vec{p}_e$ of $e$ and the average over all profiles except $e$.

**Runtime of the Profiling Phase**

Measuring the cache-hit ratio is the most expensive step in our attack. To quantify the cost we give two examples. In both cases we want to profile a 1 MB library, once for a low-frequency event, e.g., a keypress, and once for a high-frequency event, e.g., an encryption. In both cases, we try to achieve a runtime which is realistic for offline and online attacks while maintaining a high accuracy.

We choose a profiling duration of $d = 0.8$ seconds for the low-frequency event. During 0.8 seconds we can trigger around 200 events, which is enough to create a highly accurate profile. Profiling each address in the library for 0.8 seconds would take 10 days. Profiling only cache-line-aligned addresses still takes 4 hours. Applying both optimizations, the full library is profiled in 17 seconds.

In case of the high-frequency event, we attack an encryption. We assume that one encryption and the corresponding Flush+Reload measurement take 520 cycles on average. As in the previous example, we profile each address 200 times and, thus, we need 40–50 microseconds per address, *i.e.*, $d = 50\mu s$. The basic attack takes less than 55 seconds to profile the full library for one event. Profiling only cache-line-aligned addresses takes less than 1 second and applying both optimizations results in a negligible runtime.

As already mentioned above, the accuracy of the resulting profile depends on how many times an event can be triggered during profiling duration $d$. In both cases we chose durations which are more than sufficient to create accurate profiles and still achieve reasonable execution times for an online attack. Our observations showed that it is necessary to profile each event at least 10 times to get meaningful results. However, profiling an event more than a few hundred times does not increase the accuracy of the profile anymore.

## 3.2. Exploitation Phase

In the exploitation phase we execute a generic spy program which performs either the Flush+Reload or the Prime+Probe algorithm. For all addresses in the Cache Template matrix resulting from the profiling phase, the cache activity is constantly monitored.

We monitor all addresses and record whether a cache hit occurred. This information is stored in a boolean vector $\vec{h}$. To determine which event occurred based on this observation, we compute the similarity $S(\vec{h}, \vec{p}_e)$ between $\vec{h}$ and each profile $\vec{p}_e$ from the Cache Template matrix. The similarity measure $S$ can be based, for example, on a mean squared error (MSE) function. Algorithm 2 summarizes the exploitation phase.

---

**Algorithm 2:** Exploitation phase.

> **Input**: Target program binary $b$,
> Cache Template matrix $T = (\vec{p}_{e_1}, \vec{p}_{e_2}, ..., \vec{p}_{e_n})$
>
> Map binary $b$ into memory
> **repeat**
> > **foreach** *address $a$ in $T$* **do**
> > > Flush+Reload attack on address $a$
> > > Store 0/1 in $\vec{h}[a]$ for cache miss/cache hit
> >
> > **end**
> > **if** *$\vec{p}_e$ equals $\vec{h}$ w.r.t. similarity measure* **then**
> > > Event $e$ detected
> >
> > **end**

---

The exploitation phase has the same requirements as the underlying attack techniques. The attacker needs to be able to execute a spy program on the attacked system. In case of Flush+Reload, the spy program needs no privileges, except opening the attacked program binary in a read-only shared memory. It is even possible to attack binaries running in a different virtual machine on the same physical machine, if the hypervisor has page deduplication enabled. In case of Prime+Probe, the spy program needs no privileges at all and it is even possible to attack binaries running in a different virtual machine on the same physical machine, as shown by Irazoqui et al. [IES15]. However, the Prime+Probe technique is more susceptible to noise and therefore the exploitation phase will produce less reliable results, making attacks on low-frequency events more difficult.

The result of the exploitation phase is a log file containing all detected events and their corresponding timestamps. The interpretation of the log file still has to be done manually by the attacker.

# 4. Attacks on Artificial Applications

Before we actually exploit cache-based vulnerabilities in real applications in Section 5, we demonstrate the basic working principle of Cache Template Attacks on two artificial victim programs. These illustrative attacks show how Cache Template Attacks automatically profile and exploit cache activity in any program. The two attack scenarios we demonstrate are: 1) an attack on lookup tables, and 2) an attack on executed instructions. Hence, our ideal victim program or library either contains a large lookup table which is accessed depending on secret information, e.g., depending on secret lookup indices, or specific portions of program code which are executed based on secret information.

**Attack on Data Accesses**

For demonstration purposes, we spy on simple events like keypresses. In our victim program, shown in Listing 1, each keypress causes a memory access in a large array called `map`. These key-based accesses are 4096 bytes apart from each other to avoid triggering the prefetcher. The array is initialized with static values in order to place it in the data segment and to guarantee that each page contains different data and, thus, is not

```
1 int map[130][1024] = {{-1U},...,{-130U}};
2 int main(int argc, char** argv) {
3   while(1) {
4     int c = getchar(); // unbuffered
5     if (map[(c % 128) + 1][0] == 0)
6       exit(-1);
7 } }
```

Listing 1: Victim program with large array on Linux

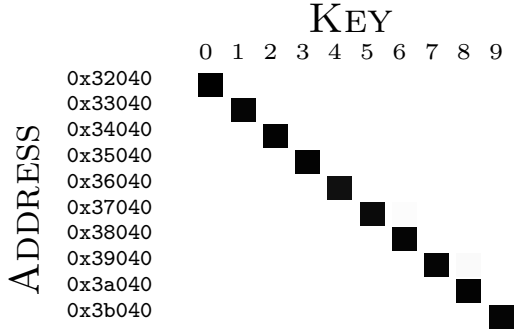

Figure 5.2.: Cache Template matrix for the artificial victim program shown in Listing 1. Dark cells indicate high cache-hit ratios.

deduplicated in any way. It is necessary to place it in the data segment in order to make it shareable with the spy program.

In the profiling phase of the Cache Template Attack, we simulate different keystroke events using the X11 automation library `libxdo`. This library can be linked statically into the spy program, *i.e.*, it does not need to be installed. The Cache Template matrix is generated as described in Section 3. Within a duration of $d = 0.8$ seconds we simulated around 700 keypress events. The resulting Cache Template matrix can be seen in Figure 5.2 for all number keys. We observe cache hits on addresses that are exactly 4 096 bytes apart, which is due to the data type and the dimension of the `map` array. In our measurements, there were less than 0.3% false positive cache hits on the corresponding addresses and less than 2% false negative cache hits. The false positive and false negative cache hits are due to the high key rate in the keypress simulation.

For verification purposes, we executed the generated keylogger for a period of 60 seconds and randomly pressed keys on the keyboard. In this setting we measured no false positives and no false negatives at all. This results

```
1 #define NOP1024 /*1024 times asm("nop");*/
2 #define CASE(X) case X:\
3   { ALIGN(0x1000) void f##X() { NOP1024 };\
4     f##X(); break; }
5 int main(int argc, char** argv) {
6   while (1) {
7     int c = getchar(); // unbuffered
8     switch (c) {
9       CASE(0);
10      // ...
11      CASE(128);
12 } } }
```

Listing 2: Victim program with long functions on Linux

from significantly lower key rates than in the profiling phase. The table is not used by any process other than the spy and the victim process and the probability that the array access happens exactly between the reload and the flush instruction is rather small, as we have longer idle periods than during the profiling phase. Thus, we are able to uniquely identify each key without errors.

**Attack on Instruction Executions**

The same attack can easily be performed on executed instructions. The source code for this example is shown in Listing 2. Each key is now processed in its own function, as defined by the `CASE(X)` macro. The functions are page aligned to avoid prefetcher activity. The `NOP1024` macro generates 1024 `nop` instructions, which is enough to avoid accidental code prefetching of function code.

Our measurements show that there is no difference between Cache Template Attacks on code and data accesses.

**Performance Evaluation**

To examine the performance limits of the exploitation phase of Cache Template Attacks, we evaluated the number of addresses which can be accurately monitored simultaneously at different key rates. At a key rate of 50 keys per second, we managed to spy on 16 000 addresses simultaneously on an Intel i5 Sandy Bridge CPU without any false positives or false negatives. The first errors occurred when monitoring 18 000 addresses simultaneously. At a key rate of 250 keys per second, which is the maximum on our system, we were able to spy on 4 000 addresses simultaneously without any errors. The first errors occurred when monitoring 5 000 addresses simultaneously. In both cases, we monitor significantly more addresses than in any practical cache attack today.

However, monitoring that many addresses is only possible if their position in virtual memory is such that the prefetcher remains inactive. Accessing several consecutive addresses on the same page causes prefetching of more data, resulting in cache hits although no program accessed the data. The limiting effect of the prefetcher on the Flush+Reload attack has already been observed by Yarom and Benger [YB14]. Based on these observations, we discuss the possibility of using the prefetcher as an effective countermeasure against cache attacks in Section 6.3.

# 5. Attacks on Real-World Applications

In this section, we consider an attack scenario where an attacker is able to execute an attack tool on a targeted machine in unprivileged mode. By executing this attack tool, the attacker extracts the cache-activity profiles which are exploited subsequently. Afterwards, the attacker collects the secret information acquired during the exploitation phase.

For this rather realistic and powerful scenario we present various case studies of attacks launched against real applications. We demonstrate the power of automatically launching cache attacks against any binary or library. First, we launch two attacks on Linux user interfaces, including GDK-based user interfaces, and an attack against a Windows user interface. In all attacks we simulate the user input in the profiling phase. Thus, the attack can be automated on the device under attack. To demonstrate the range of possible applications, we also present an automated attack on the T-table-based AES implementation of OpenSSL 1.0.2 [Ope14].

## 5.1. Attack on Linux User Interfaces

There exists a variety of software-based side-channel attacks on user input data. These attacks either measure differences in the execution time of code in other programs or libraries [Tan+08], approximate keypresses through CPU and cache activity [Ris+09], or exploit system services leaking user input data [ZW09]. In particular, Zhang et al. [ZW09] use information about other processes from `procfs` on Linux to measure inter-keystroke timings and derive key sequences. Their proposed countermeasures can be implemented with low costs and prevent their attack completely. We, however, employ Cache Template Attacks to find and exploit leaking side-channel information in shared libraries automatically in order to spy on keyboard input.

Given root access to the system, it is trivial to write a keylogger on Linux using `/dev/input/event*` devices. Furthermore, the `xinput` tool can also be used to write a keylogger on Linux, but root access is required to install it. However, using our approach of Cache Template Attacks only requires the unprivileged execution of untrusted code as well as the capability of opening the attacked binaries or shared libraries in a read-only shared memory. In the exploitation phase one round of Flush+Reload on a single address takes less than 100 nanoseconds. If we measure the average latency between keypress and cache hit, we can determine the actual keypress timing up to a few hundred nanoseconds. Compared to the existing attacks mentioned above, our attack is significantly more accurate in terms of both event detection (detection rates near 100%) and timing deviations.

In all attacks presented in this section we compute time-independent cache-hit ratios.

### Attack on the GDK Library

Launching the Cache Template profiling phase on different Linux applications revealed thousands of addresses in different libraries, binaries, and data files showing cache activity upon keypresses. Subsequently, we targeted different keypress events in order to find addresses distinguishing the different keys. Figure 5.3 shows the Cache Template of a memory area in the GDK library `libgdk-3.so.0.1000.8`, a part of the GTK framework which is the default user-interface framework on many Linux distributions.

Figure 5.3 shows several addresses that yield a cache hit with a high accuracy if and only if a certain key is pressed. For instance, every keypress on key `n` results in cache hit on address `0x7c800`, whereas the same address reacts in only 0.5% of our tests on other keypresses. Furthermore, we found a high cache-hit ratio on some addresses when a key is pressed

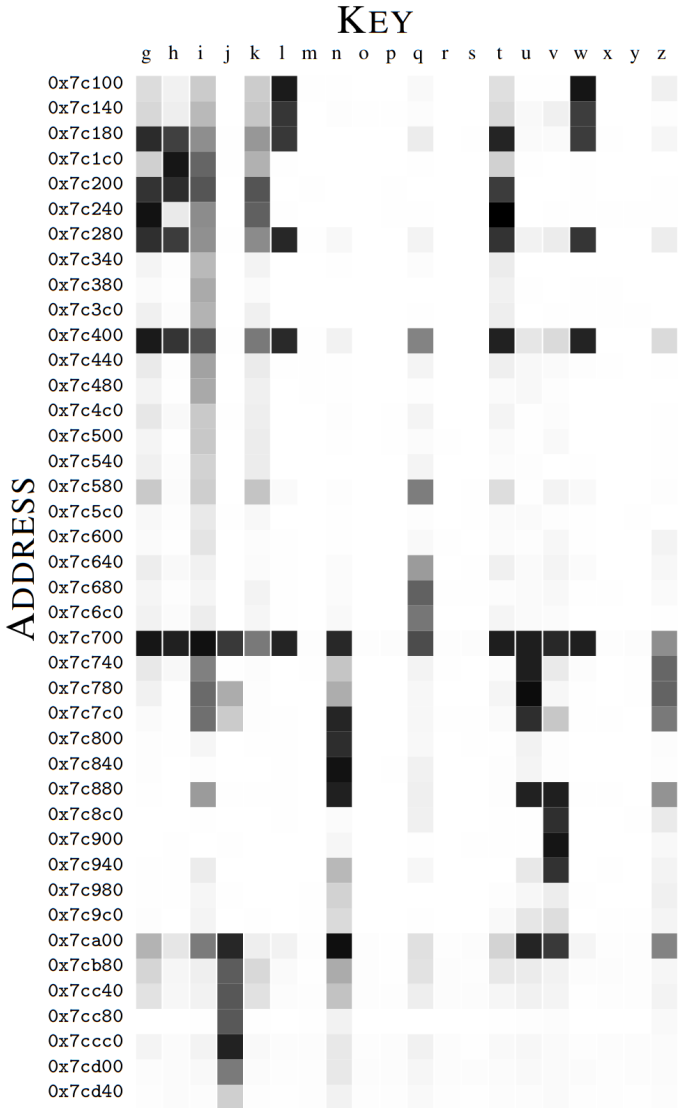

Figure 5.3.: Excerpt of the GDK Cache Template. Dark cells indicate key-address-pairs with high cache-hit ratios.

(*i.e.*, `0x6cd00` in `libgdk`), the mouse is moved (*i.e.*, `0x28760` in `libgdk`) or a modifier key is pressed (*i.e.*, `0x72fc0` in `libgdk`). We also profiled the range of keys a–f but it is omitted from Figure 5.3 because no high cache-hit ratios have been observed for the shown addresses.

We use the spy tool described in Section 3.2 in order to spy on events based on the Cache Template. We are able to accurately determine the following sets of pressed keys: $\{i\},\{j\},\{n\},\{q\},\{v\},\{l,w\},\{u,z\},\{g,h,k,t\}$. That is, we cannot distinguish between keys in the same set, but keys in one set from keys in other sets. Similarly, we can deduce whether a key is contained in none of these sets.

Not as part of our attack, but in order to understand how keyboard input is processed in the GDK library, we analyzed the binary and the source code. In general, we found out that most of the addresses revealed in the profiling phase point to code executed while processing keyboard input. The address range discussed in this section contains the array `gdk_keysym_to_unicode_tab` which is used to translate key symbols to unicode special characters. The library performs a binary search on this array, which explains why we can identify certain keys accurately, namely the leaf nodes in the binary search.

As the corresponding array is used for keyboard input in all GDK user-interface components, including password fields, our spy tool works for all applications that use the GDK library. This observation allows us to use Cache Template Attacks to build powerful keyloggers for GDK-based user interfaces automatically. Even if we cannot distinguish all keys from each other, Cache Template Attacks allow us to significantly reduce the complexity of cracking a password. In this scenario, we are able to identify 3 keys reliably, as well as the total number of keypresses. Thus, in case of a lower-case password we can reduce the entropy per character from $\log_2(26) = 4.7$ to 4.0 bits. Attacking more than 3 addresses in order to identify more keys adds a significant amount of noise to the results, as it triggers the prefetcher. First experiments demonstrated the feasibility of attacking the lock screen of Linux distributions. However, further evaluation is necessary in order to reliably determine the effectiveness of this approach.

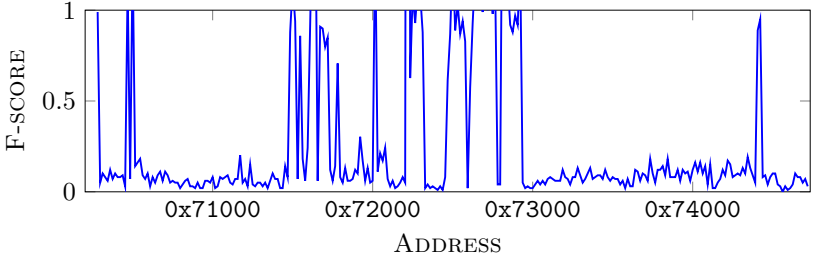

Figure 5.4.: Excerpt of the F-score plot for the address range of the `gdk_keys_by_keyval` table. High values reveal addresses that can be exploited.

**Attack on GDK Key Remapping**

If an attacker has additional knowledge about the attacked system or software, more efficient and more powerful attacks are possible. Inspired by Tannous et al. [Tan+08] who performed a timing attack on GDK key remapping, we demonstrate a more powerful attack on the GDK library, by examining how the remapping of keys influences the sets of identifiable keypresses. The remapping functionality uses a large key-translation table `gdk_keys_by_keyval` which spreads over more than four pages.

Hence, we repeated the Cache Template Attack on the GDK library with a small modification. Before measuring cache activity for an address during an event, we remapped one key to the key code at that address, retrieved from the `gdk_keys_by_keyval` table. We found significant cache activity for some address and key-remapping combinations.

When profiling each key remapping for $d = 0.8$ seconds, we measured cache activity in 52 cache-line-sized memory regions. In verification scans, we found 0.2-2.5% false positive cache hits in these memory regions. Thus, we have found another highly accurate side channel for specific key remappings. The results are shown in the F-score graph in Figure 5.4. High values allow accurate detection of keypresses if the key is remapped to this address. Thus, we find more accurate results in terms of timing in our automated attack than Tannous et al. [Tan+08].

We can only attack 8 addresses in the profiled memory area simultaneously, since it spreads over 4 pages and we can only monitor 2 or 3 addresses without triggering the prefetcher. Thus, we are able to remap any 8 keys to these addresses and reliably distinguish them. In combination with the 3 addresses of our previous results, we are able to distinguish at least 11 keys and observe the timestamp of any keystroke in the system based on cache accesses simultaneously.

It is also possible to remap more than one key to the same key code. Hence, it is possible to distinguish between groups of keys. If we consider a lower-case password again, we can now reduce the entropy per character from $\log_2(26) = 4.7$ to 1.4 bits.

We also profiled keypresses on capslock and shift. Although we were able to log keypresses on both keys, we did not consider upper case or mixed case input. The exploitation phase automatically generates a log file containing the information observed through the cache side channel. However, interpretation of these results, such as deriving a program state from a sequence of events (shift key pressed or capslock active) and the influence of the program state on subsequent events is up to analysis of the results after the attack has been performed.

Tannous et al. [Tan+08] also described a login-detection mechanism in order to avoid remapping keys unless the user types in a password field. The spy program simply watches `/proc` to see whether a login program is running. Then the keys are remapped. As soon as the user pauses, the original key mappings are restored. The user will then notice a password mismatch, but the next password entry will work as expected.

Our completely automated password keylogger is a single binary which runs on the attacked system. It maps the GDK library into its own address space and performs the profiling phase. The profiling of each keypress requires the simulation of the keypress into a hidden window. Furthermore, some events require the key remapping we just described. Finally, the keylogger switches into the exploit mode. As soon as a logon screen is detected, for instance, after the screensaver was active or the screen was locked, the keys are remapped and all keypresses are logged into a file accessible by the attacker. Thus, all steps from the deployment of the keylogger to the final log file are fully automated.

## 5.2. Attacks on other Linux Applications

We also found leakage of accurate keypress timings in other libraries, such as the `ncurses` library (*i.e.*, offset `0xbf90` in `libncurses.so`), and in files used to cache generated data related to user text input, such as

`/usr/lib/locale/locale-archive`. The latter one is used to translate keypresses into the current locale. It is a generated file which differs on each system and which changes more frequently than the attacked libraries. In consequence, it is not possible to perform an offline attack, *i.e.*, to use a pre-generated Cache Template in the exploitation phase on another system. Still, our concept of Cache Template Attacks allows us to perform an online attack, as profiling is fully automated by generating keystrokes through `libxdo` or comparable libraries. Thus, keystroke side channels are found within a few seconds of profiling. All keypress-timing side channels we found have a high accuracy and a timing deviation of less than 1 microsecond to the actual keypress.

In order to demonstrate Cache Template Attacks on a low-frequency event which is only indirectly connected to keypresses, we attacked `sshd`, trying to detect when input is sent over an active ssh connection. The received characters are unrelated to the local user input. When profiling for a duration of $d = 0.8$ seconds per address, we found 428 addresses showing cache activity when a character was received. We verified these results for some addresses manually. None of these checked addresses showed false positive hits within a verification period of 60 seconds. Thus, by exploiting the resulting Cache Template matrix, we are able to gain accurate timings for the transmitted characters (significantly less than 1 microsecond deviation to the transmission of the character). These timings can be used to derive the transmitted letters as shown by Zhang et al. [ZW09].

## 5.3. Attack on Windows User Interfaces

We also performed Cache Template Attacks on Windows applications. The attack works on Windows using MinGW identically to Linux. Even the implementation is the same, except for the keystroke simulation which is now performed using the Windows API instead of the `libxdo` library, and the file under attack is mapped using `LoadLibrary` instead of `mmap`. We performed our attack on Windows 7 and Windows 8.1 systems with the same results on three different platforms, namely Intel Core 2 Duo, Intel i5 Sandy Bridge, and Intel i5 Ivy Bridge. As in the attacks on Linux user interfaces, address space layout randomization has been activated during both profiling and exploitation phase.

In an automated attack, we found cache activity upon keypresses in different libraries with reasonable accuracy. For instance, the Windows 7 common control library `comctl32.dll` can be used to detect keypresses on different addresses. Probing `0xc5c40` results in cache hits on every keypress and mouse click within text fields accurately. Running the generated keypress logger in a verification period of 60 seconds with keyboard input by a real user, we found only a single false positive event detection based on this address. Address `0xc6c00` reacts only on keypresses and not on mouse clicks, but yields more false positive cache hits in general. Again, we can apply the attack proposed by Zhang et al. [ZW09] to recover typed words from inter-keystroke timings.

We did not disassemble the shared library and therefore do not know which function or data accesses cause the cache hit. The addresses were found by starting the Cache Template Attack with the same parameters as on Linux, but on a Windows shared library instead of a Linux shared library. As modern operating systems like Windows 7 and Windows 8.1 employ an immense number of shared libraries, we profiled only a few of these libraries. Hence, further investigations might even reveal addresses for a more accurate identification of keypresses.

## 5.4. Attack on a T-table-based AES

Cache attacks have been shown to enable powerful attacks against cryptographic implementations. Thus, appropriate countermeasures have already been suggested for the case of AES [Gue10; KS09; RSD06; Kön08]. Nevertheless, in order to compare the presented approach of Cache Template Attacks to related attacks, we launched an efficient and automated access-driven attack against the AES T-table implementation of OpenSSL 1.0.2, which is known to be insecure and susceptible to cache attacks [WHS12; Ber04; OST06; GBK11; Bog+10; Ira+14a; AK06; Ira+14b]. Recall that the T-tables are accessed according to the plaintext $p$ and the secret key $k$, *i.e.*, $T_j[p_i \oplus k_i]$ with $i \equiv j \mod 4$ and $0 \leq i < 16$, during the first round of the AES encryption. For the sake of brevity, we omit the full details of an access-driven cache attack against AES and refer the interested reader to the work of Osvik et al. [OST06; TOS10].

**Attack of Encryption Events**

In a first step, we profiled the two events "no encryption" and "encryption with random key and random plaintext". We profiled each cache-line-aligned address in the OpenSSL library during 100 encryptions. On our test system, one encryption takes around 320 cycles, which is very fast compared to a latency of at least 200 cycles caused by a single cache miss. In order to make the results more deterministically reproducible, we measure whether a cache line was used only after the encryption has finished. Thus, the profiling phase does not run in parallel and only one cache hit or miss is measured per triggered event.

This profiling step takes less than 200 seconds. We detected cache activity on 0.2%-0.3% of the addresses. Only 82 addresses showed a significant difference in cache activity depending on the event. For 18 of these addresses, the cache-hit ratio was 100% for the encryption event. Thus, our generated spy tool is able to accurately detect whenever an encryption is performed.

For the remaining 64 addresses the cache-hit ratio was around 92% for the encryption event. Thus, not each of these addresses is accessed in every encryption, depending on key and plaintext. Since we attack a T-table-based AES implementation, we know that these 64 addresses must be the T-tables, which occupy 4 KB respectively 64 cache lines. Although this information is not used in the first generated spy tool, it encourages performing a second attack to target specific key-byte values.

**Attack on Specific Key-Byte Values**

Exploiting the knowledge that we attack a T-table implementation, we enhance the attack by profiling over different key-byte values for a fixed plaintext, *i.e.*, the set of events consists of the different key-byte values. Our attack remains fully automated, as we change only the values with which the encryption is performed. The result is again a log file containing the accurate timestamp of each event monitored. The interpretation of the log file, of course, involves manual work and is specific to the targeted events, *i.e.*, key bytes in this case.

For each key byte $k_i$, we profile only the upper 4 bits of $k_i$ as the lower 4 bits cannot be distinguished because of the cache-line size of 64 bytes. This means that we need to profile only 16 addresses for each key byte $k_i$.

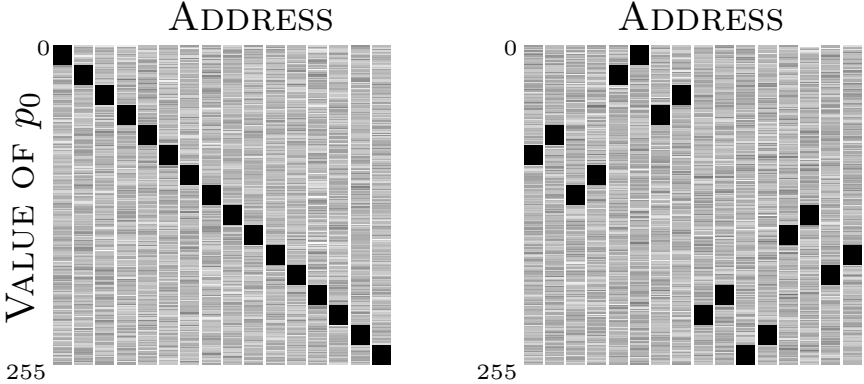

Figure 5.5.: Excerpt of the Cache Template (address range of the first T-table). The plot is transposed to match [OST06]. In the left trace $k_0 = \texttt{0x00}$, in the right trace $k_0 = \texttt{0x51}$.

Furthermore, on average 92% of these addresses are already in the cache and the Reload step of the Flush+Reload attack is unlikely to trigger the prefetcher. Thus, we can probe all addresses after a single encryption. Two profiles for different values of $k_0$ are shown in Figure 5.5. The two traces were generated using 1 000 encryptions per key byte and address to show the pattern more clearly. According to Osvik et al. [OST06] and Spreitzer et al. [SP13] these plots (or patterns) reveal at least the upper 4 bits of a key byte and, hence, attacking the AES T-table implementation works as expected. In our case, experiments showed that 1 to 10 encryptions per key byte are enough to infer these upper 4 bits correctly.

In a T-table-based AES implementation, the index of the T-table is determined by $p_i \oplus k_i$. Therefore, the same profiles can be generated by iterating over the different plaintext byte values while encrypting with a fixed key. Osvik et al. [OST06] show a similar plot, generated using the Evict+Time attack. However, in our attack the profiles are aggregated into the Cache Template matrix, as described in Section 3.1.

In the exploitation phase, the automatically generated spy tool monitors cache hits on the addresses from the Cache Template in order to determine secret key-byte values. We perform encryptions using chosen plaintexts. We attack the 16 key bytes $k_i$ sequentially. In each step $i = 0, \ldots, 15$, the plaintext is random, except for the upper 4 bits of $p_i$, which are fixed to the same chosen value as in the profiling phase. Hence, the encryption is performed over a chosen plaintext. The spy tool triggers an encryption, detects when the encryption actually happens and after each encryption, reports the set of possible values for the upper 4 bits of key byte $k_i$. As soon as only one candidate for the upper 4 bits of key byte $k_i$ remains, we continue with the next key byte.

Using Cache Template Attacks, we are able to infer 64 bits of the secret key with only 16–160 encryptions in a chosen-plaintext attack. Compared to the work of Osvik et al. [OST06] who require several hundred or thousands encryptions (depending on the measurement approach) targeting the L1 cache, and the work of Spreitzer and Plos [SP13] who require millions of encryptions targeting the L1 cache on the ARM platform, we clearly observe a significant performance improvement. More recent work shows that full key recovery is possible with less than 30 000 encryptions [Gül+15] using Flush+Reload.

The benefit of our approach, compared to existing cache attacks against AES, is that our attack is fully automated. Once the binary is deployed on the target system, it performs both profiling and exploitation phase automatically and finally returns a log file containing the key byte candidates to the attacker. Moreover, we do not need prior knowledge of the attacked system or the attacked executable or library.

AES T-table implementations are already known to be insecure and countermeasures have already been integrated, e.g., in the AES implementation of OpenSSL. Performing our attack on a non-T-table implementation (e.g., by employing AES-NI instructions) did not show key dependent information leakage, but still, we can accurately determine the start and end of the encryption through the cache behavior. However, we leave it as an interesting open issue to employ the presented approach of cache template attacks for further investigations of vulnerabilities in already protected implementations.

**Trace-Driven Attack on AES**

When attacking an insecure implementation of a cryptographic algorithm, an attacker can often gain significantly more information if it is possible to perform measurements during the encryption [AK06; GKT10], *i.e.*, in case the exact trace of cache hits and cache misses can be observed. Even if we cannot increase the frequency of the Flush+Reload attack, we are able to slow down the encryption by constantly flushing the 18 addresses which showed cache activity in every profile. We managed to increase the encryption time from 320 cycles to 16 000–20 000 cycles. Thus, a more fine-grained trace of cache hits and cache misses can be obtained which might even allow the implementation of trace-driven cache attacks purely in software.

# 6. Countermeasures

We have demonstrated in Section 5 that Cache Template Attacks are applicable to real-world applications without knowledge of the system or the application. Therefore, we emphasize the need for research on effective countermeasures against cache attacks. In Section 6.1, we discuss several countermeasures which have been proposed so far. Subsequently, in Section 6.2, we discuss how Cache Template Attacks can be employed by developers to detect and eliminate cache-based information leakage and also by users to detect and prevent cache attacks running actively on a system. Finally, in Section 6.3, we propose changes to the prefetcher to build a powerful countermeasure against cache attacks.

## 6.1. Discussion of Countermeasures

### Removal of the clflush Instruction is not Effective

The restriction of the `clflush` instruction has been suggested as a possible countermeasure against cache attacks in [YB14; YF14; Zha+14]. However, by adapting our spy tool to evict the cache line without using the `clflush` instruction (Evict+Reload instead of Flush+Reload), we demonstrate that this countermeasure is not effective at all. Thereby, we show that cache attacks can be launched successfully even without the `clflush` instruction.

Instead of using the `clflush` instruction, the eviction is done by accessing physically congruent addresses in a large array which is placed in large pages by the operating system. In order to compute physically congruent addresses we need to determine the lowest 18 bits of the physical address to attack, which can then be used to evict specific cache sets.

The actual mapping of virtual to physical addresses can be retrieved from `/proc/self/pagemap`. Even if such a mapping is not available, methods to find congruent addresses have been developed—simultaneously to this work—by Irazoqui et al. [IES15] by exploiting large pages, Oren et al.

[Ore+15] by exploiting timing differences in JavaScript, and Liu et al. [Liu+15] by exploiting timing differences in native code.

The removal of the `clflush` instruction has also been discussed as a countermeasure to protect against DRAM disturbance errors (denoted as rowhammer bug). These disturbance errors have been studied by Kim et al. [Kim+14] and, later on, exploited by Seaborn et al. [SD15] to gain kernel privileges. Several researchers have already claimed to be able to exploit the rowhammer bug without the `clflush` instruction [Goo15], This can be done by exploiting the Sandy Bridge cache mapping function, which has been reverse engineered by Hund et al. [HWH13], to find congruent addresses.

Our eviction strategy only uses the lowest 18 bits and therefore, we need more than 12 accesses to evict a cache line. With 48 accessed addresses, we measured an eviction rate close to 100%. For performance reasons we use write accesses, as the CPU does not have to wait for data fetches from the physical memory. In contrast to the `clflush` instruction, which takes only 41 cycles, our eviction function takes 325 cycles. This is still fast enough for most Flush+Reload attacks.

While `clflush` always evicts the cache line, our eviction rate is only near 100%. Therefore, false positive cache hits occur if the line has not been evicted. Using Flush+Reload, there is a rather low probability for a memory access on the monitored address to happen exactly between the Reload step and the point where the `clflush` takes effect. This probability is much higher in the case of Evict+Reload, as the eviction step takes 8 times longer than the `clflush` instruction.

We compare the accuracy of Evict+Reload to Flush+Reload using previously found cache vulnerabilities. For instance, as described in Section 5.1, probing address `0x7c800` of `libgdk-3.so.0.1000.8` allows us to detect keypresses on key `n`. The Flush+Reload spy tool detects on average 98% of the keypresses on key `n` with a 2% false positive rate (keypresses on other keys). Using Evict+Reload, we still detect 90% of the keypresses on key `n` with a 5% false positive rate. This clearly shows that the restriction of `clflush` is not sufficient to prevent this type of cache attack.

**Disable Cache-Line Sharing**

One prerequisite of Flush+Reload attacks is shared memory. In cloud scenarios, shared memory across virtual machine borders is established through page deduplication. Page deduplication between virtual machines is commonly disabled in order to prevent more coarse-grained attacks like fingerprinting operating systems and files [Suz+11; OW11] as well as Flush+Reload. Still, as shown by Irazoqui et al. [IES15], it is possible to use Prime+Probe as a fallback. However, attacking low-frequency events like keypresses becomes infeasible, because Prime+Probe is significantly more susceptible to noise.

Flush+Reload can also be prevented on a system by preventing cache-line sharing, *i.e.*, by disabling shared memory. Unfortunately, operating systems make heavy use of shared memory, and without modifying the operating system it is not possible for a user program to prevent its own memory from being shared with an attacker, even in the case of static linkage as discussed in Section 2.2.

With operating-system modifications, it would be possible to disable shared memory in all cases where a victim program cannot prevent an attack, *i.e.*, shared program binaries, shared libraries, shared generated files (for instance, `locale-archive`). Furthermore, it would be possible to provide a system call to user programs to mark memory as "do-not-share."

A hardware-based approach is to change cache tags. Virtually tagged caches are either invalidated on context switches or the virtual tag is combined with an address space identifier. Therefore, shared memory is not shared in the cache. Thus, Flush+Reload is not possible on virtually tagged caches.

We emphasize that as long as shared cache lines are available to an attacker, Flush+Reload or Evict+Reload cannot be prevented completely.

**Cache Set Associativity**

Prime+Probe, Evict+Time and Evict+Reload exploit set-associative caches. In all three cases, it is necessary to fill all ways of a cache set, either for eviction or for the detection of evicted cache sets. Based on which cache set was reloaded (respectively evicted), secret information is deduced. Fully associative caches have better security properties, as such information deduction is not possible and cache eviction can only be enforced by filling the whole cache. However, a timing attack would still be possible, e.g., due to internal cache collisions [Bog+10] leading to different execution times. As fully associative caches are impractical for larger caches, new cache architectures have been proposed to provide similar security properties [WL07; Kon+08; WL08]. However, even fully associative caches only prevent attacks which do not exploit cache-line sharing. Thus, a combination of countermeasures is necessary to prevent most types of cache attacks.

## 6.2. Proactive Prevention of Cache Attacks

Instrumenting cache attacks to detect co-residency [Zha+11] with another virtual machine on the same physical machine, or even to detect cache attacks [Zha+14] and cache-based side channels in general [Doy+13] has already been proposed in the past. Moreover, Brumley and Hakala [BH09] even suggested that developers should use their attack technique to detect and eliminate cache vulnerabilities in their programs. Inspired by these works, we present defense mechanisms against cache attacks which can be improved by using Cache Template Attacks.

### Detect Cache Vulnerabilities as a Developer

Similar to Brumley and Hakala [BH09], we propose the employment of Cache Template Attacks to find cache-based vulnerabilities automatically. Compared to [BH09], Cache Template Attacks allow developers to detect potential cache side channels for specifically chosen events automatically, which can subsequently be fixed by the developer. A developer only needs to select the targeted events (e.g., keystrokes, window switches, or encryptions) and to trigger these events automatically during the profiling phase, which significantly eases the evaluation of cache side channels. Ultimately, our approach even allows developers to find such cache vulnerabilities in third party libraries.

### Detect and Impede Ongoing Attacks as a User

Zhang et al. [Zha+14] stated the possibility to detect cache attacks by performing a cache attack on one of the vulnerable addresses or cache sets. We propose running a Cache Template Attack as a system service to detect code and data under attack. If Flush+Reload prevention is sufficient, we simply disable page sharing for all pages with cache lines under attack. Otherwise, we disable caching for these pages as proposed by Aciiçmez et al. [ABG10] and, thus, prevent all cache attacks. Only the performance for critical code and data parts is reduced, as the cache is only disabled for specific pages in virtual memory.

Furthermore, cache attacks can be impeded by performing additional memory accesses, unrelated to the secret information, or random cache flushes. Such obfuscation methods on the attacker's measurements have already been proposed by Zhang et al. [ZR13]. The idea of the proposed obfuscation technique is to generate random memory accesses, denoted as cache cleansing. However, it does not address the shared last-level cache. In contrast, Cache Template Attacks can be used to identify possible cache-based information leaks and then to specifically add noise to these specific locations by accessing or flushing the corresponding cache lines.

## 6.3. Enhancing the Prefetcher

During our experiments, we found that the prefetcher influences the cache activity of certain access patterns during cache attacks, especially due to the spatial locality of addresses, as also observed in other work [OST06; YB14; GBK11]. However, we want to discuss the prefetcher in more detail as it is crucial for the success of a cache attack.

Although the profiling phase of Cache Template Attacks is not restricted by the prefetcher, the spy program performing the exploitation phase might be unable to probe all leaking addresses simultaneously. For instance, we found 255 addresses leaking side-channel information about keypresses in the GDK library but we were only able to probe 8 of them simultaneously in the exploitation phase, because the prefetcher loads multiple cache lines in advance and, thus, generates numerous false positive cache hits.

According to the Intel 64 and IA-32 Architectures Optimization Reference Manual [Int12], the prefetcher loads multiple memory addresses in advance if "two cache misses occur in the last level cache" and the corresponding memory accesses are within a specific range (the so-called trigger distance). Depending on the CPU model this range is either 256 or 512 bytes, but does not exceed a page boundary of 4 KB. Due to this, we are able to probe at least 2 addresses per page.

We suggest increasing the trigger distance of the prefetcher beyond the 4 KB page boundary if the corresponding page already exists in the translation lookaside buffer. The granularity of the attack will then be too high for many practical targets, especially attacks on executed instructions will then be prevented.

As cache attacks constantly reaccess specific memory locations, another suggestion is to adapt the prefetcher to take temporal spatiality into consideration. If the prefetcher were to prefetch data based on that temporal distance, most existing attacks would be prevented.

Just as we did in Section 4, an attacker might still be able to establish a communication channel targeted to circumvent the prefetcher. However, the presented countermeasures would prevent most cache attacks targeting real-world applications.

# 7. Conclusion

In this paper, we introduced Cache Template Attacks, a novel technique to find and exploit cache-based side channels easily. Although specific knowledge of the attacked machine and executed programs or libraries helps, it is not required for a successful attack. The attack is performed on closed-source and open-source binaries in exactly the same way.

We studied various applications of Cache Template Attacks. Our results show that an attacker is able to infer highly accurate keystroke timings on Linux as well as Windows. For Linux distributions we even demonstrated a fully automatic keylogger that significantly reduces the entropy of passwords. Hence, we conclude that cache-based side-channel attacks are an even greater threat for today's computer architectures than assumed so far. In fact, even sensitive user input, like passwords, cannot be considered secure on machines employing CPU caches.

We argue that fundamental concepts of computer architectures and operating systems enable the automatic exploitation of cache-based vulnerabilities. We observed that many of the existing countermeasures do not prevent such attacks as expected. Still, the combination of multiple countermeasures can effectively mitigate cache attacks. However, the fact that cache attacks can be launched automatically marks a change of perspective, from a more academic interest towards practical attacks, which can be launched by less sophisticated attackers. This shift emphasizes the need to develop and integrate effective countermeasures immediately. In particular, it is not sufficient to protect only specific cryptographic algorithms like AES. More general countermeasures will be necessary to counter the threat of automated cache attacks.

# 8. Acknowledgments

We would like to thank the anonymous reviewers and our shepherd, Ben Ransford, for their valuable comments and suggestions. The research 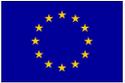 leading to these results has received funding from the European Union's Horizon 2020 research and innovation programme under grant agreement No 644052 (HECTOR). Furthermore, this work has been supported by the Austrian Research Promotion Agency (FFG) and the Styrian Business Promotion Agency (SFG) under grant number 836628 (SeCoS).

# 6

# Practical Memory Deduplication Attacks in Sandboxed Javascript

## Publication Data

D. Gruss, D. Bidner, and S. Mangard. Practical Memory Deduplication Attacks in Sandboxed JavaScript. In: 20th European Symposium on Research in Computer Security (ESORICS'15). 2015

## Contributions

Main author.

# Practical Memory Deduplication Attacks
# in Sandboxed Javascript

Daniel Gruss, David Bidner, and Stefan Mangard

Graz University of Technology, Austria

## Abstract

Page deduplication is a mechanism to reduce the memory footprint of a system. Identical physical pages are identified across borders of virtual machines and programs and merged by the operating system or the hypervisor. However, this enables side-channel information leakage through cache or memory access time. Therefore, it is considered harmful in public clouds today, but it is still considered safe to use in a private environment, *i.e.*, private clouds, personal computers, and smartphones.

We present the first memory-disclosure attack in sandboxed JavaScript which exploits page deduplication. Unlike previous attacks, our attack does not require the victim to execute an adversary's program, but simply to open a website which contains the adversary's JavaScript code. We are not only able to determine which applications are running, but also specific user activities, for instance, whether the user has specific websites currently opened. The attack works on servers, personal computers and smartphones, and across the borders of virtual machines.

## 1. Introduction

Software-based timing attacks are side-channel attacks which exploit differences in the execution time to derive secret values used during the computation. These timing differences arise from the attacked software itself, different memory types or optimizations implemented in modern computers. For instance, cache attacks exploit the timing difference between a cache access and a memory access caused by a cache miss. An attacker process can measure whether a victim process has evicted one of the attacker's cache lines [OST06] or whether a victim program has reloaded a cache line the attacker previously evicted from the cache [GBK11; YF14].

A similar timing difference can be observed between a regular memory access and a pagefault. Upon a pagefault, the operating system loads the data to the given location in virtual memory and returns control to the process. Apart from the difference in the memory access time, pagefault handling is transparent to the user process. This timing difference can be exploited to build a covert channel [War+96].

Suzaki et al. [Suz+11] presented page-deduplication attacks, which enable an attacker on the same physical machine, to determine whether specific programs are running, even across the borders of virtual machines. This is possible, because identical physical pages are merged by the operating system or the hypervisor. After a page is merged write accesses to this page cause a pagefault which is then resolved by the operating system. The timing difference the pagefault causes can be observed by the attacker program. Thus, the attacker learns that somewhere on the same physical machine another instance of this page exists.

JavaScript-based timing attacks have first been described by Felten et al. [FS00]. They were able to identify recently visited websites if website elements are fetched from the local browser cache instead of the network. A similar attack has been presented by Bortz et al. [BB07]. More recently Stone [Sto13] presented attacks which exploit timing differences caused by the modification of HTML5 elements. Using their approach, an attacker is able to determine whether specific websites have been visited, and even read pixels from other websites.

In this paper, we present the first page-deduplication attack mounted in sandboxed JavaScript. This allows a remote attacker to collect private information, such as whether a program or website is currently opened by a user. In contrast to existing JavaScript-based timing attacks, we do not exploit any weaknesses in JavaScript or the browser, but timing differences caused by optimizations in the operating system or hypervisor.

JavaScript is a scripting language implemented in modern browsers to create interactive elements on websites. It is strictly sandboxed, so it is not possible to access files or system services. The language has no representation of pointers or the virtual address space layout, and less accurate timing information than native code. Oren et al. [Ore+15] already demonstrated in a JavaScript-based cache attack that timer accuracy is high enough to distinguish cache hits from cache misses. Our attack is possible with less accurate timers, on a microsecond or millisecond basis.

To demonstrate the power of our attack, we show that we can accurately determine whether the user has opened specific websites. Our attack can be applied in a generic way to any system which employs page deduplication, independently of the CPU architecture and in particular independently of the CPU cache structure. This is a significant share of modern personal computers and smartphones.

With our attack, an adversary is not only able to perform the attack remotely through a website, on an arbitrary number of victims, but an adversary is also able to attack a variety of different devices in the same way. Thus, page-deduplication attacks no longer target one specific system, but instead target large numbers of internet users simultaneously. For instance, a website can detect which other websites a user has opened and thereby add more valuable information to user profiles. Furthermore, the attack causes negligible CPU and memory utilization and is thus, hard to detect if placed in a large JavaScript framework.

We show that page deduplication must be considered a security threat on any system and not only on public cloud servers. Therefore, we conclude that the only effective countermeasure is to disable page deduplication.

**Outline**

The remaining paper is organized as follows. In Section 2, we provide background information on shared memory and page deduplication, as well as existing attacks. We describe the implementation of our attack in Section 3. In Section 4.1, we present the performance of our attack in a private cloud and in Section 4.2, we present results of our attack on personal computers and smartphones. We discuss countermeasures against page-deduplication attacks in Section 5. Finally, we conclude in Section 6.

# 2. Background

## 2.1. Shared Memory

Operating systems and hypervisors use shared memory to reduce physical memory utilization. Libraries which are used by several programs are loaded into physical memory only once, and are then shared among the processes using it. Thus, multiple programs access the same physical pages mapped within their own virtual address space.

The operating system makes use of shared memory in more cases. When forking a process, the memory is first shared between the parent process and the child process. As soon as one of the processes writes into the shared memory area, a copy-on-write page fault occurs and the operating system creates a copy of the according memory region. Note that write accesses into non-shared memory areas do not incur page faults and thus are significantly faster.

Shared memory is not only used when forking a process, but when starting instances of an already running program, or if a user program explicitly requests shared memory using system calls like `mmap` or `dlopen`. Mapping a file using one of these methods results in a memory region shared with all other processes mapping the same file.

The form of shared memory we target in this paper is content-based page deduplication. The hypervisor or operating system scans the physical memory for pages with identical content. If identical pages are found, they are remapped to one of the pages, while the other pages are marked as free. Thus, memory is shared between completely unrelated and possibly sandboxed processes, and even between processes running in different virtual machines. If a process modifies its shared data, a copy-on-write page fault occurs and the hypervisor or operating system creates a copy of the memory region. Although searching for identical pages costs CPU time, page deduplication can increase the system performance, by reducing the number of block device accesses, as more data can be held in memory. Therefore, it is especially relevant in small systems like smartphones, besides the primary application in cloud systems.

## 2.2. Page-Deduplication Attacks

Page-deduplication attacks are a specific type of side-channel attacks, which exploit timing differences in write accesses on deduplicated pages. The first attack on page deduplication was presented by Suzaki et al. [Suz+11]. They were able to determine whether specific applications are running in a co-located virtual machine in the cloud. Furthermore, they described the possibility of building covert communication channels between virtual machines by exploiting page deduplication.

In the basic attack scheme, an attacker is able to run a spy program on the victim's system. However, the spy program may be sandboxed or even run in a virtual machine. The spy program fills a page with data it suspects to find in the memory of the victim machine. The hypervisor or operating system constantly deduplicates identical physical pages. When the spy program tries to write to the page again, it can measure the elapsed time and infer whether a copy-on-write page fault occurs or not. Thus, the attacker can determine whether some other process on the same physical machine has an identical page in memory. Such attacks can be performed on both, binary code and static data as well as dynamically generated data.

Owens et al. [OW11] demonstrated that it is possible to efficiently fingerprint operating systems in co-located virtual machines by exploiting page deduplication. Since then, covert channels based on page deduplication [Xia+13; Xia+12] have been constructed and evaluated.

At the same time, researchers were able to build more efficient cache attacks if attacker and victim process share memory [GBK11]. Page deduplication introduces a way to share memory with a victim process in a co-located virtual machine in the cloud. The possibility of performing a cache attack on a victim process across virtual machine borders has first been described by Yarom et al. [YF14]. Since then, several page-deduplication-based cache attacks have been demonstrated [Ira+14a; Ira+14b].

# 3. Description of our JavaScript-based Attack

Our attack follows the same methodology as the page-deduplication attack presented by Suzaki et al. [Suz+11], which was implemented in native code. As our attack is implemented in JavaScript, we face several new challenges, such as setting the content of a whole page in physical memory or detecting whether and when page deduplication has occurred.

As described in Section 2.2, the first step of a page-deduplication attack is to fill a page with data we expect to find on the system under attack. In native code, this done by filling a page-aligned region in an array with the according data. We found that JavaScript engines in common browsers (Firefox and Chrome) perform a call to their own internal `malloc` implementation when creating a large array in JavaScript. As a means of optimization, these `malloc` implementations align large memory allocations to page borders. Therefore, creating and filling a large array in JavaScript works as in native code, in terms of our attack.

The second step is to wait until the operating system or hypervisor deduplicates our array. In our attacker model, the adversary performs the attack through a website on every visitor. Therefore, we cannot make assumptions about how long it takes until page deduplication has been performed. Instead, we repeatedly write the same value to the same position on the target page and measure the time the write access took. We observed no influence of these repeated writes on whether the page is considered for deduplication. Thus, we can perform the deduplication check in a regular frequency.

The third step is the measurement of the write-access time, to infer whether a page has been deduplicated. This is done by measuring the time a write access on our own page takes. Based on the access time, we decide whether a copy-on-write page fault occurred. In native x86 code, we use the `rdtsc` assembly instruction for this purpose. In JavaScript, we can use the function `performance.now()`. The accuracy of this function varies from sub-microsecond to millisecond range. If checking for deduplication of a single page in memory, our attack requires accurate microsecond measurements. However, usually more pages are attacked and thus less accurate timers are sufficient. For instance, when checking for deduplication of a 600 kilobyte image, even an accurate millisecond-based timer can be used to implement our attack. Thus, `performance.now()` is sufficient to distinguish copy-on-write page faults from regular write accesses. Furthermore, `performance.now()` is available independently of the underlying hardware. Therefore, we can attack systems with a variety of different processors using the same JavaScript code, such as personal computers or smartphones.

The only remaining question to perform our attack is how to know the data we want to fill the page with. Neither static code and data nor dynamically generated data is necessarily page-aligned. However, if the attacker knows the content of 8192 bytes contiguous in virtual memory, we can fill 4096 pages with data from these 8192 bytes, with every possible offset from the page alignment. Although this allows us to attack systems and programs with random offsets for the targeted data, we found that this is hardly necessary for most cases. For instance, we observed that images and CSS style sheets in websites are page-aligned in memory. This greatly facilitates our attack, as we can trivially extract the page content from a file and include it in our JavaScript code.

The resulting attack applies to a wide range of scenarios, from mobile phone usage, over personal computers, to multi-tenant cloud systems. A user on a targeted system accesses a website, which contains the adversary's JavaScript code. The JavaScript code is then executed. After a few minutes, the JavaScript code transmits the results back to the adversary. Our attack not only extracts sensitive information, like the browsing behavior of a user, but it is also extremely powerful due to its scalability. Once the JavaScript code is deployed on a website, it automatically attacks anyone who accesses the website. We will demonstrate the attack in different scenarios in the following section.

# 4. Practical Attacks and Evaluation

In this section, we demonstrate our attack on a KVM-based private cloud server, on Windows 8 personal computers and finally on Android smartphones. In all scenarios, we use the same JavaScript source code.

## 4.1. Cross-VM Attack on Private Clouds

Existing page-deduplication attacks have been demonstrated on public IaaS (Infrastructure-as-a-Service) cloud systems [Suz+11; Xia+13; Xia+12]. In this attack scenario, an adversary tries to be co-located on the same physical server with a targeted virtual machine. Once the adversary is co-located, the adversary extracts sensitive information from other virtual machines, e.g., whether vulnerable versions of specific server applications are running, or whether specific files are currently open.

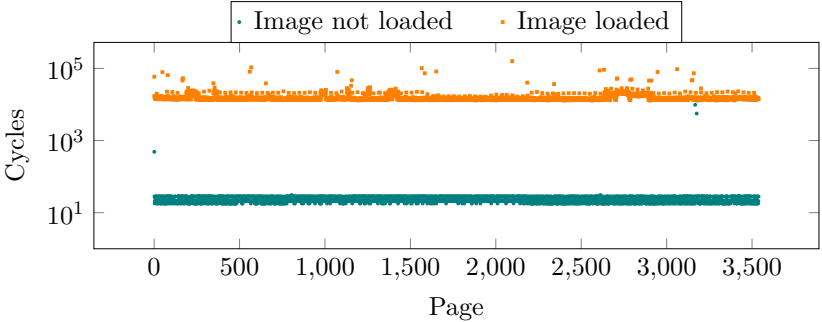

Figure 6.1.: Timings measured in native code on an otherwise idle Linux KVM virtual machine. The graph shows write-access times on an array containing an image file.

Although public cloud providers reacted and now disable page deduplication in public IaaS clouds [IES14], we found that page deduplication is not yet considered a security problem on private cloud systems and servers. Popular Linux server distributions enable page deduplication, either by default, or automatically when reaching a certain memory usage level. For instance, we observed this behavior on Proxmox VE, Redhat Server and Ubuntu Server if configured as a KVM host.

Therefore, we demonstrate our attack in a private IaaS cloud. This is a realistic scenario, for instance in companies where users work on thin clients, connected to a virtual machine in the private IaaS cloud. In this scenario, a victim working in one virtual machine opens a website containing the malicious JavaScript code, which is then continuously run in the background in a browser tab. Compared to existing attacks, our attack is possible even if the system does not allow users to start arbitrary programs, or if the user is well-educated to avoid executing programs from an untrusted origin. Furthermore, we want to emphasize that our attack is doubly sandboxed in this scenario, by running in the JavaScript sandbox in the virtual machine separated from the targeted program in another virtual machine. That is, the adversary is able to extract sensitive information from the victim's virtual machine and other virtual machines on the same server.

The malicious JavaScript code has to stay in memory until page deduplication has been performed. Depending on the system configuration, this can be between 30 seconds and several hours. During our tests with 4 gigabytes of physical memory and the default system configuration, we found that our memory is deduplicated after 3 minutes.

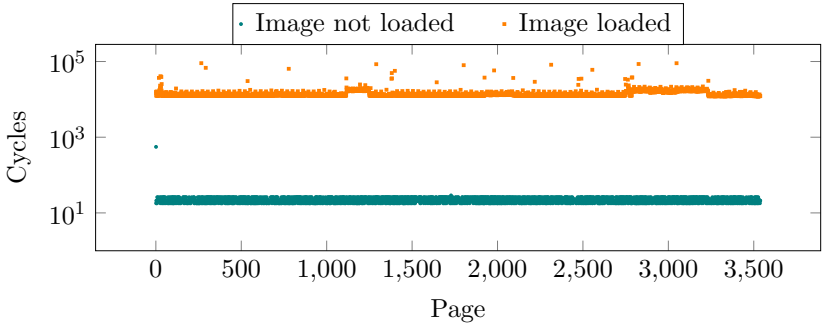

Figure 6.2.: Timings measured in native code on a Linux KVM virtual machine under high CPU load. The graph shows write-access times on an array containing an image file.

In order to evaluate the accuracy of our attack in JavaScript code, we first perform the same attack in native x86 code. Figure 6.1 and Figure 6.2 show write-access times on an array containing a 14 megabyte image file as measured by our native-code spy program within the same virtual machine, with low and high system load. This is equivalent to loading 3584 small images (2–4 kilobytes) and measuring the deduplication of each of them. These write-access times quantify the accuracy of our page deduplication detection. When the image is loaded, we found no measurements to be lower than the expected copy-on-write access time. When having the image not loaded in the browser, we found less than $0.1\,\%$ of the measurements to be significantly above the expected regular write-access time. These $0.1\,\%$ can lead to false positive copy-on-write detection. However, as there is a timing difference of at least a factor of $10^3$, we found an even smaller number of peaks to be above the lowest copy-on-write access times. We subsequently tested our attack using native code in the cross-VM setting and achieved the same accuracy. Therefore, we can accurately determine whether an image has been deduplicated and thus, has been loaded by a user.

Subsequently, we measured the performance of our JavaScript-based attack. In Figure 6.3 and Figure 6.4, the write-access times on an array containing the same 14 megabyte image file are shown, but this time measured by our JavaScript spy program. Even in with full system load and the browser under attack running in a different virtual machine, page deduplication was detected correctly in all of our measurements. However, in contrast to the native-code implementation of our spy program, we found up to 0.3 % of the pages to be falsely detected as deduplicated when having low system load and 1.1 % on average when having a high system load.

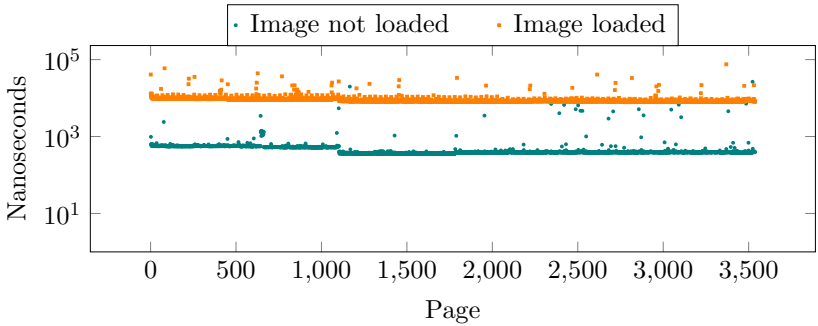

Figure 6.3.: Timings measured in JavaScript on an otherwise idle Linux KVM virtual machine. The graph shows write-access times on an array containing an image file.

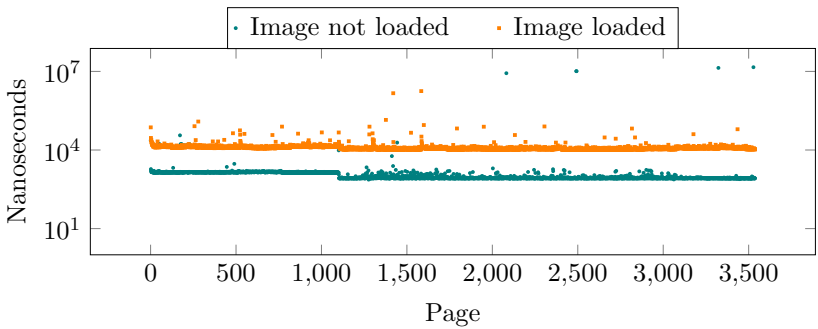

Figure 6.4.: Timings measured in JavaScript on a Linux KVM virtual machine under high CPU load. The graph shows write-access times on an array containing an image file.

We performed this attack on recent versions of the most commonly used browsers, Chrome 40 and Firefox 31. As both browsers load the image file to a page-aligned location in memory, the attack works in exactly the same way and gives the same results for both browsers. Furthermore, we performed the same attack on a browser in a different virtual machine. Even in this setting, we did not find more false positives, and all deduplicated pages were detected successfully.

In order to demonstrate our attack on a real-world scenario, we determine the websites currently opened by a user. In this scenario, the adversary creates arrays containing image data of the websites to detect on the targeted machine. For demonstration purposes, we examined the 10 most-visited websites [Ale15] and chose an image or style sheet file from each website, to determine whether it is currently open in a web browser on the same machine. Furthermore, we generate several pages filled with zeros and several pages filled with random data, to measure reference timings for deduplicated and non-deduplicated pages. When the operating system or hypervisor has tried to deduplicate our pages, the zero-filled pages will have high write-access times, as they are deduplicated. The random-filled pages still have low write-access times, as each random-filled page is unique in the system and therefore not deduplicated. Some websites only contain very small or very few images. In these cases we combine several images to perform the attack more reliably. In all cases we had at least 24 kilobytes of data to measure deduplication.

Figure 6.5 shows the write-access times to arrays containing image data from these websites, as well as the zero-filled pages and random-filled pages. We can clearly see which websites are currently opened in the browser, because of the higher write-access times, due to the copy-on-write page-fault handling. Based on such measurements, an adversary is able to spy on users' browsing behavior through malicious JavaScript code, even across browsers and virtual machine borders.

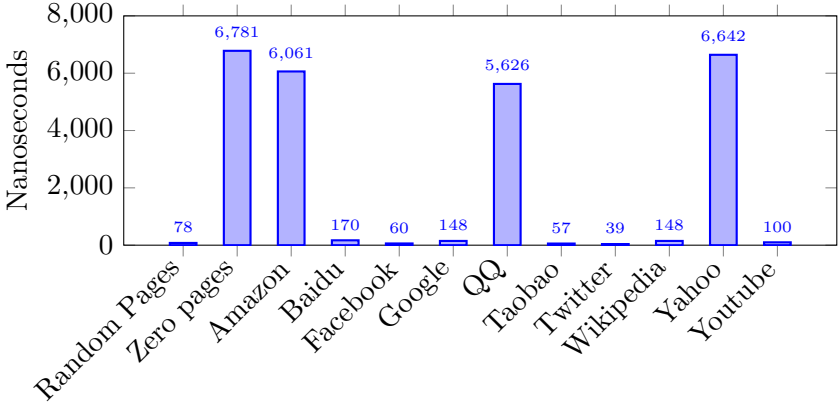

Figure 6.5.: Write-access times measured in JavaScript inside a Linux KVM virtual machine, for images from frequented websites as well as random-filled an zero-filled pages. We measured high access times only for the currently opened websites: Amazon, QQ, Taobao, Wikipedia and Yahoo.

## 4.2. Attack on Personal Computers and Smartphones

Our attack is even more precise if performed on a personal computer or smartphone, as the device under attack is only used by a single user at a time. Therefore, we can create accurate profiles of single users. As in the cross-VM attack, the victim merely needs to access a website containing the malicious JavaScript code.

This scenario is not only very simple and realistic, but moreover, it has a huge impact, as it can be applied to popular operating systems like Windows 8 on personal computers, or Android on smartphones. Windows 8 and 8.1 have a market share of around 15 % [Net15] on personal computers and have page deduplication enabled by default [KS11]. Android has a market share of 81.5 % [Int15] on smartphones, but it is device-specific whether page deduplication is enabled by default or not. However, Google recommends [Goo15] that manufacturers enable page deduplication by default on memory-constrained devices, and many manufacturers follow this recommendation. Therefore, we assume that the number of smartphones having page deduplication enabled, and thus vulnerable to this attack, is significant.

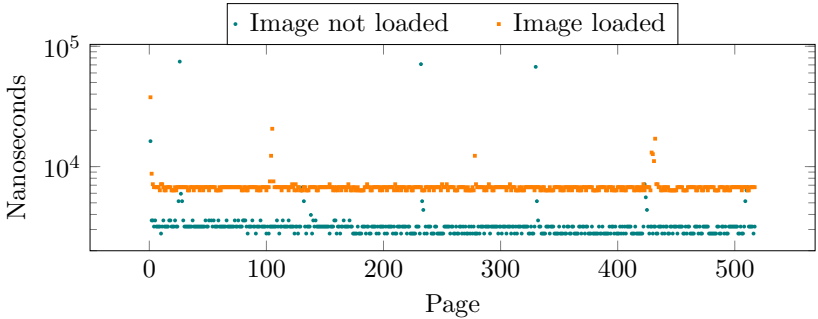

Figure 6.6.: Timings measured in JavaScript on Windows 8.1. The graph shows write-access times on an array containing an image file.

In our attack, the malicious JavaScript code runs continuously in the background in a browser tab. We found that on our Windows 8.1 test machine, page deduplication has been performed after 15 minutes on average. On our Android 4.4.4 test device, page deduplication has been performed after 45 minutes on average. As in the cloud scenario, we can then detect running applications and which specific version of an application is running, or even detect which specific websites are opened by a user. To evaluate the side channel, we again measure the deduplication detection rate for an image loaded in a browser. As we encountered problems with browsers on smartphones loading the 14 megabyte image file, we now use a 2 megabyte image file. This is equivalent to performing the same test with 512 small images (2–4 kilobytes).

Figure 6.6 shows our JavaScript-based measurements for the image file, using Firefox 36 on Windows. We can detect page deduplication almost as reliably as in the private-cloud scenario, with less than 2 % false positives. We are able to perform the attack without changes in Internet Explorer 11 and Firefox 36 on Windows, as both return micro- or nanosecond accurate timings via the `window.performance.now()` function. However, Chrome 41 on Windows only allows measuring time in milliseconds. Thus, we cannot measure the timing difference for each single page. Instead, we have to measure the time over a large number of pages at once. When measuring time over 150 write accesses at once, we are able to distinguish whether these 150 pages were deduplicated or not, with only millisecond timer accuracy.

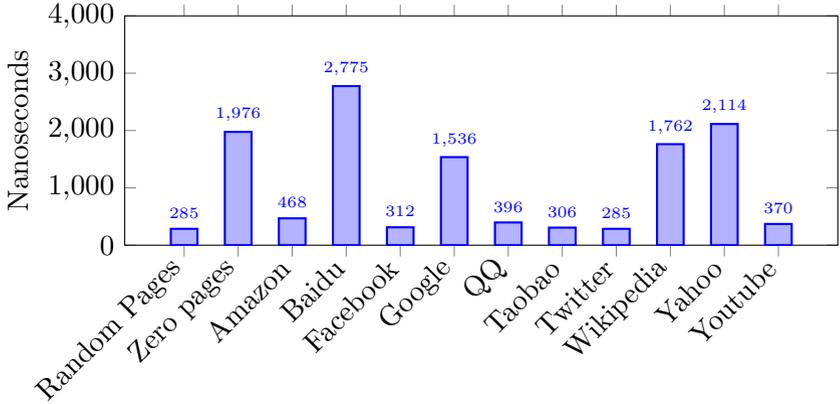

Figure 6.7.: Write-access times measured in JavaScript on Windows 8.1, for images from frequented websites as well as random an zero-filled pages. We measured high access times only for the currently opened websites: Baidu, Google, Wikipedia and Yahoo.

When targeting website usage as a real-world scenario, the adversary creates arrays containing image data of the websites to detect on the targeted machine. As in the private-cloud scenario, we examined the 10 most-visited websites [Ale15]. Figure 6.7 shows the write-access times to arrays containing image data from these websites, as well as the zero-filled pages and random-filled pages. Again, we clearly see which websites are opened, based on the higher write-access time.

When attacking Android smartphones, we found that although it takes up to one hour until deduplication is performed, the accuracy is not much worse than in the other scenarios we tested. We measured up to 0.8 % of false positives when having the image file not loaded in a browser and up to 0.5 % false negatives when having the image file loaded in a browser. This is slightly less accurate than in the other scenarios. Figure 6.8 shows the timing difference with and without the image loaded by a browser. Again, we examined the 10 most-visited websites [Ale15]. Figure 6.9 shows the write-access times to arrays containing image data from these websites, as well as the zero-filled pages and random-filled pages. As in all other scenarios, we also see on Android which websites are opened, based on the higher write-access time.

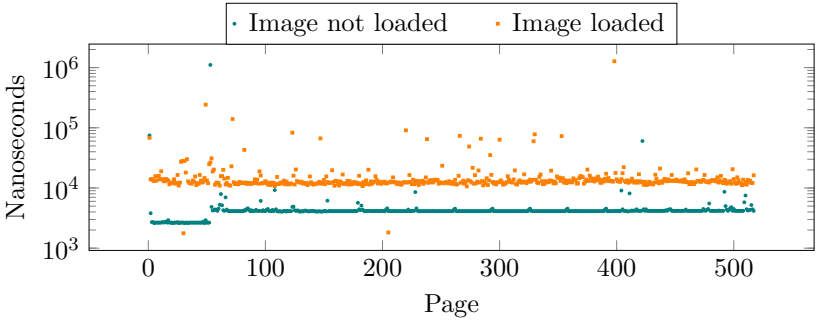

Figure 6.8.: Timings measured in JavaScript on Android 4.4.4. The graph shows write-access times on an array containing an image file.

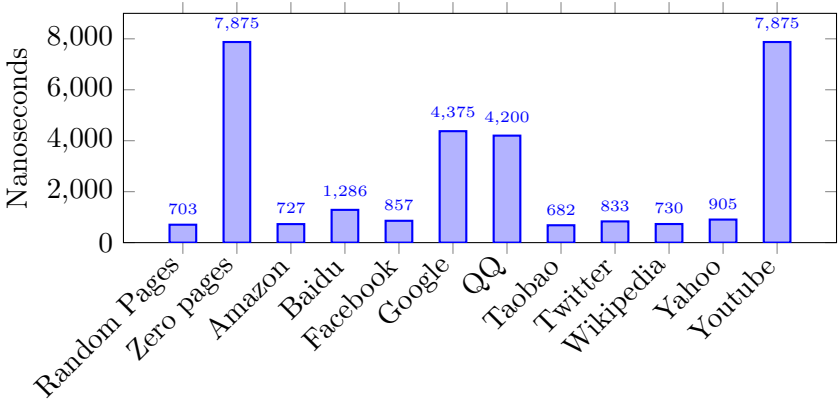

Figure 6.9.: Write-access times measured in JavaScript on Android 4.4.4, for images from frequented websites as well as random an zero-filled pages. We measured high access times only for the currently opened websites: Google, QQ and Youtube.

# 5. Countermeasures

Our attack shows that even in sandboxed JavaScript code, an adversary is able to extract significant sensitive information from real-world applications if the underlying system employs page deduplication. Our specific attack can be prevented on application level, *i.e.*, in the browser executing the adversary's code, or in the applications under attack. However, countermeasures on this level incur limitation of functionality. Disabling page deduplication is the only generic effective countermeasure against page-deduplication attacks.

It is possible prevent or at least weaken our specific attack in JavaScript runtime environments by changing the way data is stored in memory, reducing the accuracy of timers, or disabling JavaScript execution for untrusted code completely.

Our attack benefits from the fact that we are able to allocate page-sized physically contiguous memory areas. Thus, we are able to define the value of each byte on a physical page. JavaScript engines could prevent this by adding small offsets to array indices, so that a few bytes per page cannot be controlled by the attacker. Consequently, the attacker-controlled memory will not be deduplicated. This would cause a small performance impact while impeding page-deduplication attacks in JavaScript.

Another optimization we exploit is page alignment of large data, like images, as performed by modern web browsers. However, adding a random offset to the page alignment would not prevent our attack. The adversary can create 4096 copies of a targeted page, and thereby perform the same attack with only a small overhead. Furthermore, such a countermeasure would require manual modification of existing software, and would incur a performance penalty at the same time.

Oren et al. [Ore+15] suggested reducing the accuracy of JavaScript timers as a countermeasure against JavaScript-based cache attacks. However, a reduced timer accuracy would not prevent our attack. It is easily possible to measure the timing over a large number of pages and thereby invoke several copy-on-write page faults, resulting in timing differences in a millisecond range, which can be detected even with coarse-grained timers.

Our attack could also be prevented by disabling the execution of untrusted JavaScript, *i.e.*, disable JavaScript on websites completely. However, this imposes a significant drawback on functionality of modern browsers and websites. In any case, the attack is still possible if implemented in a browser plugin or smartphone application, where JavaScript-level countermeasures do not apply.

However, we think that any form of content-based page deduplication implies a security problem. As writable pages can be generated in any script language, sandboxed or not, and furthermore, we only require coarse-grained timer accuracy, we consider it insecure to perform page deduplication on writable pages. Considering only read-only pages has already been suggested by Suzaki et al. [Suz+11] as a countermeasure. Apparently, this countermeasure has not been implemented on the systems we attacked. We assume that one of the reasons is that the hypervisor or operating system is not able to distinguish between read-only pages and writable pages within virtual machines, one of the core applications of page deduplication.

However, not considering writeable pages would prevent page-deduplication attacks in JavaScript or other script languages which do not support read-only data. Still, even in case that only read-only pages are merged, an attack could still be possible through browser plugins or smartphone applications on code and static data of targeted binaries, as they are able to load read-only pages or even execute native code. Thus, disabling page deduplication completely is the only way to effectively prevent page-deduplication attacks as presented in this paper.

# 6. Conclusion

In this paper, we presented the first page-deduplication attack in sandboxed JavaScript. In particular, the attack can be launched from any website. We show how the attack can be used to determine whether specific images or websites are currently opened by a user. We demonstrated the attack on private clouds, personal computers and smartphones. In all scenarios, it is even possible to mount the attack across the borders of virtual machines. Thus, we conclude that page deduplication must always be considered vulnerable to attacks as presented in this paper. Systems which have page deduplication enabled cannot be considered secure anymore.

The fact that page-deduplication attacks can be launched through websites marks a paradigm shift, from a targeted attack on a specific system towards large-scale practical attacks launched on a huge number of devices simultaneously. Therefore, we strongly recommend to disable page deduplication.

# 7. Acknowledgments

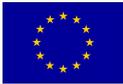 The research leading to these results has received funding from the European Union's Horizon 2020 research and innovation programme under grant agreement No 644052 (HECTOR). Furthermore, this work has been supported by the Austrian Research Promotion Agency (FFG) and the Styrian Business Promotion Agency (SFG) under grant number 836628 (SeCoS).

# 7

# Rowhammer.js: A Remote Software-Induced Fault Attack in JavaScript

## Publication Data

## Contributions

Main author.

# Rowhammer.js: A Remote Software-Induced Fault Attack in JavaScript

Daniel Gruss, Clémentine Maurice[†], and Stefan Mangard

Graz University of Technology, Austria

## Abstract

A fundamental assumption in software security is that a memory location can only be modified by processes that may write to this memory location. However, a recent study has shown that parasitic effects in DRAM can change the content of a memory cell without accessing it, but by accessing other memory locations in a high frequency. This so-called Rowhammer bug occurs in most of today's memory modules and has fatal consequences for the security of all affected systems, e.g., privilege escalation attacks.

All studies and attacks related to Rowhammer so far rely on the availability of a cache flush instruction in order to cause accesses to DRAM modules at a sufficiently high frequency. We overcome this limitation by defeating complex cache replacement policies. We show that caches can be forced into fast cache eviction to trigger the Rowhammer bug with only regular memory accesses. This allows to trigger the Rowhammer bug in highly restricted and even scripting environments.

We demonstrate a fully automated attack that requires nothing but a website with JavaScript to trigger faults on remote hardware. Thereby we can gain unrestricted access to systems of website visitors. We show that the attack works on off-the-shelf systems. Existing countermeasures fail to protect against this new Rowhammer attack.

## 1. Introduction

Hardware-fault attacks have been a security threat since the first attacks in 1997 by Boneh et al. [BDL97] and Biham et al. [BS97]. Fault attacks

---

typically require physical access to the device to expose it to physical conditions which are outside the specification. This includes high or low temperature, radiation, as well as laser on dismantled microchips. However, software-induced hardware faults are also possible, if the device can be brought to the border or out of the specified operation conditions using software. Kim et al. [Kim+14] showed that frequently accessing specific memory locations can cause random bit flips in DRAM chips. 85% of the DDR3 modules they examined are vulnerable. The number of bit flips varies from one module to another, *i.e.*, some modules can be more vulnerable than others. More recently, DDR4 modules have been found to be vulnerable as well [Pes+15]. Bit flips can be triggered by software by flushing a memory location from the cache and reloading it. Seaborn [SD15] demonstrated that an attacker can exploit such bit flips for privilege escalation. These exploits are written in native code and use special instructions to flush data from the cache.

We show that it is possible to trigger hardware faults by performing fast cache eviction on all architectures, if the DRAM modules are vulnerable. Compared to previous work, we do not use any specific instruction, but only regular memory accesses to evict data from the cache. The attack technique is thus generic and can be applied to any architecture, programming language and runtime environment that allows producing a fast stream of memory accesses. Therefore, proposed countermeasures such as removing the `clflush` instruction cannot prevent attacks. Even more severe, we show that on vulnerable modules, we can also perform remote JavaScript-based Rowhammer attacks.

Since an attack through a website can be performed on millions of victim machines simultaneously and stealthily, it poses an enormous security threat. Rowhammer.js is independent of the instruction set of the CPU. It is the first remote software-induced hardware-fault attack. As a proof of concept, we implemented a JavaScript version that as of today runs in all recent versions of Firefox and Google Chrome.

For a Rowhammer attack in JavaScript we perform the following steps:

1. Find 2 addresses in different rows
2. Evict and reload the 2 addresses in a high frequency
3. Search for an exploitable bit flip
4. Exploit the bit flip (e.g., manipulate page tables, remote code execution)

| Platform | CPU | Architecture | RAM |
|----------|-----|--------------|-----|
| Lenovo T420 | i5-2540M | Sandy Bridge | Corsair DDR3-1333 8 GB and Samsung DDR3-1600 4 GB (2×) |
| Lenovo x230 | i5-3320M | Ivy Bridge | Samsung DDR3-1600 4 GB (2×) |
| Asus H97-Pro | i7-4790 | Haswell | Kingston DDR3-1600 8 GB |
| ASRock Z170 ITX | i7-6700K | Skylake | G.Skill DDR4-3200 8 GB (2×) and Crucial DDR4-2133 8 GB (2×) |

Table 7.1.: Experimental setups.

Steps 3 and 4 have already been solved in previous work [SD15], but step 1 and 2 remain open challenges.

The challenge in step 1 is to retrieve information on the physical addresses from JavaScript. It is strictly sandboxed and provides no possibility to retrieve virtual or physical addresses. To tackle this challenge, we determine parts of the physical addresses using large arrays that are allocated by operating systems on large pages. We thus do not exploit any weaknesses in JavaScript or the browser, but only OS-level optimizations.

The challenge in step 2 is to find fast cache eviction strategies to replace the `clflush` instruction. On older CPUs, simply accessing $n+1$ addresses is sufficient to evict lines for an $n$-way cache [Liu+15; Ore+15]. On Intel CPUs produced in the last 4 years, *i.e.*, post Sandy Bridge, the replacement policy has changed and is undocumented. Consequently, known eviction strategies have a low eviction rate or a high execution time, which is not suitable for Rowhammer attacks. To tackle this challenge, we present a novel generic method for finding cache eviction strategies that achieve the best performance in both timing and eviction rate by comprehensively exploring the parameter space. We present the best eviction strategies so far, outperforming previous ones on all recent Intel architectures. Based on this method, we build a two-phase online attack for remote systems with unknown hardware configuration.

We compare the different implementations of the Rowhammer attacks on a fixed set of configurations (see Table 7.1), some vulnerable in default settings, others at decreased refresh rates.

As of today, software countermeasures against Rowhammer native code attacks only target specific exploits, and, as we show, do not protect sufficiently against attacks from JavaScript. Hardware countermeasures are harder to deploy, since they do not affect legacy hardware including recent vulnerable DDR4 modules. BIOS updates can be used to solve the problem on commodity systems, however it is only a practical solution for very advanced users.

Summarizing, our key contributions are:

- We provide the first comprehensive exploration of the cache eviction parameter space on all recent Intel CPUs. This also benefits broader domains, e.g., cache attacks, cache-oblivious algorithms, cache replacement policies.

- We build a native code implementation of the Rowhammer attack that only uses memory accesses. The attack is successful on Sandy Bridge, Ivy Bridge, Haswell and Skylake, in various DDR3 and DDR4 configurations.

- We build a pure JavaScript Rowhammer implementation, showing that an attacker can trigger Rowhammer bit flips remotely, through a web browser.

The remainder of this paper is organized as follows. In Section 2, we provide background information on DRAM, the Rowhammer bug, CPU caches, and cache attacks. In Section 3, we describe a two-phase automated attack to trigger bit flips on unknown systems. In Section 4, we demonstrate the Rowhammer bug without `clflush` in native code and in JavaScript. In Section 5, we provide a discussion of our proof-of-concept exploit, limitations, and countermeasures. Finally, we discuss future work in Section 6 and provide conclusions in Section 7.

# 2. Background

## 2.1. DRAM

Modern memory systems have multiple *channels* of DRAM memory connected to the memory controller. A channel consists of multiple *Dual Inline Memory Modules* (*DIMMs*), that are the physical modules on the motherboard. Each DIMM has one or two *ranks*, that are the sides of the physical module. Each rank is a collection of *chips*, that are further composed of *banks*. Accesses to different banks can be served concurrently. Each bank is an array of capacitor cells that are either in a charged or discharged state, representing a binary data value. The bank is represented as a collection of rows, typically $2^{14}$ to $2^{17}$.

The charge from the cells is read into a *row buffer* on request and written back to the cells as soon as another row is requested. Thus, access to the DRAM is done in three steps: 1. opening a row, 2. accessing the data in the row buffer, 3. closing the row before opening a new row, writing data back to the cells.

DRAM is volatile memory and discharges over time. The *refresh interval* defines when the cell charge is read and restored to sustain the value. DDR3 and DDR4 specifications require refreshing all rows at least once within 64ms [Kim+14; Aic15a].

The selection of channel, rank, bank and row is based on physical address bits. The mapping for Intel CPUs has recently been reverse engineered [Sea15; Pes+15].

## 2.2. The Rowhammer Bug

The increase of DRAM density has led to physically smaller cells, thus capable of storing smaller charges. As a result, cells have a lower noise margin, and cells can interact electrically with each other although they should be isolated. The so called *Rowhammer bug* consists in the corruption of data, not in rows that are directly accessed, but rather in rows nearby the accessed one.

DRAM and CPU manufacturers have known the Rowhammer bug since at least 2012 [BH14; Bai+14]. Hammering DRAM chips is a quality assurance tests applied to modules [AlA05]. As refreshing DRAM cells consumes time, DRAM manufacturers optimize the refresh rate to the lowest rate that still works reliably.

The Rowhammer bug has recently been studied [Kim+14; Hua+12; Par+14] and the majority of off-the-shelf DRAM modules has been found vulnerable to bit flips using the `clflush` instruction. The `clflush` instruction flushes data from the cache, forcing the CPU to serve the next memory access from DRAM. Their proof-of-concept implementation frequently accesses and flushes two memory locations in a loop, causing bit flips in a third memory location.

Seaborn implemented Rowhammer exploits [SD15] in native code with the `clflush` instruction: a privilege escalation on a Linux system caused by a bit flip in a page table and an escape from the Google Native Client sandbox caused by a bit flip in indirect jumps. As a countermeasure, the

| | | Address Bit | | | | | | | | | | | | | | | | | | | | | | | | | |
|---|---|---|---|---|---|---|---|---|---|---|---|---|---|---|---|---|---|---|---|---|---|---|---|---|---|---|---|---|
| | | 3 2 | 3 1 | 3 0 | 2 9 | 2 8 | 2 7 | 2 6 | 2 5 | 2 4 | 2 3 | 2 2 | 2 1 | 2 0 | 1 9 | 1 8 | 1 7 | 1 6 | 1 5 | 1 4 | 1 3 | 1 2 | 1 1 | 1 0 | 0 9 | 0 8 | 0 7 | 0 6 |
| 2 cores | $o_0$ | ⊕ | | ⊕ | | ⊕ | ⊕ | ⊕ | ⊕ | ⊕ | | ⊕ | | | | ⊕ | ⊕ | ⊕ | | ⊕ | | ⊕ | | | | | | ⊕ |
| 4 cores | $o_0$ | ⊕ | | ⊕ | | ⊕ | ⊕ | ⊕ | ⊕ | | ⊕ | | ⊕ | | ⊕ | ⊕ | ⊕ | | ⊕ | | ⊕ | | ⊕ | | | | | ⊕ |
| | $o_1$ | | ⊕ | | ⊕ | ⊕ | | ⊕ | | ⊕ | ⊕ | ⊕ | ⊕ | ⊕ | ⊕ | | ⊕ | | ⊕ | | ⊕ | | ⊕ | | | | ⊕ | |

Table 7.2.: Complex addressing function from [Mau+15a].

`clflush` instruction was removed from the set of allowed instructions in Google Chrome Native Client [SD15].

## 2.3. CPU Caches

A CPU cache is a small and fast memory inside the CPU hiding the latency of main memory by keeping copies of frequently used data. Modern Intel CPUs have three levels of cache, where L1 is the smallest and fastest cache and L3 the slowest and largest cache. The L3 cache is an inclusive cache, *i.e.*, all data in L1 and L2 cache is also present in the L3 cache. It is divided into one slice per CPU core, but shared, *i.e.*, cores can access all slices. The undocumented *complex addressing* function that maps physical addresses to slices was recently reverse engineered [Mau+15a; Inc+15; Yar+15]. We used the results published by Maurice et al. [Mau+15a], shown in Table 7.2. The table shows how address bits 6 to 32 are xor'd into one or two output bits $o_0$ and $o_1$. In case of a dual-core CPU, output bit $o_0$ determines to which of the two cache slices the physical address maps. In case of a quad-core CPU, output bits $o_1$ and $o_0$ determine the slice.

Caches are organized in sets of multiple lines. The mapping from physical addresses to sets is fixed. Addresses that map to the same set and slice are called *congruent*. To load a new line from memory, the *replacement policy* decides which line to evict. Intel has not disclosed the cache replacement policy of their CPUs. However, the replacement policies for some architectures have been reverse-engineered: Sandy Bridge has a pseudo-LRU replacement policy and Ivy Bridge a modification of the pseudo-LRU replacement policy [Won]. Moreover, Ivy Bridge, Haswell and Skylake use adaptive cache replacement policies which only behave as pseudo-LRU in some situations [Qur+07]. These CPUs can switch the cache replacement policy frequently.

## 2.4. Cache Attacks and Cache Eviction

Cache side-channel attacks exploit timing differences between cache hits and cache misses. Practical attacks on cryptographic algorithms have been explored thoroughly [Ber05; Per05]. There are two main types of cache attacks called Prime+Probe and Flush+Reload. The Prime+Probe attack has been introduced by Percival [Per05] and Osvik et al. [OST06]. It determines activities of a victim process by repeatedly measuring the duration to access once every address in a set of congruent addresses, *i.e.*, a so-called eviction set. Prime+Probe on the last-level cache enables cross-core cache attacks such as cross-VM attacks without shared memory [IES15; Liu+15], covert channels [Mau+15b] and attacks from within sandboxed JavaScript [Ore+15]. Oren et al. [Ore+15] and Liu et al. [Liu+15] compute the eviction set by adding addresses to the eviction set until eviction works. Flush+Reload has been introduced by Gullasch et al. [GBK11] and Yarom and Falkner [YF14]. It exploits shared memory between attacker and victim and is very fine-grained. Cache lines are flushed with the `clflush` instruction or using cache eviction [GSM15].

Evicting data from the cache is just as crucial to cache attacks as it is for the Rowhammer attack. Previous work either uses the `clflush` instruction or hand-crafted eviction loops. Hund et al. [HWH13] showed that data can be evicted by filling a large memory buffer the size of the cache. However, this is very slow and thus not applicable to fine-grained cache attacks or Rowhammer attacks. Using the reverse-engineered complex addressing function solves the problem of finding addresses that are congruent in the cache, but it leaves the non-trivial problem of finding access sequences to achieve high eviction rates while maintaining a low execution time.

# 3. Cache Eviction Strategies

In this section, we describe how to find cache eviction strategies in a fully automated way for microarchitectures post Sandy Bridge. An *eviction strategy* accesses addresses from an eviction set in a specific access pattern and can ideally be used as a replacement for `clflush`. *Eviction set* is commonly defined as a set of congruent addresses. The access pattern defines in which order addresses from the eviction set are accessed, including multiple accesses per address.

An efficient eviction strategy can replace the `clflush` instruction in any cache attack and significantly improves cache attacks based on Prime+Probe, like JavaScript-based attacks [Ore+15] or cross-VM cache attacks [Liu+15]. It also allows to replace the `clflush` instruction in a Rowhammer attack (see Section 4).

The replacement policy of the CPU influences the size of the eviction set and the access pattern necessary to build an efficient eviction strategy. For a pseudo-LRU replacement policy, accessing as many congruent locations as the number of ways of the L3 cache (for instance 12 or 16) once, evicts the targeted address with a high probability. For adaptive cache replacement policies, an eviction strategy that is effective for one policy is likely to be ineffective for the other. Thus it is necessary to craft an eviction strategy that causes eviction for both policies and ideally does not introduce a significant timing overhead.

We distinguish between the following ways to generate an eviction strategy:

1. *Static eviction set and static access pattern*: uses information on cache slice function and physical addresses, and generates a predefined pattern in negligible time. Sections 3.2 and 3.3 describe new efficient eviction strategies computed this way.
2. *Dynamic eviction set and static access pattern*: computes the eviction set in an automated way, without any knowledge of the system, e.g., the number of cores. A good access pattern that matches the replacement policy of the targeted system is necessary for a successful attack. Section 3.3 describes this approach.
3. *Dynamic eviction set and dynamic access pattern*: automatically computes the eviction set and the access pattern based on randomness. This comes at the cost of performing a huge number of eviction tests, but it has the advantage to require almost no information on the system, and allows to implement fully automated online attacks for unknown systems. Section 3.3 describes this approach.
4. *Static eviction set and dynamic access pattern*: uses a pre-defined eviction set, but a random pattern that is computed in an automated way. This is possible in theory, but it has no advantage over automatically testing static access patterns. We thus do not further investigate this approach.

We first describe a model to represent access patterns, given several parameters. To find a good eviction strategy for a given system, we define an offline and an online phase. In the offline phase, the attacker explores the parameter space to find the best eviction strategies for a set of controlled systems. The goal is to find a eviction strategy that matches the undocumented replacement policy the closest, including the possibility of policy switches. In the online phase, the attacker targets an unknown system, with no privileges.

## 3.1. Cache Eviction Strategy Model

The success of a cache eviction strategy is measured by testing whether the targeted memory address is not cached anymore over many experiments, *i.e.*, average success rate. For such cases, we made the following three observations.

First, only cache hits and cache misses to addresses in the same cache set have a non-negligible influence on the cache, apart from cache maintenance and prefetching operations to the same cache set. We verified this by taking an eviction algorithm and randomly adding memory accesses that are not congruent. The eviction rate is the average success rate of the eviction function. It does not change by adding non-congruent accesses to an eviction strategy as long as the timing does not deviate. Thus, the eviction set only contains congruent addresses and the effectiveness of the eviction strategy depends on the *eviction set size*.

Second, addresses are indistinguishable with respect to the cache. Thus, we represent access patterns as sequences of address labels $a_i$, e.g., $a_1 a_2 a_3 \ldots$. Each address label is set to a different address and thus for each time frame the sequence defines which address to access. A pattern $a_1 a_2 a_3$ is equivalent to any pattern $a_k a_l a_m$ where $k \neq l \neq m$. If run in a loop, the number of *different memory addresses* has an influence on the effectiveness on the eviction strategy.

Third, repeated accesses to the same address are necessary to keep it in the cache, as replacement policies can prefer to evict recently added cache lines over older ones. Changing the eviction sequence from $a_1 a_2 \ldots a_{17}$ to $a_1 a_1 a_2 a_2 \ldots a_{17} a_{17}$ reduces the execution time by more than 33% on Haswell, and increases the eviction rate significantly if executed repeatedly, as the cache remains filled with our eviction set. However, we observed a diminishing marginal utility for the number of accesses to the same address. For all addresses we observed that after a certain number of accesses, further accesses do not increase and can even decrease the eviction rate. Thus, we describe eviction strategies as a loop over an eviction set of size

```
1 for (s = 0; s <= S-D; s += L)
2   for (c = 0; c < C; c += 1)
3     for (d = 0; d < D; d += 1)
4       *a[s+d];
```

Listing 3: Eviction loop for pattern testing.

$S$, where only a subset of $D$ addresses is *accessed per round*. A parameter $L$ allows to make accesses *overlap* for repeated accesses.

While testing all possible sequences even for very small sequence lengths is not possible in practical time (cf. Stirling numbers of second kind as a good estimate), a systematic exploration of influential parameters is possible. In theory, better eviction strategies may lie outside of this reduced search space. However using this method, we found eviction strategies that allowed us to successfully trigger bit flips using eviction-based Rowhammer (see Section 4). To discuss and compare eviction strategies systematically, we use the following naming scheme in this paper to describe parametrized eviction strategies as depicted in Listing 3. The eviction strategy name has the form $\mathcal{P}$-$C$-$D$-$L$-$S$, with $C$, the number of accesses to each memory address per loop round, $D$, the number of different memory addresses accessed per loop round, $L$, the step size/increment of the loop (for overlapping accesses), and $S$, the eviction set size. For instance, LRU-eviction is $\mathcal{P}$-1-1-1-$S$ with an access sequence of $a_1 a_2 a_3 \ldots a_S$.

## 3.2. Offline Phase

In the offline phase, the attacker has at his disposal a set of machines and tries to learn the eviction strategy that matches the replacement policy the closest for each machine. While it is not strictly a reverse engineering of the replacement policy, by knowing the best eviction strategy, the attacker gains knowledge on the systems. In this phase, the attacker has no time constraints.

We discuss the evaluation in detail for the Haswell platform with a single DIMM in single channel mode. We explored the parameter space up to degree 6 in the dimensions of $C$, $D$ and $L$ and 23 different eviction set sizes each, in order to find eviction strategies that are fast and effective enough to perform Rowhammer attacks. Including the equivalent eviction strategies we evaluated a total of 18293 eviction strategies on 3 of our test platforms. We tested each eviction strategy in 20 double-sided Rowhammer tests with 2 million hammering rounds (*i.e.*, 80 million evictions per eviction strategy) and evaluated them using different evaluation criteria including eviction rate, runtime, number of cache hits and misses. The runtime was more than 6 days. The hammering was performed on a fixed set of physical addresses congruent to one specific cache set to allow for a fair comparison of the eviction strategies. Half of the evictions, *i.e.*, 40 millions, were used to measure eviction rate, cache hits and cache misses. The other half was used to measure the average execution time per eviction. We verified that the sample size is high enough to get reproducible measurements.

The number of bit flips is not suitable for the evaluation of a single eviction strategy, but only to determine whether and how cache hits, cache misses, the execution time and the eviction rate influence the probability of a bit flip. Bit flips are reproducible in terms of the memory location, but the time and the number of memory accesses until a bit flip occurs again varies widely. In order to measure the average number of bit flips for a eviction strategy, we would have to test every eviction strategy for several hours instead of minutes. This would increase the test time per machine to several weeks, and even then, it would not yield reproducible results, as it has been observed that the DRAM cells get permanently damaged if hammered for a long time [Kim+14].

High execution times are too slow to trigger bit flips and low execution times are useless without a good eviction rate. The execution time of the eviction strategy is directly related to the number of memory accesses to the two victim addresses. Hence, it influences the probability of a bit flip directly. On our default configured Ivy Bridge notebook we observed bit flips even with execution times of 1.5 microseconds per hammering round, that is approximately 21,500 accesses per address within the specified total refresh interval of 64ms. This maps to the average periodic refresh interval `tREFI` by dividing 64ms by 8192 [Mic03]. Double-sided rowhammering using `clflush` takes only 60 nanoseconds on our Haswell test system, that is approximately 0.6 million accesses per address in 64ms. Figure 7.1a shows how bit flips are correlated with the eviction execution time.

The eviction rate has to be very high to trigger bit flips. Figure 7.1b shows how many bit flips occurred at which eviction rate. We observe that 81% of the bit flips occurred at an eviction rate of 99.75% or higher and thus use this as a threshold for good eviction strategies on our Haswell system. Even though a bit flip may occur at lower eviction rates, the probability is significantly lower.

| $C$ | $D$ | $L$ | $S$ | Accesses | Hits | Misses | Time (ns) | Eviction |
|---|---|---|---|---|---|---|---|---|
| - | - | - | - | - | 2 | 2 | 60 | 99.9999% |
| 5 | 2 | 2 | 18 | 90 | 34 | 4 | 179 | 99.9624% |
| 2 | 2 | 1 | 17 | 64 | 35 | 5 | 180 | 99.9820% |
| 2 | 1 | 1 | 17 | 34 | 47 | 5 | 191 | 99.8595% |
| 6 | 2 | 2 | 18 | 108 | 34 | 5 | 216 | 99.9365% |
| 1 | 1 | 1 | 17 | 17 | 96 | 13 | 307 | 74.4593% |
| 4 | 2 | 2 | 20 | 80 | 41 | 23 | 329 | 99.7800% |
| 1 | 1 | 1 | 20 | 20 | 187 | 78 | 934 | 99.8200% |

Table 7.3.: The fastest 5 eviction strategies with an eviction rate above 99.75% compared to `clflush` and LRU eviction on the Haswell test system.

The eviction loop contributes to a high number of cache hits and cache misses, apart from the two addresses we want to hammer. We measure the number of cache hits and cache misses that occur during our test run using hardware performance counters through the Linux syscall interface `perf_event_open`. Cache hits have a negligible influence on the execution time and no effect on the DRAM. Cache misses increase the execution time and, if performed on a different row but in the same channel, rank and bank, additional DRAM accesses. However, Figures 7.1c and 7.1d show that both cache hits and cache misses do not impact the number of bit flips significantly, as the average for all eviction strategies is in the range of the eviction strategies that triggered a bit flip.

Thus, we thus use the eviction rate as a criteria for good eviction strategies, and among those eviction strategies, we prefer those with a lower average execution time. This method requires no access to any system interfaces and can be implemented in any language and execution environment that allows to measure time and perform arbitrary memory accesses, such as JavaScript.

Table 7.3 shows a comparison of the fastest 5 of these eviction strategies with an eviction rate above 99.75% (see Figure 7.1b) and `clflush` based rowhammering as well as the fastest LRU ($\mathcal{P}$-1-1-1-20) eviction strategy that achieves the same eviction rate. The best two eviction strategies are $\mathcal{P}$-5-2-2-18 and $\mathcal{P}$-2-2-1-17, both with an execution time around 180 nanoseconds.

Accessing each address in the eviction set only once (LRU eviction) is far from optimal for cache attacks and impractical for Rowhammer. Although counterintuitive, adding more accesses to the eviction loop will lower the overall execution time. We can observe this for instance by comparing the eviction strategies $\mathcal{P}$-1-1-1-20 and $\mathcal{P}$-4-2-2-20. While both access the same set of 20 addresses, the latter one performs 4 times as many memory accesses, yet its execution time is only one third. Comparing the best eviction strategy we found to LRU eviction as described in previous work, performs only as good if the set size is at least $S = 25$, increasing the average execution time 9 times higher than the one of the best eviction strategy we found. On the other hand, the eviction set size in previous work is typically specified as $S = 17$. For $\mathcal{P}$-1-1-1-17 we measured an eviction rate of 74.5% and even then a 1.7 times higher execution time than with the best eviction strategy we found. This shows that the eviction strategies we found are a significant improvement over previously published eviction methods.

We performed the same evaluation for the other architectures. The distribution of bit flips on our Ivy Bridge test system relative to eviction rate and execution time is shown in Figure 7.2. Most bit flips occurred at eviction rates above 99%. The fastest 5 of these eviction strategies are shown in Table 7.4 in comparison with `clflush` and the fastest LRU ($\mathcal{P}$-1-1-1-15) eviction strategy.

According to our measurements the complex addressing function on Skylake is not the same as in Haswell, but it can be trivially derived from the reverse engineered 8-core function. We again found that LRU eviction performs much worse than the best eviction strategy we found as shown in Table 7.4.

## 3.3. Online Phase

In the online phase, the attacker targets an unknown system. In particular, microarchitecture and number of CPU cores are unknown to the attacker. The attacker has the knowledge gained from the offline phase at his disposal. However, he has no privilege on the victim's machine and no time to run the extensive search from the offline phase. The online phase consists in two attacks: an assumption-based attack, and a fall-back attack in case the first one does not work. In both cases the attack is based on a series of timing attacks and no access to specific system interfaces is necessary.

| $C$ | $D$ | $L$ | $S$ | Acc. | Hits | Misses | Time (ns) | Eviction |
|---|---|---|---|---|---|---|---|---|
| - | - | - | - | - | 2 | 2 | 40 | 100.000% |
| 4 | 5 | 5 | 20 | 80 | 43 | 35 | 327 | 99.514% |
| 1 | 1 | 1 | 13 | 13 | 52 | 33 | 333 | 72.145% |
| 3 | 1 | 1 | 17 | 51 | 46 | 41 | 341 | 99.081% |
| 4 | 5 | 5 | 17 | 60 | 45 | 37 | 345 | 99.604% |
| 3 | 1 | 1 | 19 | 57 | 50 | 47 | 369 | 99.267% |
| 3 | 2 | 2 | 18 | 54 | 48 | 43 | 376 | 99.412% |
| 1 | 1 | 1 | 15 | 15 | 97 | 84 | 632 | 99.085% |
| $C$ | $D$ | $L$ | $S$ | Acc. | Hits | Misses | Time (ns) | Eviction |
| - | - | - | - | - | 2 | 2 | 47 | 100.000% |
| 3 | 1 | 1 | 22 | 66 | 48 | 45 | 218 | 99.937% |
| 2 | 2 | 1 | 22 | 84 | 47 | 45 | 222 | 99.932% |
| 3 | 3 | 3 | 24 | 72 | 50 | 45 | 222 | 99.938% |
| 3 | 3 | 3 | 21 | 63 | 51 | 45 | 223 | 99.937% |
| 4 | 3 | 3 | 24 | 96 | 49 | 45 | 225 | 99.905% |
| 1 | 1 | 1 | 17 | 17 | 240 | 36 | 240 | 82.959% |
| 1 | 1 | 1 | 21 | 21 | 145 | 87 | 495 | 99.970% |

Table 7.4.: `clflush` and LRU eviction compared to the fastest 5 eviction strategies above 99% eviction rate on the Ivy Bridge test system (top) and compared to the fastest 5 eviction strategies above 99.9% eviction rate on the Skylake DDR4 test system (bottom).

**Assumption-based Attack**

The attacker first tests whether the targeted system resembles a system tested in the offline phase, by performing timing attacks. No access to syscalls or system interfaces is required for this step. The attacker defines a threshold eviction rate based on the results from the offline phase (for instance 99.75%) and searches for eviction strategies above this threshold on the system under attack. By testing a set of eviction strategies from the offline phase, the attacker learns whether the architecture of the system under attack resembles an architecture from the offline phase. In this case the best eviction strategy for the system under attack is within the set of eviction strategies previously tested. The number of eviction strategies to test is as low as the number of targeted CPU architectures and thus it only takes a few seconds to compute.

The eviction set can be computed in a static or dynamic way. Without any further assumptions we can run modified versions of the algorithms by Oren et al. [Ore+15] or Liu et al. [Liu+15]. Instead of the $\mathcal{P}$-1-1-1 access pattern they implement, we use one of the suspected eviction strategies to build a dynamic assumption-based algorithm. This improves the success rate of their algorithms on recent architectures. However, we make additional assumptions to reduce the execution time to a minimum and build a static assumption-based algorithm. One assumption is that large arrays are allocated on large pages, as has been observed before [GBM15]. Based on this assumption we can use the complex addressing function from Table 7.2 to determine the slice patterns for 4KB and 2MB pages as shown in Figure 7.3. These distinct patterns in the mapping from physical addresses to cache slices depend only on the number of cache slices and are the same for Intel CPUs since the Sandy Bridge architecture. The algorithm by Oren et al. [Ore+15] or Liu et al. [Liu+15] finds only addresses in the same cache slice and cache set. We use it to build an eviction set of 2MB-aligned congruent addresses in the same slice. Subsequent eviction set computations are performed statically based on the complex addressing function and the identified 2MB offsets.

**Fall-back Attack**

If the assumption-based phase does not work on a system under attack, e.g., because the unknown system is none of the systems tested in the offline phase, the attacker runs a fall-back phase to find an eviction strategy that is sufficient to trigger a bit flip with Rowhammer.

Oren et al. [Ore+15] and Liu et al. [Liu+15] compute a dynamic eviction set with a static access pattern $\mathcal{P}$-1-1-1. We extend their algorithms to compute eviction strategies with dynamic eviction sets and dynamic access patterns. In the first step, we continuously add addresses to the eviction strategy multiple times to create eviction strategies with multiple accesses to the same address. We know that the eviction strategy is large enough as soon as we can clearly measure the eviction of the target physical address. In a second step, when the eviction rate is above the attacker chosen threshold, eviction addresses that do not lower the eviction rate are removed by replacing them with other addresses that are still in the eviction set. Thus, the number of memory accesses does not decrease, but the eviction set is minimized. This decreases the number of cache misses and thus the execution time. Finally, we randomly remove accesses that do not decrease the eviction rate and do not increase the execution time.

This again decreases the number of unnecessary cache hits and thus the execution time.

The resulting eviction strategy can neither access less addresses nor can any duplicate accesses be removed without lowering the eviction rate. They thus perform similarly to statically computed eviction strategies. The result of the algorithm is a series of accesses that fulfill the eviction rate threshold chosen by the attacker and that has a low execution time on the system under attack. If the threshold was set high enough so that bit flips are likely to occur in practice, the eviction strategy found by the fall-back algorithm can be used for an attack.

The algorithm uses a function `cached(p)` that tries to evict a target address `p` using the current eviction strategy and set and decides whether `p` is cached or not based on the access time. The quality of the solution depends on the number of tests that are performed in this function. The function only returns true, if an eviction rate below the attacker defined threshold is measured. A higher number of tests increases the execution time and the accuracy of this binary decision. Figure 7.4 shows how the number of tests influences the eviction rate and the execution time of the resulting eviction strategy. If a high eviction rate is necessary, the execution time of the algorithm is can exceed 40 minutes. Thus, our algorithm can precompute a working eviction strategy once and subsequent eviction set computations are done with the fixed eviction strategy within seconds.

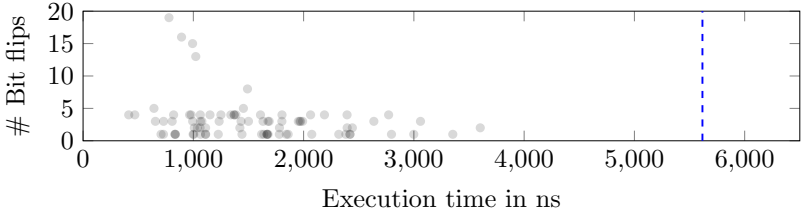

(a) Low execution time is better.

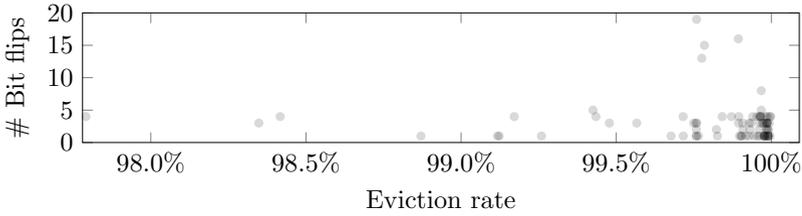

(b) High eviction rate is better. Average over all eviction strategies is 73.96%.

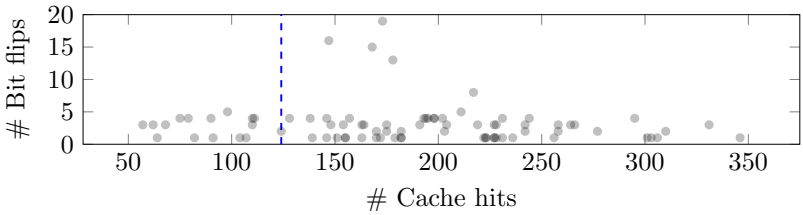

(c) Number of cache hits is not a good criteria for bit flips.

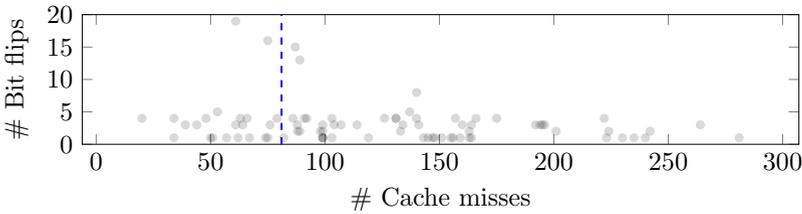

(d) Number of cache misses is not a good criteria for bit flips.

Figure 7.1.: Relation between the number of bit flips and average execution time, cache hits and cache misses per eviction and the eviction rate of the corresponding eviction strategy measured in 40 million samples. One point per eviction strategy that caused a bit flip, others are omitted. The darker the more points overlay. Average over all eviction strategies shown as dashed line. Good eviction strategies have high eviction rates and low execution times.

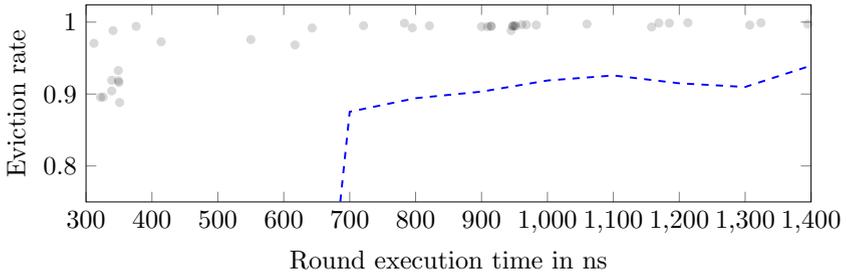

Figure 7.2.: Average execution time and eviction rate per eviction strategy on Ivy Bridge measured in 40 million samples per eviction strategy. One point per eviction strategy that caused a bit flip, others are omitted. The darker the more points overlay. Average over all eviction strategies shown as dashed line.

```
0123  0123  0123  0123  1032  1032  1032  1032  2301  2301  2301  2301  3210  3210  3210  3210
1032  1032  1032  1032  0123  0123  0123  0123  3210  3210  3210  3210  2301  2301  2301  2301
2301  2301  2301  2301  3210  3210  3210  3210  0123  0123  0123  0123  1032  1032  1032  1032
3210  3210  3210  3210  2301  2301  2301  2301  1032  1032  1032  1032  0123  0123  0123  0123
```

Figure 7.3.: Slice patterns for 64-byte offsets on 4KB pages on a 4-core system. An attacker can derive which addresses map to the same cache slice. Substituting 2 by 0 and 3 by 1 gives the slice pattern for 2-core systems.

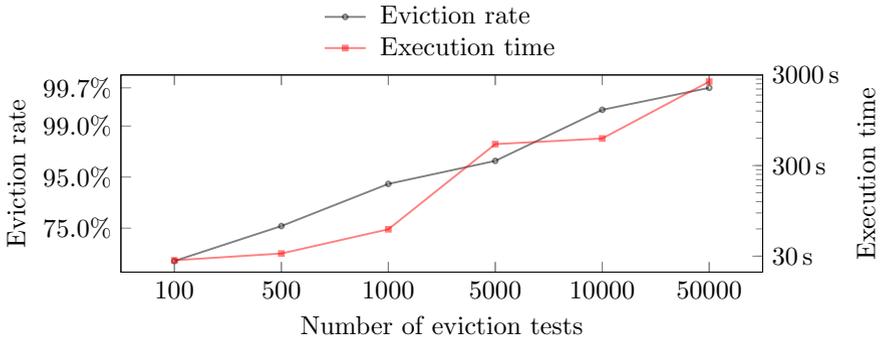

Figure 7.4.: The eviction rate and execution time of the dynamic eviction strategy when implementing the `cached(p)` function with $n$ eviction tests.

# 4. Implementation of eviction-based Rowhammer

We now perform Rowhammer attacks using the eviction strategies from Section 3 instead of `clflush` in different scenarios. First, we demonstrate that it is possible to trigger bit flips in the same conditions as in the existing attacks where an attacker is able to execute native code on the system under attack. We then show that given knowledge about the physical addresses, it is possible to trigger bit flips even from a remote website using JavaScript. In a third step, we show that the full Rowhammer attack is possible from a remote website using JavaScript without any additional information on the system.

## 4.1. Rowhammer in Native Code

We extended the `double_sided_rowhammer` program by Dullien [SD15] by using the best eviction strategy we have found. The two `clflush` instructions were first replaced by the eviction code described in Section 3.1, with parameters for a $\mathcal{P}$-2-2-1 eviction strategy. The eviction sets are either precomputed statically using the physical address mapping and the complex addressing function in Table 7.2, or using a dynamic eviction strategy computation algorithm.

This way, we were able to reproducibly flip bits on our Sandy Bridge and Ivy Bridge test machine using different eviction strategies when running with the Samsung DDR3 RAM and our Skylake test machine when running with the Crucial DDR4 RAM. The machines were operated in default configuration.

On our Haswell test machine we were not able to reproducibly flip bits with the default settings, not even with the `clflush` instruction. However, the BIOS configuration allows setting a custom refresh rate by setting the average periodic refresh interval `tREFI`. We had to increase the `tREFI` value from 6,549 to over 19,000 just to be able to trigger bit flips *with* the `clflush` instruction. The refresh interval is a typical parameter used by computer gaming enthusiasts and the overclocking community to increase system performance. However, while this might also be an interesting target group, we rather want to analyze the influence of the refresh interval on the applicability of the Rowhammer attack using cache eviction and the Rowhammer attack in JavaScript. Kim et al. [Kim+14] observed that the refresh interval directly influences the number of bit flips that occur and that below a module dependent `tREFI` value no bit flips occur. We will show that their observation also applies to Rowhammer with cache eviction and Rowhammer in JavaScript.

Lowering the refresh interval is not part of an actual attack. Existing work has already examined the prevalence of the Rowhammer and found that 85% of the DDR3 modules examined are susceptible to Rowhammer bit flips [Kim+14]. Also in our case only the modules of the Haswell test system and the G.Skill DIMMs in the Skylake test system were not susceptible to Rowhammer bit flips at default settings, whereas it was possible to induce Rowhammer bit flips in the other three DIMMs at default settings. Thus, our results do not contradict previous estimates and we must assume that millions of systems are still vulnerable.

Rowhammer with eviction in native code revives the Google Native Client exploit [SD15] that allows privilege escalation in Google Chrome. The `clflush` instruction has been blacklisted to solve this vulnerability, however, this is ineffective and a sandbox escape is still possible, as we can trigger bit flips in Google Native Client based on eviction.

## 4.2. Rowhammer in JavaScript

Triggering the Rowhammer bug from JavaScript is more difficult as JavaScript has no concept of virtual addresses or pointers and no access to physical address mappings. We observed that large typed arrays in JavaScript in all recent Firefox and Google Chrome versions on Linux are allocated 1MB aligned and use anonymous 2MB pages when possible. The reason for this lies in the memory allocation mechanism implemented by the operating system. Any memory allocation in a comparable scripting language and environment will also result in the allocation of anonymous 2MB pages for large arrays.

By performing a timing attack similar to the one performed by Gruss et al. [GBM15], we can determine the 2MB page frames in the browser. In this attack we iterate over an array and measure the access latency. The latency peaks during memory initialization are caused by the pagefaults that occur with the start of each new 2MB page, as shown in Figure 7.5. This also works in recent browser versions with a reduced timer resolution as suggested by Oren et al. [Ore+15] and added to the HTML5 standard by the W3C [W3C15]. Thus, we know the lowest 21 bits of the virtual and physical address by knowing the offset in the array.

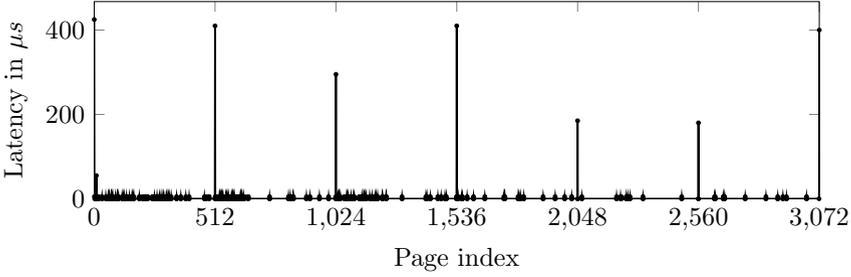

Figure 7.5.: Access latency of 4KB aligned addresses in a large array in JavaScript. Pagefaults cause the latency peaks at the start of the 2MB pages.

As a first proof-of-concept we reproduced bit flips in JavaScript in Firefox by hammering the exact physical addresses as in native code. In order to do this we built a tool to translate physical to virtual addresses for another process. To compute the eviction sets we use the assumption-based algorithm from Section 3.2. We observed that simple memory accesses as in our native code implementation are not optimized out by the just-in-time-compiler.

The final JavaScript-based attack does not require any outside computation and thus, runs entirely without user interaction in the browser. It exploits the fact that large typed arrays are allocated on 2MB pages. Thus, we know that each 2MB region of our array is divided into 16 row offsets of size 128KB (depends on the lowest row index bit). We can now perform double-sided hammering in these 2MB regions to trigger a bit flip within the 2MB region or amplified single-sided hammering on the outer two rows of every 2MB pages to induce a bit flip in another physical 2MB region. The result is the first hardware-fault attack implemented in JavaScript on a remote website.

## 4.3. Attack Evaluation

As described by Kim et al. [Kim+14] not all addresses in a DRAM are equally susceptible to bit flips. Therefore, to provide a fair comparison of the different techniques, we measured the number of bit flips for a fixed address pair already known to be susceptible. Figure 7.6 shows how different refresh rates influence the number of bit flips for a fixed time interval in different setups. The system was under slight usage during the tests (browsing, typing in an editor, etc.). We see that the `clflush` instruction yields the highest number of bit flips. If the refresh interval was set to a value where bit flips can be triggered using `clflush`, they can be triggered using native code eviction as well. To trigger bit flips in JavaScript, a slightly higher refresh interval was necessary. Again, it depends on the particular DIMM whether the refresh interval is chosen correctly so that no bit flips occur.

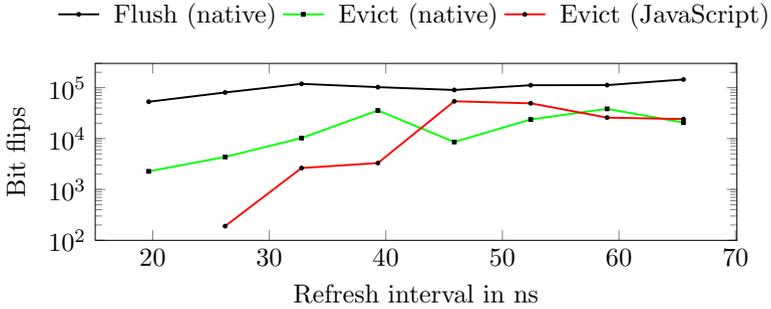

Figure 7.6.: Number of bit flips within 15 minutes on a fixed address pair for different values for the average periodic refresh interval `tREFI` on Haswell in three different setups.

The probability for bit flips in JavaScript is slightly lower than in native code, as native code is slightly faster. However, if a machine is vulnerable to our native code implementation it is likely vulnerable using our JavaScript implementation as well. While these plots were obtained on the Haswell machine, we were also able to trigger bit flips on our Ivy Bridge laptop with default settings from JavaScript. However, as the Laptop BIOSes did not allow to set the refresh interval `tREFI` directly, we could not obtain a comparable plot.

While DDR4 was assumed to have countermeasures against rowhammering, countermeasures are not part of the final DDR4 standard [Aic15a]. Using the Crucial DDR4 DIMMs we even were able to induce bit flips at default system settings and with the most recent BIOS version, after applying the functions reverse engineered by Pessl et al. [Pes+15]. On the G.Skill DDR4 DIMMs we could only induce bit flips at an increased refresh interval. Thus, even on these very recent and up-to-date systems

Rowhammer countermeasures have not been implemented in hardware and those implemented in software are ineffective. Whether a system is vulnerable to Rowhammer-based attacks still crucially depends on the refresh interval chosen by DIMM.

# 5. Discussion and Related Work

## 5.1. Building an Exploit With Rowhammer.js

Existing exploits assume that a page table is mapped in a row between two rows occupied by the attacker. However, we observed that this situation rarely occurs in practice. The operating system prefers to use large pages to reduce the pressure on the TLB. To make the organization and changes to physical address mappings easier the operating system will also group small pages into the same organizational physical frames. Page tables are only allocated between two user pages in a near-out-of-memory situation. Thus, the exploits allocate almost all system memory to enforce such a situation [SD15]. However, swapping is enabled by default in all major operating systems and thus the system will be severely unresponsive due to swapping. In our proof-of-concept exploit, we perform "amplified single-sided hammering". By hammering two adjacent rows we increase the probability for a bit flip in a surrounding row significantly compared to single-sided hammering. This allows to induce bit flips even across the borders of physically coherent 2MB regions with a high probability. As we already have been able to trigger bit flips in JavaScript we will only focus on how to manipulate a page table similar to previous exploits [SD15]. The attacker can repeat any step of the attack as long as necessary to be successful.

In the first step, the exploit locates an exploitable bit flip as described in Section 4.2, *i.e.*, a bit flip in the $\frac{1}{3}$ of the page table bits that are used for physical addresses. An exploitable bit flip changes an address bit in a page table that is in an adjacent 2MB region. We have found such bit flips on our all our test machines. In the second step, the exploit script releases all pages but the two that have previously been hammered and the ones that are required for cache eviction. Thus, also the page that contained the bit flip is released. Allocating arrays requires the browser to reserve virtual memory regions and to map them to physical memory upon the first access. The attacker determines the largest array size that still triggers the allocation of a page table in a timing attack (see 4.2). The array size was 1MB on all our test systems. We only access and thus allocate one 4KB page per 1MB array and thus 2 user pages per page table. The probability to place a group of page tables in the targeted 2MB region is $\approx \frac{1}{3}$. In the third step, the exploit script triggers the bit flip again and may find that its own memory mappings changed. With a chance of $\approx \frac{1}{3}$ the memory mapped is now one of the attackers page tables. The attacker can now change mapped addresses in that page table and if successful, has gained full access to the physical memory of the system. Our proof-of-concept works on recent Linux systems with all recent versions of Firefox and it does not require a near-out-of-memory situation. It does not work in Google Chrome due to the immediate allocation of all physical memory for an allocated 1MB array after a single access.

## 5.2. Limitations

In JavaScript we use 2MB pages to find congruent addresses and adjacent rows efficiently. If the operating system does not provide 2MB pages, we cannot perform double-sided or amplified single-sided hammering. However, the probability of a bit flip with single-sided hammering is significantly lower. Exploiting double-sided hammering with 2MB pages is not possible because we can then only induce bit flips in our own memory. Thus, an attack is only possible with amplified single-sided hammering to induce a bit flip in an adjacent row in an adjacent 2MB page. There is only a limited number of such rows in a system. Still the search for an exploitable bit flip can easily take several hours, especially as the probability of a bit flip in JavaScript is lower than in native code. Furthermore, if we cannot guess the best eviction strategy for the system, it will take up to an hour of precomputations to find a good eviction strategy. The victim has to stay on the website for the duration of the attack. While this was the case in our proof-of-concept attack it is less realistic for a real-world attack.

## 5.3. Countermeasures

The operating system allocates memory in large physical memory frames (often 2MB) for reasons of optimization. Page tables, kernel pages and user pages are not allocated in the same memory frame, unless the system is close to out-of-memory (*i.e.*, allocating the last few kilobytes of physical memory). Thus, the most efficient Rowhammer attack (double-sided hammering) would not possible if the operating system memory allocator was less aggressive in near-out-of-memory situations. Preventing (amplified) single-sided hammering is more difficult, as hammering across the boundaries of a 2MB region is possible.

To fully close the attack vector for double-sided hammering, we also have to deal with read-only shared code and data, *i.e.*, shared libraries. If the attacker hammers on a shared library, a fault can be induced in this library. Therefore, shared libraries should not be shared over processes that run at different privilege levels or under different users. As a consequence, the attacker would be unable to escape from a sandbox or gain access to a higher privilege level using `clflush` or eviction-based Rowhammer.

Kim et al. [Kim+14] proposed several countermeasures which should be implemented for new DRAM modules, including increasing the refresh rate. However, this would cause significant performance impacts. BIOS updates supplied so far only double the refresh rate, which is insufficient to prevent attacks on all DRAM modules. Moreover, many users to not update the BIOS unless it is unavoidable.

Pseudo Target Row Refresh (pTRR) and Target Row Refresh (TRR) are features that refresh neighboring rows when the number of accesses to one row exceeds a threshold. They have less overhead compared to double the refresh rate. Although TRR has been announced as implemented in all DDR4 modules it has been removed from the final DDR4 standard. Manufacturers can still choose to implement it in their devices, but if the memory controller does not support it, it has no effect.

Error-correcting code (ECC) memory is often mentioned as a countermeasure against Rowhammer attacks. However, recent work shows that it cannot reliably protect against Rowhammer attacks.cases [Aic15b; Lan16].

At the software level, one proposed countermeasure is the detection using hardware performance counters [HF15; Pay16; Gru+16; Awe+16]. The excessive number of cache references and cache hits allows to detect on-going attacks. However, this countermeasure can suffer from false positives, so it needs further evaluation before it can be brought to practice.

## 5.4. Related Work

The initial work by Kim et al. [Kim+14] and Seaborn's [SD15] root exploit made the scientific community aware of the security implications of a Rowhammer attack. However, to date, there have been very few other publications, focusing on different aspects than our work. Barbara Aichinger [Aic15a] analyzed Rowhammer faults in server systems where the problem exists in spite of ECC memory. She remarks that it will be difficult to fix the problem in the millions or even billions of DDR3 DRAMs in server systems. Rahmati et al. [Rah+15] have shown that bit flips can be used to identify a system based on the unique and repeatable error pattern that occurs at a significantly increased refresh interval. Our paper is the first to examine how to perform Rowhammer attacks based on cache eviction.[1] Our cache eviction techniques facilitated cache side-channel attacks on ARM CPUs [Lip+15]. Concurrent and independent work by Aweke et al. [Awe+16] have also demonstrated bit flips without `clflush` on an old Sandy Bridge laptop. They focus on countermeasures, whereas we focus on attacking a wider range of architectures and environments. Qiao and Seaborn [QS16] implemented a Rowhammer attack with non-temporal memory accesses.

# 6. Future Work

While we only investigated the possibility of a JavaScript Rowhammer attack in Firefox and Google Chrome on Linux, the attack exploits fundamental concepts that are inbuilt in the way hardware and operating system work. Whenever the operating system uses 4KB pages, page tables are required and at latest allocated when one of the 4KB pages belonging to this page table is accessed. Thus, the operating system cannot prevent that $\frac{1}{3}$ of memory is allocated for page tables. The same attack approach could be applied to hypervisors that allocate 4KB pages to virtual machines, even if they applies similar allocation mechanisms as the Linux kernel. While it might seem unreasonable and not realistic that hypervisors allocate 4KB pages, it in fact makes cross-VM page deduplication easier. According to Barresi et al. [Bar+15], page deduplication is in fact still widely used in public clouds. Our work opens the possibility for further investigation on whether page deduplication in fact is not only a problem for security and privacy of virtual machines, but a security problem for the hypervisor itself.

---

[1]A draft of this paper was published online since July 24, 2015.

# 7. Conclusion

In this paper, we presented Rowhammer.js, an implementation of the Rowhammer attack using fast cache eviction to trigger the Rowhammer bug with only regular memory accesses. It is the first work to investigate eviction strategies to defeat complex cache replacement policies. This does not only enable to trigger Rowhammer in JavaScript, it also benefits research on cache attacks as it allows to perform attacks on recent and unknown CPUs fast and reliably. Our fully automated attack runs in JavaScript through a remote website and can gain unrestricted access to systems. The attack technique is independent of CPU microarchitecture, programming language and execution environment.

The majority of DDR3 modules are vulnerable and DDR4 modules can be vulnerable too. Thus, it is important to discover all Rowhammer attack vectors. Automated attacks through websites pose an enormous threat as they can be performed on millions of victim machines simultaneously.

# 8. Acknowledgments

We would like to thank our shepherd Stelios Sidiroglou-Douskos and our anonymous reviewers for their valuable comments and suggestions. We would also like to thank Mark Seaborn, Thomas Dullien, Yossi Oren, Yuval Yarom, Barbara Aichinger, Peter Pessl and Raphael Spreitzer for feedback and advice.

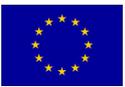 Supported by the EU Horizon 2020 programme under GA No. 644052 (HECTOR), the EU FP7 programme under GA No. 610436 (MATTHEW), the Austrian Research Promotion Agency (FFG) and Styrian Business Promotion Agency (SFG) under GA No. 836628 (SeCoS), and Cryptacus COST Action IC1403.

# 8

# Flush+Flush: A Fast and Stealthy Cache Attack

## Publication Data

## Contributions

Main author.

# Flush+Flush
# A Fast and Stealthy Cache Attack

Daniel Gruss, Clémentine Maurice[†], Klaus Wagner, and Stefan Mangard

Graz University of Technology, Austria

## Abstract

Research on cache attacks has shown that CPU caches leak significant information. Proposed detection mechanisms assume that all cache attacks cause more cache hits and cache misses than benign applications and use hardware performance counters for detection.

In this article, we show that this assumption does not hold by developing a novel attack technique: the *Flush+Flush* attack. The *Flush+Flush* attack only relies on the execution time of the flush instruction, which depends on whether data is cached or not. *Flush+Flush* does not make any memory accesses, contrary to any other cache attack. Thus, it causes no cache misses at all and the number of cache hits is reduced to a minimum due to the constant cache flushes. Therefore, *Flush+Flush* attacks are stealthy, *i.e.*, the spy process cannot be detected based on cache hits and misses, or state-of-the-art detection mechanisms. The *Flush+Flush* attack runs in a higher frequency and thus is faster than any existing cache attack. With 496 KB/s in a cross-core covert channel it is 6.7 times faster than any previously published cache covert channel.

## 1. Introduction

The CPU cache is a microarchitectural element that reduces the memory access time of recently-used data. It is shared across cores in modern processors, and is thus a piece of hardware that has been extensively studied in terms of information leakage. Cache attacks include covert and cryptographic side channels, but caches have also been exploited in other

---

types of attacks, such as bypassing kernel ASLR [HWH13], detecting cryptographic libraries [Ira+15a], or keystroke logging [GSM15]. Hardware performance counters have been proposed recently as an OS-level detection mechanism for cache attacks and Rowhammer [CSY15; HF15; Pay16]. This countermeasure is based on the assumption that all cache attacks cause significantly more cache hits and cache misses than benign applications. While this assumption seems reasonable, it is unknown whether there are cache attacks that do not cause a significant number of cache hits and cache misses.

In this article, we present the *Flush+Flush* attack. *Flush+Flush* exploits the fact that the execution time of the `clflush` instruction is shorter if the data is not cached and higher if the data is cached. At the same time, the `clflush` instruction evicts the corresponding data from all cache levels. *Flush+Flush* exploits the same hardware and software properties as *Flush+Reload* [YF14]: it works on read-only shared memory, cross-core attack and in virtualized environments. In contrast to *Flush+Reload*, *Flush+Flush* does not make any memory accesses and thus does not cause any cache misses at all and only a minimal number of cache hits. This distinguishes *Flush+Flush* from any other cache attack. However, with both *Flush+Reload* and *Flush+Flush* the victim process experiences an increased number of cache misses.

We evaluate *Flush+Flush* both in terms of *performance* and *detectability* in three scenarios: a covert channel, a side-channel attack on user input, and a side-channel attack on AES with T-tables. We implement a detection mechanism that monitors cache references and cache misses of the last-level cache, similarly to state of the art [CSY15; HF15; Pay16]. We show that existing cache attacks as well as Rowhammer attacks can be detected using performance counters. However, we demonstrate that this countermeasure is non-effective against the *Flush+Flush* attack, as the fundamental assumption fails. The *Flush+Flush* attack is thus more stealthy than existing cache attacks, *i.e.*, a *Flush+Flush* spy process cannot be detected based on cache hits and cache misses. Thus, it cannot be detected by state-of-the-art detection mechanisms.

The *Flush+Flush* attack runs in a higher frequency and thus is faster than any existing cache attack in side-channel and covert channel scenarios. It achieves a cross-core transmission rate of 496 KB/s, which is 6.7 times faster than any previously published cache covert channel. The *Flush+Flush* attack does not trigger prefetches and thus allows to monitor multiple addresses within a 4 KB memory range in contrast to *Flush+Reload* that fails in these scenarios [GSM15].

Our key contributions are:

- We detail a new cache attack technique that we call *Flush+Flush*. It relies only on the difference in timing of the `clflush` instruction between cached and non-cached memory accesses.

- We show that in contrast to all other attacks, *Flush+Flush* is stealthy, *i.e.*, it cannot be detected using hardware performance counters. We show that *Flush+Flush* also outperforms all existing cache attacks in terms of speed.

The remainder of this paper is organized as follows. Section 2 provides background information on CPU caches, shared memory, and cache attacks. Section 3 describes the *Flush+Flush* attack. Section 4 investigates how to leverage hardware performance counters to detect cache attacks. We compare the performance and detectability of *Flush+Flush* attacks compared to state-of-the-art attacks in three scenarios: a covert channel in Section 5, a side-channel attack on keystroke timings in Section 6, and on cryptographic algorithms in Section 7. Section 8 discusses implications and countermeasures. Section 9 discusses related work. Finally, we conclude in Section 10.

# 2. Background

## 2.1. CPU Caches

CPU caches hide the memory accesses latency to the slow physical memory by buffering frequently used data in a small and fast memory. Modern CPU architectures implement $n$-way set-associative caches, where the cache is divided into cache sets, and each cache set comprises several cache lines. A line is loaded in a set depending on its address, and each line can occupy any of the $n$ ways.

On modern Intel processors, there are three cache levels. The L3 cache, also called last-level cache, is shared between all CPU cores. The L3 cache is inclusive, *i.e.*, all data within the L1 and L2 caches is also present in the L3 cache. Due to these properties, executing code or accessing data on one core has immediate consequences even for the private caches of the other cores. This can be exploited in so called cache attacks. The last-level cache is divided into as many slices as cores, interconnected by a ring bus. Since the Sandy Bridge microarchitecture, each physical address is mapped to a slice by an undocumented so-called *complex-addressing* function, that has recently been reversed-engineered [Mau+15a].

A cache replacement policy decides which cache line to replace when loading new data in a set. Typical replacement policies are least-recently used (LRU), variants of LRU and bimodal insertion policy where the CPU can switch between the two strategies to achieve optimal cache usage [Qur+07]. The unprivileged `clflush` instruction evicts a cache line from all the cache hierarchy. However, a program can also evict a cache line by accessing enough memory.

## 2.2. Shared Memory

Operating systems and hypervisors instrument shared memory to reduce the overall physical memory utilization and the TLB utilization. Shared libraries are loaded into physical memory only once and shared by all programs using them. Thus, multiple programs access the same physical pages mapped within their own virtual address space.

The operating system similarly optimizes mapping of files, forking a process, starting a process twice, or using `mmap` or `dlopen`. All cases result in a memory region shared with all other processes mapping the same file.

On personal computers, smartphones, private cloud systems and even in public clouds [Bar+15], another form of shared memory can be found, namely content-based page deduplication. The hypervisor or operating system scans the physical memory for byte-wise identical pages. Identical pages are remapped to the same physical page, while the other page is marked as free. This technique can lower the use of physical memory and TLB significantly. However, sharing memory between completely unrelated and possibly sandboxed processes, and between processes running in different virtual machines brings up security and privacy concerns.

## 2.3. Cache Attacks and Rowhammer

Cache attacks exploit timing differences caused by the lower latency of CPU caches compared to physical memory. Access-driven cache attacks are typically devised in two types: *Prime+Probe* [OST06; Per05; TOS10] and *Flush+Reload* [GBK11; YF14].

In *Prime+Probe* attacks, the attacker occupies a cache set and measures whenever a victim replaces a line in that cache set. Modern processors have a physically indexed last-level cache, use complex addressing, and undocumented replacement policies. Cross-VM side-channel attacks [IES15; Liu+15] and covert channels [Mau+15b] that tackle these challenges have been presented in the last year. Oren et al. [Ore+15] showed that a *Prime+Probe* cache attack can be launched from within sandboxed JavaScript in a browser, allowing a remote attacker to eavesdrop on network traffic statistics or mouse movements through a website.

*Flush+Reload* is a two phase attack that works on a single cache line. First, it *flushes* a cache line using the `clflush` instruction, then it measures the time it takes to *reload* the data. Based on the time measurement, the attacker determines whether a targeted address has been reloaded by another process in the meantime. In contrast to *Prime+Probe*, *Flush+Reload* exploits the availability of shared memory and especially shared libraries between the attacker and the victim program. Applications of *Flush+Reload* have been shown to be reliable and powerful, mainly to attack cryptographic algorithms [Ira+15a; Ira+15b; Gül+15; Zha+14].

Rowhammer is not a typical cache attack but a DRAM vulnerability that causes random bit flips by repeatedly accessing a DRAM row [Kim+14]. It however shares some similarities with caches attacks since the accesses must bypass all levels of caches to reach DRAM and trigger bit flips. Attacks exploiting this vulnerability have already been demonstrated to gain root privileges and to evade a sandbox [SD15]. Rowhammer causes a significant number of cache hits and cache misses, that resemble a cache attack.

# 3. The *Flush+Flush* Attack

The *Flush+Flush* attack is a faster and stealthier alternative to existing cache attacks that also has fewer side effects on the cache. In contrast to other cache attacks, it does not perform any memory accesses. For this reason it causes no cache misses and only a minimal number of cache hits. Thus, proposed detection mechanisms based on hardware performance counters fail to detect the *Flush+Flush* attack. *Flush+Flush* exploits the same hardware and software properties as *Flush+Reload*. It runs across cores and in virtualized environments if read-only shared memory with the victim process can be acquired.

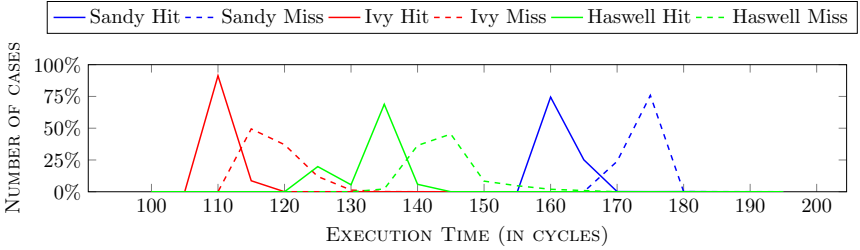

Figure 8.1.: Execution time of the `clflush` instruction on cached and uncached memory on different CPU architectures

Our attack builds upon the observation that the `clflush` instruction can abort early in case of a cache miss. In case of a cache hit, it has to trigger eviction on all local caches. This timing difference can be exploited in form of a cache attack, but it can also be used to derive information on cache slices and CPU cores as each core can access its own cache slice faster than others.

The attack consists of only one phase, that is executed in an endless loop. It is the execution of the `clflush` instruction on a targeted shared memory line. The attacker measures the execution time of the `clflush` instruction. Based on the execution time, the attacker decides whether the memory line has been cached or not. As the attacker does not load the memory line into the cache, this reveals whether some other process has loaded it. At the same time, `clflush` evicts the memory line from the cache for the next loop round of the attack.

The measurement is done using the `rdtsc` instruction that provides a sub-nanosecond resolution timestamp. It also uses `mfence` instructions, as `clflush` is only ordered by `mfence`, but not by any other means.

Figure 8.1 shows the execution time histogram of the `clflush` instruction for cached and non-cached memory lines, run on the three setups with different recent microarchitectures: a Sandy Bridge i5-2540M, an Ivy Bridge i5-3320M and a Haswell i7-4790. The timing difference of the peaks is 12 cycles on Sandy Bridge, 9 cycles on Ivy Bridge, and 12 cycles on Haswell. If the address maps to a remote core, another penalty of 3 cycles is added to the minimum execution time for cache hits. The difference is enough to be observed by an attacker. We discuss this timing difference and its implications in Section 9.1. In either case the execution time is less than the access time for both memory cached in the last-level cache and memory accesses that are not cached. Therefore, *Flush+Flush* is significantly faster than any other last-level cache attack.

The *Flush+Flush* attack inherently has a slightly lower accuracy than the *Flush+Reload* technique in some cases, due to the lower timing difference between a hit and a miss and because of a lower access time on average. Nevertheless, the same amount of information is extracted faster using the *Flush+Flush* attack due to the significantly lower execution time. Furthermore, the reload-step of the *Flush+Reload* attack can trigger the prefetcher and thus destroy measurements by fetching data into the cache. This is the case especially when monitoring more than one address within a physical page [GSM15]. As the *Flush+Flush* attack never performs any memory accesses, this problem does not exist and the *Flush+Flush* attack achieves an even higher accuracy here. For the same reason, the *Flush+Flush* attack causes no cache misses and only a minimal number of cache hits. Thus, recently proposed detection mechanisms using cache references and cache misses fail to detect *Flush+Flush*.

# 4. Detecting Cache Attacks with Hardware Performance Counters

Cache attacks can lead to an increased number of cache hits or cache misses in the attacker process or in other processes. Thus, it may be possible to detect abnormal behavior on a system level. However, to stop or prevent an attack, it is necessary to identify the attacking process. Therefore, we consider an attack *stealthy* if the attacking spy process cannot be identified.

Hardware performance counters are special-purpose registers that are used to monitor special hardware-related events. Events that can be monitored include cache references and cache misses on the last-level cache. They are mostly used for performance analysis and fine tuning, but have been found to be suitable to detect Rowhammer and the *Flush+Reload* attack [HF15; CSY15; Pay16]. The focus of our work is to show that detection of existing attacks is straightforward, but detection of the *Flush+Flush* attack using these performance counters is infeasible, due to the absence of cache misses and the minimal number of cache references.

| Name | Description |
|---|---|
| BPU_RA/_RM | Branch prediction unit read accesses/misses |
| BRANCH_INSTRUCTIONS/_MISSES | Retired branch instructions/mispredictions |
| BUS_CYCLES | Bus cycles |
| CACHE_MISSES/_REFERENCES | Last-level cache misses/references |
| UNC_CBO_CACHE_LOOKUP | C-Box events incl. `clflush` (all slices) |
| CPU_CYCLES/REF_CPU_CYCLES | CPU cycles with/without scaling |
| DTLB_RA/_RM/_WA/_WM | Data TLB read/write accesses/misses |
| INSTRUCTIONS | Retired instructions |
| ITLB_RA/_RM | Instruction TLB read/write accesses |
| L1D_RA/_RM/_WA/_WM | L1 data cache read/write accesses/misses |
| L1I_RM | L1 instruction cache read misses |
| LL_RA/_WA | Last-level cache read/write accesses |

Table 8.1.: List of hardware performance events we use.

We analyze the feasibility of such detection mechanisms using the Linux `perf_event_open` syscall interface that provides userspace access to a subset of all available performance counters on a per-process basis. The actual accesses to the model specific registers are performed in the kernel. The same information can be used by a system service to detect ongoing attacks. During our tests we ran the performance monitoring with system service privileges.

We analyzed all 23 hardware and cache performance events available with the Linux syscall interface on our system. Additionally, we analyzed the so called *uncore* [Int14] performance monitoring units and found one called C-Box that is influenced by cache hits, misses and `clflush` instructions directly. The UNC_CBO_CACHE_LOOKUP event of the C-Box allows monitoring a last-level cache lookups per cache slice, including by the `clflush` instruction. The C-Box monitoring units are not available through a generic interface but only through model specific registers. Table 8.1 lists all events we evaluated. We found that there are no other performance counters documented to monitor cache hits, misses or `clflush` instructions specifically. Furthermore, neither the hypervisor nor the operating system can intercept the `clflush` instruction or monitor the frequency of `clflush` instructions being executed using performance counters.

The number of performance events that can be monitored simultaneously is limited by hardware. On all our test systems it is possible to monitor up to 4 events simultaneously. Thus, any detection mechanism can only use 4 performance events simultaneously.

We evaluated the 24 performance counters for the following scenarios:

1. Idle: idle system,

2. Firefox: user scrolling down a chosen Twitter feed in Firefox,

3. OpenTTD: user playing a game

4. stress -m 1: loop reading and writing in dynamically allocated 256 MB arrays,

5. stress -c 1: loop doing a CPU computation with almost no memory,

6. stress -i 1: loop calling the I/O `sync()` function,

7. *Flush+Reload*: cache attack on the GTK library to spy on keystroke events,

8. Rowhammer: Rowhammer attack.

The first 3 scenarios are casual computer usage scenarios, the next 3 cause a benign high load situation and the last 2 perform an attack. A good detection mechanism classifies as benign the scenarios 1 to 6 and as attacks 7 and 8.

We use the instruction TLB (ITLB) performance counters (`ITLB_RA` + `ITLB_WA`) to normalize the performance counters to make cache attacks easier to detect, and prevent scenarios 2 and 3 from being detected as malicious. Indeed, the main loop that is used in the *Flush+Reload* and Rowhammer attacks causes a high number of last-level cache misses while executing only a small piece of code. Executing only a small piece of code causes a low pressure on the ITLB.

Table 8.2 shows a comparison of performance counters for the 8 scenarios tested over 135 seconds. These tests were performed in multiple separate runs as the performance monitoring unit can only monitor 4 events simultaneously. Not all cache events are suitable for detection. The `UNC_CBO_CACHE_LOOKUP` event that counts cache slice events including `clflush` operations shows very high values in case of `stress -i`. It would thus lead to false positives. Similarly, the `INSTRUCTIONS` event used by Chiappetta et al. [CSY15] has a significantly higher value in case of `stress`

| Event / Test | Idle | Firefox | OTTD | stress -m | stress -c | stress -i | F+R | Rowhammer |
|---|---|---|---|---|---|---|---|---|
| BPU_RA | 4.35 | 14.73 | 67.21 | 92.28 | 6 109 276.79 | 3.23 | 127 443.28 | 23 778.66 |
| BPU_RM | 0.36 | 0.32 | 1.87 | 0.00 | 12 320.23 | 0.36 | 694.21 | 25.53 |
| BRANCH_INST. | 4.35 | 14.62 | 74.73 | 92.62 | 6 094 264.03 | 3.23 | 127 605.71 | 23 834.59 |
| BRANCH_MISS. | 0.36 | 0.31 | 2.06 | 0.00 | 12 289.93 | 0.35 | 693.97 | 25.85 |
| BUS_CYCLES | 4.41 | 1.94 | 12.39 | 52.09 | 263 816.26 | 6.20 | 30 420.54 | 98 406.44 |
| CACHE_MISSES | 0.09 | 0.15 | 2.35 | 58.53 | 0.06 | 1.92 | 693.67 | 13 766.65 |
| CACHE_REFER. | 0.40 | 0.98 | 6.84 | 61.05 | 0.31 | 2.28 | 693.92 | 13 800.01 |
| UNC_CBO_LOO | 432.99 | 3.88 | 18.66 | 4 166.71 | 0.31 | 343 224.44 | 2 149.72 | 50 094.17 |
| CPU_CYCLES | 38.23 | 67.45 | 449.23 | 2 651.60 | 9 497 363.56 | 237.62 | 1 216 701.51 | 3 936 969.93 |
| DTLB_RA | 5.11 | 19.19 | 123.68 | 31.78 | 6 076 031.42 | 3.04 | 47 123.44 | 25 459.36 |
| DTLB_RM | 0.07 | 0.09 | 1.67 | 0.05 | 0.05 | 0.04 | 0.05 | 0.03 |
| DTLB_WA | 1.70 | 11.18 | 54.88 | 30.97 | 3 417 764.10 | 1.13 | 22 868.02 | 25 163.03 |
| DTLB_WM | 0.01 | 0.01 | 0.03 | 2.50 | 0.01 | 0.01 | 0.01 | 0.16 |
| INSTRUCTIONS | 20.24 | 66.04 | 470.89 | 428.15 | 20 224 639.96 | 11.77 | 206 014.72 | 132 896.65 |
| ITLB_RA | 0.95 | 0.97 | 0.98 | 1.00 | 0.96 | 0.97 | 0.96 | 0.97 |
| ITLB_RM | 0.05 | 0.03 | 0.02 | 0.00 | 0.04 | 0.03 | 0.04 | 0.03 |
| L1D_RA | 5.11 | 18.30 | 128.75 | 31.53 | 6 109 271.97 | 3.01 | 47 230.08 | 26 173.65 |
| L1D_RM | 0.37 | 0.82 | 8.47 | 61.63 | 0.51 | 0.62 | 695.22 | 15 630.85 |
| L1D_WA | 1.70 | 10.69 | 57.66 | 30.72 | 3 436 461.82 | 1.13 | 22 919.77 | 25 838.20 |
| L1D_WM | 0.12 | 0.19 | 1.50 | 30.57 | 0.16 | 0.44 | 0.23 | 10.01 |
| L1I_RM | 0.12 | 0.65 | 0.21 | 0.03 | 0.65 | 1.05 | 1.17 | 1.14 |
| LL_RA | 0.14 | 0.39 | 5.61 | 30.73 | 0.12 | 0.47 | 695.35 | 9 067.77 |
| LL_WA | 0.01 | 0.02 | 0.74 | 30.30 | 0.01 | 0.01 | 0.02 | 4 726.97 |
| REF_CPU_CYC. | 157.70 | 69.69 | 445.89 | 1 872.05 | 405 922.02 | 223.08 | 1 098 534.32 | 3 542 570.00 |

Table 8.2.: Comparison of performance counters normalized to the number of ITLB events in different cache attacks and normal scenarios over 135 seconds in separate runs.

-c than in the attack scenarios and would cause false positives in the case of benign CPU intensive activities. The `REF_CPU_CYCLES` is the unscaled total number of CPU cycles consumed by the process. Divided by the TLB events, it shows how small the executed loop is. The probability of false positive matches is high, for instance in the case of `stress -c`.

Thus, 4 out of 24 events allow detecting both *Flush+Reload* and Rowhammer without causing false positives for benign applications. The rationale behind these events is as follows:

1. `CACHE_MISSES` occur after data has been flushed from the last-level cache,

2. `CACHE_REFERENCES` occur when reaccessing memory,

3. `L1D_RM` occur because flushing from last-level cache also flushes from the lower cache levels,

4. `LL_RA` are a subset of the `CACHE_REFERENCES` counter, they occur when reaccessing memory,

Two of the events are redundant: `L1D_RM` is redundant with `CACHE_MISSES`, and `LL_RA` with `CACHE_REFERENCES`. We will thus focus on the `CACHE_MISSES` and `CACHE_REFERENCES` events as proposed in previous work [HF15; CSY15; Pay16].

We define that a process is considered malicious if more than $k_m$ cache miss or $k_r$ cache reference per ITLB event are observed. The attack is detected if

$$\frac{C_{\text{CACHE\_MISSES}}}{C_{\text{ITLB\_RA}} + C_{\text{ITLB\_WA}}} \geq k_m, \quad \text{or} \quad \frac{C_{\text{CACHE\_REFERENCES}}}{C_{\text{ITLB\_RA}} + C_{\text{ITLB\_WA}}} \geq k_r,$$

with $C$ the value of the corresponding performance counter. The operating system can choose the frequency in which to run the detection checks.

The thresholds for the cache reference and cache hit rate are determined based on a set of benign applications and malicious applications. It is chosen to have the maximum distance to the minimum value for any malicious application and the maximum value for any benign application. In our case this is $k_m = 2.35$ and $k_r = 2.34$. Based on these thresholds, we perform a classification of processes into malicious and benign processes. We tested this detection mechanism against various cache attacks and found that it is suitable to detect different *Flush+Reload*, *Prime+Probe* and Rowhammer attacks as malicious. However, the focus of our work is not the evaluation of detection mechanisms based on performance counters, but to show that such detection mechanisms cannot reliably detect the *Flush+Flush* attack due to the absence of cache misses and a minimal number of cache references.

In the following sections, we evaluate the performance and the detectability of *Flush+Flush* compared to the state-of-the-art cache attacks *Flush+Reload* and *Prime+Probe* in three scenarios: a covert channel, a side channel on user input and a side channel on AES with T-tables.

# 5. Covert Channel Comparison

In this section, we describe a generic low-error cache covert channel framework. In a covert channel, an attacker runs two unprivileged applications on the system under attack. The processes are cooperating to communicate with each other, even though they are not allowed to by the security policy. We show how the two processes can communicate using the *Flush+Flush*,

*Flush+Reload*, and *Prime+Probe* technique. We compare the performance and the detectability of the three implementations. In the remainder of the paper, all the experiments are performed on a Haswell i7-4790 CPU.

## 5.1. A Low-error Cache Covert Channel Framework

In order to perform meaningful experiments and obtain comparable and fair results, the experiments must be reproducible and tested in the same conditions. This includes the same hardware setup, and the same protocols. Indeed, we cannot compare covert channels from published work [Mau+15b; Liu+15] that have different capacities and error rates. Therefore, we build a framework to evaluate covert channels in a reproducible way. This framework is generic and can be implemented over any covert channel that allows bidirectional communication, by implementing the `send()` and `receive()` functions.

The central component of the framework is a simple transmission protocol. Data is transmitted in packets of $N$ bytes, consisting of $N - 3$ bytes payload, a 1 byte sequence number and a CRC-16 checksum over the packet. The sequence number is used to distinguish consecutive packets. The sender retransmits packets until the receiver acknowledges it. Packets are acknowledged by the receiver if the checksum is valid.

Although errors are still possible in case of a false positive CRC-16 checksum match, the probability is low. We choose the parameters such that the effective error rate is below 5%. The channel capacity measured with this protocol is comparable and reproducible. Furthermore, it is close to the effective capacity in a real-world scenario, because error-correction cannot be omitted. The number of transmitted bits is the minimum of bits sent and bits received. The transmission rate can be computed by dividing the number of transmitted bits by the runtime. The error rate is given by the number of all bit errors between the sent bits and received bits, divided by the number of transmitted bits.

## 5.2. Covert Channel Implementations

We first implemented the *Flush+Reload* covert channel. By accessing fixed memory locations in a shared library the a 1 is transmitted, whereas a 0 is transmitted by omitting the access. The receiver performs the actual *Flush+Reload* attack to determine whether a 1 or a 0 was transmitted. The bits retrieved are then parsed as a data frame according to the transmission protocol. The sender also monitors some memory locations using *Flush+Reload* for cache hits too, to receive packet acknowledgments.

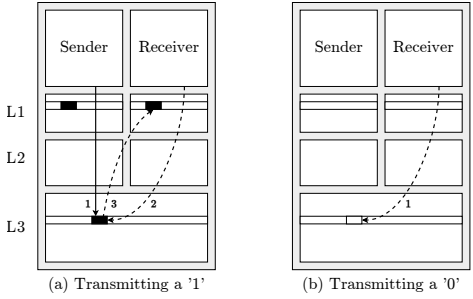

(a) Transmitting a '1'     (b) Transmitting a '0'

Figure 8.2.: Illustration of the *Flush+Flush* covert channel.

The second implementation is the *Flush+Flush* covert channel, illustrated by Figure 8.2. It uses the same sender process as the *Flush+Reload* covert channel. To transmit a 1 (Figure 8.2-a), the sender accesses the memory location, that is cached (step 1). This time, the receiver only flushes the shared line. As the line is present in the last-level cache by inclusiveness, it is flushed from this level (step 2). A bit also indicates that the line is present in the L1 cache, and thus must also be flushed from this level (step 3). To transmit a 0 (Figure 8.2-b), the sender stays idle. The receiver flushes the line (step 1). As the line is not present in the last-level cache, it means that it is also not present in the lower levels, which results in a faster execution of the `clflush` instruction. Thus only the sender process performs memory accesses, while the receiver only flushes cache lines. To send acknowledgment bytes the receiver performs memory accesses and the sender runs a *Flush+Flush* attack.

The third implementation is the *Prime+Probe* covert channel. It uses the same attack technique as Liu et al. [Liu+15], Oren et al. [Ore+15], and Maurice et al. [Mau+15b]. The sender transmits a 1 bit by priming a cache set. The receiver probes the same cache set. Again the receiver determines whether a 1 or a 0 was transmitted. We make two adjustments for convenience and to focus solely on the transmission part. First, we compute a static eviction set by using the complex addressing function [Mau+15a] on physical addresses. This avoids the possibility of errors introduced by timing-based eviction set computation. Second, we map the shared library into our address space to determine the physical address to attack to make an agreement on the cache sets in sender and receiver. Yet, the shared library is never accessed and unmapped even before the *Prime+Probe* attack is started. We assume that the sender and receiver have agreed on the cache sets in a preprocessing step. This is practical even for a timing-based approach.

| Technique | Packet size | Capacity in KB/s | Error rate | Sender refer- ences | Sender misses | Sender stealth | Receiver refer- ences | Receiver misses | Receiver stealth |
|---|---|---|---|---|---|---|---|---|---|
| *Flush+Flush* | 28 | 496 | 0.84% | 1809.26 | 96.66 | ✗ | 1.75 | 1.25 | ✓ |
| *Flush+Reload* | 28 | 298 | 0.00% | 526.14 | 56.09 | ✗ | 110.52 | 59.16 | ✗ |
| *Flush+Reload* | 5 | 132 | 0.01% | 6.19 | 3.20 | ✗ | 45.88 | 44.77 | ✗ |
| *Flush+Flush* | 5 | 95 | 0.56% | 425.99 | 418.27 | ✗ | 0.98 | 0.95 | ✓ |
| *Prime+Probe* | 5 | 67 | 0.36% | 48.96 | 31.81 | ✗ | 4.64 | 4.45 | ✗ |
| *Flush+Reload* | 4 | 54 | 0.00% | 0.86 | 0.84 | ✓ | 2.74 | 1.25 | ✗ |
| *Flush+Flush* | 4 | 52 | 1.00% | 0.06 | 0.05 | ✓ | 0.59 | 0.59 | ✓ |
| *Prime+Probe* | 4 | 34 | 0.04% | 55.57 | 32.66 | ✗ | 5.23 | 5.01 | ✗ |

Table 8.3.: Comparison of capacity and detectability of the three cache covert channels with different parameters. *Flush+Flush* and *Flush+Reload* use the same sender process.

## 5.3. Performance Evaluation

Table 8.3 compares the capacity and the detectability of the three covert channels in different configurations. The *Flush+Flush* covert channel is the fastest of the three covert channels. With a packet size of 28 bytes the transmission rate is 496 KB/s. At the same time the effective error rate is only 0.84%. The *Flush+Reload* covert channel also achieved a good performance at a packet size of 28 bytes. The transmission rate then is 298 KB/s and the error rate $< 0.005\%$. With a packet size of 4 bytes, the performance is lower in all three cases.

A *Prime+Probe* covert channel with a 28-byte packet size is not realistic. First, to avoid triggering the hardware prefetcher we do not access more than one address per physical page. Second, for each eviction set we need 16 addresses. Thus we would require $28B \cdot 4096 \cdot 16 = 14\,GB$ of memory only for the eviction sets. For *Prime+Probe* we achieved the best results with a packet size of 5 bytes. With this configuration the transmission rate is 68 KB/s at an error rate of 0.14%, compared to 132 KB/s using *Flush+Reload* and 95 KB/s using *Flush+Flush*.

The *Flush+Flush* transmission rate of 496 KB/s is significantly higher than any other state-of-the-art cache covert channels. It is 6.7 times as fast as the fastest cache covert channel to date [Liu+15] at a comparable error rate. Our covert channel based on *Flush+Reload* is also faster than previously published cache covert channels, but still much slower than the *Flush+Flush* covert channel. Compared to our *Prime+Probe* covert channel, *Flush+Flush* is 7.3 times faster.

## 5.4. Detectability

Table 8.3 shows the evaluation of the detectability for packet sizes that yielded the highest performance in one of the cases. *Flush+Reload* and *Flush+Flush* use the same sender process, the reference and miss count is mainly influenced by the number of retransmissions and executed program logic. *Flush+Reload* is detected in all cases either because of its sender or its receiver, although its sender process with a 4-byte packet size stays below the detection threshold. The *Prime+Probe* attack is always well above the detection threshold and therefore always detected as malicious. All *Flush+Flush* receiver processes are classified as benign. However, only the sender process used for the *Flush+Flush* and the *Flush+Reload* covert channels with a 4-byte packet size is classified as benign.

The receiver process performs most of the actual cache attack. If it is sufficient to keep the receiver process stealthy, *Flush+Flush* clearly outperforms all other cache attacks. If the sender has to be stealthy as well, the sender process used by *Flush+Flush* and *Flush+Reload* performs better than the *Prime+Probe* sender process. However, due to the high number of cache hits it is difficult to keep the sender process below the detection threshold. An adversary could choose to reduce the transmission rate in order to be stealthier in either case.

# 6. Side-Channel Attack on User Input

Another cache attack that has been demonstrated recently using *Flush+Reload*, is eavesdropping on keystroke timings. We attack an address in the GTK library invoked when processing keystrokes. The attack is implemented as a program that constantly flushes the address, and derives when a keystroke occurred, based on memory access times or the execution time of the `clflush` instruction.

| Technique | Cache references | Cache misses | Stealthy |
|-----------|------------------|--------------|----------|
| *Flush+Reload* | 5.140 | 5.138 | <span style="color:red">✗</span> |
| ***Flush+Flush*** | 0.002 | 0.000 | <span style="color:green">✓</span> |

Table 8.4.: Comparison of performance counters normalized to the number of ITLB events for cache attacks on user input.

## 6.1. Performance Evaluation

We compare the three attacks *Flush+Flush*, *Flush+Reload*, and *Prime+Probe*, based on their performance in this side-channel attack scenario. During each test we simulate a user typing a 1000-character text into an editor. Each test takes 135 seconds. As expected, *Flush+Reload* has a very high accuracy of 96.1%. This allows direct logging of keystroke timings. *Flush+Flush* performs notably well, with 74.7% correctly detected keystrokes. However, this makes a practical attack much harder than with *Flush+Reload*. The attack with *Prime+Probe* yielded no meaningful results at all due to the high noise level. In case of *Flush+Reload* and *Flush+Flush* the accuracy can be increased significantly by attacking 3 addresses that are used during keystroke processing simultaneously. The decision whether a keystroke was observed is then based on these 3 addresses increasing the accuracy significantly. Using this technique reduces the error rate in case of *Flush+Reload* close to 100% and above 92% in case of *Flush+Flush*.

## 6.2. Detectability

To evaluate the detectability we again monitored the cache references and cache misses events, and compared the three cache attacks with each other and with an idle system. Table 8.4 shows that *Flush+Reload* generates a high number of cache references, whereas *Flush+Flush* causes a negligible number of cache references. We omitted *Prime+Probe* in this table as it was not sufficiently accurate to perform the attack.

*Flush+Reload* yields the highest accuracy in this side-channel attack, but it is easily detected. The accuracy of *Flush+Flush* can easily be increased to more than 92% and it still is far from being detected. Thus, *Flush+Flush* is a viable and stealthy alternative to the *Flush+Reload* attack as it is not classified as malicious based on the cache references or cache misses performance counters.

# 7. Side-Channel Attack on AES with T-Tables

To round up our comparison with other cache attacks, we compare *Flush+Flush*, *Flush+Reload*, and *Prime+Probe* in a high frequency side-channel attack scenario. Finding new cache attacks is out of scope of our work. Instead, we try to perform a fair comparison between the different attack techniques by implementing a well known cache attack using the three techniques on a vulnerable implementation of a cryptographic algorithm. We attack the OpenSSL T-Table-based AES implementation that is known to be susceptible to cache attacks [Ber05; OST06]. This AES implementation is disabled by default for security reasons, but still exists for the purpose of comparing new and existing side-channel attacks.

The AES algorithm uses the T-tables to compute the ciphertext based on the secret key $k$ and the plaintext $p$. During the first round, table accesses are made to entries $T_j[p_i \oplus k_i]$ with $i \equiv j \mod 4$ and $0 \leq i < 16$. Using a cache attack it is possible to derive values for $p_i \oplus k_i$ and thus, possible key-byte values $k_i$ in case $p_i$ is known.

## 7.1. Attack Implementation Using Flush+Flush

The implementation of the chosen-plaintext attack side-channel attacks for the three attack techniques is very similar. The attacker triggers an encryption, choosing $p_i$ while all $p_j$ with $i \neq j$ are random. One cache line holds 16 T-Table entries. The cache attack is now performed on the first line of each T-Table. The attacker repeats the encryptions with new random plaintext bytes $p_j$ until only one $p_i$ remains to always cause a cache hit. The attacker learns that $p_i \oplus k_i \equiv_{\lceil 4 \rceil} 0$ and thus $k_i \equiv_{\lceil 4 \rceil} p_i$. After performing the attack for all 16 key bytes, the attacker has derived 64 bits of the secret key $k$. As we only want to compare the three attack techniques, we do not extend this attack to a full key recovery attack.

## 7.2. Performance Evaluation

Figure 8.3 shows a comparison of cache templates generated with *Flush+Reload*, *Flush+Flush*, and *Prime+Probe* using 1 000 000 encryptions to create a visible pattern in all three cases. Similar templates can be found in previous work [OST06; SP13; GSM15]. Table 8.5 shows how many encryptions are necessary to determine the upper 4 bits correctly. We performed encryptions until the correct guess for the upper 4 bits of key byte $k_0$ had a 5% margin over all other key candidates. *Flush+Flush* requires around 1.4 times as many encryptions as *Flush+Reload*, but 13.7 times less than *Prime+Probe* to achieve the same accuracy.

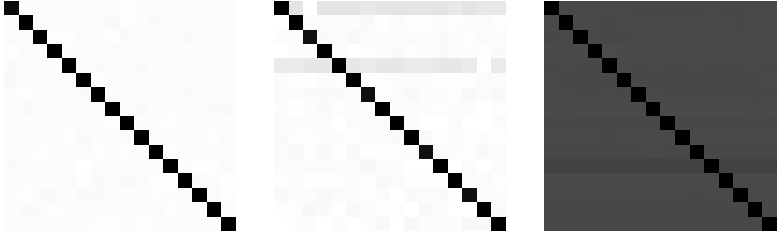

Figure 8.3.: Comparison of Cache Templates (address range of the first T-table) generated using *Flush+Reload* (left), *Flush+Flush* (middle), and *Prime+Probe* (right). In all cases $k_0 = \texttt{0x00}$.

| Technique | Number of encryptions |
|---|---|
| *Flush+Reload* | 250 |
| ***Flush+Flush*** | 350 |
| *Prime+Probe* | 4 800 |

Table 8.5.: Number of encryptions to determine the upper 4 bits of a key byte.

*Flush+Flush* is the only attack that does not trigger the prefetcher. Thus, we can monitor multiple adjacent cache sets. By doing this we double the number of cache references, but increase the accuracy of the measurements so that 275 encryptions are sufficient to identify the correct key byte with a 5% margin. That is only 1.1 times as many encryptions as *Flush+Reload* and 17.5 times less than *Prime+Probe*. Thus, *Flush+Flush* on multiple addresses is faster at deriving the same information as *Flush+Reload*.

## 7.3. Detectability

Table 8.6 shows a comparison of the performance counters for the three attacks over 256 million encryptions. The *Flush+Flush* attack took only 163 seconds whereas *Flush+Reload* took 215 seconds and *Prime+Probe* 234 seconds for the identical attack. On a system level, it is possible to notice ongoing cache attacks on AES in all three cases due to the high number of cache misses caused by the AES encryption process. However, to stop or prevent the attack, it is necessary to detect the spy process. *Prime+Probe* exceeds the detection threshold by a factor of 468 and *Flush+Reload* exceeds the threshold by a factor of 1070. To stay below the detection threshold, slowing down the attack by at least the same factor would be necessary. In contrast, *Flush+Flush* is not detected based on our classifier and does not have to be slowed down to be stealthy.

| Technique | Cache references | Cache misses | Execution time in s | References (norm.) | Misses (norm.) | Stealthy |
|---|---|---|---|---|---|---|
| *Flush+Reload* | $1\,024 \cdot 10^6$ | $19\,284\,602$ | 215 | $2\,513.43$ | 47.33 | ✗ |
| *Prime+Probe* | $4\,222 \cdot 10^6$ | $294\,897\,508$ | 234 | $1\,099.63$ | 76.81 | ✗ |
| **Flush+Flush** | $768 \cdot 10^6$ | $1\,741$ | 163 | 1.40 | 0.00 | ✓ |

Table 8.6.: Comparison of the performance counters when performing 256 million encryptions with different cache attacks and without an attack.

# 8. Discussion

## 8.1. Using `clflush` to Detect Cores and Cache Slices

The *Flush+Flush* attack can be used to determine on which CPU core a process is running or to which cache slice an address maps. Indeed, a `clflush` on a remote cache slice takes longer than a `clflush` on a local cache slice, as shown in Figure 8.4. This is due to the ring bus architecture connecting remote slices. Knowing the physical address of a memory access on a local slice, we can then use the complex addressing function [Mau+15a] to determine on which core the process runs. However, this would require high privileges. Yet, it is possible to determine to which slice an address maps without knowing the physical address by performing a timing attack. This can be done by an unprivileged process, as pinning a thread to a CPU core requires no privileges.

This can be exploited to detect colocation on the same CPU, CPU core or hyperthreading core in restricted environments even if the `cpuid` instructions is virtualized. It is more difficult to determine which CPU core a thread runs on based on memory access timings because of the influence of lower level caches. Such an attack has also not been demonstrated yet. The information on the executing CPU core can be used to enhance cache attacks and other attacks such as the Rowhammer attack [Kim+14;

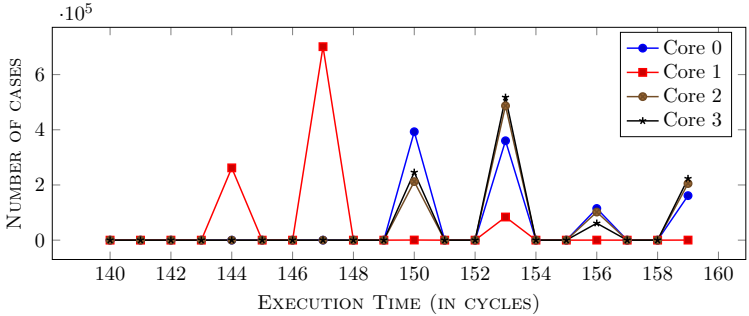

Figure 8.4.: Excerpt of the `clflush` histogram for an address in slice 1 on different cores. The lower execution time on core 1 shows that this address maps to slice 1.

GMM16]. Running `clflush` on a local slice lowers the execution time of each Rowhammer loop round by a few cycles. The probability of bit flips increases as the execution time lowers, thus we can leverage the information whether an address maps to a local slice to improve this attack.

A similar timing difference also occurs upon memory accesses that are served from the local or a remote slice respectively. The reason again is the direct connection to the local cache slice while remote cache slices are connected via a ring bus. However, as memory accesses will also be cached in lower level caches, it is more difficult to observe the timing difference without `clflush`. The `clflush` instruction directly manipulates the last-level cache, thus lower level caches cannot hide the timing difference.

While the operating system can restrict access on information such as the CPU core the process is running on and the physical address mapping to make efficient cache attacks harder, it cannot restrict access to the `clflush` instruction. Hence, the effect of such countermeasures is lower than expected.

## 8.2. Countermeasures

We suggest modifying the `clflush` instruction to counter the wide range of attacks that it can be used for. The difference in the execution time of `clflush` is 3 cycles depending on the cache slice and less than 12 cycles depending on whether it is a cache miss. In practice the `clflush` instruction is used only in rare situations and not in a high frequency. Thus, a hypothetical performance advantage cannot justify introducing these exploitable timing differences. We propose making `clflush` a constant-time instruction. This would prevent the *Flush+Flush* attack completely, as well as information leakage on cache slices and CPU cores.

*Flush+Flush* is the only cache attack that does not perform any memory accesses and thus causes no cache misses and only a minimal number of cache references. One theoretical way to detect our attack would be to monitor each load, e.g., by timing, and to stop when detecting too many misses. However, this solution is currently not practical, as a software-based solution that monitors each load would cause a significant performance degradation. A similar hardware-based solution called *informing loads* has been proposed by Kong et al. [Kon+09], however it needs a change in the instruction set. Without hardware modifications it would be possible to enable the `rdtsc` instruction only in privileged mode as can be done using seccomp on Linux [lwn08] since 2008. Fogh [Fog15] proposed to simulate the `rdtsc` in an interrupt handler, degrading the accuracy of measurements far enough to make cache attacks significantly harder.

*Flush+Reload* and *Flush+Flush* both require shared memory. If shared memory is not available, an attacker would have to resort to a technique that even works without shared memory such as *Prime+Probe*. Furthermore, making the `clflush` instruction privileged would prevent *Flush+Reload* and *Flush+Flush* as well. However, this would require changes in hardware and could not be implemented in commodity systems.

# 9. Related work

## 9.1. Detecting and Preventing Cache Attacks

Zhang et al. [Zha+11] proposed HomeAlone, a system-level solution that uses a *Prime+Probe* covert channel to *detect* the presence of a foe co-resident virtual machine. The system monitors random cache sets so that friendly virtual machines can continue to operate if they change their workload, and that foe virtual machines are either detected or forced to be silent. Cache Template Attacks [GSM15] can be used to detect attacks on shared libraries and binaries as a user. However, such a permanent scan increases the system load and can only detect attacks in a small address range within a reasonable response time.

Herath and Fogh [HF15] proposed to monitor cache misses to detect *Flush+Reload* attacks and Rowhammer. The system would slow down or halt all attacker processes. With the detection mechanism we implemented, we show that this technique is feasible for previous attacks but not for the *Flush+Flush* attack. Chiappetta et al. [CSY15] proposed to build a trace of cache references and cache misses over the number of executed instructions to detect *Flush+Reload* attacks. They then proposed three methods to analyze this trace: a correlation-based method, and two other ones based on machine learning techniques. However, a learning phase is needed to detect malicious programs that are either from a set of known malicious programs or resemble a program from this set. They are thus are less likely to detect new or unknown cache attacks or Rowhammer attacks, in contrast to our ad-hoc detection mechanism. Payer [Pay16] proposed a system called HexPADS to use cache references, cache misses, but also other events like page faults to detect cache attacks and Rowhammer at runtime.

Cache attacks can be *prevented* at three levels: at the hardware level, at the system level, and finally, at the application level. At the hardware level, several solutions have been proposed to prevent cache attacks, either by removing cache interferences, or randomizing them. The solutions include new secure cache designs [WL07; WL08; LL14] or altering the prefetcher policy [FL15]. However, hardware changes are not applicable to commodity systems. At the system level, page coloring provides cache isolation in software [Raj+09; KPM12]. Zhang et al. [ZR13] proposed a more relaxed isolation like repeated cache cleansing. These solutions cause performance issues, as they prevent optimal use of the cache. Application-level countermeasures seek to find the source of information leakage and patch it [Bri+06]. However, application-level countermeasures are bounded and cannot prevent cache attacks such as covert channels and Rowhammer. In contrast with prevention solutions that incur a loss of performance, using performance counters does not prevent attacks but rather detect them without overhead.

## 9.2. Usage of Hardware Performance Counters in Security

Hardware performance counters are made for performance monitoring, but security researchers found other applications. In defensive cases, performance counters allow detection of malware [Dem+13], integrity checking of programs [MZK11], control flow integrity [Xia+12], and binary analysis [Wil+12]. In offensive scenarios, it has been used for side-channel attacks against AES [UGV08] and RSA [BM15]. Performance counters have also been used by Maurice et al. [Mau+15a] to reverse engineer the complex addressing function of the last-level cache of modern Intel CPUs.

## 9.3. Cache Covert Channels

Cache covert channels are a well-known problem, and have been studied relatively to the recent evolutions in microarchitecture. The two main types of access-driven attacks can be used to derive a covert channel. Covert channels using *Prime+Probe* have already been demonstrated in [Mau+15b; Liu+15]. *Flush+Reload* has been used for side-channels attacks [YF14], thus a covert channel can be derived easily. However, to the best of our knowledge, there was no study of the performance of such a covert channel.

In addition to building a covert channel with our new attack *Flush+Flush*, we re-implemented *Prime+Probe* and implemented *Flush+Reload*.[1] We thus provide an evaluation and a fair comparison between these different covert channels, in the same hardware setup and with the same protocol.

## 9.4. Side-Channel Attacks on User Inputs

Section 6 describes a side channel to eavesdrop on keystrokes. If an attacker has root access to a system, there are simple ways to implement a keylogger. Without root access, software-based side-channel attacks have already proven to be a reliable way to eavesdrop on user input. Attacks exploit the execution time [Tan+08], peaks in CPU and cache activity graphs [Ris+09], or system services [ZW09]. Zhang et al. [ZW09] showed that it is possible to derive key sequences from inter-keystroke timings obtained via `procfs`. Oren et al. [Ore+15] demonstrated that cache attacks in sandboxed JavaScript inside a browser can derive user activities, such as mouse movements. Gruss et al. [GSM15] showed that auto-generated *Flush+Reload* attacks can be used to measure keystroke timings as well as identifying keys.

---

[1]After public disclosure of the *Flush+Flush* attack on November 14, 2015, *Flush+Flush* has also been demonstrated on ARM-based mobile devices [Lip+15].

# 10. Conclusion

In this paper we presented *Flush+Flush*, a novel cache attack that, unlike any other, performs no memory accesses. Instead, it relies only on the execution time of the flush instruction to determine whether data is cached. *Flush+Flush* does not trigger prefetches and thus is applicable in more situations than other attacks. The *Flush+Flush* attack is faster than any existing cache attack. It achieves a transmission rate of 496 KB/s in a covert channel scenario, which is 6.7 times faster than any previous cache covert channel. As it performs no memory accesses, the attack causes no cache misses at all. For this reason, detection mechanisms based on performance counters to monitor cache activity fail, as their underlying assumption is incorrect.

While the *Flush+Flush* attack is significantly harder to detect than existing cache attacks, it can be prevented with small hardware modifications. Making the `clflush` instruction constant-time has no measurable impact on today's software and does not introduce any interface changes. Thus, it is an effective countermeasure that should be implemented.

Finally, the experiments led in this paper broaden the understanding of the internals of modern CPU caches. Beyond the adoption of detection mechanisms, the field of cache attacks benefits from these findings, both to discover new attacks and to be able to prevent them.

# 11. Acknowledgments

We would like to thank Mathias Payer, Anders Fogh, and our anonymous reviewers for their valuable comments and suggestions.

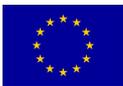 Supported by the EU Horizon 2020 programme under GA No. 644052 (HECTOR), the EU FP7 programme under GA No. 610436 (MATTHEW), the Austrian Research Promotion Agency (FFG) and Styrian Business Promotion Agency (SFG) under GA No. 836628 (SeCoS), and Cryptacus COST Action IC1403.

# ARMageddon: Cache Attacks on Mobile Devices

## Publication Data

M. Lipp, D. Gruss, R. Spreitzer, C. Maurice, and S. Mangard. ARMageddon: Cache Attacks on Mobile Devices. In: USENIX Security Symposium. 2016

## Contributions

Idea. 50% of Text.

# ARMageddon:
# Cache Attacks on Mobile Devices

Moritz Lipp, Daniel Gruss, Raphael Spreitzer, Clémentine Maurice, and Stefan Mangard

Graz University of Technology, Austria

# Abstract

In the last 10 years, cache attacks on Intel x86 CPUs have gained increasing attention among the scientific community and powerful techniques to exploit cache side channels have been developed. However, modern smartphones use one or more multi-core ARM CPUs that have a different cache organization and instruction set than Intel x86 CPUs. So far, no cross-core cache attacks have been demonstrated on non-rooted Android smartphones. In this work, we demonstrate how to solve key challenges to perform the most powerful cross-core cache attacks *Prime+Probe*, *Flush+Reload*, *Evict+Reload*, and *Flush+Flush* on non-rooted ARM-based devices without any privileges. Based on our techniques, we demonstrate covert channels that outperform state-of-the-art covert channels on Android by several orders of magnitude. Moreover, we present attacks to monitor tap and swipe events as well as keystrokes, and even derive the lengths of words entered on the touchscreen. Eventually, we are the first to attack cryptographic primitives implemented in Java. Our attacks work across CPUs and can even monitor cache activity in the ARM TrustZone from the normal world. The techniques we present can be used to attack hundreds of millions of Android devices.

# 1. Introduction

Cache attacks represent a powerful means of exploiting the different access times within the memory hierarchy of modern system architectures. Until recently, these attacks explicitly targeted cryptographic implementations, for instance, by means of cache timing attacks [Ber04] or the well-known

*Evict+Time* and *Prime+Probe* techniques [OST06]. The seminal paper by Yarom and Falkner [YF14] introduced the so-called *Flush+Reload* attack, which allows an attacker to infer which specific parts of a binary are accessed by a victim program with an unprecedented accuracy and probing frequency. Recently, Gruss et al. [GSM15] demonstrated the possibility to use *Flush+Reload* to automatically exploit cache-based side channels via cache template attacks on Intel platforms. *Flush+Reload* does not only allow for efficient attacks against cryptographic implementations [Ben+14; Ira+14; PSY15], but also to infer keystroke information and even to build keyloggers on Intel platforms [GSM15]. In contrast to attacks on cryptographic algorithms, which are typically triggered multiple times, these attacks require a significantly higher accuracy as an attacker has only one single chance to observe a user input event.

Although a few publications about cache attacks on AES T-table implementations on mobile devices exist [Bog+10; WHS12; SP13b; SP13a; SG14], the more efficient cross-core attack techniques *Prime+Probe*, *Flush+Reload*, *Evict+Reload*, and *Flush+Flush* [Gru+16] have not been applied on smartphones. In fact, there was reasonable doubt [YF14] whether these cross-core attacks can be mounted on ARM-based devices at all. In this work, we demonstrate that these attack techniques are applicable on ARM-based devices by solving the following key challenges systematically:

1. *Last-level caches are not inclusive on ARM and thus cross-core attacks cannot rely on this property.* Indeed, existing cross-core attacks exploit the inclusiveness of shared last-level caches [Liu+15; Mau+15b; Mau+15a; Ore+15; GSM15; YF14; IES15; Gru+16; Gül+15] and, thus, no cross-core attacks have been demonstrated on ARM so far. We present an approach that exploits coherence protocols and L1-to-L2 transfers to make these attacks applicable on mobile devices with non-inclusive shared last-level caches, irrespective of the cache organization.[1]

2. *Most modern smartphones have multiple CPUs that do not share a cache.* However, cache coherence protocols allow CPUs to fetch cache lines from remote cores faster than from the main memory. We utilize this property to mount both cross-core and cross-CPU attacks.

---

[1]Simultaneously to our work on ARM, Irazoqui et al. [IES16] developed a technique to exploit cache coherence protocols on AMD x86 CPUs and mounted the first cross-CPU cache attack.

3. *Except ARMv8-A CPUs, ARM processors do not support a flush instruction.* In these cases, a fast eviction strategy must be applied for high-frequency measurements. As existing eviction strategies are too slow, we analyze more than 4 200 eviction strategies for our test devices, based on Rowhammer attack techniques [GMM16].

4. *ARM CPUs use a pseudo-random replacement policy* to decide which cache line to replace within a cache set. This introduces additional noise even for robust time-driven cache attacks [SP13b; SG14]. For the same reason, *Prime+Probe* has been an open challenge [SP13a] on ARM, as an attacker needs to predict which cache line will be replaced first and wrong predictions destroy measurements. We design re-access loops that interlock with a cache eviction strategy to reduce the effect of wrong predictions.

5. *Cycle-accurate timings require root access on ARM* [ARM12] and alternatives have not been evaluated so far. We evaluate different timing sources and show that cache attacks can be mounted in any case.

Based on these building blocks, we demonstrate practical and highly efficient cache attacks on ARM.[2] We do not restrict our investigations to cryptographic implementations but also consider cache attacks as a means to infer other sensitive information—such as inter-keystroke timings or the length of a swipe action—requiring a significantly higher measurement accuracy. Besides these generic attacks, we also demonstrate that cache attacks can be used to monitor cache activity caused within the ARM TrustZone from the normal world. Nevertheless, we do not aim to exhaustively list possible exploits or find new attack vectors on cryptographic algorithms. Instead, we aim to demonstrate the immense attack potential of the presented cross-core and cross-CPU attacks on ARM-based mobile devices based on well-studied attack vectors. Our work allows to apply existing attacks to millions of off-the-shelf Android devices without any privileges. Furthermore, our investigations show that Android still employs vulnerable AES T-table implementations.

**Contributions.** The contributions of this work are:

---

[2]Source code for ARMageddon attack examples can be found at `https://github.com/IAIK/armageddon`.

- We demonstrate the applicability of highly efficient cache attacks like *Prime+Probe*, *Flush+Reload*, *Evict+Reload*, and *Flush+Flush* on ARM.
- Our attacks work irrespective of the actual cache organization and, thus, are the first last-level cache attacks that can be applied cross-core and also cross-CPU on off-the-shelf ARM-based devices. More specifically, our attacks work against last-level caches that are instruction-inclusive and data-non-inclusive as well as caches that are instruction-non-inclusive and data-inclusive.
- Our cache-based covert channel outperforms all existing covert channels on Android by several orders of magnitude.
- We demonstrate the power of these attacks by attacking cryptographic implementations and by inferring more fine-grained information like keystrokes and swipe actions on the touchscreen.

**Outline.**    The remainder of this paper is structured as follows. In Section 2, we provide information on background and related work. Section 3 describes the techniques that are the building blocks for our attacks. In Section 4, we demonstrate and evaluate fast cross-core and cross-CPU covert channels on Android. In Section 5, we demonstrate cache template attacks on user input events. In Section 6, we present attacks on cryptographic implementations used in practice as well the possibility to observe cache activity of cryptographic computations within the TrustZone. We discuss countermeasures in Section 7 and conclude this work in Section 8.

# 2.  Background and Related Work

In this section, we provide the required preliminaries and discuss related work in the context of cache attacks.

## 2.1.  CPU Caches

Today's CPU performance is influenced not only by the clock frequency but also by the latency of instructions, operand fetches, and other interactions with internal and external devices. In order to overcome the latency of system memory accesses, CPUs employ caches to buffer frequently used data in small and fast internal memories.

Modern caches organize cache lines in multiple sets, which is also known as set-associative caches. Each memory address maps to one of these cache sets and addresses that map to the same cache set are considered congruent. Congruent addresses compete for cache lines within the same set and a predefined replacement policy determines which cache line is replaced. For instance, the last generations of Intel CPUs employ an undocumented variant of least-recently used (LRU) replacement policy [GMM16]. ARM processors use a pseudo-LRU replacement policy for the L1 cache and they support two different cache replacement policies for L2 caches, namely round-robin and pseudo-random replacement policy. In practice, however, only the pseudo-random replacement policy is used due to performance reasons. Switching the cache replacement policy is only possible in privileged mode. The implementation details for the pseudo-random policy are not documented.

CPU caches can either be virtually indexed or physically indexed, which determines whether the index is derived from the virtual or physical address. A so-called tag uniquely identifies the address that is cached within a specific cache line. Although this tag can also be based on the virtual or physical address, most modern caches use physical tags because they can be computed simultaneously while locating the cache set. ARM typically uses physically indexed, physically tagged L2 caches.

CPUs have multiple cache levels, with the lower levels being faster and smaller than the higher levels. ARM processors typically have two levels of cache. If all cache lines from lower levels are also stored in a higher-level cache, the higher-level cache is called *inclusive*. If a cache line can only reside in one of the cache levels at any point in time, the caches are called *exclusive*. If the cache is neither inclusive nor exclusive, it is called *non-inclusive*. The last-level cache is often shared among all cores to enhance the performance upon transitioning threads between cores and to simplify cross-core cache lookups. However, with shared last-level caches, one core can (intentionally) influence the cache content of all other cores. This represents the basis for cache attacks like *Flush+Reload* [YF14].

In order to keep caches of multiple CPU cores or CPUs in a coherent state, so-called coherence protocols are employed. However, coherence protocols also introduce exploitable timing effects, which has recently been exploited by Irazoqui et al. [IES16] on x86 CPUs.

In this paper, we demonstrate attacks on three smartphones as listed in Table 9.1. The Krait 400 is an ARMv7-A CPU, the other two processors are ARMv8-A CPUs. However, the stock Android of the Alcatel One Touch Pop 2 is compiled for an ARMv7-A instruction set and thus ARMv8-A instructions are not used. We generically refer to ARMv7-A and ARMv8-A as "ARM architecture" throughout this paper. All devices have a shared L2 cache. On the Samsung Galaxy S6, the flush instruction is unlocked by default, which means that it is available in userspace. Furthermore, all devices employ a cache coherence protocol between cores and on the Samsung Galaxy S6 even between the two CPUs [ARM14].

| Device | SoC | CPU (cores) | L1 caches | L2 cache | Inclusiveness |
|--------|-----|-------------|-----------|----------|---------------|
| OnePlus One | Qualcomm Snapdragon 801 | Krait 400 (2) 2.5 GHz | 2× 16 KB, 4-way, 64 sets | 2048 KB, 8-way, 2048 sets | non-inclusive |
| Alcatel One Touch Pop 2 | Qualcomm Snapdragon 410 | Cortex-A53 (4) 1.2 GHz | 4× 32 KB, 4-way, 128 sets | 512 KB, 16-way, 512 sets | instruction-inclusive, data-non-inclusive |
| Samsung Galaxy S6 | Samsung Exynos 7 Octa 7420 | Cortex-A53 (4) 1.5 GHz | 4× 32 KB, 4-way, 128 sets | 256 KB, 16-way, 256 sets | instruction-inclusive, data-non-inclusive |
| | | Cortex-A57 (4) 2.1 GHz | 4× 32 KB, 2-way, 256 sets | 2048 KB, 16-way, 2048 sets | instruction-non-inclusive, data-inclusive |

Table 9.1.: Test devices used in this paper.

## 2.2. Shared Memory

Read-only shared memory can be used as a means of memory usage optimization. In case of shared libraries it reduces the memory footprint and enhances the speed by lowering cache contention. The operating system implements this behavior by mapping the same physical memory into the address space of each process. As this memory sharing mechanism is independent of how a file was opened or accessed, an attacker can map a binary to have read-only shared memory with a victim program. A similar effect is caused by content-based page deduplication where physical pages with identical content are merged.

Android applications are usually written in Java and, thus, contain self-modifying code or just-in-time compiled code. This code would typically not be shared. Since Android version 4.4 the Dalvik VM was gradually replaced by the Android Runtime (ART). With ART, Java byte code is compiled to native code binaries [And15] and thus can be shared too.

## 2.3. Cache Attacks

Initially, cache timing attacks were performed on cryptographic algorithms [Koc96; Kel+00; Pag02; Tsu+03; Ber04; Nev06; NSW06]. For example, Bernstein [Ber04] exploited the total execution time of AES T-table implementations. More fine-grained exploitations of memory accesses to the CPU cache have been proposed by Percival [Per05] and Osvik et al. [OST06]. More specifically, Osvik et al. formalized two concepts, namely *Evict+Time* and *Prime+Probe*, to determine which specific cache sets were accessed by a victim program. Both approaches consist of three basic steps.

***Evict+Time*:**

1. Measure execution time of victim program.
2. Evict a specific cache set.
3. Measure execution time of victim program again.

***Prime+Probe*:**

1. Occupy specific cache sets.
2. Victim program is scheduled.
3. Determine which cache sets are still occupied.

Both approaches allow an adversary to determine which cache sets are used during the victim's computations and have been exploited to attack cryptographic implementations [OST06; TOS10; IES15; Liu+15] and to build cross-VM covert channels [Mau+15b]. Yarom and Falkner [YF14] proposed *Flush+Reload*, a significantly more fine-grained attack that exploits three fundamental concepts of modern system architectures. First, the availability of shared memory between the victim process and the adversary. Second, last-level caches are typically shared among all cores. Third, Intel platforms use inclusive last-level caches, meaning that the eviction of information from the last-level cache leads to the eviction of this data from all lower-level caches of other cores, which allows any program to evict data from other programs on other cores. While the basic idea of this attack has been proposed by Gullasch et al. [GBK11], Yarom and Falkner extended this idea to shared last-level caches, allowing cross-core attacks. *Flush+Reload* works as follows.

***Flush+Reload*:**

1. Map binary (e.g., shared object) into address space.
2. Flush a cache line (code or data) from the cache.

3. Schedule the victim program.
4. Check if the corresponding line from step 2 has been loaded by the victim program.

Thereby, *Flush+Reload* allows an attacker to determine which specific instructions are executed and also which specific data is accessed by the victim program. Thus, rather fine-grained attacks are possible and have already been demonstrated against cryptographic implementations [Ira+15a; Ira+15b; Gül+15]. Furthermore, Gruss et al. [GSM15] demonstrated the possibility to automatically exploit cache-based side-channel information based on the *Flush+Reload* approach. Besides attacking cryptographic implementations like AES T-table implementations, they showed how to infer keystroke information and even how to build a keylogger by exploiting the cache side channel. Similarly, Oren et al. [Ore+15] demonstrated the possibility to exploit cache attacks on Intel platforms from JavaScript and showed how to infer visited websites and how to track the user's mouse activity.

Gruss et al. [GSM15] proposed the *Evict+Reload* technique that replaces the flush instruction in *Flush+Reload* by eviction. While it has no practical application on x86 CPUs, we show that it can be used on ARM CPUs. Recently, *Flush+Flush* [Gru+16] has been proposed. Unlike other techniques, it does not perform any memory access but relies on the timing of the flush instruction to determine whether a line has been loaded by a victim. We show that the execution time of the ARMv8-A flush instruction also depends on whether or not data is cached and, thus, can be used to implement this attack.

While the attacks discussed above have been proposed and investigated for Intel processors, the same attacks were considered not applicable to modern smartphones due to differences in the instruction set, the cache organization [YF14], and in the multi-core and multi-CPU architecture. Thus, only same-core cache attacks have been demonstrated on smartphones so far. For instance, Weiß et al. [WHS12] investigated Bernstein's cache-timing attack [Ber04] on a Beagleboard employing an ARM Cortex-A8 processor. Later on, Weiß et al. [Wei+14] investigated this timing attack in a multi-core setting on a development board. As Weiß et al. [WHS12] claimed that noise makes the attack difficult, Spreitzer and Plos [SP13b] investigated the applicability of Bernstein's cache-timing attack on different ARM Cortex-A8 and ARM Cortex-A9 smartphones running Android. Both investigations [WHS12; SP13b] confirmed that timing information is leaking, but the attack takes several hours due to the high number of measurement samples that are required, *i.e.*, about $2^{30}$ AES encryptions. Later on, Spreitzer and Gérard [SG14] improved upon these results and managed to reduce the key space to a complexity which is practically relevant.

Besides Bernstein's attack, another attack against AES T-table implementations has been proposed by Bogdanov et al. [Bog+10], who exploited so-called wide collisions on an ARM9 microprocessor. In addition, power analysis attacks [GKT10] and electromagnetic emanations [GK11] have been used to visualize cache accesses during AES computations on ARM microprocessors. Furthermore, Spreitzer and Plos [SP13a] implemented *Evict+Time* [OST06] in order to attack an AES T-table implementation on Android-based smartphones. However, so far only cache attacks against AES T-table implementations have been considered on smartphone platforms and none of the recent advances have been demonstrated on mobile devices.

# 3. ARMageddon Attack Techniques

We consider a scenario where an adversary attacks a smartphone user by means of a malicious application. This application *does not require any permission* and, most importantly, it can be executed in unprivileged userspace and *does not require a rooted device.* As our attack techniques do not exploit specific vulnerabilities of Android versions, they work on stock Android ROMs as well as customized ROMs in use today.

## 3.1. Defeating the Cache Organization

In this section, we tackle the aforementioned challenges 1 and 2, *i.e.*, the last-level cache is not inclusive and multiple processors do not necessarily share a cache level.

When it comes to caches, ARM CPUs are very heterogeneous compared to Intel CPUs. For example, whether or not a CPU has a second-level cache can be decided by the manufacturer. Nevertheless, the last-level cache on ARM devices is usually shared among all cores and it can have different inclusiveness properties for instructions and data. Due to cache coherence, shared memory is kept in a coherent state across cores and CPUs. This is of importance when measuring timing differences between cache accesses and memory accesses (cache misses), as fast remote-cache accesses are performed instead of slow memory accesses [ARM14]. In case of a non-coherent cache, a cross-core attack is not possible but an attacker can run the spy process on all cores simultaneously and thus fall back to a same-core attack. However, we observed that caches are coherent on all our test devices.

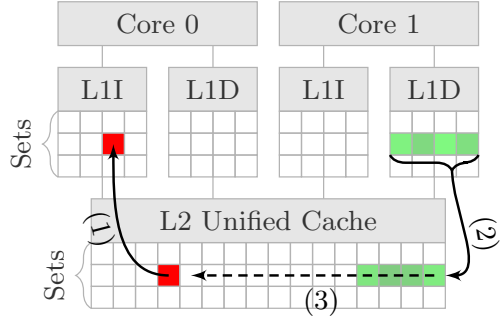

Figure 9.1.: Cross-core instruction cache eviction through data accesses.

To perform a cross-core attack we load enough data into the cache to fully evict the corresponding last-level cache set. Thereby, we exploit that we can fill the last-level cache directly or indirectly depending on the cache organization. On the Alcatel One Touch Pop 2, the last-level cache is instruction-inclusive and thus we can evict instructions from the local caches of the other core. Figure 9.1 illustrates such an eviction. In step 1, an instruction is allocated to the last-level cache and the instruction cache of one core. In step 2, a process fills its core's data cache, thereby evicting cache lines into the last-level cache. In step 3, the process has filled the last-level cache set using only data accesses and thereby evicts the instructions from instruction caches of other cores as well.

We access cache lines multiple times to perform transfers between L1 and L2 cache. Thus, more and more addresses used for eviction are cached in either L1 or L2. As ARM CPUs typically have L1 caches with a very low associativity, the probability of eviction to L2 through other system activity is high. Using an eviction strategy that performs frequent transfers between L1 and L2 increases this probability further. Thus, this approach also works for other cache organizations to perform cross-core and cross-CPU cache attacks. Due to the cache coherence protocol between the CPU cores [Lal13; ARM14], remote-core fetches are faster than memory accesses and thus can be distinguished from cache misses. For instance,

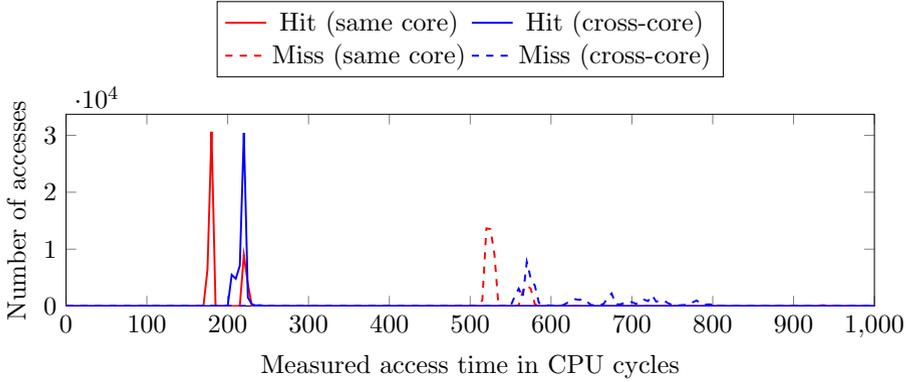

Figure 9.2.: Histograms of cache hits and cache misses measured same-core and cross-core on the OnePlus One.

Figure 9.2 shows the cache hit and miss histogram on the OnePlus One. The cross-core access introduces a latency of 40 CPU cycles on average. However, cache misses take more than 500 CPU cycles on average. Thus, cache hits and misses are clearly distinguishable based on a single threshold value.

## 3.2. Fast Cache Eviction

In this section, we tackle the aforementioned challenges 3 and 4, *i.e.*, not all ARM processors support a flush instruction, and the replacement policy is pseudo-random.

There are two options to evict cache lines: (1) the flush instruction or (2) evict data with memory accesses to congruent addresses, *i.e.*, addresses that map to the same cache set. As the flush instruction is only available on the Samsung Galaxy S6, we need to rely on eviction strategies for the other devices and, therefore, to defeat the replacement policy. The L1 cache in Cortex-A53 and Cortex-A57 has a very small number of ways and employs a least-recently used (LRU) replacement policy [ARM13b]. However, for a full cache eviction, we also have to evict cache lines from the L2 cache, which uses a pseudo-random replacement policy.

**Eviction strategies.** Previous approaches to evict data on Intel x86 platforms either have too much overhead [HWH13] or are only applicable to caches implementing an LRU replacement policy [Liu+15; Mau+15b; Ore+15]. Spreitzer and Plos [SP13a] proposed an eviction strategy for ARMv7-A CPUs that requires to access more addresses than there are cache lines per cache set, due to the pseudo-random replacement policy. Recently, Gruss et al. [GMM16] demonstrated how to automatically find fast eviction strategies on Intel x86 architectures. We show that their algorithm is applicable to ARM CPUs as well. Thereby, we establish eviction strategies in an automated way and significantly reduce the overhead compared to [SP13a]. We evaluated more than 4 200 access patterns on our smartphones and identified the best eviction strategies. Even though the cache employs a random replacement policy, average eviction rate and average execution time are reproducible. Eviction sets are computed based on physical addresses, which can be retrieved via `/proc/self/pagemap` as current Android versions allow access to these mappings to any unprivileged app without any permissions. Thus, eviction patterns and eviction sets can be efficiently computed.

We applied the algorithm of Gruss et al. [GMM16] to a set of physically congruent addresses. Table 9.2 summarizes different eviction strategies, *i.e.*, loop parameters, for the Krait 400. $N$ denotes the total eviction set size (length of the loop), $A$ denotes the shift offset (loop increment) to be applied after each round, and $D$ denotes the number of memory accesses in each iteration (loop body). The column *cycles* states the average execution time in CPU cycles over 1 million evictions and the last column denotes the average eviction rate. The first line in Table 9.2 shows the average execution time and the average eviction rate for the privileged flush instruction, which gives the best result in terms of average execution time (549 CPU cycles). We evaluated 1 863 different strategies and our best identified eviction strategy ($N = 11$, $A = 2$, $D = 2$) also achieves an average eviction rate of 100% but takes 1 578 CPU cycles. Although a strategy accessing every address in the eviction set only once ($A = 1$, $D = 1$, also called LRU eviction) performs significantly fewer memory accesses, it consumes more CPU cycles. For an average eviction rate of 100%, LRU eviction requires an eviction set size of at least 16. The average execution time then is 3 026 CPU cycles. Considering the eviction strategy used in [SP13a] that takes 4 371 CPU cycles, clearly demonstrates the advantage of our optimized eviction strategy that takes only 1 578 CPU cycles.

We performed the same evaluation with 2 295 different strategies on the ARM Cortex-A53 in our Alcatel One Touch Pop 2 test system and summarize them in Table 9.3. For the best strategy we found ($N = 21$, $A = 1$, $D = 6$), we measured an average eviction rate of 99.93% and an average execution time of 4 275 CPU cycles. We observed that LRU eviction ($A = 1$, $D = 1$) on the ARM Cortex-A53 would take 28 times more CPU cycles to achieve an average eviction rate of only 99.10%, thus it is not suitable for attacks on the last-level cache as used in previous work [SP13a]. The reason for this is that data can only be allocated to L2 cache by evicting it from the L1 cache on the ARM Cortex-A53. Therefore, it is better to reaccess the data that is already in the L2 cache and gradually add new addresses to the set of cached addresses instead of accessing more different addresses.

| N | A | D | Cycles | Eviction rate |
|---|---|---|--------|---------------|
| - | - | - | 549 | 100.00% |
| 11 | 2 | 2 | 1 578 | 100.00% |
| 12 | 1 | 3 | 2 094 | 100.00% |
| 13 | 1 | 5 | 2 213 | 100.00% |
| 16 | 1 | 1 | 3 026 | 100.00% |
| 24 | 1 | 1 | 4 371 | 100.00% |
| 13 | 1 | 2 | 2 372 | 99.58% |
| 11 | 1 | 3 | 1 608 | 80.94% |
| 11 | 4 | 1 | 1 948 | 58.93% |
| 10 | 2 | 2 | 1 275 | 51.12% |

Table 9.2.: Different eviction strategies on the Krait 400.

| N | A | D | Cycles | Eviction rate |
|---|---|---|--------|---------------|
| - | - | - | 767 | 100.00% |
| 23 | 2 | 5 | 6 209 | 100.00% |
| 23 | 4 | 6 | 16 912 | 100.00% |
| 22 | 1 | 6 | 5 101 | 99.99% |
| 21 | 1 | 6 | 4 275 | 99.93% |
| 20 | 4 | 6 | 13 265 | 99.44% |
| 800 | 1 | 1 | 142 876 | 99.10% |
| 200 | 1 | 1 | 33 110 | 96.04% |
| 100 | 1 | 1 | 15 493 | 89.77% |
| 48 | 1 | 1 | 6 517 | 70.78% |

Table 9.3.: Different eviction strategies on the Cortex-A53.

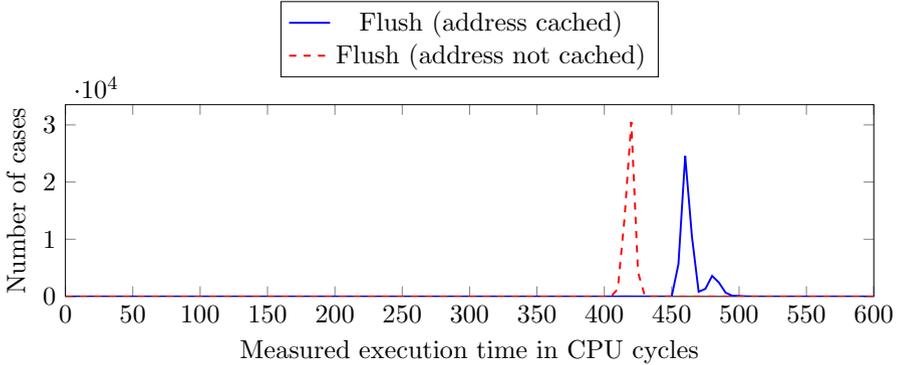

Figure 9.3.: Histograms of the execution time of the flush operation on cached and not cached addresses measured on the Samsung Galaxy S6.

On the ARM Cortex-A57 the userspace flush instruction was significantly faster in any case. Thus, for *Flush+Reload* we use the flush instruction and for *Prime+Probe* the eviction strategy. Falling back to *Evict+Reload* is not necessary on the Cortex-A57. Similarly to recent Intel x86 CPUs, the execution time of the flush instruction on ARM depends on whether or not the value is cached, as shown in Figure 9.3. The execution time is higher if the address is cached and lower if the address is not cached. This observation allows us to distinguish between cache hits and cache misses depending on the timing behavior of the flush instruction, and therefore to perform a *Flush+Flush* attack. Thus, in case of shared memory between the victim and the attacker, it is not even required to evict and reload an address in order to exploit the cache side channel.

**A note on *Prime+Probe*.**   Finding a fast eviction strategy for *Prime+ Probe* on architectures with a random replacement policy is not as straightforward as on Intel x86. Even in case of x86 platforms, the problem of cache trashing has been discussed by Tromer et al. [TOS10]. Cache trashing occurs when reloading (probing) an address evicts one of the addresses that are to be accessed next. While Tromer et al. were able to overcome this problem by using a doubly-linked list that is accessed forward during the prime step and backwards during the probe step, the random replacement policy on ARM also contributes to the negative effect of cache trashing.

We analyzed the behavior of the cache and designed a prime step and a probe step that work with a smaller set size to avoid set thrashing. Thus, we set the eviction set size to 15 on the Alcatel One Touch Pop 2. As we run the *Prime+Probe* attack in a loop, exactly 1 way in the L2 cache will not be occupied after a few attack rounds. We might miss a victim access in $\frac{1}{16}$ of the cases, which however is necessary as otherwise we would not be able to get reproducible measurements at all due to set thrashing. If the victim replaces one of the 15 ways occupied by the attacker, there is still one free way to reload the address that was evicted. This reduces the chance of set thrashing significantly and allows us to successfully perform *Prime+Probe* on caches with a random replacement policy.

## 3.3. Accurate Unprivileged Timing

In this section, we tackle the aforementioned challenge 5, *i.e.*, cycle-accurate timings require root access on ARM.

In order to distinguish cache hits and cache misses, timing sources or dedicated performance counters can be used. We focus on timing sources, as cache misses have a significantly higher access latency and timing sources are well studied on Intel x86 CPUs. Cache attacks on x86 CPUs employ the unprivileged `rdtsc` instruction to obtain a sub-nanosecond resolution timestamp. The ARMv7-A architecture does not provide an instruction for this purpose. Instead, the ARMv7-A architecture has a performance monitoring unit that allows to monitor CPU activity. One of these performance counters—denoted as *cycle count register* (PMCCNTR)—can be used to distinguish cache hits and cache misses by relying on the number of CPU cycles that passed during a memory access. However, these performance counters are not accessible from userspace by default and an attacker would need root privileges.

We broaden the attack surface by exploiting timing sources that are accessible without any privileges or permissions. We identified three possible alternatives for timing measurements.

**Unprivileged syscall.** The `perf_event_open` syscall is an abstract layer to access performance information through the kernel independently of the underlying hardware. For instance, `PERF_COUNT_HW_CPU_CYCLES` returns an accurate cycle count including a minor overhead due to the syscall. The availability of this feature depends on the Android kernel configuration, e.g., the stock kernel on the Alcatel One Touch Pop 2 as well as the OnePlus One provide this feature by default. Thus, in contrast to previous work [SP13a], the attacker does not have to load a kernel module to access this information as the `perf_event_open` syscall can be accessed without any privileges or permissions.

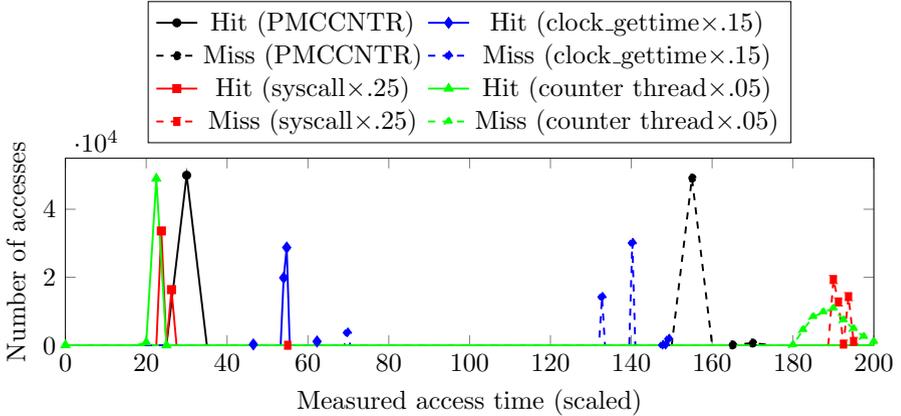

Figure 9.4.: Histogram of cross-core cache hits/misses on the Alcatel One Touch Pop 2 using different methods. X-values are scaled for visual representation.

**POSIX function.** Another alternative to obtain sufficiently accurate timing information is the POSIX function `clock_gettime()`, with an accuracy in the range of microseconds to nanoseconds. Similar information can also be obtained from `/proc/timer_list`.

**Dedicated thread timer.** If no interface with sufficient accuracy is available, an attacker can run a thread that increments a global variable in a loop, providing a fair approximation of a cycle counter. Our experiments show that this approach works reliably on smartphones as well as recent x86 CPUs. The resolution of this threaded timing information is as high as with the other methods.

In Figure 9.4 we show the cache hit and miss histogram based on the four different methods, including the cycle count register, on a Alcatel One Touch Pop 2. Despite the latency and noise, cache hits and cache misses are clearly distinguishable with all approaches. Thus, all methods can be used to implement cache attacks. Determining the best timing method on the device under attack can be done in a few seconds during an online attack.

| Work | Type | Bandwidth [bps] | Error rate |
|------|------|-----------------|------------|
| Ours (Samsung Galaxy S6) | *Flush+Reload*, cross-core | **1 140 650** | 1.10% |
| Ours (Samsung Galaxy S6) | *Flush+Reload*, cross-CPU | **257 509** | 1.83% |
| Ours (Samsung Galaxy S6) | *Flush+Flush*, cross-core | **178 292** | 0.48% |
| Ours (Alcatel One Touch Pop 2) | *Evict+Reload*, cross-core | **13 618** | 3.79% |
| Ours (OnePlus One) | *Evict+Reload*, cross-core | **12 537** | 5.00% |
| Marforio et al. [Mar+12] | Type of Intents | 4 300 | – |
| Marforio et al. [Mar+12] | UNIX socket discovery | 2 600 | – |
| Schlegel et al. [Sch+11] | File locks | 685 | – |
| Schlegel et al. [Sch+11] | Volume settings | 150 | – |
| Schlegel et al. [Sch+11] | Vibration settings | 87 | – |

Table 9.4.: Comparison of covert channels on Android.

# 4. High Performance Covert Channels

To evaluate the performance of our attacks, we measure the capacity of cross-core and cross-CPU cache covert channels. A covert channel enables two unprivileged applications on a system to communicate with each other without using any data transfer mechanisms provided by the operating system. This communication evades the sandboxing concept and the permission system (cf. collusion attacks [Mar+12]). Both applications were running in the background while the phone was mostly idle and an unrelated app was running as the foreground application.

Our covert channel is established on addresses of a shared library that is used by both the sender and the receiver. While both processes have read-only access to the shared library, they can transmit information by loading addresses from the shared library into the cache or evicting (flushing) it from the cache, respectively.

The covert channel transmits packets of $n$-bit data, an $s$-bit sequence number, and a $c$-bit checksum that is computed over data and sequence number. The sequence number is used to distinguish consecutive packets and the checksum is used to check the integrity of the packet. The receiver acknowledges valid packets by responding with an $s$-bit sequence number and an $x$-bit checksum. By adjusting the sizes of checksums and sequence numbers the error rate of the covert channel can be controlled.

Each bit is represented by one address in the shared library, whereas no two addresses are chosen that map to the same cache set. To transmit a bit value of 1, the sender accesses the corresponding address in the library. To transmit a bit value of 0, the sender does not access the corresponding address, resulting in a cache miss on the receiver's side. Thus, the receiving process observes a cache hit or a cache miss depending on the memory access performed by the sender. The same method is used for the acknowledgements sent by the receiving process.

We implemented this covert channel using *Evict+Reload*, *Flush+Reload*, and *Flush+Flush* on our smartphones. The results are summarized in Table 9.4. On the Samsung Galaxy S6, we achieve a cross-core transmission rate of 1 140 650 bps at an error rate of 1.10%. This is 265 times faster than any existing covert channel on smartphones. In a cross-CPU transmission we achieve a transmission rate of 257 509 bps at an error rate of 1.83%. We achieve a cross-core transition rate of 178 292 bps at an error rate of 0.48% using *Flush+Flush* on the Samsung Galaxy S6. On the Alcatel One Touch Pop 2 we achieve a cross-core transmission rate of 13 618 bps at an error rate of 3.79% using *Evict+Reload*. This is still 3 times faster than previous covert channels on smartphones. The covert channel is significantly slower on the Alcatel One Touch Pop 2 than on the Samsung Galaxy S6 because the hardware is much slower, *Evict+Reload* is slower than *Flush+Reload*, and retransmission might be necessary in 0.14% of the cases where eviction is not successful (cf. Section 3.2). On the older OnePlus One we achieve a cross-core transmission rate of 12 537 bps at an error rate of 5.00%, 3 times faster than previous covert channels on smartphones. The reason for the higher error rate is the additional timing noise due to the cache coherence protocol performing a high number of remote-core fetches.

# 5. Attacking User Input on Smartphones

In this section we demonstrate cache side-channel attacks on Android smartphones. We implement cache template attacks [GSM15] to create and exploit accurate cache-usage profiles using the *Evict+Reload* or *Flush+Reload* attack. Cache template attacks have a profiling phase and an exploitation phase. In the profiling phase, a template matrix is computed that represents how many cache hits occur on a specific address when triggering a specific event. The exploitation phase uses this matrix to infer events from cache hits.

To perform cache template attacks, an attacker has to map shared binaries or shared libraries as read-only shared memory into its own address space. By using shared libraries, the attacker bypasses any potential countermeasures taken by the operating system, such as restricted access to runtime data of other apps or address space layout randomization (ASLR). The attack can even be performed online on the device under attack if the event can be simulated.

Triggering the actual event that an attacker wants to spy on might require either (1) an offline phase or (2) privileged access. For instance, in case of a keylogger, the attacker can gather a cache template matrix offline for a specific version of a library, or the attacker relies on privileged access of the application (or a dedicated permission) in order to be able to simulate events for gathering the cache template matrix. However, the actual exploitation of the cache template matrix to infer events neither requires privileged access nor any permission.

## 5.1. Attacking a Shared Library

Just as Linux, Android uses a large number of shared libraries, each with a size of up to several megabytes. We inspected all available libraries on the system by manually scanning the names and identified libraries that might be responsible for handling user input, e.g., the `libinput.so` library. Without loss of generality, we restricted the set of attacked libraries since testing all libraries would have taken a significant amount of time. Yet, an adversary could exhaustively probe all libraries.

We automated the search for addresses in these shared libraries and after identifying addresses, we monitored them in order to infer user input events. For instance, in the profiling phase on `libinput.so`, we simulated events via the android-debug bridge (adb shell) with two different methods. The first method uses the `input` command line tool to simulate user input events. The second method is writing event messages to `/dev/input/event*`. Both methods can run entirely on the device for instance in idle periods while the user is not actively using the device. As the second method only requires a `write()` statement it is significantly faster, but it is also more device specific. Therefore, we used the `input` command line except when profiling differences between different letter keys. While simulating these events, we simultaneously probed all addresses within the `libinput.so` library, *i.e.*, we measured the number of cache hits that occurred on each address when triggering a specific event. As already mentioned above, the simulation of some events might require either an offline phase or specific privileges in case of online attacks.

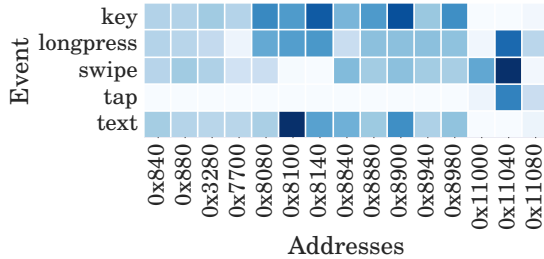

Figure 9.5.: Cache template matrix for `libinput.so`.

Figure 9.5 shows part of the cache template matrix for `libinput.so`. We triggered the following events: key events including the power button (*key*), long touch events (*longpress*), *swipe* events, touch events (*tap*), and text input events (*text*) via the `input` tool as often as possible and measured each address and event for one second. The cache template matrix clearly reveals addresses with high cache-hit rates for specific events. Darker colors represent addresses with higher cache-hit rates for a specific event and lighter colors represent addresses with lower cache-hit rates. Hence, we can distinguish different events based on cache hits on these addresses.

We verified our results by monitoring the identified addresses while operating the smartphone manually, *i.e.*, we touched the screen and our attack application reliably reported cache hits on the monitored addresses. For instance, address `0x11040` of `libinput.so` can be used to distinguish tap actions and swipe actions on the screen of the Alcatel One Touch Pop 2. Tap actions cause a smaller number of cache hits than swipe actions. Swipe actions cause cache hits in a high frequency as long as the screen is touched. Figure 9.6 shows a sequence of 3 tap events, 3 swipe events, 3 tap events, and 2 swipe events. These events can be clearly distinguished due to the fast access times. The gaps mark periods of time where our program was not scheduled on the CPU. Events occurring in those periods can be missed by our attack.

Swipe input allows to enter words by swiping over the soft-keyboard and thereby connecting single characters to form a word. Since we are able to determine the length of swipe movements, we can correlate the length of the swipe movement with the actual word length in any Android application or system interface that uses swipe input without any privileges. Furthermore, we can determine the actual length of the unlock pattern for the pattern-unlock mechanism.

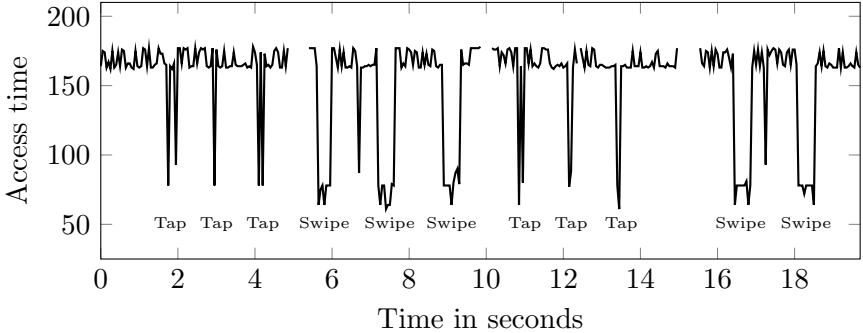

Figure 9.6.: Monitoring address `0x11040` of `libinput.so` on the Alcatel One Touch Pop 2 reveals taps and swipes.

Figure 9.7 shows a user input sequence consisting of 3 tap events and 3 swipe events on the Samsung Galaxy S6. The attack was conducted using *Flush+Reload*. An attacker can monitor every single event. Taps and swipes can be distinguished based on the length of the cache hit phase. The length of a swipe movement can be determined from the same information. Figure 9.8 shows the same experiment on the OnePlus One using *Evict+Reload*. Thus, our attack techniques work on coherent non-inclusive last-level caches.

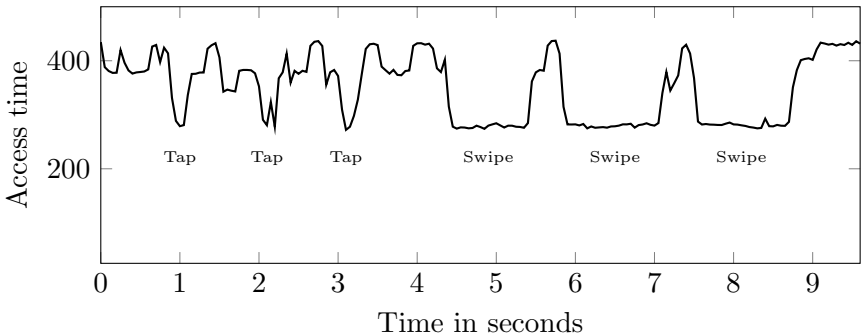

Figure 9.7.: Monitoring address `0xDC5C` of `libinput.so` on the Samsung Galaxy S6 reveals tap and swipe events.

## 5.2. Attacking ART Binaries

Instead of attacking shared libraries, it is also possible to apply this attack to ART (Android Runtime) executables [And15] that are compiled ahead of time. We used this attack on the default AOSP keyboard and evaluated the number of accesses to every address in the optimized executable that responds to an input of a letter on the keyboard. It is possible to find addresses that correspond to a key press and more importantly to distinguish between taps and key presses. Figure 9.9 shows the corresponding cache template matrix. We summarize the letter keys in one line (*alphabet*) as they did not vary significantly. These addresses can be used to monitor key presses on the keyboard. We identified an address that corresponds only to letters on the keyboard and hardly on the space bar or the return button. With this information it is possible to precisely determine the length of single words entered using the default AOSP keyboard.

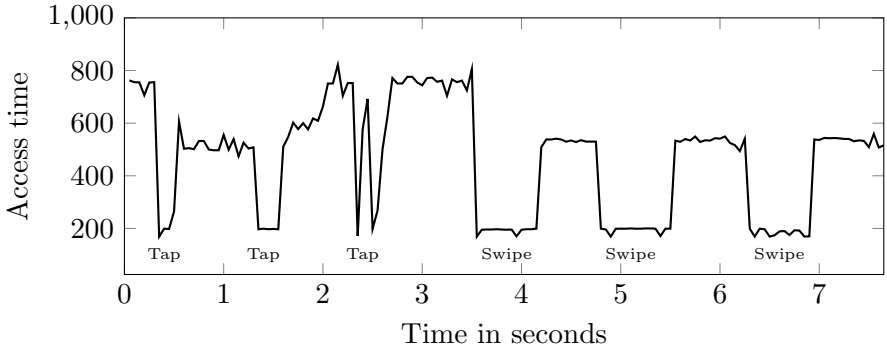

Figure 9.8.: Monitoring address `0xBFF4` of `libinput.so` on the OnePlus One reveals tap and swipe events.

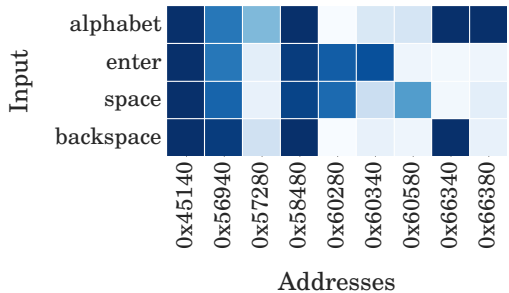

Figure 9.9.: Cache template matrix for the default AOSP keyboard.

We illustrate the capability of detecting word lengths in Figure 9.10. The blue line shows the timing measurements for the address identified for keys in general, the red dots represent measurements of the address for the space key. The plot shows that we can clearly determine the length of entered words and monitor user input accurately over time.

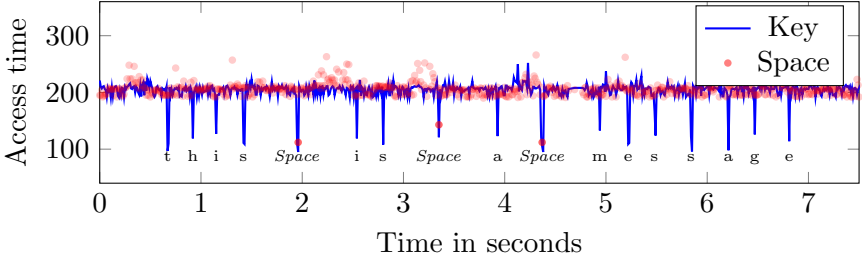

Figure 9.10.: *Evict+Reload* on 2 addresses in `custpack@app@` `withoutlibs@LatinIME.apk@classes.dex` on the Alcatel One Touch Pop 2 while entering the sentence "this is a message".

## 5.3. Discussion and Impact

Our proof-of-concept attacks exploit shared libraries and binaries from Android apk files to infer key strokes. The cache template attack technique we used for these attacks is generic and can also be used to attack any other library. For instance, there are various libraries that handle different hardware modules and software events on the device, such as GPS, Bluetooth, camera, NFC, vibrator, audio and video decoding, web and PDF viewers. Each of these libraries contains code that is executed and data that is accessed when the device is in use. Thus, an attacker can perform a cache template attack on any of these libraries and spy on the corresponding device events. For instance, our attack can be used to monitor activity of the GPS sensor, bluetooth, or the camera. An attacker can record such user activities over time to learn more about the user.

We can establish inter-keystroke timings at an accuracy as high as the accuracy of cache side-channel attacks on keystrokes on x86 systems with a physical keyboard. Thus, the inter-keystroke timings can be used to infer entered words, as has been shown by Zhang et al. [ZW09]. Our attack even has a higher resolution than [ZW09], *i.e.*, it is sub-microsecond accurate. Furthermore, we can distinguish between keystrokes on the soft-keyboard and generic touch actions outside the soft-keyboard. This information can be used to enhance sensor-based keyloggers that infer user input on mobile devices by exploiting, e.g., the accelerometer and the gyroscope [CC11; CC12; Avi+12; Mil+12; XBZ12] or the ambient-light sensor [Spr14]. However, these attacks suffer from a lack of knowledge when exactly a user touches the screen. Based on our attack, these sensor-based keyloggers can be improved as our attack allows to infer (1) the exact time when the user touches the screen, and (2) whether the user touches the soft-keyboard or any other region of the display.

Our attacks only require the user to install a malicious app on the smartphone. However, as shown by Oren et al. [Ore+15], *Prime+Probe* attacks can even be performed from within browser sandboxes through remote websites using JavaScript on Intel platforms. Gruss et al. [GBM15] showed that JavaScript timing measurements in web browsers on ARM-based smartphones achieve a comparable accuracy as on Intel platforms. Thus, it seems likely that *Prime+Probe* through a website works on ARM-based smartphones as well. We expect that such attacks will be demonstrated in future work. The possibility of attacking millions of users shifts the focus of cache attacks to a new range of potential malicious applications.

In our experiments with the predecessor of ART, the Dalvik VM, we found that the just-in-time compilation effectively prevents *Evict+Reload* and *Flush+Reload* attacks. The just-in-time compiled code is not shared and thus the requirements for these two attacks are not met. However, *Prime+Probe* attacks work on ART binaries and just-in-time compiled Dalvik VM code likewise.

# 6. Attack on Cryptographic Algorithms

In this section we show how *Flush+Reload*, *Evict+Reload*, and *Prime+Probe* can be used to attack AES T-table implementations that are still in use on Android devices. Furthermore, we demonstrate the possibility to infer activities within the ARM TrustZone by observing the cache activity using *Prime+Probe*. We perform all attacks cross-core and in a synchronized setting, *i.e.*, the attacker triggers the execution of cryptographic algorithms by the victim process. Although more sophisticated attacks are possible, our goal is to demonstrate that our work enables practical cache attacks on smartphones.

## 6.1. AES T-Table Attacks

Many cache attacks against AES T-table implementations have been demonstrated and appropriate countermeasures have already been proposed. Among these countermeasures are, e.g., so-called bit-sliced implementations [RSD06; Kön08; KS09]. Furthermore, Intel addressed the problem by adding dedicated instructions for AES [Gue10] and ARM also follows the same direction with the ARMv8 instruction set [ARM13a]. However, our investigations showed that Bouncy Castle, a crypto library widely used in Android apps such as the WhatsApp messenger[App16], still uses a T-table implementation. Moreover, the OpenSSL library, which is the default crypto provider on recent Android versions, uses T-table implementations until version 1.0.1.[3] This version is still officially supported and commonly used on Android devices, e.g., the Alcatel One Touch Pop 2. T-tables contain the precomputed AES round transformations, allowing to perform encryptions and decryptions by simple XOR operations. For instance, let $p_i$ denote the plaintext bytes, $k_i$ the initial key bytes, and $s_i = p_i \oplus k_i$ the initial state bytes. The initial state bytes are used to retrieve precomputed T-table elements for the next round. If an attacker knows a plaintext byte $p_i$ and the accessed element of the T-table, it is possible to recover the key bytes $k_i = s_i \oplus p_i$. However, it is only possible to derive the upper 4 bits of $k_i$ through our cache attack on a device with a cache line size of 64 bytes. This way, the attacker can learn 64 key bits. In second-round and last-round attacks the key space can be reduced further. For details about the basic attack strategy we refer to the work of Osvik et al. [OST06; TOS10]. Although we successfully mounted an *Evict+Reload* attack on the Alcatel One Touch Pop 2 against the OpenSSL AES implementation, we do not provide further insights as we are more interested to perform the first cache attack on a Java implementation.

**Attack on Bouncy Castle.** Bouncy Castle is implemented in Java and provides various cryptographic primitives including AES. As Bouncy Castle 1.5 still employs AES T-table implementations by default, all Android devices that use this version are vulnerable to our presented attack. To the best of our knowledge, we are the first to show an attack on a Java implementation.

---

[3]Later versions use a bit-sliced implementation if ARM NEON is available or dedicated AES instructions if ARMv8-A instructions are available. Otherwise, a T-table implementation is used. This is also the case for Google's BoringSSL library.

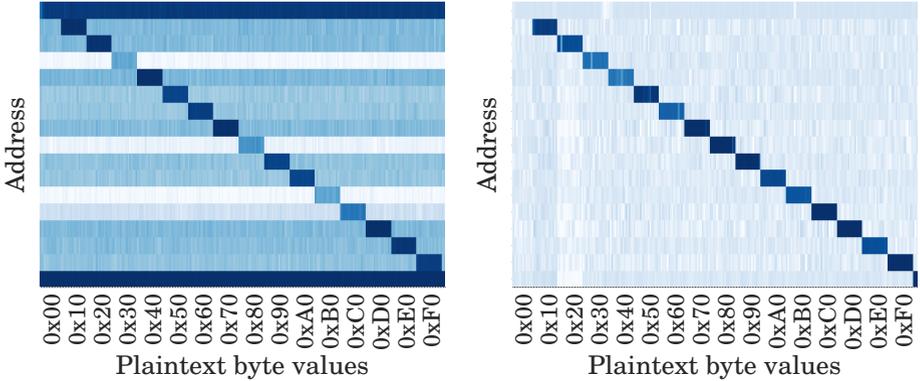

Figure 9.11.: Attack on Bouncy Castle's AES using *Evict+Reload* on the Alcatel One Touch Pop 2 (left) and *Flush+Reload* on the Samsung Galaxy S6 (right).

During the initialization of Bouncy Castle, the T-tables are copied to a local private memory area. Therefore, these copies are not shared among different processes. Nevertheless, we demonstrate that *Flush+Reload* and *Evict+Reload* are efficient attacks on such an implementation if shared memory is available. Further, we demonstrate a cross-core *Prime+Probe* attack without shared memory that is applicable in a real-world scenario.

Figure 9.11 shows a template matrix of the first T-table for all 256 values for plaintext byte $p_0$ and a key that is fixed to 0 while the remaining plaintext bytes are random. These plots reveal the upper 4 key bits of $k_0$ [OST06; SP13a]. Thus, in our case the key space is reduced to 64 bits after 256–512 encryptions. We consider a first-round attack only, because we aim to demonstrate the applicability of these attacks on ARM-based mobile devices. However, full-key recovery is possible with the same techniques by considering more sophisticated attacks targeting different rounds [TOS10; SY15], even for asynchronous attackers [Ira+14; Gül+15].

We can exploit the fact that the T-tables are placed on a different boundary every time the process is started. By restarting the victim application we can obtain arbitrary disalignments of T-tables. Disaligned T-tables allow to reduce the key space to 20 bits on average and for specific disalignments even full-key recovery without a single brute-force computation is possible [SP13a; Tak+13]. We observed not a single case where the T-tables were aligned. Based on the first-round attack matrix in Figure 9.11, the expected number of encryptions until a key byte is identified is $1.81 \cdot 128$.

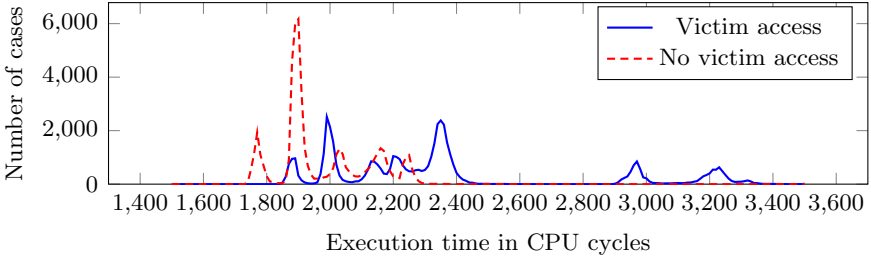

Figure 9.12.: Histogram of *Prime+Probe* timings depending on whether the victim accesses congruent memory on the ARM Cortex-A53.

Thus, full key recovery is possible after $1.81 \cdot 128 \cdot 16 = 3\,707$ encryptions by monitoring a single address during each encryption.

**Real-world cross-core attack on Bouncy Castle.** If the attacker has no way to share a targeted memory region with the victim, *Prime+ Probe* instead of *Evict+Reload* or *Flush+Reload* can be used. This is the case for dynamically generated data or private memory of another process. Figure 9.12 shows the *Prime+Probe* histogram for cache hits and cache misses. We observe a higher execution time if the victim accesses a congruent memory location. Thus, *Prime+Probe* can be used for a real-world cross-core attack on Bouncy Castle and also allows to exploit disaligned T-tables as mentioned above.

In a preprocessing step, the attacker identifies the cache sets to be attacked by performing random encryptions and searching for active cache sets. Recall that the cache set (index) is derived directly from the physical address on ARM, *i.e.*, the lowest $n$ bits determine the offset within a $2^n$-byte cache line and the next $s$ bits determine one of the $2^s$ cache sets. Thus, we only have to find a few cache sets where a T-table maps to in order to identify all cache sets required for the attack. On x86 the replacement policy facilitates this attack and allows even to deduce the number of ways that have been replaced in a specific cache set [OST06]. On ARM the random replacement policy makes *Prime+Probe* more difficult as cache lines are replaced in a less predictable way. To launch a *Prime+ Probe* attack, we apply the eviction strategy and the crafted reaccess patterns we described in Section 3.2.

Figure 9.13 shows an excerpt of the cache template matrix resulting from a *Prime+Probe* attack on one T-table. For each combination of plaintext byte and offset we performed $100\,000$ encryptions for illustration purposes. We only need to monitor a single address to obtain the upper 4 bits of $s_i$ and, thus, the upper 4 bits of $k_i = s_i \oplus p_i$. Compared to the *Evict+ Reload* attack from the previous section, *Prime+Probe* requires 3 times as many measurements to achieve the same accuracy. Nevertheless, our results show that an attacker can run *Prime+Probe* attacks on ARM CPUs just as on Intel CPUs.

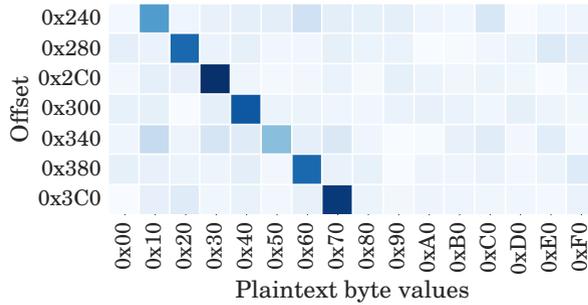

Figure 9.13.: Excerpt of the attack on Bouncy Castle's AES using *Prime+ Probe*.

## 6.2. Spy on TrustZone Code Execution

The ARM TrustZone is a hardware-based security technology built into ARM CPUs to provide a secure execution environment [ARM13a]. This trusted execution environment is isolated from the *normal world* using hardware support. The TrustZone is used, e.g., as a hardware-backed credential store, to emulate secure elements for payment applications, digital rights management as well as verified boot and kernel integrity measurements. The services are provided by so-called trustlets, *i.e.*, applications that run in the secure world.

Since the secure monitor can only be called from the supervisor context, the kernel provides an interface for the userspace to interact with the TrustZone. On the Alcatel One Touch Pop 2, the TrustZone is accessible through a device driver called QSEECOM (Qualcomm Secure Execution Environment Communication) and a library `libQSEEComAPI.so`. The key master trustlet on the Alcatel One Touch Pop 2 provides an interface to generate hardware-backed RSA keys, which can then be used inside the TrustZone to sign and verify signatures.

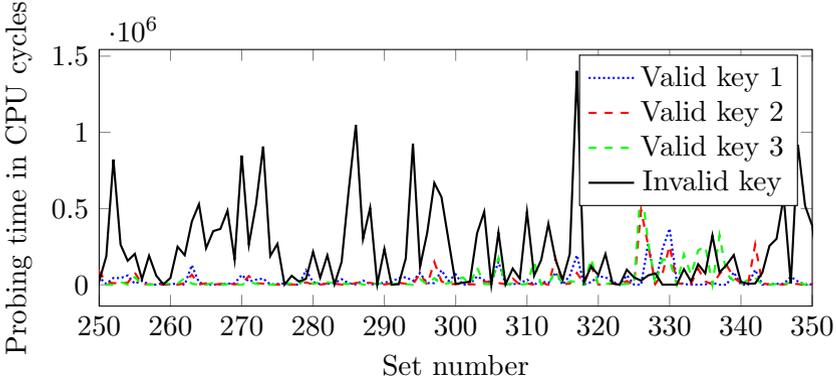

Figure 9.14.: Mean squared error between the average *Prime+Probe* timings of valid keys and invalid keys on the Alcatel One Touch Pop 2.

Our observations showed that a *Prime+Probe* attack on the TrustZone is not much different from a *Prime+Probe* attack on any application in the normal world. However, as we do not have access to the source code of the TrustZone OS or any trustlet, we only conduct simple attacks.[4] We show that *Prime+Probe* can be used to distinguish whether a provided key is valid or not. While this might also be observable through the overall execution time, we demonstrate that the TrustZone isolation does not protect against cache attacks from the normal world and any trustlet can be attacked.

We evaluated cache profiles for multiple valid as well as invalid keys. Figure 9.14 shows the mean squared error over two runs for different valid keys and one invalid key compared to the average of valid keys. We performed *Prime+Probe* before and after the invocation of the corresponding trustlet, *i.e.*, prime before the invocation and probe afterwards. We clearly see a difference in some sets (cache sets 250–320) that are used during the signature generation using a valid key. These cache profiles are reproducible and can be used to distinguish whether a valid or an invalid key has been used in the TrustZone. Thus, the secure world leaks information to the non-secure world.

On the Samsung Galaxy S6, the TrustZone flushes the cache when entering or leaving the trusted world. However, by performing a *Prime+Probe* attack in parallel, *i.e.*, multiple times while the trustlet performs the corresponding computations, the same attack can be mounted.

---

[4]More sophisticated attacks would be possible by reverse engineering these trustlets.

# 7. Countermeasures

Although our attacks exploit hardware weaknesses, software-based countermeasures could impede such attacks. Indeed, we use unprotected access to system information that is available on all Android versions.

As we have shown, the operating system cannot prevent access to timing information. However, other information supplied by the operating system that facilitates these attacks could be restricted. For instance, we use `/proc/pid/` to retrieve information about any other process on the device, e.g., `/proc/pid/pagemap` is used to resolve virtual addresses to physical addresses. Even though access to `/proc/pid/pagemap` and `/proc/self/pagemap` has been restricted in Linux in early 2015, the Android kernel still allows access to these resources. Given the immediately applicable attacks we presented, we stress the urgency to merge the corresponding patches into the Android kernel. Furthermore, we use `/proc/pid/maps` to determine shared objects that are mapped into the address space of a victim. Restricting access to procfs to specific privileges or permissions would make attacks harder. We recommend this for both the Linux kernel as well as Android.

We also exploit the fact that access to shared libraries as well as `dex` and `art` optimized program binaries is only partially restricted on the file system level. While we cannot retrieve a directory listing of `/data/dalvik-cache/`, all files are readable for any process or Android application. We recommend to allow read access to these files to their respective owner exclusively to prevent *Evict+Reload*, *Flush+Reload*, and *Flush+Flush* attacks through these shared files.

In order to prevent cache attacks against AES T-tables, hardware instructions should be used. If this is not an option, a software-only bit-sliced implementation must be employed, especially when disalignment is possible, as it is the case in Java. Since OpenSSL 1.0.2 a bit-sliced implementation is available for devices capable of the ARM NEON instruction set and dedicated AES instructions are used on ARMv8-A devices. Cryptographic algorithms can also be protected using cache partitioning [Liu+16]. However, cache partitioning comes with a performance impact and it can not prevent all attacks, as the number of cache partitions is limited.

We responsibly disclosed our attacks and the proposed countermeasures to Google and other development groups prior to the publication of our attacks. Google has applied upstream patches preventing access to

`/proc/pid/pagemap` in early 2016 and recommended installing the security update in March 2016 [Goo16].

## 8. Conclusion

In this work we demonstrated the most powerful cross-core cache attacks *Prime+Probe*, *Flush+Reload*, *Evict+Reload*, and *Flush+Flush* on default configured unmodified Android smartphones. Furthermore, these attacks do not require any permission or privileges. In order to enable these attacks in real-world scenarios, we have systematically solved all challenges that prevented highly accurate cache attacks on ARM so far. Our attacks are the first cross-core and cross-CPU attacks on ARM CPUs. Furthermore, our attack techniques provide a high resolution and a high accuracy, which allows monitoring singular events such as touch and swipe actions on the screen, touch actions on the soft-keyboard, and inter-keystroke timings. In addition, we show that efficient state-of-the-art key-recovery attacks can be mounted against the default AES implementation that is part of the Java Bouncy Castle crypto provider and that cache activity in the ARM TrustZone can be monitored from the normal world.

The presented example attacks are by no means exhaustive and launching our proposed attack against other libraries and apps will reveal numerous further exploitable information leaks. Our attacks are applicable to hundreds of millions of today's off-the-shelf smartphones as they all have very similar if not identical hardware. This is especially daunting since smartphones have become the most important personal computing devices and our techniques significantly broaden the scope and impact of cache attacks.

## Acknowledgment

We would like to thank our anonymous reviewers for their valuable comments and suggestions.

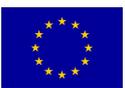
Supported by the EU Horizon 2020 programme under GA No. 644052 (HECTOR), the EU FP7 programme under GA No. 610436 (MATTHEW), and the Austrian Research Promotion Agency (FFG) under grant number 845579 (MEMSEC).

# Prefetch Side-Channel Attacks: Bypassing SMAP and Kernel ASLR

## Publication Data

## Contributions

Main author.

# Prefetch Side-Channel Attacks: Bypassing SMAP and Kernel ASLR

Daniel Gruss*, Clémentine Maurice*, Anders Fogh†, Moritz Lipp*, and Stefan Mangard*

*Graz University of Technology     † G DATA Advanced Analytics

## Abstract

Modern operating systems use hardware support to protect against control-flow hijacking attacks such as code-injection attacks. Typically, write access to executable pages is prevented and kernel mode execution is restricted to kernel code pages only. However, current CPUs provide no protection against code-reuse attacks like ROP. ASLR is used to prevent these attacks by making all addresses unpredictable for an attacker. Hence, the kernel security relies fundamentally on preventing access to address information.

We introduce Prefetch Side-Channel Attacks, a new class of generic attacks exploiting major weaknesses in prefetch instructions. This allows unprivileged attackers to obtain address information and thus compromise the entire system by defeating SMAP, SMEP, and kernel ASLR. Prefetch can fetch inaccessible privileged memory into various caches on Intel x86. It also leaks the translation-level for virtual addresses on both Intel x86 and ARMv8-A. We build three attacks exploiting these properties. Our first attack retrieves an exact image of the full paging hierarchy of a process, defeating both user space and kernel space ASLR. Our second attack resolves virtual to physical addresses to bypass SMAP on 64-bit Linux systems, enabling ret2dir attacks. We demonstrate this from unprivileged user programs on Linux and inside Amazon EC2 virtual machines. Finally, we demonstrate how to defeat kernel ASLR on Windows 10, enabling ROP attacks on kernel and driver binary code. We propose a new form of strong kernel isolation to protect commodity systems incuring an overhead of only 0.06–5.09%.

# 1. Introduction

The exploitation of software bugs imperils the security of modern computer systems fundamentally. Especially, buffer overflows can allow an attacker to overwrite data structures that are used in the control flow of the program. These attacks are not limited to user space software but are also possible on operating system kernels [HHF09]. Modern computer hardware provides various features to prevent exploitation of software bugs. To protect against control-flow hijacking attacks, the operating system configures the hardware such that write access to executable pages is prevented. Furthermore, the hardware is configured such that in kernel mode, the instruction pointer may not point into the user space, using a mechanism called supervisor mode execution prevention (SMEP). Data accesses from kernel mode to user space virtual addresses are prevented by operating system and hardware, using a mechanism called supervisor mode access prevention (SMAP). To close remaining attack vectors, address-space layout randomization (ASLR) is used to make all addresses unpredictable for an attacker and thus make return-oriented-programming (ROP) attacks infeasible. All major operating systems employ kernel ASLR (KASLR) [PaX03; Lev12; RSI12]. Information on where objects are located in the kernel address space is generally not available to user programs.

Knowledge of virtual address information can be exploited by an attacker to defeat ASLR [Sha+04; HWH13]. Knowledge of physical address information can be exploited to bypass SMEP and SMAP [KPK14], as well as in side-channel attacks [Liu+15; IES15; Gru+16; Mau+15; Pes+16] and Rowhammer attacks [SD15b; Kim+14; GMM16; Kir15]. Thus, the security of user programs and the kernel itself relies fundamentally on preventing access to address information. Address information is often leaked directly through system interfaces such as `procfs` [KPK14] or indirectly through side channels such as double page faults [HWH13]. However, operating system developers close these information leaks through security patches [Kir15]. In this paper, we show that even if the operating system itself does not leak address information, recent Intel and ARM systems leak this information on the microarchitectural level.

We introduce Prefetch Side-Channel Attacks, a new class of generic attacks that allow an unprivileged local attacker to completely bypass access control on address information. This information can be used to compromise the entire physical system by bypassing SMAP and SMEP in

ret2dir attacks or defeating KASLR and performing ROP attacks in the kernel address space. Our attacks are based on weaknesses in the hardware design of prefetch instructions. Indeed, prefetch instructions leak timing information on the exact translation level for every virtual address. More severely, they lack a privilege check and thus allow fetching inaccessible privileged memory into various CPU caches. Using these two properties, we build two attack primitives: the translation-level oracle and the address-translation oracle. Building upon these primitives, we then present three different attacks. Our first attack infers the translation level for every virtual address, effectively defeating ASLR. Our second attack resolves virtual addresses to physical addresses on 64-bit Linux systems and on Amazon EC2 PVM instances in less than one minute per gigabyte of system memory. This allows an attacker to perform ret2dir-like attacks. On modern systems, this mapping can only be accessed with root or kernel privileges to prevent attacks that rely on knowledge of physical addresses. Prefetch Side-Channel Attacks thus render existing approaches to KASLR ineffective. Our third attack is a practical KASLR exploit. We provide a proof-of-concept on a Windows 10 system that enables return-oriented programming on Windows drivers in memory. We demonstrate our attacks on recent Intel x86 and ARM Cortex-A CPUs, on Windows and Linux operating systems, and on Amazon EC2 virtual machines.

We present a countermeasure against Prefetch Side-Channel Attacks on commodity systems, that involves reorganizing the user and kernel address space to protect KASLR. Our countermeasure requires only a small number of changes to operating system kernels and comes with a performance impact of 0.06–5.09%.

Our key contributions are:

1. We present two generic attack primitives leveraging the prefetch instructions: the translation-level oracle and the address-translation oracle. We then use these primitives in three different attacks.
2. We present a generic attack to infer the translation level for every virtual address to defeat ASLR.
3. We demonstrate generic unprivileged virtual-to-physical address translation attack in the presence of a physical direct map in kernel or hypervisor, on Linux and in a PVM on Amazon EC2. This allows bypassing SMAP and SMEP, enabling ret2dir attacks.
4. We present a generic attack to circumvent KASLR, which enables ROP attacks inside the kernel. We demonstrate our attack on a Windows 10 system.

5. We propose a new form of strong kernel isolation to mitigate Prefetch Side-Channel Attacks and double page fault attacks on kernel memory.

**Outline**  This paper is structured as follows. Section 2 provides background on caches and address spaces. Section 3 presents the settings and two novel attack primitives leveraging the prefetch instructions: the translation-level oracle and the address-translation oracle. The translation-level oracle is used in Section 4 to perform a translation-level recovery attack to defeat ASLR. The address-translation oracle is used in Section 5 to perform unprivileged virtual-to-physical address translation as the basis of ret2dir attacks. Both oracles are used in Section 6 to defeat KASLR. Section 7 shows how to perform cache side-channel and Rowhammer attacks on inaccessible kernel memory. Section 8 presents countermeasures against our attacks. Section 9 discusses related work, and Section 10 concludes this article.

# 2. Background and Related Work

## 2.1. Address translation

To isolate processes from each other, CPUs support virtual address spaces. For this purpose, they typically use a multi-level translation table. Which translation table is used is determined by a value stored in a CPU register. This register value is exchanged upon a context switch. Thus, each process has its own address mappings and only access to its own address space. The kernel is typically mapped into every address space but protected via hardware-level access control. When a thread performs a system call it switches to an operating system controlled stack and executes a kernel-level system call handler. However, it still has the same translation table register value.

In the case of recent Intel CPUs, this translation table has 4 levels. On each level, translation table entries define the properties of this virtual memory region, e.g., whether the memory region is present (*i.e.*, mapped to physical memory), or whether it is accessible to user space. The uppermost level is the page map level 4 (PML4). It divides the 48-bit virtual address space into 512 memory regions of each 512 GB (PML4 entries). Each PML4 entry maps to page directory pointer table (PDPT) with

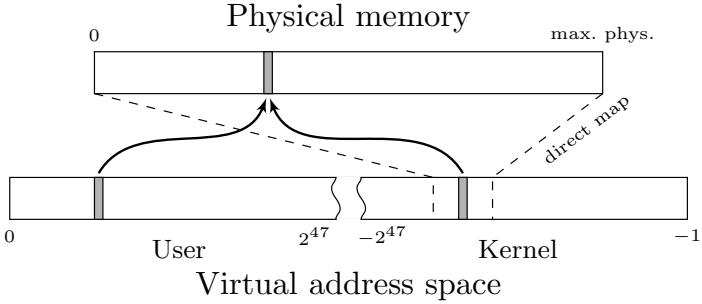

Figure 10.1.: Direct mapping of physical memory. A physical address is mapped multiple times, once accessible for user space and once in the kernel space.

512 entries each controlling a 1 GB memory region that is either 1 GB of physical memory directly mapped (a so-called 1 GB page), or to a page directory (PD). The PD again has 512 entries, each controlling a 2 MB region that is either 2 MB of physical memory directly mapped (a so-called 2 MB page), or a page table (PT). The PT again has 512 entries, each controlling a 4 KB page. The lowest level that is involved in the address translation is called the *translation level*. The CPU has special caches and so-called translation-lookaside buffers for different translation levels, to speed up address translation and privilege checks.

A second, older mechanism that is used on x86 CPUs in virtual-to-physical address translation is segmentation. User processes can be isolated from each other and especially from the kernel by using different code and data segments. Segments can have a physical address offset and a size limit, as well as access control properties. However, these features are widely redundant with the newer translation table mechanism. Thus, most of these features are not available in 64-bit mode on x86 CPUs. In particular, all general purpose segments are required to have the offset set to physical address 0 and the limit to the maximum value. Thus, the CPU can ignore these values at runtime and does not have to perform runtime range checks for memory accesses.

## 2.2. Virtual address space

The virtual address space of every process is divided into user address space and kernel address space. The user address space is mapped as user-accessible, unlike the kernel space that can only be accessed when the CPU is running in kernel mode.The user address space is divided into memory regions for code, data, heap, shared libraries and stack. Depending on the operating system, the user address space may look entirely different in different processes with respect to the absolute virtual offsets of the regions and also the order of the regions. In contrast, the kernel address space looks mostly identical in all processes.

To perform context switches, the hardware requires mapping parts of the kernel in the virtual address space of every process. When a user thread performs a syscall or handles an interrupt, the hardware simply switches into kernel mode and continues operating in the same address space. The difference is that the privileged bit of the CPU is set and kernel code is executed instead of the user code. Thus, the entire user and kernel address mappings remain generally unchanged while operating in kernel mode. As sandboxed processes also use a regular virtual address space that is primarily organized by the kernel, the kernel address space is also mapped in an inaccessible way in sandboxed processes.

Many operating systems have a physical memory region or the whole physical memory directly mapped somewhere in the kernel space [ker09; Lev12]. This mapping is illustrated in Figure 10.1. It is used to organize paging structures and other data in physical memory. The mapping is located at a fixed and known location, even in the presence of KASLR. Some hypervisors also employ a direct map of physical memory [xen09]. Thus, every user page is mapped at least twice, once in the user address space and once in the kernel direct map. When performing operations on either of the two virtual addresses, the CPU translates the corresponding address to the same physical address in both cases. The CPU then performs the operation based on the physical address.

Physical direct maps have been exploited in ret2dir attacks [KPK14]. The attacker prepares a code page to be used in the kernel in the user space. Exploiting a kernel vulnerability, code execution in the kernel is then redirected to the same page in the physical direct map. Hence, the attacker has obtained arbitrary code execution in the kernel.

## 2.3. Address-space layout randomization

Modern CPUs protect against code injection attacks (e.g., NX-bit, $W \oplus X$ policy), code execution in user space memory in privileged mode (e.g.,

SMEP, supervisor mode execution protection), and data accesses in user space memory regions in privileged mode (e.g., SMAP, supervisor mode access protection). However, by chaining return addresses on the stack it is possible to execute small code gadgets that already exist in the executable memory regions, e.g., return-to-libc and ROP attacks. In an ROP attack, the attacker injects return addresses into the stack and in some cases modifies the stack pointer to a user-controlled region, in order to chain the execution of so-called gadgets. These gadgets are fragments of code already existing in the binary, typically consisting of a few useful instructions and a return instruction.

ASLR is a countermeasure against these control flow hijacking attacks. Every time a process is started, its virtual memory layout is randomized. ASLR can be applied on a coarse-grained level or a fine-grained level. In the case of coarse-grained ASLR, only the base addresses of different memory regions are randomized, e.g., code, data, heap, libraries, stack. This is mostly performed on a page-level granularity. An attacker cannot predict addresses of code and data and thus cannot inject modified code or manipulate data accesses. In particular, an attacker cannot predict the address of gadgets to be used in an ROP attack. All modern operating systems implement coarse-grained ASLR. Fine-grained ASLR randomizes even the order of functions, variables, and constants in memory on a sub-page-level granularity. However, it incurs performance penalties, and can be bypassed [Sno+13] and thus is rarely used in practice.

User space ASLR primarily protects against remote attackers that only have restricted access to the system and thus cannot predict addresses for ROP chains. KASLR primarily protects against local attackers as they cannot predict addresses in the kernel space for ROP chains. In particular, invalid accesses cause a crash of the application under attack or the entire system. On Windows, the start offsets of the kernel image, drivers and modules, are randomized.

## 2.4. CPU caches

CPU caches hide slow memory access latencies by buffering frequently used data in smaller and faster internal memory. Modern CPUs employ set-associative caches, where addresses are mapped to cache sets and each cache set consists of multiple equivalent cache lines (also called ways). The index to determine the cache set for an address can be based on the virtual or physical address. The last-level cache is typically physically indexed and shared among all cores. Thus executing code or accessing data on one core has immediate consequences for all other cores.

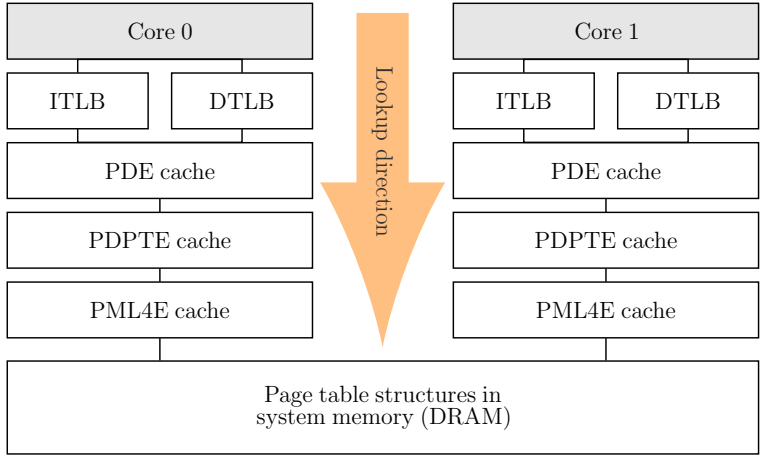

Figure 10.2.: Paging caches are used to speed-up address translation table lookups.

Address translation structures are stored in memory and thus will also be cached by the regular data caches [Int14c]. In addition to that, address translation table entries are stored in special caches such as the translation-lookaside buffers to allow the CPU to work with them. When accessing virtual addresses these buffers are traversed to find the corresponding physical address for the requested memory area. The caches of the different table lookups are represented in Figure 10.2. These caches are typically fully-associative.

As CPUs are getting faster, they rely on speculative execution to perform tasks before they are needed. Data prefetching exploits this idea to speculatively load data into the cache. This can be done in two different ways: hardware prefetching, that is done transparently by the CPU itself, and software prefetching, that can be done by a programmer. Recent Intel CPUs have five instructions for software prefetching: `prefetcht0`, `prefetcht1`, `prefetch2`, `prefetchnta`, and `prefetchw`. These instructions are treated like hints to tell the processor that a specific memory location is likely to be accessed soon. The different instructions allow hinting future repeated accesses to the same location or write accesses.

Similarly, recent ARMv8-A CPUs supply the prefetch instruction `PRFM`. Both on Intel and ARM CPUs, the processor may ignore prefetch hints.

## 2.5. Cache attacks

Cache attacks are side-channel attacks exploiting timing differences introduced by CPU caches. Cache attacks have first been studied theoretically [Koc96; Kel+00], but practical attacks on cryptographic algorithms followed since 2002 [Pag02; TSS03; Ber04].

In the last ten years, fine-grained cache attacks have been proposed, targeting single cache sets. In an *Evict+Time* attack [OST06], the attacker measures the average execution time of a victim process, e.g., running an encryption. The attacker then measures how the average execution time changes when evicting one specific cache set before the victim starts its computation. If the average execution time is higher, then this cache set is probably accessed by the victim.

A *Prime+Probe* attack [OST06; Per05] consists of two steps. In the Prime step, the attacker occupies one specific cache set. After the victim program has been scheduled, the Probe step is used to determine whether the cache set is still occupied. A new generation of *Prime+Probe* attacks have recently been used to perform attacks across cores and virtual machine borders [IES15; Liu+15; Mau+15] as well as from within sandboxes [Ore+15].

Gullasch et al. [GBK11] built a significantly more accurate attack that exploits the fact that shared memory, e.g., shared libraries, is loaded into the same cache set for different processes running on the same CPU core. Yarom and Falkner [YF14] presented an improvement over this attack, called *Flush+Reload* that targets the last-level cache and thus works across cores. *Flush+Reload* attacks work on a single cache line granularity. These attacks exploit shared inclusive last-level caches. An attacker frequently flushes a targeted memory location using the `clflush` instruction. By measuring the time it takes to reload the data, the attacker determines whether data was loaded into the cache by another process in the meantime. Applications of *Flush+Reload* are more reliable and powerful in a wide range of attacks [GSM15; Gül+15; Zha+14; Ira+15a; Ira+15b].

*Flush+Reload* causes a high number of cache misses due to the frequent cache flushes. This has recently also been used to perform a memory corruption attack called Rowhammer [Kim+14]. In a Rowhammer attack, an attacker causes random bit flips in inaccessible and higher privileged memory regions. These random bit flips occur in DRAM memory and the *Flush+Reload* loop is only used to bypass all levels of caches to reach DRAM in a high frequency. Proof-of-concept exploits to gain root privileges and to evade a sandbox have been demonstrated [SD15a]. For the attack to succeed, an attacker must hammer memory locations that map to different rows in the same bank. However, the mapping from addresses to rows and banks is based on physical addresses. Thus, Rowhammer attacks are substantially faster and easier if physical address information is available as an attacker can directly target the comparably small set of addresses that map to different rows in the same bank. As a countermeasure, operating systems have recently restricted access to physical address information to privileged processes [Kir15].

# 3. Setting and Attack Primitives

In this section, we describe the prefetch side channel and two primitives that exploit this side channel. We build a translation-level oracle, that determines whether a page is present and which translation table level is used for the mapping. This primitive is the basis for our translation-level recovery attack described in Section 4 to defeat ASLR. We build an address-translation oracle that allows verifying whether a specific virtual address maps to a specific physical address. We use this to resolve the mapping of arbitrary virtual addresses to physical addresses to mount ret2dir attacks, defeating SMAP and SMEP, in Section 5. We use both attack primitives in our the KASLR exploit described in Section 6.

## 3.1. Attack setting and attack vector

**Attack setting**    In our attacks, we consider a local attack scenario where user space and KASLR are in place. The attacker can run arbitrary code on the system under attack, but does not have access to the kernel or any privileged interfaces such as `/proc/self/pagemap` providing user space address information. This includes settings such as an unprivileged process in a native environment, an unprivileged process in a virtual machine, and a sandboxed process.

To exploit a potential vulnerability in kernel code, an attacker cannot inject code into a writable memory region in the kernel, or directly jump into code located in the user address space as this is prevented by modern CPUs with features like the NX-bit, SMEP, and SMAP. Thus, an attacker can only reuse existing code in a so-called code reuse attack, e.g., ROP attacks. However, building an ROP payload requires exact knowledge of the addresses space layout. Even if the operating system does not leak any address space information and ASLR is employed and effective, we show that the hardware leaks a significant amount of address space information.

The information gained allows an attacker to conduct cache side-channel attacks and Rowhammer attacks, as well as to defeat KASLR and bypass SMAP and SMEP in a ret2dir-like attack.

**Attack vector** Prefetch Side-Channel Attacks are novel and generic side-channel attacks. We exploit the following two properties:

**Property 1** The execution time of prefetch instructions varies depending on the state of various CPU internal caches.

**Property 2** Prefetch instructions do not perform any privilege checks.

The execution time (Property 1) of a prefetch instruction can be directly measured. It is independent of privilege levels and access permissions. We exploit this property in our translation-level oracle. Intel states that prefetching "addresses that are not mapped to physical pages" can introduce non-deterministic performance penalties [Int14c]. ARM states that the prefetch instructions are guaranteed not to cause any effect that is not equivalent to loading the address directly from the same user process[ARM13]. In both cases, we found timing differences to be deterministic enough to be exploitable. That is, Property 1 can be observed on all our test platforms shown in Table 10.1, *i.e.*, all Intel microarchitectures since Sandy Bridge as well as the ARMv8-A microarchitecture. Thus, attacks based on Property 1 are applicable to the vast majority of systems used in practice. We demonstrate our translation-level recovery attack on all platforms.

The timing difference caused by the lack of privilege checks (Property 2) can only be measured indirectly using one of the existing cache attack techniques. The combination of the prefetch side channel with different cache attack techniques yields different properties. Intel states that software prefetches should not be used on addresses that are not "managed or owned" by the user process [Int14a], but in practice does not prevent it, thus letting us do this in our attack. Property 2 can be observed on all

| CPU / SoC | Microarchitecture | System type |
|---|---|---|
| i5-2530M, i5-2540M | Sandy Bridge | Laptop |
| i5-3230M | Ivy Bridge | Laptop |
| i7-4790 | Haswell | Desktop |
| i3-5005U, i5-5200U | Broadwell | Laptop |
| i7-6700K | Skylake | Desktop |
| Xeon E5-2650 | Sandy Bridge | Amazon EC2 VM |
| Exynos 7420 | ARMv8-A | Smartphone |

Table 10.1.: Experimental setups.

Intel test platforms shown in Table 10.1, *i.e.*, all microarchitectures since Sandy Bridge. Thus, attacks based on Property 2 are applicable to the vast majority of desktop, server, and cloud systems.

**Measurements**   Measuring the execution time of instructions or memory accesses is typically necessary to perform micro-benchmarks. On ARM CPUs we experienced no difficulties with out-of-order execution. We used `clock_gettime()` to measure time in nanoseconds, as in previous work [Lip+16], and surrounded the *target instr.* with a memory and instruction barrier consisting of `DSB SY; ISB`. Depending on the attack we used a memory access or the `PRFM` instruction as *target instruction*.

On Intel CPUs micro-benchmark measurements are significantly harder, due to out-of-order execution. The instructions `rdtsc` and `rdtscp` both provide a sub-nanosecond timestamp. `rdtscp` also waits for all memory load operations to be processed before retrieving the timestamp. The `cpuid` instruction can be used to serialize the instruction stream. To perform accurate measurements, Intel recommends using a sequence of `cpuid; rdtsc` before executing the code to measure and `rdtscp; cpuid` afterward [Int10]. In cache side-channel attacks memory fences like `mfence` can be used to ensure that memory store operations are also serialized. However, the `prefetch` instruction is not serialized by `rdtscp` or any memory fence, but only by `cpuid` [Int14b]. Due to these serialization issues we crafted instruction sequences to measure exactly a *target instruction* in different scenarios:

1. In cases of long measurements and measurements of memory access times, the target instruction is unlikely to be reordered before a preceding `rdtscp` instruction. We thus use:
   `mfence cpuid rdtscp` *target instr.* `rdtscp cpuid mfence`.

2. When measuring prefetch instructions repeatedly, correct values for minimum and median latency are important. Thus, noise introduced by `cpuid` is tolerable but reordering the target instruction is not, because it could lead to a lower measurement for the minimum latency. In this case we use:

   `mfence rdtscp cpuid` *target instr.* `cpuid rdtscp mfence`.

Depending on the attack we used a memory access, or the prefetch instructions `prefetchnta` and `prefetcht2` as *target instruction.*

## 3.2. Translation-level oracle

In the translation-level oracle, we exploit differences in the execution time of prefetch instructions (Property 1). Prefetch instructions resolve virtual addresses to physical addresses to enqueue the prefetching request. Intel CPUs follow a defined procedure to find a cache entry or a physical address for a specific virtual address (cf. Section 4.10.3.2 of Intel Manual Vol. 3A [Int14c]):

1. Cache lookup (requires TLB lookup)
2. TLB lookup
3. PDE cache lookup
4. PDPTE cache lookup
5. PML4E cache lookup

The procedure aborts as early as possible omitting all subsequent steps. Step 1 and 2 can be executed in parallel for the L1 cache and thus the latency of step 2 is hidden in this case. However, in case of the L2 or L3 cache, step 2 needs to be executed before to complete the cache lookup in step 1. If no entry is found in any TLB, step 3 needs to be executed to complete step 2 and the same applies for steps 4 and 5. Depending on the specific CPU, some caches may not be present and the corresponding steps are omitted. Every step of the lookup procedure introduces a timing differences that can be measured. For ARM CPUs, the same mechanism applies to the corresponding translation tables. However, on all CPUs tested, we found at least 4 distinct average execution times for different cases.

The translation-level oracle works in two steps. First, we calibrate the execution time of a prefetch instruction on the system under attack. Second, we measure the execution time of a prefetch instruction on an arbitrary virtual address. Based on the execution time, we can now derive on which level a prefetch instruction finished the search for a cache entry or a physical address.

Prefetch instructions on Intel CPUs ignore privilege levels and access permissions (Property 2). Thus, it is possible to prefetch execute-only pages, as well as inaccessible kernel memory. When running the procedure over the whole address space, we now also obtain information on all kernel pages. Note that even a process with root privileges could not obtain this information without loading a kernel module on modern operating systems.

## 3.3. Address-translation oracle

The lack of privilege checks (Property 2) is the basis for our second oracle, as well as other privilege checks that are not active in 64-bit mode on x86 CPUs (cf. Section 2.1). An attacker can execute prefetch on any virtual address including kernel addresses and non-mapped addresses. Thus, we can use prefetch as an oracle to verify whether two virtual addresses $p$ and $\bar{p}$ map to the same physical address. The address-translation oracle works in three steps:

1. Flush address $p$
2. Prefetch (inaccessible) address $\bar{p}$
3. Reload $p$

If the two addresses map to the same physical address, the prefetch of $\bar{p}$ in step 2 leads to a cache hit in step 3 with a high probability. Thus, the access time in step 3 is lower than for a cache miss. By repeating this measurement, the confidence level can be increased to the desired value. One measurement round takes 100–200 nanoseconds on our test systems. Thus, an attacker can run up to 10 million such measurements per second. Figure 10.3 shows the minimum access time from step 3, over a set of inaccessible addresses $\bar{p}$ measured on an i5-3320M. The peak shows the single address $\bar{p}$ that maps to the same physical address as $p$.

Similarly, we can also perform a microarchitectural timing attack on prefetch instructions directly. Based on the execution time of prefetch instructions (Property 1), we can measure whether a targeted address $p$ is in the cache. In this *Evict+Prefetch*-variant of the address-translation oracle, we exploit both properties of prefetch instructions (cf. Section 3.1).

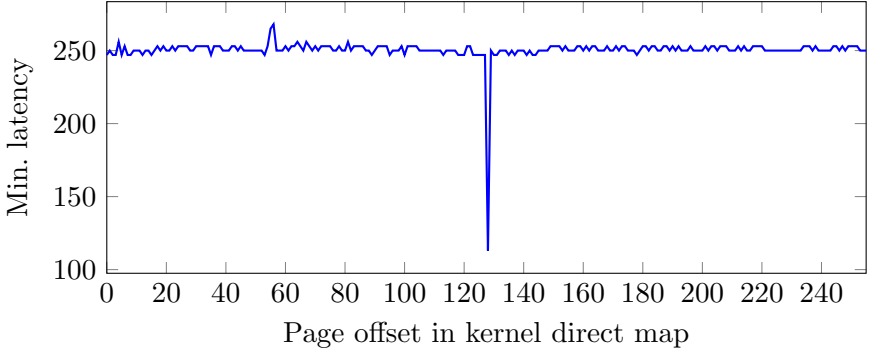

Figure 10.3.: Minimum memory access time for an address $p$ after prefetching different inaccessible addresses, on an i5-3320M. Peak shows the single address $\bar{p}$ mapping to the same physical address as $p$.

As the target address might be inaccessible, we evict the address instead of flushing it in the first step. The prefetching replaces the reload step and checks whether the inaccessible address is already in the cache:

1. Evict address $p$
2. Execute function or system call
3. Prefetch $p$

If the function or system call in step 2 accesses any address $\bar{p}$ that maps to the same physical address as address $p$, we will observe a lower timing in step 3 with a high probability. Thus, as in the regular address-translation oracle, we determine whether an address $\bar{p}$ and an address $p$ map to the same physical address. The difference is that in the *Evict+Prefetch*-variant the address $\bar{p}$ is unknown to the attacker. Instead, the attacker learns that a targeted address $p$ is used by the function or system call.

## 4. Translation-level recovery attack

In this section, we describe how to determine the translation level from an unprivileged user space process based on the translation-level oracle described in Section 3.2. Processes with root privileges can normally obtain system information to derive the translation level, for instance on Linux using the `pagemap` file in `procfs`. However, even here the information provided by the operating system is incomplete and to obtain the translation-level information for kernel memory, it is necessary to install a kernel module for this purpose. We instead only rely on the timing difference observed when running a prefetch attack on an address.

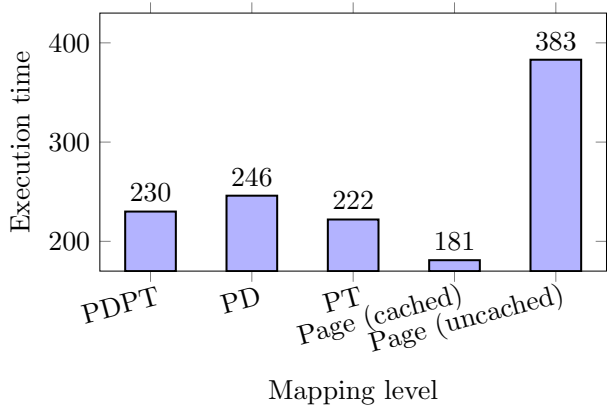

Figure 10.4.: Median prefetch execution time in cycles compared to the actual address mapping level, measured on an i5-2540M.

Figure 10.4 shows the median prefetch execution time for 5 different cases measured on an i5-2540M compared to the actual mapping level. We measured the execution time of `prefetchnta` and `prefetcht2` in 0.5 million tests. The lowest median timing can be observed when the address is valid and cached, with a median of 181 cycles. It is interesting to observe that prefetching a non-cached address when all TLB entries are present has a median execution time of 383 cycles. Thus, we can use prefetch instructions to distinguish cache hits and misses for valid addresses. In case the targeted address is not valid, we observe different execution times depending the mapping level where the address resolution ends. If the memory region has a page directory but the page table is not valid, the median execution time is 222 cycles. If the memory region does not have a page directory but a PDPT, the median execution time is 246 cycles. If the memory region does not have a PDPT, the median execution time is 230 cycles. Note that these timings strongly depend on the measurement techniques in Section 3.1.

We perform a breadth-first search starting with the PML4 and going down to single pages as illustrated in Figure 10.5. We start the recovery attack with the top-level PML4 recursively going down to the lowest level (cf. Section 2.1). We eliminate measurement noise by checking multiple addresses in each of the 512 regions on each layer. The median execution time of a prefetch instruction sequence is used to decide whether a PDPT is mapped or not. On the PDPT level we thus obtain information on 1 GB pages, on the PD level we obtain information on 2 MB pages and on the lowest level (PT) we obtain information on 4 KB pages. On each level we learn whether the address is mapped directly from this level, or a lower level, or whether it is marked as invalid.

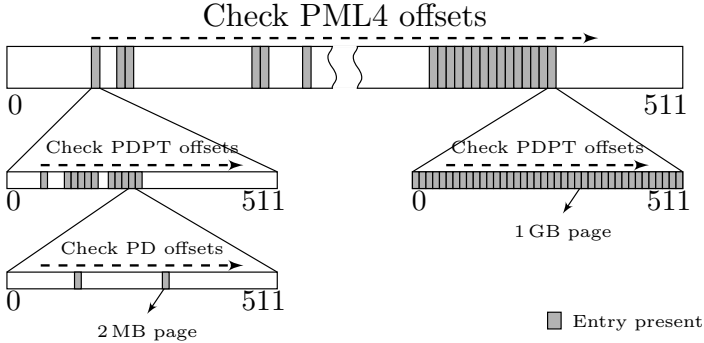

Figure 10.5.: Breadth-first search through the page translation entries that are present. The attacker obtains an accurate map of the page translation level for user and kernel address space.

For a single check, we perform $2^8$ tests that in total take less than 4ms on the i5-2540M. We check 4 addresses per region and thus require less than 16ms per memory region. Thus, for every translation table that is present, our attack has a runtime of approximately 8 seconds. Programs on Linux typically use at least 8 translations tables (1 PML4, 2 PDPTs, 2 page directories, 3 page tables). The total runtime to only recover the translation levels for the user space here is approximately 1 minute. However, recovering the translation levels for the kernel can take several minutes if 1 GB pages are used by the kernel, or even several hours if 2 MB pages are used for the physical direct map. If more addresses are mapped in the user space, the execution time can increase to several minutes or hours, depending on the target process.

In either case, our attack successfully recovers the translation level which is normally only accessible for processes with root privileges. Obtaining this information effectively defeats ASLR, as the attacker can now accurately determine which addresses are mapped to physical memory by locating libraries or drivers in inaccessible memory regions. Finally, our attack defeats recently proposed countermeasures [Cra+15] that employ execute-only mappings, as prefetch instructions ignore access permissions.

**Translation-level recovery from Android apps**   Similarly to 64-bit x86, 64-bit ARMv8-A has a 4-level page translation mechanism. This is for instance the case on our Samsung Galaxy S6 with an Exynos 7420 system-on-chip with a non-rooted stock Android system. On this system, we use the unprivileged `PRFM PLDL1KEEP` instruction to prefetch memory and the unprivileged `DC CIVAC` instruction to flush memory from user space.

The basic translation-level recovery attack on our ARMv8-A CPU is the same as on Intel x86 CPUs. The timing measurement by `clock_gettime` provides a measurement on a nanosecond scale. The timing measurement is significantly faster than on Intel x86 CPUs as there is no `cpuid` instruction consuming a significant amount of cycles. We performed $2^8$ tests per address from an Android app. In total this takes less than $50\mu s$ on our ARMv8-A system. Thus, the translation-level recovery attack runs on ARM-based devices successfully.

# 5. Address-translation attack

In this section, we describe how to mount ret2dir-like attacks without knowledge of physical addresses based on our address-translation oracle. We also build an efficient attack to resolve virtual addresses to physical addresses. This attack exploits the physical direct map in the kernel. Many operating systems and hypervisors use such a mapping to read and write on physical memory [ker09; Lev12; xen09]. The physical direct map has been exploited by Kemerlis et al. [KPK14] to bypass SMEP in their attack called ret2dir. Similarly, this map can also be used for return stacks in an ROP attack if the attacker has knowledge of the physical address of user-accessible memory. However, our address-translation attack is not restricted to ret2dir-like attacks, it also provides physical address information that is necessary in many side-channel attacks and fault attacks [Liu+15; IES15; Gru+16; Pes+16; SD15b; Kim+14; Kir15].

The attack does not require prior knowledge of the virtual offset of the physical direct map in the kernel. This offset is typically fixed, but it can also be determined by using a translation-level recovery attack. We demonstrate our attack on a native Ubuntu Linux system, and from within an Amazon EC2 instance.

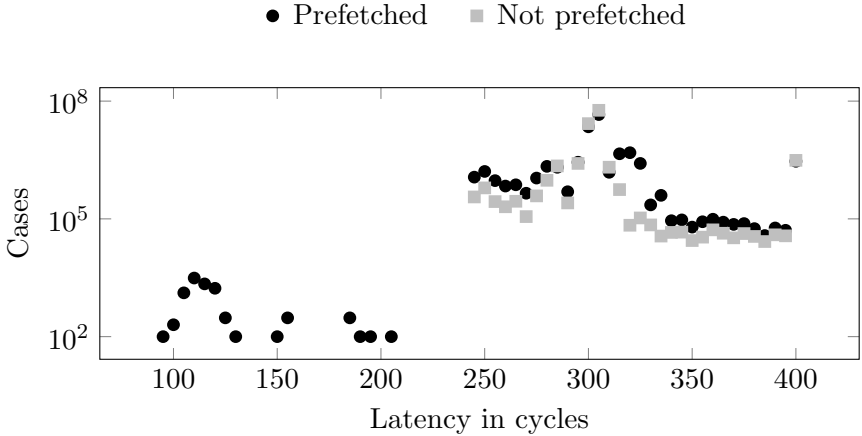

Figure 10.6.: Access latency that has (or has not) been prefetched through a kernel address. Measurements performed on Linux on an Intel i5-3320M.

The attacker runs the address-translation oracle on one address $p$ and one address $\bar{p}$ in the virtual memory area of the kernel that is directly mapped to physical memory. The timing difference resulting from prefetching $\bar{p}$ is shown in Figure 10.6. Note that $p$ and $\bar{p}$ have the same page offset, as the page offset is identical for physical and virtual pages. By only checking the possible offsets based on the known page size, the attacker reduces the number of addresses to check to a minimum. The search space for 2 MB pages is only 512 possibilities per 1 GB of physical memory and for 4 KB pages only 262 144 possibilities per 1 GB of physical memory. The attacker performs a brute-force search for the correct kernel physical direct-map address by trying all possible values for $\bar{p}$. Finding an address $\bar{p}$ means that the attacker has obtained an address that maps user space memory in the kernel address space, thus providing a bypass for SMAP and SMEP. Thus it is possible to mount ret2dir-like attacks using this address. However, the correct physical address can be obtained by subtracting the virtual address of the start of the physical direct map. On Linux, the physical direct map is located at virtual address `0xffff 8800 0000 0000`.

Figure 10.7 shows the average brute-force search time per gigabyte of physical memory, when searching for a 2 MB page. In this search, we ran our address-translation oracle $2^{14}$ to $2^{17}$ times, depending on the architecture. Increasing the number of runs of the address-translation oracle decreases the false negative rate but at the same time increases the execution time. We did not find any false positives in any of the native attack settings. Depending on the architecture, delays were introduced to lower the pressure on the prefetcher. The highest accuracy was achieved on the i5-3230M (Ivy Bridge), the i7-4790 (Haswell), and the i5-5200U (Broadwell) systems where the false negative rate was between 7% and 13%. The accuracy was significantly lower on the i5-2540M (Sandy Bridge) test system. However, the false-negative rate remained at $\geq 25\%$ even with a higher number of address-translation oracle runs. For the expected execution time per gigabyte of physical memory, we computed how long the attacks have to be repeated until the physical address is found with a probability of more than 99%. The i5-2540M (Sandy Bridge) test system had the highest false negative rate and thus the expected search time is the highest here. Similarly, on the Skylake system, the attack needs to be executed 3 times to find a physical address with a probability of more than 99%. However, as the execution time per round of the attack on the Skylake system was much lower than on the other systems, the expected execution time is close to the other systems.

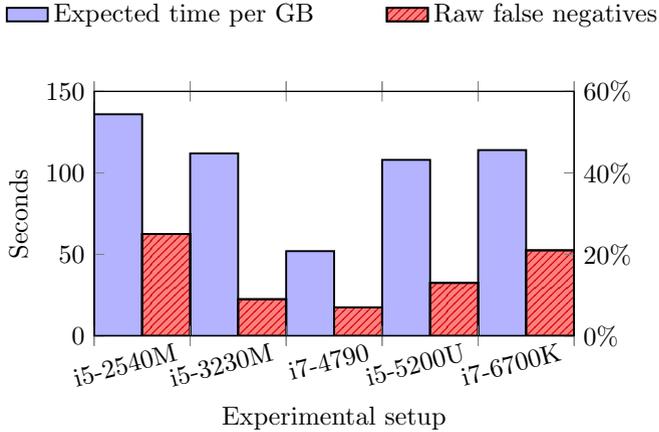

Figure 10.7.: Expected brute-force search time per GB of physical memory, searching for a 2 MB page. Raw false negative rate after a single run of the attack. On all platforms the whole system memory can be searched exhaustively within minutes to hours.

**Getting host physical addresses on Amazon EC2**   On the Amazon EC2 instance running Linux, we exploit the Xen PVM physical direct map, located at virtual address `0xffff 8300 0000 0000`. Apart from this, the basic attack remains the same. To perform the attack on an Amazon EC2 instance, we compensated the noise by checking multiple 4 KB offsets per 2 MB page. Our machine was scheduled on an Intel Xeon E5-2650 (Sandy Bridge). In a dual CPU configuration, it can manage up to 768 GB of physical memory. To compensate for this huge amount of potentially addressable physical memory, we reduced the number of address-translation oracle runs from $2^{15}$ to $2^{13}$. Our attack speed is thus reduced from 154 seconds to 46 seconds per gigabyte on average, limiting the total attack time to less than 10 hours. While this is significantly more than in a native environment with a smaller amount of physical memory, it is still practical to use this attack to translate a small number of virtual addresses to physical addresses.

As we had no direct access to the real address translation information, we verified our results based on the technique from Section 3.3. Translations are considered correct if multiple consecutive verification loops confirm that the hypervisor physical direct-map addresses indeed allow prefetching the targeted user virtual address, and if the mappings of multiple addresses from the same virtual page can be confirmed as well using the address-translation oracle. We obtained an accuracy of the attack in the cloud that is comparable to the accuracy of the attack in a native environment. The presumably correct physical address is always found, *i.e.*, no false negatives. When searching through the maximum 768 GB of address space, we consistently found 1 false positive match (*i.e.*, a 2 MB page) that was later eliminated in the verification loop.

**Other operating systems**   Many other operating systems, such as BSD or OSX, maintain a physical direct map. However, we found no such mapping on Windows. Thus, our address-translation oracle can not directly be applied to Windows systems.

Although 64-bit Android has a physical direct map located at virtual address `0xffff ffc0 0000 0000` and 32-bit Android at virtual address `0xc000 0000`, we were not able to build an address-translation oracle on Android. As the prefetch instructions do not prefetch kernel addresses mapped through the second translation-table base register, the attack is mitigated. However, an attack could be possible on systems where user space and kernel space share a translation-table base register, while the kernel would still be inaccessible. Similarly, the attack does not work on today's Google NaCl sandbox as it uses a 32-bit address space using 32-bit segmentation. The sandboxed process therefore only partially shares an address space with the non-sandboxed code and thus the attack is mitigated. However, we verified that a Prefetch Side-Channel Attack using cache eviction instead of `clflush` within the Google NaCl sandbox works on the lowest 4 GB of virtual memory. Thus, when Google NaCl introduces support for 64-bit address spaces in the NaCl sandbox, 32-bit segmentation cannot be used anymore and our attack is likely to succeed on all virtual addresses and thus to leak physical addresses to sandboxed processes.

# 6. Kernel ASLR exploit

In this section, we demonstrate how to defeat KASLR by using prefetch instructions. We demonstrate our attack on Windows 10 and Windows 7 systems. Similarly as in the previous attack, we try to locate mapped memory regions in address space regions that are not accessible from user space. Again, we exploit the omission of privilege checks by prefetch instructions (Property 2). As described in Section 3, we use prefetch instructions in combination with code execution to identify the load address of drivers in kernel mode in this first stage of the attack. In the second stage of the attack, we determine addresses used by a specific driver. By locating the driver, we effectively defeat KASLR.

Similarly, on Windows 7, kernel and hardware-abstraction layer are located between virtual address `0xffff f800 0000 0000` and `0xffff f87f ffff ffff` and system drivers are located between virtual address `0xffff f880 0000 0000` and `0xffff f88f ffff ffff`. On Windows 10, the address range is extended to the region from `0xffff 8000 0000 0000` to `0xffff 9fff ffff ffff`. Which drivers are present in the driver area depends on the system configuration. Furthermore, the order in virtual address space directly depends on the order the drivers are loaded, which again depends on the system configuration. To exploit a kernel vulnerability and build an ROP chain in the code of a known driver, an attacker has to know the exact address offset of the driver. However, the exact address offsets are randomized and can normally not be retrieved from user processes. Our attack exploits that KASLR does not randomize the offset of drivers on a sub-page level. Kernel and hardware-abstraction layer are loaded on consecutive 2 MB pages with a random 4 KB start offset on Windows 7. Thus, we cannot attack this memory region directly using a translation-level recovery attack. However, an *Evict+Prefetch* attack is possible on any kernel memory region. To build the most efficient attack, we target the driver memory area where we can first perform a translation-level recovery attack and an *Evict+Prefetch* attack afterward. Windows 10 uses 4 KB pages instead, adding entropy to the randomized driver location.

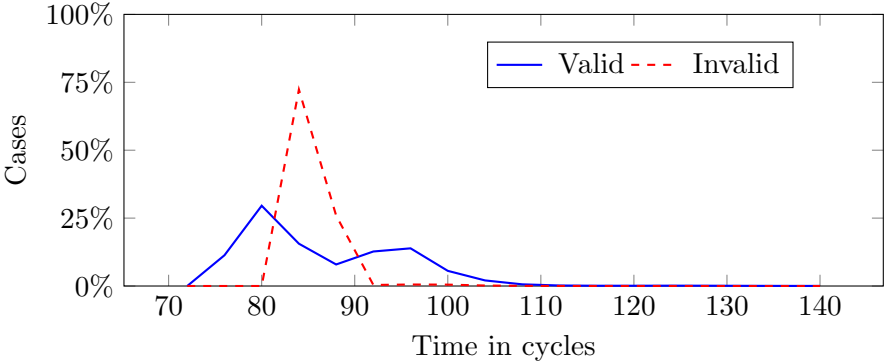

Figure 10.8.: Timing difference of a prefetch sequence on valid and invalid addresses in kernel space, from unprivileged user space process. Measurements performed on Windows 7 on an Intel i3-5005U.

In the first stage of our attack, we locate addresses mapped to physical memory in the driver memory area using our translation-level recovery attack. Figure 10.8 illustrates the timing difference between valid and invalid addresses in the driver region on Windows 7 on an Intel i3-5005U. As drivers are loaded consecutively in the virtual address space, we found it to be sufficient for our attack to search through the address space in 2 MB steps and measure where pages are mapped to physical memory. On Windows 7, the average runtime for the first stage of the attack, mapping both the kernel region and the driver region, is 7 ms on an idle system. On Windows 10, the first stage runs in 64 KB steps and takes 101 ms on average. As Windows 10 maps 4 KB pages we scan the address range in 4 KB steps in an intermediate step taking 180 ms on average. At a high system load, the attack requires several hundred repetitions to perform the first stage of the attack reliably, having an average runtime below 2 seconds on Windows 7.

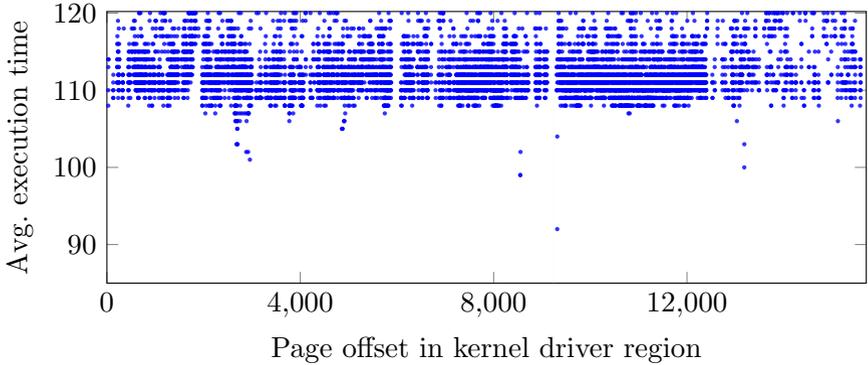

Figure 10.9.: Second stage: driver region is searched by measuring average execution times of prefetching addresses in the driver memory area from the first stage. Lowest average execution time is measured on an address in the memory of the targeted driver.

In the second stage of our attack, we use the *Evict+Prefetch* variant of the address-translation oracle. Instead of searching for pages that are mapped to physical memory, we now determine whether a target address $p$ is used by a syscall. Therefore, we perform the *Evict+Prefetch* attack over all potentially used addresses in a random order. We run the following three steps:

1. *We evict all caches.* For this purpose, we access a buffer large enough to evict all driver addresses from all TLBs, page translation caches, code and data caches.
2. *We perform a syscall to the targeted driver.* If the target address $p$ is used by the targeted driver, the CPU fetches it into the caches while executing the syscall.
3. *We measure the timing of a prefetch instruction sequence.* This reveals whether the target address $p$ was loaded into the cache by the driver in the second step.

In order to verify the measurement, we perform a control run where we omit the system call to the targeted driver. If the execution time of a prefetch instruction sequence on the target address $p$ is higher without the system call, we learn that $p$ is in fact used by the driver and not loaded into the cache by other driver activity on the system. The attack can be repeated multiple times to increase the accuracy.

By determining the lowest virtual address in the driver region that is used by the targeted driver, we learn where the driver starts. As we know the driver version we can now use the virtual addresses from this kernel driver in return-oriented-programming attacks.

The average runtime for the second stage of the attack is 490 seconds on Windows 7. Thus, the total average runtime is below 500 seconds on Windows 7 on our i3-5005U. On Windows 10 we narrowed down the potential addresses in the first stage more than in Windows 7. Thus, the average runtime of the second stage is also lower on Windows 10, requiring only 12 seconds on average to locate a driver.

# 7. Other applications

In this section, we discuss how prefetch instructions can be used in other cache attacks. First, we implemented modified variant of *Flush+Reload* called *Flush+Prefetch*. The measurement accuracy of this cache attack is comparable to *Prime+Probe* while the spatial accuracy is the same as in a *Flush+Reload* attack. We verified the feasibility of this attack by implementing a cross-core covert channel. On a Haswell i7-4790 we achieved a performance of 146 KB/s at an error rate of < 1%. This is in the same order of magnitude as the fastest state-of-the-art cache covert channels [Gru+16].

Second, the *Evict+Prefetch* variant of the address-translation oracle can be used to perform a *Flush+Reload*-style attack on privileged kernel addresses. Indeed, we demonstrated such an attack in Section 6 to detect whether specific virtual addresses are used by a driver. However, an attacker could also spy on the usage of known virtual addresses in kernel code and drivers. This would allow monitoring activity on system and specific hardware interfaces.

Third, the *Evict+Prefetch* attack also allows performing Rowhammer attacks on privileged addresses. An attacker could directly target kernel page tables or any other kernel data structure. As the execution time is lower than that of *Evict+Reload*, an attack is likely possible. We verified that bit flips can be induced by this attack on a system running at a refresh rate reduced to 25%. However, we leave examinations on the prevalence of this problem on default configured systems and the study of practical Rowhammer exploits using *Evict+Prefetch* open to future work.

# 8. Countermeasures

In this section, we discuss countermeasures against Prefetch Side-Channel Attacks. First, we propose a new form of strong kernel isolation, that effectively prevents all Prefetch Side-Channel Attacks on the kernel address space. Second, we will discuss countermeasures that have been proposed against other side-channel attacks and hardware modifications to mitigate Prefetch Side-Channel Attacks.

**Stronger kernel isolation**  Removing the identity mapping would help against our virtual-to-physical address translation attack and completely prevent ret2dir-like attacks, however, it would not protect against our KASLR or translation-level recovery attacks.

We propose *stronger kernel isolation*, a new form of strong kernel isolation, to provide security against a wide range of attacks. Strong kernel isolation ensures that no address is mapped in both user space and kernel space. This mechanism has initially been proposed by Kemerlis et al. [KPK14]. Their approach unmaps pages from the kernel physical direct map when they are mapped in user space. This only introduces a performance penalty of 0.18–2.91%. However, this is not sufficient to protect against our attacks. Instead, *stronger kernel isolation* does not run syscalls and unrelated kernel threads in the same address space as user threads. We propose to switch the address translation tables immediately after the context switch into the kernel. Thus, only short and generic interrupt dispatching code would need to be mapped in the same address space used by the user program. The remainder of the kernel and also the direct mapping of physical memory would thus not be mapped in the address translation tables of the user program. This layout is illustrated in Figure 10.10.

Stronger kernel isolation also eliminates the double page fault side channel [HWH13], as no virtual address in the user program is valid in both user space and kernel space. This countermeasure can be implemented on commodity hardware and existing operating systems and it only requires a few modifications in operating system kernels. The performance impact is comparably small as switching the address translation tables has to be done once per context switch into the kernel and once per context switch from the kernel back to the user space. This is done by replacing the value in the `cr3` register on Intel x86 CPUs once per context switch. We implemented a proof-of-concept to measure the overhead of updating the

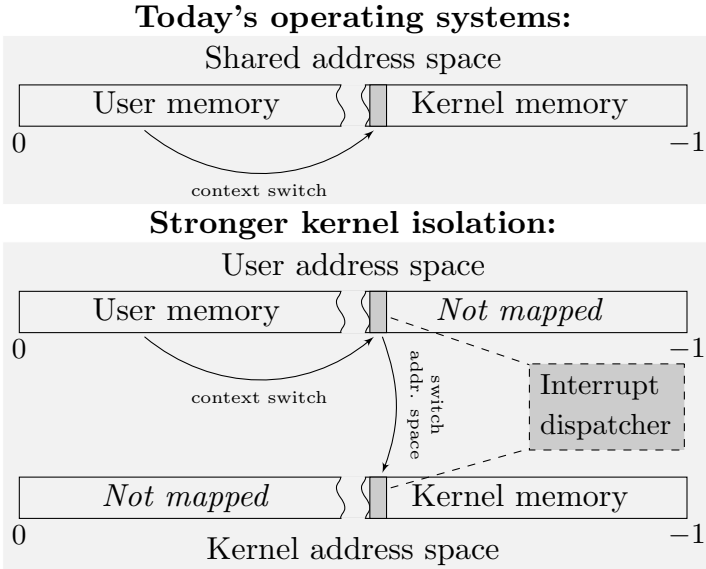

Figure 10.10.: Currently kernel and user memory are only separated through privilege levels. With stronger kernel isolation, the kernel switches from the user space to a dedicated kernel space, directly after a context switch into privileged mode. Thus, only a negligible portion of interrupt dispatcher code is mapped in both address spaces.

`cr3` as an estimate for the performance penalty of stronger kernel isolation. Table 10.2 shows the overhead in different benchmarks. We observe that for benchmarks that perform a small number of syscalls, the performance overhead is negligible, e.g., 0.06%. For other benchmarks the overhead can be higher, e.g., up to 5.09% in the case of pgbench.

**State-of-the-art countermeasures** While there have been recent advances in detecting cache attacks using performance counters [HF15; CSY15; Gru+16; Pay16] it is less clear whether this is also applicable to Prefetch Side-Channel Attacks. Prefetch Side-Channel Attacks can indeed cause an increased number of DTLB misses and thus could be detected using hardware performance counters. We observe approximatively 4 billion DTLB hits/minute while browsing in Firefox, and approximatively 47 billion while running our virtual-to-physical attack. A more thorough evaluation is needed to assess false positives. While there are numerous events related to prefetching that can be monitored with performance counters, to the best of our knowledge, since Nehalem micro-architecture it is not possible anymore to monitor software prefetching but only hardware prefetching [Int14c]. Future work has to show whether performance counters can indeed be used for a reliable detection mechanism. We also note that while it is possible to disable hardware prefetching, it is not possible to disable software prefetching.

| Benchmark | Baseline | Stronger kernel isolation | Overhead |
|-----------|----------|---------------------------|----------|
| apache | 37578.83 req./s | 37205.16 req./s | +1.00% |
| pgbench | 146.81 trans./s | 139.70 trans./s | +5.09% |
| pybench | 1552 ms | 1553 ms | +0,06% |
| x264 | 96.20 fps | 96.14 fps | +0.06% |

Table 10.2.: Estimation of overhead.

**Hardware modifications**   Complete protection against Prefetch Side-Channel Attacks could also be achieved through microarchitectural modifications. We think that prefetch instructions need to be modified in two ways to completely close this attack vector. First, if prefetch instructions performed privilege checks just as other memory referencing instructions, prefetching kernel addresses would trigger a segmentation fault and the process would be killed. It would also prevent measuring the translation table levels over the whole address space as the process would be killed after accessing the first invalid address. Second, prefetch instructions leak timing information on the cache state. The timing difference on our ARM-based smartphones was even higher than on our Intel x86 test system. Eliminating this timing difference would only introduce a small performance overhead, as prefetch instruction are not used by most software. This would prevent cache attacks based on prefetch instructions completely.

# 9. Related Work

Hund et al. [HWH13] demonstrated three timing side channel attacks to obtain address information. The first is a cache attack searching for cache collisions with kernel addresses. The second performs double page faults to measure timing differences for valid and invalid memory regions introduced by the TLB. The third exploits page fault timing differences due to the TLB and address translation caches. The first attack is mitigated on current operating systems by preventing access to physical addresses, and the second and third attacks can be prevented at the operating system level by preventing excessive use of page faults leading to segmentation faults. In contrast, our attack exploits the TLB and address translation caches without triggering any page faults. Furthermore, as our approach leaks the timing more directly through prefetch instructions, it is faster and retrieves information on a finer granularity, *i.e.*, we can obtain the exact virtual-to-physical address translation. Our approach is also more generic as it bypasses the operating system.

Kemerlis et al. [KPK14] presented two methods providing a basis of ret2dir attacks. First, they use the `procfs` interface to obtain physical addresses, now mitigated on current operating systems by preventing access to this interface. Second, they perform a memory spraying attack where they can use any address in the physical direct map for their ret2dir attack. Our attack enables ret2dir-like attacks without knowledge of physical addresses and recovery of physical addresses from unprivileged user space applications, enabling ret2dir attacks. As a countermeasure, they proposed strong kernel isolation, which we extended in this paper.

Barresi et al. [Bar+15] focused on a cross-VM scenario to break ASLR in the cloud with CAIN, while our work mostly focuses on a local attack, that can also be performed on a guest VM. However, CAIN attacks assume a cloud environment that enables memory deduplication, which is already known to be nefarious and is not deployed on e.g., Amazon EC2. In contrast, our attacks do not require memory deduplication and have been performed on Amazon EC2.

Bhattacharya et al. [BRM12] showed that hardware prefetching, performed automatically by the CPU, leaks information. In contrast to this work, we exploit software prefetching which can be triggered at any time by an attacker, from user space. The hardware prefetcher has also been used by Fuchs and Lee [FL15], as a countermeasure against cache side channels.

Concurrent to our work, Jang et al. [JLK16] exploited Intel TSX transaction to defeat KASLR. TSX transactions prevent pagefaults by jumping to an alternative code path. When accessing or executing on kernel address the timing difference until reaching the alternative code path leaks information on the address translation caches. Evtyushkin et al. [EPA16] exploit the branch-target buffer to break KASLR. Finally, Chen et al.

[Che+16] proposed dynamic fine-grained ASLR during runtime to defeat KASLR attacks.

# 10. Conclusion

Prefetch Side-Channel Attacks are a new class of generic attacks exploiting fundamental weaknesses in the hardware design of prefetch instructions. These new attacks allow unprivileged local attackers to completely bypass access control on address information and thus to compromise an entire physical system by defeating SMAP, SMEP, and kernel ASLR. Our attacks work in native and virtualized environments alike. We introduced two primitives that build the basis of our attacks. First, the translation-level oracle, exploiting that prefetch leaks timing information on address translation. Second, the address-translation oracle, exploiting that prefetch does not perform any privilege checks and can be used to fetch inaccessible privileged memory into various caches. The translation-level oracle allowed us to defeat ASLR and locate libraries and drivers in inaccessible memory regions. Using the address-translation oracle, we were able to resolve virtual to physical addresses on 64-bit Linux systems and from unprivileged user programs inside an Amazon EC2 virtual machine. This is the basis for ret2dir-like attacks that bypass SMEP and SMAP. Based on both oracles, we demonstrated how to defeat kernel ASLR on Windows 10, providing the basis for ROP attacks on kernel and driver binary code. As a countermeasure against this new class of attacks, we proposed stronger kernel isolation, such that syscalls and unrelated kernel threads do not run in the same address space as user threads. This countermeasure only requires a few modifications in operating system kernels and that the performance penalty is as low as 0.06–5.09%. Therefore, we recommend that it is deployed in all commodity operating systems.

# 11. Acknowledgments

We would like to thank Klaus Wagner for help with some experiments and our anonymous reviewers for their valuable comments and suggestions.

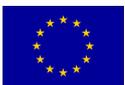 Supported by EU Horizon 2020 programme GA No. 644052 (HECTOR) and EU FP7 programme GA No. 610436 (MATTHEW).

# Statutory Declaration

I declare that I have authored this thesis independently, that I have not used other than the declared sources / resources, and that I have explicitly marked all material which has been quoted either literally or by content from the used sources.